\documentclass[aps,prd,twocolumn,epsf,groupaddress,nofootinbib]{revtex4}
\usepackage{bm}
\usepackage{epsfig}
\usepackage{amsfonts,amsmath,amsthm,amssymb}
\usepackage{latexsym}
\usepackage{mathrsfs}
\usepackage[makeroom]{cancel}
\newcommand{\postscript}[2]{\setlength{\epsfxsize}{#2\hsize}
   \centerline{\epsfbox{#1}}}
\usepackage[titletoc,toc]{appendix}
\usepackage{multirow}
\usepackage{slashbox}
\usepackage[usenames,dvipsnames]{xcolor}
\definecolor{orange}{cmyk}{0,0.5,1,0}
\definecolor{rossoCP3}{cmyk}{0,.88,.77,.40}
\definecolor{graa}{rgb}{0.8,0.8,0.8}
\definecolor{blaa}{rgb}{0.2,0.2,0.6}
\usepackage{dsfont}
\newcommand{\be}{\begin{equation}}
\newcommand{\ee}{\end{equation}}

\newcommand{\bea}{\begin{eqnarray}}
\newcommand{\eea}{\end{eqnarray}}

\def\vev#1{\langle #1\rangle_0}

\def\ket#1{\left| #1\right>}

\def\amp#1#2#3{ \left<#1\right| #2\left|#3\right>}

\def\tfrac#1#2{{\textstyle\frac{#1}{#2}}}
\def\thalf{\tfrac{1}{2}}

\def\Lm{\Lambda}
\def\lm{\lambda}
\def\Lmp{\Lm'}
\def\lme{\lm_e}
\def\lmnu{\lm_{\bar \nu_e}}

\begin{document}

\title{\color{rossoCP3} Symmetries and the fundamental forces of Nature II}

\author{Luis A. Anchordoqui}
\affiliation{Department of Physics and Astronomy,  Lehman College, City University of New York, NY 10468, USA \\
Department of Physics, Graduate Center, City University  of New York, 365 Fifth Avenue, NY 10016, USA\\
Department of Astrophysics, American Museum of Natural History, Central Park West  79 St., NY 10024, USA
}

\date{Fall 2015}
\begin{abstract}
 \noindent  
 Lecture script of a one-semester course that aims to develop an understanding and appreciation of fundemental concepts in modern physics for students who are comfortable with calculus. This document contains the last part of the course where we introduce the concept of the {\it quantum unit} to describe the dynamic properties of subatomic particles and the interactions of matter and radiation. 

\end{abstract}
\maketitle

\setcounter{section}{8}
\setcounter{equation}{205}
\setcounter{figure}{28}
\setcounter{footnote}{9}

\section{Origins of Quantum Mechanics}

Quantum mechanics was born in the beginning of the 20th century due to an apparent collapse of the deterministic classical mechanics driven by Euler-Lagrange equations~\cite{Euler,Lagrange}. The collapse resulted from the discovery of various phenomena which are inexplicable with classical physics~\cite{Faraday,Hertz,Thomson,Lenard:1900,Lenard:1902}. The historical path to quantum mechanics invariably begins with Planck and his analysis of blackbody spectral data~\cite{Planck:1900a,Planck:1900b,Planck:1901tja}. It is this that we now turn to study.

\subsection{Blackbody radiation}

The rate at which an object radiates energy has been found to be proportional to the fourth power of the Kelvin temperature $T$ and to the area $A$ of the emitting object, i.e., $L \propto AT^4$. At normal temperatures ($\approx 300~{\rm K}$) we are not aware of this electromagnetic radiation because of its low intensity. At higher temperatures, there is sufficient infrared radiation that we can feel heat if we are close to the object. At still higher temperatures (on the order of $1000~{\rm K}$), objects actually glow, such as a red-hot electric stove burner. At temperatures above $2000~{\rm K}$, objects glow with a yellow or whitish color, such as the filament of a lightbulb. 

For an idealized object that absorbs all incident electromagnetic radiation  regardless of frequency or angle of incidence (a.k.a. blackbody~\cite{Kirchhoff:1860}),  the bolometric luminosity becomes 
\begin{equation}
L = \sigma \, A \, T^4 \,,
\end{equation}
where $\sigma = 5.67 \times 10^{-8}~{\rm W \, m^{-2} \,  K^{-4}}$ is the Stefan-Boltzmann constant~\cite{Stefan:1879,Boltzmann:1884}. 
The total power leaving $1~{\rm m}^2$ of a blackbody surface -- that is the radiant flux $F$ measured in ${\rm W}/{\rm m}^2$ --  at absolute temperature $T$ is then
\begin{equation}
F (T) = L/A = \sigma T^4 \, .
\label{umizoomi2}
\end{equation}
Another important characteristic of the blackbody radiation is  the experimentally obtained Wien's displacement law~\cite{Wien:1894}, which states that
the wavelength $\lambda_{\rm max}$ at which the spectral emittance reaches its maximum value decreases as the temperature is increased, in inverse proportion to the temperature
\begin{equation}
\lambda_{\rm max} T = 2.90 \times 10^{-3}~{\rm m \, K} \, .
\end{equation}
The term ``displacement'' refers to the way the peak is moved or displaced as the temperature is varied. Wien's law is qualitatively consistent with our previous observation that heated objects first begin to glow with a red color, and at higher temperatures the color becomes more yellow. As the temperature is increased, the wavelength at which most of the radiation is emitted moves from the longer- wavelength (red) part of the visible region toward medium (yellow or whitish) wavelengths.

\begin{figure}[tbp] \postscript{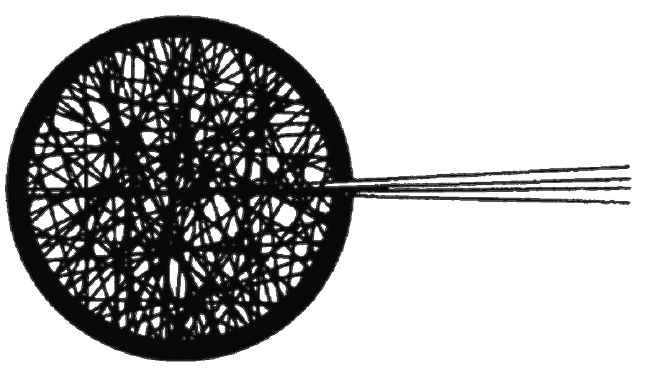}{0.9} \caption{A cavity filled with electromagnetic radiation in thermal equilibrium with its walls at temperature $T$. Some radiation escapes through the hole, which represents an ideal blackbody.}
\label{fig:blackbody}
\end{figure}

An approximate realization of a blackbody surface is as follows. Consider a hollow metal box whose walls are in thermal equilibrium at temperature $T$. The cavity is filled with electromagnetic radiation forming standing waves (normal modes) with nodes at the walls, as shown in Fig.~\ref{fig:blackbody}. Suppose there is a small hole in one wall of the box which allows some of the radiation to escape. It is the hole, and not the box itself, that is the blackbody. Radiation from outside that is incident on the hole gets lost inside the box and has a negligible chance of reemerging from the hole; thus no reflections occur from the blackbody (the hole). The radiation that emerges from the hole is just a sample of the radiation inside the box, so understanding the nature of the radiation inside the box allows us to understand the radiation that leaves through the hole.

\begin{figure}[tbp] \postscript{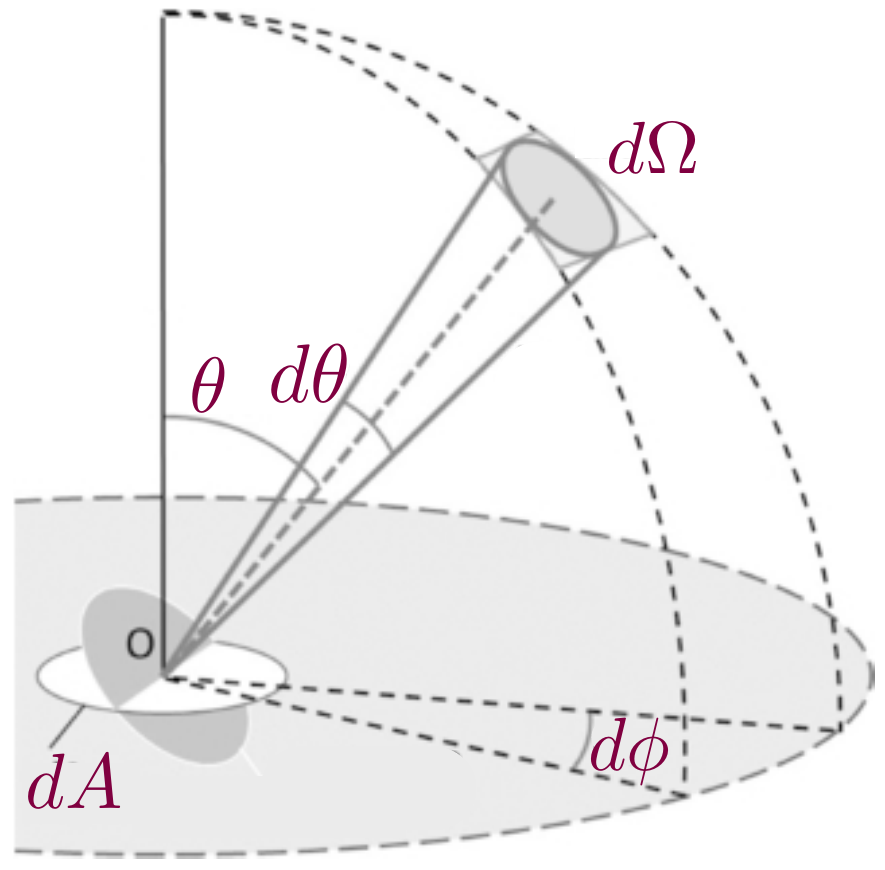}{0.85} \caption{Geometry for an obliquely emerging solid angle $d\Omega$ from a unit surface area $dA$, with unitary normal vector $\hat n$.}
\label{fig:solid-angle}
\end{figure}

Let us consider the radiation inside the box. It has an energy density (that is energy per unit volume) $u(T)$ measured in ${\rm J} \, {\rm m}^{-3}$, and a spectral energy density $du/d\lambda \equiv u_\lambda (\lambda,T)$, measured in ${\rm J} \, {\rm m}^{-3} \, {\rm nm}^{-1}$. That is, if we could look into the interior of the box and measure the spectral energy density of the electromagnetic radiation with wavelengths between $\lambda$ and $\lambda + d\lambda$ in a small volume element, the result would be $u_\lambda(\lambda, T) \, d\lambda$. 
The surface brightness (or spectral emittance) $B_\lambda (\lambda, T)$ is defined as the spectral radiant flux  per sterradian emitted from a unit surface that lies normal to the view direction. The surface element is thus different for each view direction. Because photons of all wavelengths travel at speed $c$, the wavelength dependence of the spectral energy density is the same as that of the surface brightness. It does not matter whether the radiation sampled is that in $1~{\rm m^3}$ at fixed time, or that impinging on a 1~${\rm m}^2$ in 1~s. The relation between the two quantities, $B_\lambda$ and $u_\lambda$, follows from their definitions. Consider first the spectral energy density $u_\lambda$. It includes photons moving at speed $c$ isotropically in all directions into all $4~\pi~{\rm sr}$. Divide by $4\pi$ to obtain the energy per unit volume  flowing into 1~sr, i.e. $u_\lambda/4\pi$. Multiply this by the speed $c$ of the photons to obtain the spectral radiant flux per steradian passing through $1~{\rm m}^2$ in 1~s, which is the surface brightness, 
\begin{equation}
B_\lambda (\lambda,T) = \frac{c}{4 \pi} u_\lambda (\lambda, T) \, .
\end{equation}
It is useful to know the total radiant flux passing through a unit area of a fixed surface immersed within the  blackbody cavity. Consider the radiation field we have inside the box (i.e. a bundle of rays flowing in all directions)  and construct a small element of area $dA$ at some arbitrary orientation $\hat n$ as shown in Fig.~\ref{fig:solid-angle}. The differential amount of flux from the solid angle $d\Omega$ is (reduced by the lowered effective area $\cos \theta dA$), 
\begin{equation}
d F_\lambda (\lambda,T) = B_\lambda(\lambda, T) \cos \theta d \Omega \, ,
\label{umizoomi1}
\end{equation}
where the solid angle element is $d\Omega = \sin \theta d\theta d \phi$. The spectral radiant flux in the direction $\hat n$, is obtained by integrating (\ref{umizoomi1}) over the upper hemisphere shown in Fig.~\ref{fig:solid-angle}. Note that if $B_\lambda$ is an isotropic radiation field (not a function of angle), then the {\it net} flux is zero, since $\int \cos \theta d\Omega =0$ over all solid angles. That is, there is as much energy crossing $dA$ in the $\hat n$ direction as in the $-\hat n$ direction. The total power leaving $1~{\rm m}^2$ of the surface (i.e. the  radiant flux) is the surface brightness integrated over all frequencies and all angles of the upper hemisphere
\begin{equation}
F(T) = \int_0^\infty \int_{\theta =0}^{\pi/2} \int_{\phi=0}^{2 \pi} B_\lambda(\lambda,T) \frac{dA \cos \theta}{dA} \ d\Omega d\lambda \, .
\label{Pumpkin1}
\end{equation}
The integration limits for $\theta$ specify that the integration is only over the upper hemisphere, yielding
\begin{equation}
\int_{\theta=0}^{\pi/2} \int_{\phi=0}^{2 \pi} \cos \theta \sin \theta d\theta d\phi = \pi \, .
\end{equation}
On the other hand, the total energy $u(T)$ in a unit volume is simply $u_\lambda (\lambda, T)$ integrated over all wavelengths. From (\ref{umizoomi2}) it follows that
\begin{equation}
u(T) = \int_0^\infty u_\lambda (\lambda, T) d \lambda = a T^4 \,,
\label{yogabagaba1}
\end{equation}
where $a = 4 \sigma/c = 7.566 \times 10^{-16}~{\rm J} \, {\rm m}^{-3} \, {\rm K}^{-4}$.
Our next goal is to derive an analytic expression for the spectral energy density that satisfies (\ref{yogabagaba1}).

\begin{figure}[tbp] \postscript{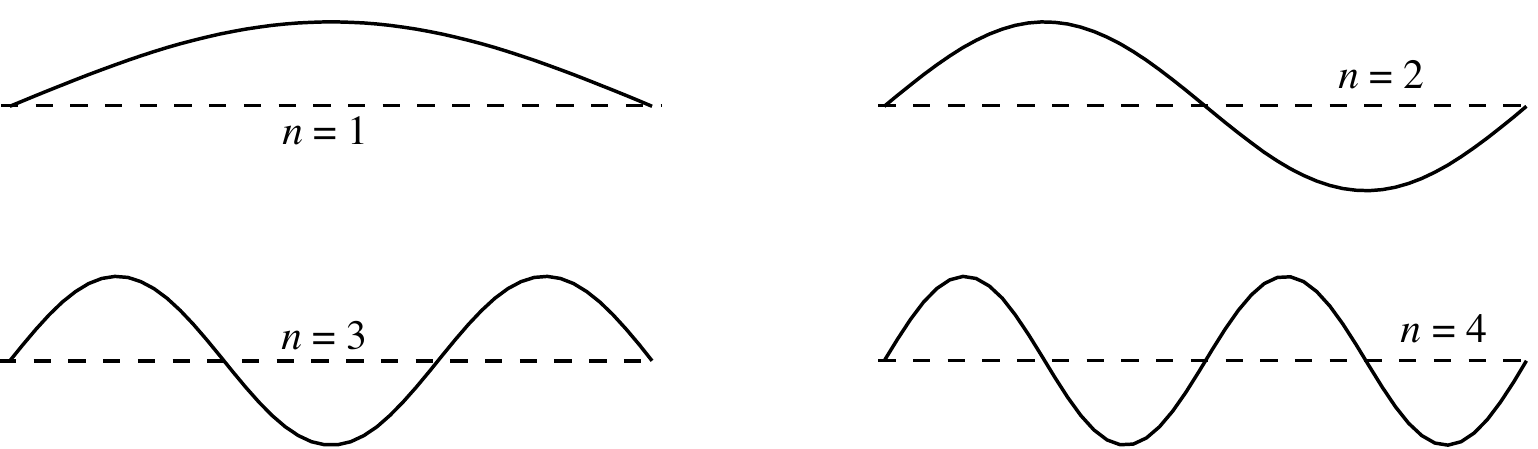}{0.95} \caption{Standing waves corresponding to $n\lambda/2 = L_x$ for $n = 1, 2, 3,4$.  Abscissa: $x$-axis of the cavity, bounded by walls at $x = 0$ and $x = L_x$. Ordinate: Electric field strengths of the longest standing waves satisfying the boundary condition $\vec E = 0$ at the walls.}
\label{fig:normal-modes}
\end{figure}

We have seen that any wave can be characterized by: the wavelength $\lambda$, the speed $c$, the period $ T= \lambda/c$, the frequency $\nu = 1/T = c/\lambda$, or else by the angular frequency 
\begin{equation}
\omega = 2 \pi \nu = 2\pi c/\lambda \, .
\label{yogabagaba2}
\end{equation}
In what follows we will refer to both the frequency $\nu$ and the angular frequency $\omega$ as simply the frequency, since this will not lead to a confusion of the two. We start by relating $u_\omega (\omega, T)$ with $u_\lambda(\lambda,T)$. This is done as follows. The spectral energy density  within the wavelength interval $(\lambda, \lambda + \Delta \lambda)$ can be written in two ways:
\begin{equation}
u_\lambda ( \lambda, T) \ d\lambda = u_\omega(\omega, T) \ d\omega \, .
\label{yogabagaba3}
\end{equation}
The left-hand side of (\ref{yogabagaba3}) is merely the definition of the spectral energy density. The right-hand side expresses the fact that $\lambda$ and $\omega$ are related by a one-to-one function (\ref{yogabagaba2}). Then, (\ref{yogabagaba2}) and (\ref{yogabagaba3}) provide the following relation 
\begin{equation}
u_\lambda (\lambda, T) = u_\omega (\omega, T) \, \left|\frac{d\omega}{d\lambda}\right| = u_\omega (\omega, T) \, \frac{2 \pi c}{\lambda^2} \, .
\label{yogabagaba4}
\end{equation}
To begin the derivation of $u_\omega (\omega, T)$ consider a rectangular   
 box of volume $V = L_x L_y L_z$, which is at thermal equlibrium at temperature $T$. The spectral energy density of the electromagnetic radiation with frequencies between $\omega$ and $\omega + d \omega$ in a small volume element is 
\begin{equation}
u_\omega (\omega, T)  d\omega = N(\omega, T) \ \langle E \rangle \,,
\label{yogabagaba5}
\end{equation}
where $N(\omega, T)$ is the number of electromagnetic modes (standing waves)  within a given frequency interval $(\omega, \omega + d \omega)$ allowed inside the box, $\langle E \rangle$ is the average energy of one radiation mode of frequency $\omega$, and $u_\omega (\omega, T)$ is the average energy per $\omega$-interval per volume; the qualifier ``average'' is used because the radiation is in equilibrium with its source on average over some macroscopic interval of time. As we show below, the difference between the (classical) Rayleigh-Jeans prediction~\cite{Rayleigh:1900,Rayleigh:1905,Jeans:1905} and the (quantum) prediction by Planck~\cite{Planck:1900a,Planck:1900b,Planck:1901tja} depends on how $\langle E \rangle$ is calculated. Before proceeding  we estimate the first term on the right-hand side of (\ref{yogabagaba5}). To this end we note that
\begin{equation}
\left(\begin{array}{c}
{\rm number \, of \, states \, inside \, the \, box} \\ {\rm within \, the \, interval} \, (\omega, \omega + d\omega) \end{array} \right) = \frac{dZ}{d \omega} d \omega \,,
\label{yogabagaba6}
\end{equation}
where $Z(\omega)$ is the number of standing waves up to $\omega$ that can exist in this box.  We assume that the allowed frequencies of the radiation propagating in any one direction are spaced evenly. With this in mind, we can write
\begin{equation}
Z(\omega) = \varkappa \left(\frac{\omega}{\omega_{{\rm min},x} }\right) \, \left(\frac{\omega}{\omega_{{\rm min},y} }\right)  \left(\frac{\omega}{\omega_{{\rm min},z} }\right)
\label{yogabagaba7}
\end{equation}
where $\omega_{{\rm min},x}$ is the minimum frequency of radiation that can propagate in the box in the $x$-direction; and similarly for $\omega_{{\rm min},y}$ and $\omega_{{\rm min},z}$. This minimum frequency exists because there is the maximum wavelength, 
\begin{equation}
\lambda_{{\rm max},x} = 2L_x
\label{yogabagaba8}
\end{equation}
that can exist between the walls located $L_x$ units apart. (The illustrating Fig.~\ref{fig:normal-modes}  assumes that the wave is zero at the walls, but a similar result can also be obtained for other boundary conditions.)
Now, from (\ref{yogabagaba2}) and (\ref{yogabagaba8}) we have
\begin{equation}
\omega_{{\rm min},j} = \frac{2 \pi c}{2L_j} = \frac{\pi c}{L_j} \,,
\label{yogabagaba9}
\end{equation}
with $j = \{x,y,z\}$.
From Fig.~\ref{fig:normal-modes} we can conclude that the next three largest wavelengths are $2L_x/2$, $2L_x/3$, and $2L_x/4$ where two, three, and four semi-periods of the wave fit between the walls, respectively. The corresponding frequencies, in analogy with (\ref{yogabagaba9}), are $2\omega_{{\rm min},x}$, $3\omega_{{\rm min},x}$, and $4\omega_{{\rm min},x}$. This justifies our assumption that the frequencies of the radiation in a box are spaced evenly. Next, from (\ref{yogabagaba7}) and (\ref{yogabagaba9}) one has:
\begin{eqnarray}
Z(\omega)  =  \varkappa \frac{\omega^3}{(\pi c)^3/(L_x L_y l_z)} 
=  \varkappa \frac{\omega^3}{(\pi c)^3} L_x L_y L_z \ .
\label{yogabagaba10}
\end{eqnarray}
Using the fact that photons have two independent polarizations (that is there are two states per wave vector $\vec k = (\omega/c) \, \hat n$ of photons of frequency $\omega$ propagating in the direction $\hat n$ inside the box) it is easily seen that $\varkappa = \pi/3$~\cite{Rybicki:1979}.
Then according to (\ref{yogabagaba10}) 
\begin{equation}
\left(\begin{array}{c}
{\rm number \, of \, states \, inside \, the \, box} \\ {\rm within \, the \, interval} \, (\omega, \omega + d\omega) \end{array} \right) = \frac{\omega^2}{\pi^2 c^3} L_x L_y L_z \,, \nonumber 
\end{equation}
and finally
\begin{equation}
\frac{\left(\begin{array}{c}
{\rm number \, of \, states \, inside \, the \, box} \\ {\rm within \, the \, interval} \, (\omega, \omega + d\omega) \end{array} \right)}{\rm volume \, of \, the \, box} = \frac{\omega^2 d\omega}{\pi^2 c^3} \, .
\label{yogabagaba11}
\end{equation}

The energy of each standing wave (normal mode) is distributed according to the Maxwell-Boltzmann distribution~\cite{Maxwell:1860a,Maxwell:1860b,Boltzmann:1877}
\begin{equation}
P(E) \, dE = \frac{e^{-E/kT}}{kT} \, dE \,,
\label{yogabagaba12}
\end{equation}
where $k= 1.38 \times 10^{-23}~{\rm J} \, {\rm K}^{-1}$ is the Boltzmann's constant and $T$ the absolute temperature; 
(\ref{yogabagaba12}) reflects Boltzmann's connection between entropy and probability~\cite{Carter}. The classical Rayleigh-Jeans prediction is arrived at assuming $E$ to be a {\it continuous} variable from $0$ to $\infty$, i.e. all energies are possible. Then,
\begin{equation}
\langle E \rangle = \frac{\int_0^\infty P(E) \, E \, dE}{\int_0^\infty P(E) \, dE}= \cdots = kT \, .
\label{Rayleigh-Jeans}
\end{equation}
It is worthwhile to point out that $P(E)$ is already normalized, i.e.  the denominator of (\ref{Rayleigh-Jeans}) equals one. We are keeping it so that we will be consistent when  considering the discrete case later.

\begin{figure}[tbp] \postscript{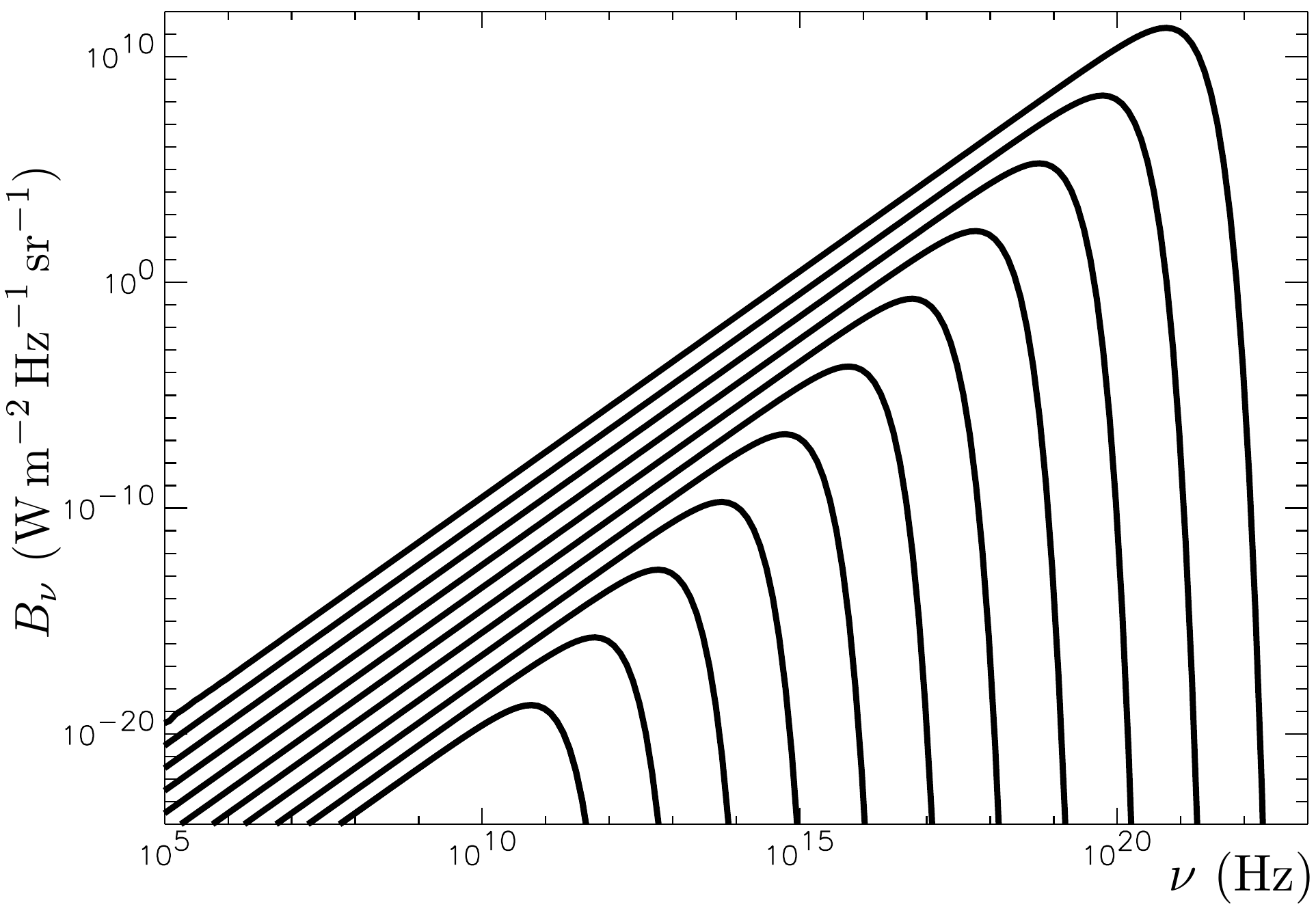}{0.9} \caption{Planck spectrum of blackbody radiation. Each curve corresponds to a certain absolute thermodynamic temperature value. Shown is the surface brightness for $10^0~{\rm K}$, $10^1~{\rm K}$, $10^2~{\rm K}$, $\cdots,\, 10^{10}~{\rm K}$.}
\label{fig:Planck}
\end{figure}

Putting all these ingredients together, we arrive at the spectral energy density of
radiation in the wavelength interval $d\lambda$ inside the cavity
\begin{equation}
u_\lambda (\lambda,T) \, d \lambda = \frac{N(\lambda) \, d \lambda}{V} \ kT = \frac{8 \pi}{\lambda^4} \ kT \ d \lambda \, .
\label{RJu}
\end{equation}
The corresponding spectral emittance per unit wavelength interval $d \lambda$ is the so-called  ``Rayleigh-Jeans formula''
\begin{equation}
B_\lambda (\lambda, T)  =  \frac{c}{4\pi} \, u_\lambda(\lambda, T) 
 = \frac{2 c}{\lambda^4} \, kT \, .
\label{yogabagabaRJ}
\end{equation}
This result, based firmly on the theories of electromagnetism and thermodynamics, represents our best attempt to apply classical physics to understanding the problem of blackbody radiation. The surface brightness  calculated with (\ref{yogabagabaRJ}) is in agreement with experimental data at long wavelengths. At short wavelengths, however, the classical theory (which predicts $u_\lambda \to \infty$ as $\lambda \to 0$) is absolutely not physical. The surface brightness of the emitted light (radiated energy) must remain finite! The failure of the Rayleigh-Jeans formula at short wavelengths is known as the {\it ultraviolet catastrophe} and represents a serious problem for classical physics, because the theories of thermodynamics and electromagnetism on which the Rayleigh-Jeans formula is based have been carefully tested in many other circumstances and found to give extremely good agreement with experiment. It is apparent in the case of blackbody radiation that the classical theories do not work, and that a new kind of physical theory is needed.

In 1901, Planck proposed a solution to this problem by considering a different way to calculate $\langle E \rangle$.  In Planck's theory, the energy is {\it not} a continuous variable. Each oscillator in the cavity can emit or absorb energy only in quantities that are integer multiples of a certain basic quantity $\Delta E$, 
 \begin{equation}
E_n = n \, \Delta E, \quad {\rm with} \quad n = 0,1,2,3, \cdots \, .
\end{equation}
Furthermore, the discrete energy increment is proportional to the frequency of the oscillator
\begin{equation}
\Delta E = h \nu \,,
\end{equation}
where $h = 6.626 \times 10^{-34}~{\rm J} \, s = 4.136 \times 10^{-15}~{\rm eV} \, {\rm s}$ is the Planck's constant.
 The average energy of an oscillator is then given by the discrete sum 
\begin{equation}
\langle E \rangle = \frac{\sum_{n=0}^\infty E_n \, P(E_n)}{\sum_{n=0}^\infty P(E_n) } = \cdots = \frac{hc/\lambda}{e^{hc/(\lambda kT)} -1} \, .
\label{Planck-E}
\end{equation}
Multiplying this result by the number of oscillators per unit volume in the interval $d \lambda$ given by (\ref{yogabagaba11}), we obtain  the spectral emittance distribution function of the radiation inside the cavity
\begin{equation}
B_\lambda (\lambda, T) = \frac{2  c}{\lambda^4} \ \langle E \rangle = \frac{2 c}{\lambda^4} \frac{hc/\lambda}{e^{hc/(\lambda kT)} - 1} \, .
\label{yogabagabaP}
\end{equation}
Planck's spectra for a range of values of $T$ is shown in Fig.~\ref{fig:Planck}.
Note that Planck's result reduces to the classical limit if $h = \Delta E = 0$.  For $h \to 0$,  the exponential in (\ref{yogabagabaP}) can be expanded using $e^x \approx 1 + x + \cdots$ for $x\ll 1$, where $x = hc/(\lambda kT)$. Then
\begin{equation}
e^{hc/(\lambda k T)} -1 \approx \frac{hc}{\lambda kT}
\label{yogabagabaexp}
\end{equation}
and so
\begin{equation}
\langle E \rangle = \frac{hc/\lambda}{e^{hc/(\lambda kT}) - 1} = kT \, .
\end{equation}
For long wavelength (low photon energy) again we use (\ref{yogabagabaexp}) to show
\begin{equation}
\lim_{\lambda \to \infty} 
u_\lambda \to \frac{8 \pi}{\lambda^4}  kT \,,
\end{equation}
which is the Rayleigh-Jeans formula (\ref{RJu}). Finally, in the quantum regime $\lambda \to 0$ (i.e. high photon energy) $e^{hc/(\lambda kT)} \to \infty$ exponentially faster than $\lambda^5 \to 0$ so
\begin{equation}
\lim_{\lambda \to 0} \frac{1}{\lambda^5 (e^{hc/(\lambda kT)} - 1)} \to 0 \, .
\end{equation}
There is no ultraviolet catastrophe in the quantum limit.\\

{\bf EXERCISE 9.1}~Calculate the integrals in (\ref{Rayleigh-Jeans}) and the sum of the series in (\ref{Planck-E}).\\

{\bf EXERCISE 9.2}~Verify that if you integrate (\ref{yogabagabaP}) over all wavelengths and solid angles you  can reproduce Stefan-Boltzmann law (\ref{umizoomi2}).\\

{\bf EXERCISE 9.3}~Show that Wien's displacement law can be derived by determining the maximum of (\ref{yogabagabaP}).\\

\begin{figure}[tbp] \postscript{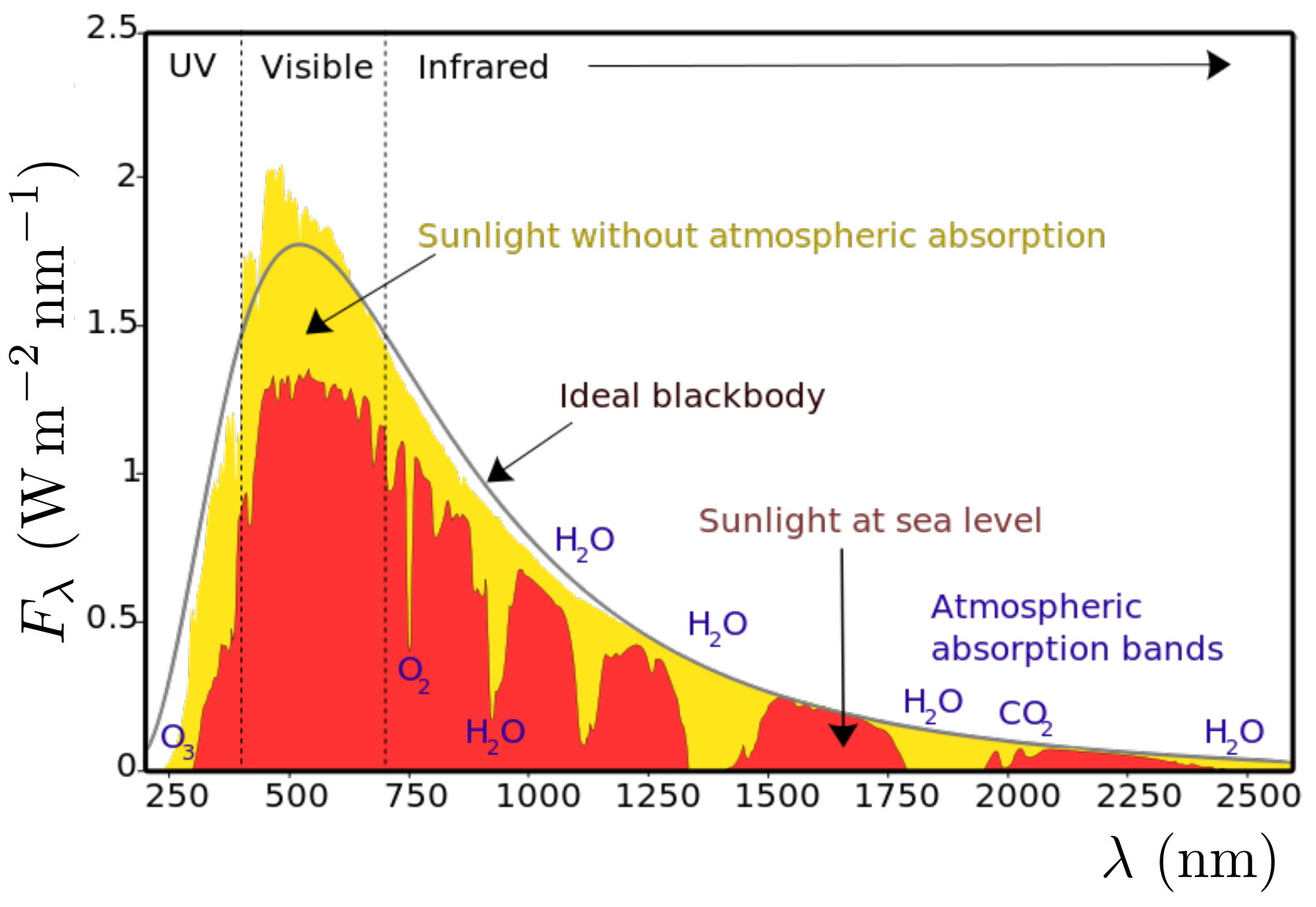}{0.9} \caption{Solar energy incident at Earth's atmosphere and surface. The yellow band is the radiation incident at the top of the atmosphere, while the red band is the radiation at Earth's surface, diminished by the atmospheric absorbers shown. The radiation approximates a blackbody curve. These data are from the American Society for Testing and Materials (ASTM) Terrestrial Reference Spectra.} 
\label{fig:sun-blackbody}
\end{figure}

{\bf EXERCISE 9.4}~{\it (i)}~Stars behave approximately like blackbodies. Use Wien's displacement formula to obtain a rough estimate of the surface temperature of the Sun, assuming that it is an ideal blackbody as suggested by the ASTM data shown in Fig.~\ref{fig:sun-blackbody} and that evolution on Earth worked well (i.e., that the human eye uses optimal the light from the Sun). {\it (ii)}~The solar constant (radiant flux at the surface of the Earth) is about 1.365~${\rm kW/m}^2$. Find the effective surface temperature of the Sun {\it (iii)}~Assuming that the surface of Neptune and the thermodynamics of its atmosphere are similar to those of the Earth estimate the surface temperature of Neptune. Neglect any possible internal source of heat. [{\it Hint:} Astronomical data which may be helpful: radius of Sun $R_\odot = 7 \times 10^5~{\rm km}$; radius of Neptune $R_{\rm N}= 2.2 \times 10^4~{\rm km}$; mean Sun-Earth distance $r_{\rm SE} = 1~{\rm AU}= 1.5 \times 10^8~{\rm
  km}$; mean Sun-Neptune distance $r_{\rm SN}= 4.5 \times 10^9~{\rm km}$.]\\

\begin{figure}[tbp] \postscript{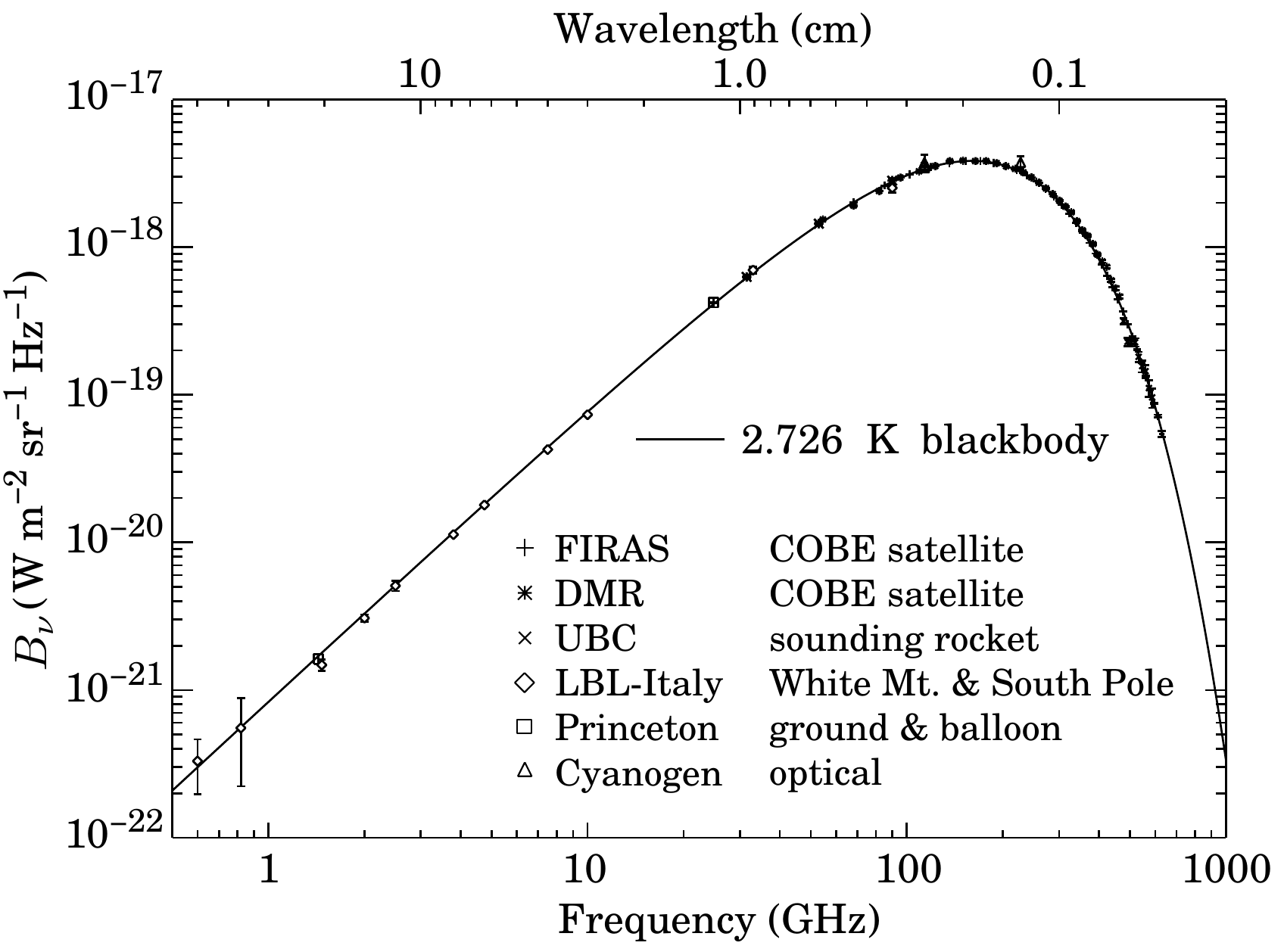}{0.9} \caption{The CMB blackbody spectrum as confirmed by measurements over a broad range of wavelengths~\cite{Smoot:1997xt}.}
\label{fig:CMB-blackbody}
\end{figure}

{\bf EXERCISE 9.5}~A compilation of experimental measurements of the CMB reveals an accurate blackbody spectrum, see Fig.~\ref{fig:CMB-blackbody}. Actually, according to the FIRAS (Far InfraRed Absolute Spectrometer) instrument aboard the COBE ({\it Cosmic Background Explorer}) satellite, which measured a temperature of $T = 2.726 \pm 0.010~{\rm K}$, the CMB is the most perfect blackbody ever seen~\cite{Mather:1993ij}. {\it (i)}~Write down an integral which determines how many photons per cubic centimeter are contained in the CMB.  Estimate the result within an order of magnitude and show that it agrees with the fiducial value adopted in exercise 8.9. {\it (ii)}~Convince yourself that the average energy of a CMB photon is $\langle E_{\gamma}^{\rm CMB} \rangle \approx 6 \times 10^{-4}~{\rm eV}${\it (iii)}~Show that a freely expanding blackbody radiation remains described by the Planck formula, but with a temperature that drops in proportion to the scale expansion.

\subsection{Photoelectric effect}

The success of Planck's idea immediately raises the question: why is it that oscillators in the walls can only emit and absorb energies in multiples of $h\nu$? The reason for this was supplied by  Einstein in 1905, in connection with his explanation of the photoelectric effect: light is composed of particles called photons, and  each photon has an energy $E_\gamma = h \nu$~\cite{Einstein:1905cc}.

\begin{figure}[tbp] \postscript{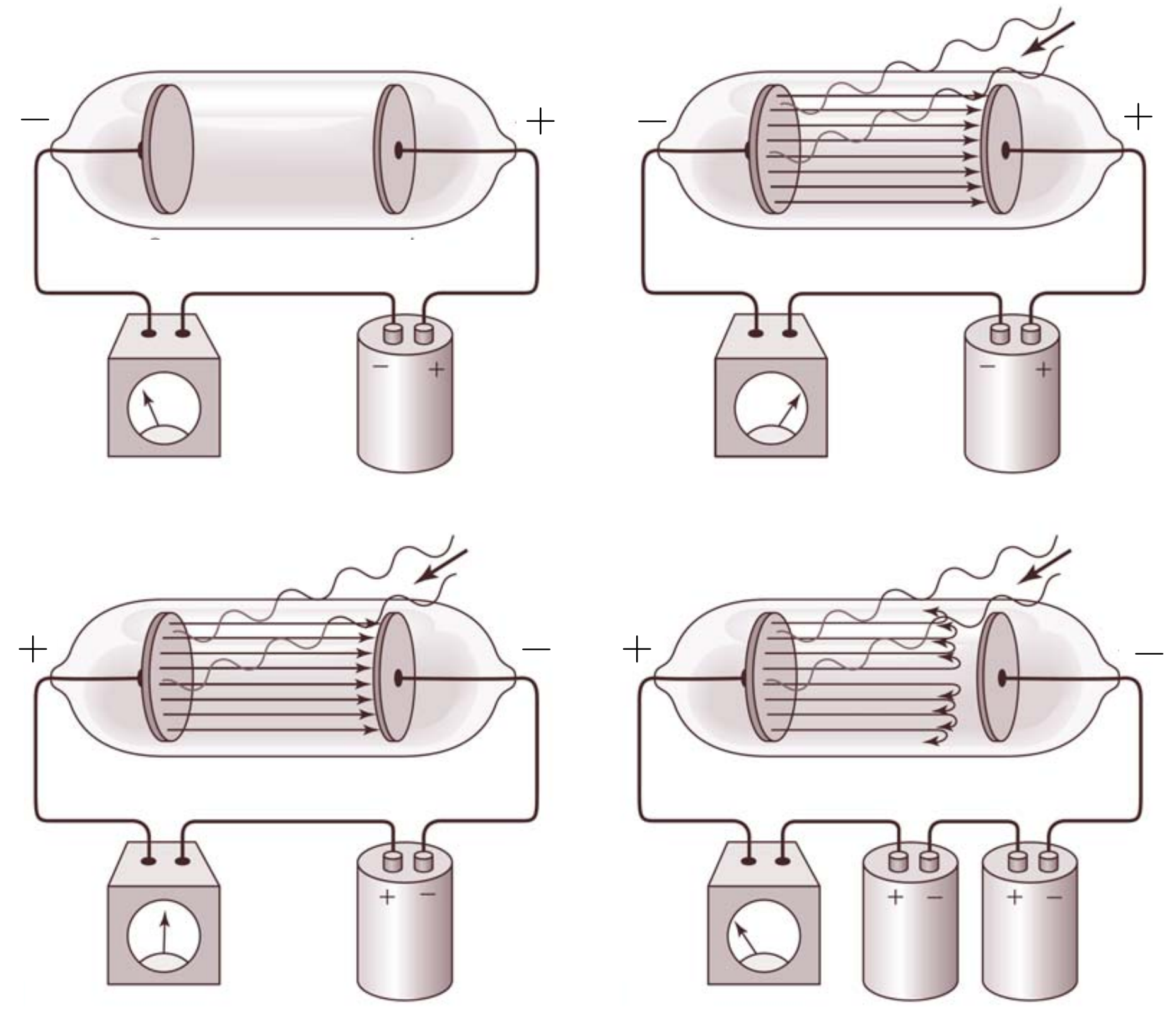}{0.85} \caption{Schematic diagram of the apparatus used  to study the photoelectric effect.}
\label{fig:photoelectric-effect}
\end{figure}

The photoelectric effect is the observation that a beam of light can knock electrons out of the surface of a metal. The electrons emitted from the surface are called photoelectrons. The phenomenon was discovered by Hertz~\cite{Hertz}  and further studied by Lenard~\cite{Lenard:1900,Lenard:1902}. What is surprising about the photoelectric effect is that the energy of the photoelectrons is independent of the intensity of the incident light. If the frequency of the light is swept, we find that there is a minimum frequency $\nu_0$ below which no electrons are emitted. The energy $\varphi = h \nu_0$ corresponding to this frequency is called the work function of the surface. In experiments used to investigate the photoelectric effect, the photoelectrons are usually collected on a metal plate detector, which forms part of an electrical circuit; see Fig.~\ref{fig:photoelectric-effect}. The current measured in the circuit is proportional to the number of electrons striking the detector plate. To measure the kinetic energy of the electrons, we can apply a static potential $V$ (often known as a retarding potential) to the plate. Only electrons with a kinetic energy greater than $eV$, where $e$ is the electron charge, will reach the plate; any electrons with a kinetic energy less than $eV$ will be repelled, and will not be detected. The stopping potential, $V_0$, is the retarding potential at which no more electrons are detected, and tells us the kinetic energy of the fastest electrons, i.e. $K_{\rm max}=eV_0$. One of the important early observations about the photoelectric effect is that for a particular metal surface and frequency of light, the stopping potential is independent of the light intensity.

\begin{figure}[tbp] \postscript{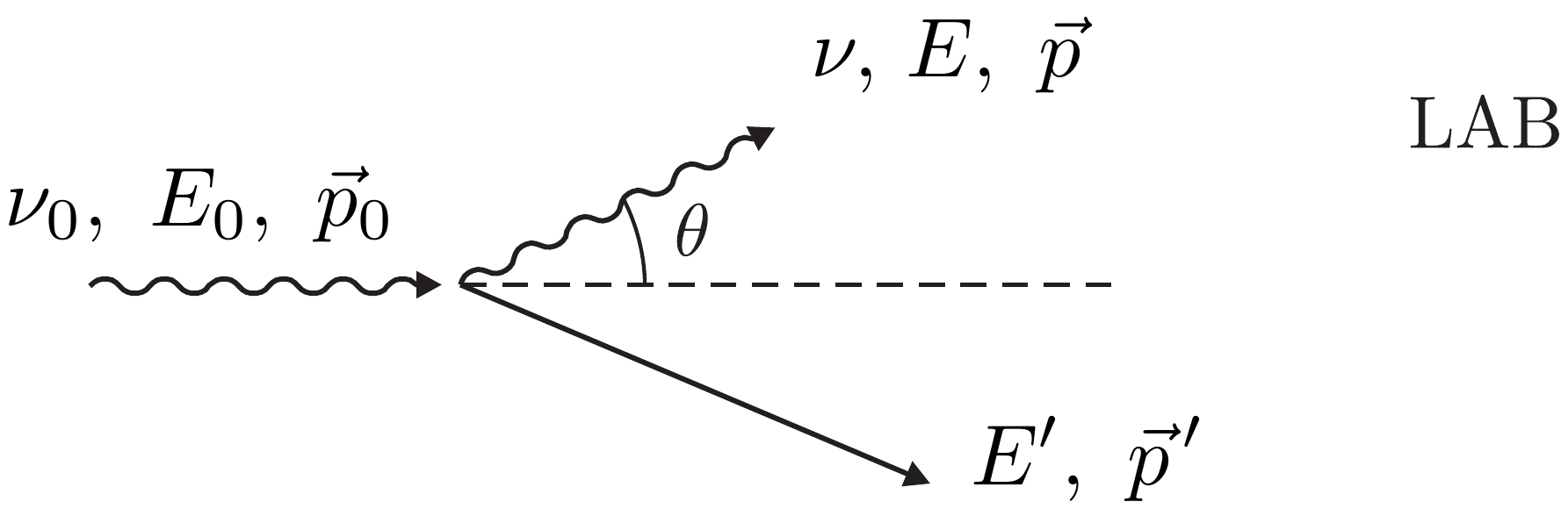}{0.85} \caption{The geometry of Compton scattering.}
\label{fig:Compton}
\end{figure}

We can now explain why classical electromagnetism fails to explain the photoelectric effect, and why all the problems are solved by quantum mechanics. {\it (i)}~In classical electromagnetism, increasing the intensity of a beam of light increases the amplitude of the oscillating electric field $\vec E$. Since the force that the incident beam exerts on an electron is $e\vec E$, the theory would predict that the energy of the photoelectrons would increase with increasing light intensity. However, this is not the case: $V_0$ is independent of light intensity. In quantum mechanics, however, doubling the intensity simply doubles the number of photons, it does not change their energy. If the photoelectric effect is interpreted as a collision between a single photon and an electron in the metal surface, then the total number of photons striking the surface is immaterial in determining the energy of the ejected electron. {\it (ii)}~According to classical electromagnetism, as long as the intensity of the light is large enough, the photoelectric effect should occur at any frequency, a direct contradiction of the experimental evidence, which shows a clear cutoff frequency below which no electrons are ejected. In quantum mechanics the frequency of the light determines the photon energy, and since it takes a certain minimum amount of energy to knock an electron out of the surface (defined by the work function $\varphi$ of the surface), photons with energy $h\nu < \varphi$ simply do not have enough energy to achieve this. {\it (iii)}~In classical electromagnetism the energy imparted to the electron must somehow be ``soaked up'' from the incident wave. Since this takes some time, if very weak light is used it would be expected that there should be a measurable time delay between the light striking the surface and the electron being emitted. This has never been observed: measurements have shown that if there is a time lag, it is less than $10^{-9}~{\rm s}$. In quantum mechanics, the photoelectric effect is viewed as a single collisional event and no time delay is predicted. {\it (iv)}~When $K_{\rm max}$ is plotted as a function of frequency  $\nu > \nu_0$, the experimental data  fit a straight line, whose slope equals Planck's constant: $K_{\rm max} = h \nu - \varphi$~\cite{Millikan:1914,Millikan:1916}.

Einstein's theory of the photon composition of light immediately explains Planck's condition that the energy of electromagnetic radiation of frequency $\nu$, in a box, is restricted to the values $E= nh\nu$. Since there can only be an integer number of photons $n$ at any given frequency, each of energy $h\nu$, the energy of the field at that frequency can only be $nh\nu$. Planck's restriction on energies is thereby explained in a very natural, appealing way.\\

{\bf EXERCISE 9.6}~For a typical case of photoemission from sodium, show that classical theory predicts that: {\it (i)}~$K_{\rm max}$ depends on the incident light intensity $I$;  {\it (ii)}~$K_{\rm max}$ does not depend on the frequency of the incident light; {\it (iii)} there is a long time lag between the start of illumination and the beginning of the photocurrent. The work function for sodium is $\varphi = 2.28~{\rm eV}$ and an absorbed power per unit area of $1.00\times 10^{-7}~{\rm  mW/cm}^2$ produces a measurable photocurrent in sodium.\\

{\bf EXERCISE 9.7}~A metal surface has a photoelectric cutoff wavelength of 325.6~nm. It is illuminated with light of wavelength 259.8~nm. What is the stopping potential?\\

{\bf EXERCISE 9.8}~Another effect that revealed the quantized nature of radiation is the (elastic) scattering of light on particles shown in Fig.~\ref{fig:Compton}, called the Compton effect~\cite{Compton:1923zz}. {\it (i)}~Using conservation of energy and momentum, derive the Compton shift formula,
\begin{equation}
\lambda - \lambda_0 = \lambda_{\rm c} ( 1 - \cos \theta) \,,
\label{Compton}
\end{equation}
where  $\lambda_{\rm c}$ is the Compton wavelength of the particle, which is equivalent to the wavelength of a photon whose energy is the same as the mass of the particle. {\it (ii)}~In the experiment by Compton, X-rays are scattered by nearly free electrons ($\lambda_{\rm c} = 2.43 \times 10^{-10}~{\rm cm}$) in carbon (graphite).  (Although no scattering target contains actual ``free'' electrons, the outer or valence electrons in many materials are very weakly attached to the atom and behave like nearly free electrons. The binding energies of these electrons in the atom are so small compared with the energies of the incident X-ray photons that they can be regarded as nearly ``free'' electrons.) A movable detector measured the energy of the scattered X rays at various angles $\theta$. At each angle, two peaks appear, corresponding to scattered X-ray photons with two different energies or wavelengths. The wavelength of one peak does not change as the angle is varied; this peak corresponds to scattering that involves ``inner'' electrons of the atom, which are more tightly bound to the atom so that the photon can scatter with no loss of energy. The wavelength of the other peak, however, varies strongly with angle. This variation is exactly as the Compton formula predicts. Show that for a maximal scattering angle the fractional change $\Delta \lambda/\lambda_0$ is about 7\%. In summary, the particle character of light is confirmed in Compton's experiment and we assign the energy of $E = \hslash \omega$ and the momentum $\vec p = \hslash \vec k$ to the (undivisible) photon, where $\hslash = h/(2\pi)$. The Compton shift formula (\ref{Compton}) reveals a proportionality to $\hslash$, a quantum mechanical property that is confirmed by experiment. Classically no change of the wavelength is to be expected.

\subsection{Line spectra of atoms}

When a gaseous element is energetically excited so that it emits radiation, the emitted radiation -- when passed through a prism-- is found to consist of a series of well-defined lines (called the spectrum of the element), each associated with a different wavelength. When the excitation is carried out by heating to incandescence in a flame, the spectra are found to be associated with neutral atoms, but if the excitation is more energetic, e.g., due to a high-voltage electrical discharge or spark, the resulting spectrum (spark spectrum) is found to be associated with ionized atoms. 

\begin{figure*}[tbp] \postscript{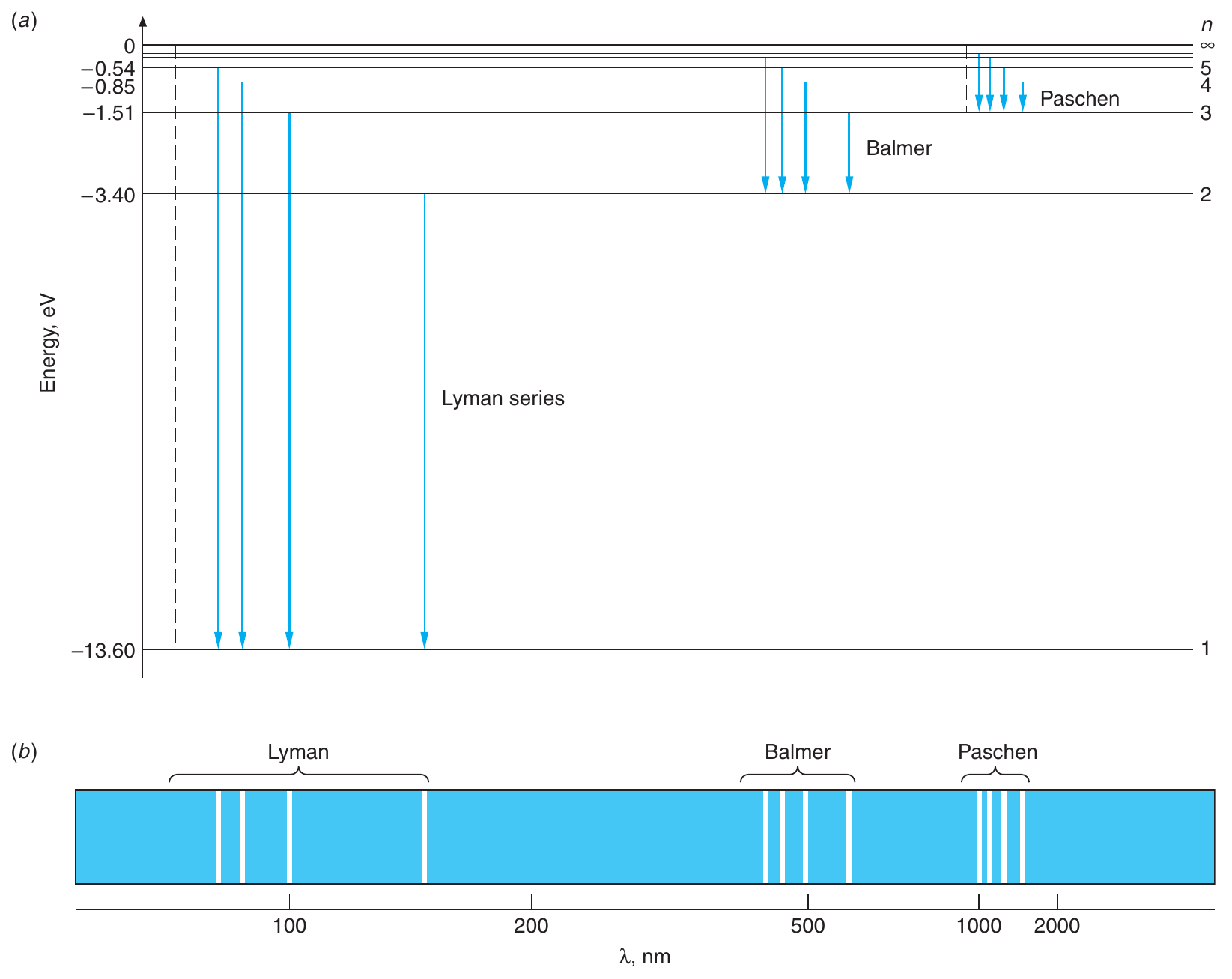}{0.85} \caption{(a)~Energy-level diagram for hydrogen showing the seven lowest stationary states and the four lowest energy
transitions each for the Lyman, Balmer, and Paschen series. There are an infinite number of levels. The dashed line shown for each series is the series limit, corresponding to the energy $n$
that would be radiated by an electron at rest far from the nucleus ($n_2 \to \infty$) in a transition to the state $n_1$ for that series. The horizontal spacing between the transitions shown for each series is proportional to the wavelength spacing between the lines of the spectrum. (b)~The spectral lines corresponding to the transitions shown for the three series. Notice the regularities within each series, particularly the short-wavelength limit and the successively smaller separation between adjacent lines as the limit is approached. The wavelength scale in the diagram is not linear~\cite{Tipler}.}
\label{fig:atomic-spectra}
\end{figure*}

When  hydrogen in a glass tube is excited by a 5000~V electrical discharge, four lines are observed in the visible part of the emission spectrum: red at 656.3~nm, blue-green at 486.1~nm, blue violet at 434.1~nm and violet at 410.2~nm. These can be explained by  Balmer's empirical formula  $\lambda = 364.56 \ n^2/(n^2 -4)~{\rm nm}$, where $n$ is a variable integer that takes on the values $n= 3, 4, 5, \cdots$~\cite{Balmer}.
Balmer's formula was generalized subsequently by Rydberg~\cite{Rydberg:1890a,Rydberg:1890b} and Ritz~\cite{Ritz:1908} to accommodate newly discovered spectral lines in the ultraviolet~\cite{Lyman:1906,Lyman:1914} and infrared~\cite{Paschen} regions; see Fig.~\ref{fig:atomic-spectra}. The modern form, called the Balmer-Rydberg-Ritz formula, gives the reciprocal wavelength as
\begin{equation}
\frac{1}{\lambda} = {\cal R} \left(\frac{1}{n_1^2} - \frac{1}{n_2^2} \right), \quad {\rm for} \quad n_2 > n_1\,,
\label{Bohr1}
\end{equation}
where $n_1$ and $n_2$ are integers and ${\cal R}$, the Rydberg constant, is the same for all series of spectral lines of the same element and varies only slightly, and in a regular way, from element to element. For hydrogen, the value of ${\cal R}$ is ${\cal R}_{\rm H} = 1.096776 \times 10^7~{\rm m}^{-1}$. The Balmer series of spectral lines in the visible region correspond to the values $n_1 = 2$, $n_2 = 3,4,5$ and 6. The lines with $n_1 = 1$ in the ultraviolet make up the Lyman series. The line with $n_2 = 2$, designated the Lyman alpha, has the longest wavelength (lowest wavenumber) in this series, with $1/\lambda = 82.258~{\rm cm}^{-1}$ or $\lambda = 121.57~{\rm nm}$. For very heavy elements, ${\cal R}$ approaches the value of ${\cal R}_\infty =1.097373 \times 10^7~{\rm m}^{-1}$.

Many attempts were made to construct a model of the atom that yielded the Balmer-Rydberg-Ritz formula.  It was known that an atom was about $10^{-10}~{\rm m}$ in diameter, that it contained electrons much lighter than the atom, and that it was electrically neutral. Thomson attempted various models consisting of electrons embedded in a fluid that contained most of the mass of the atom and had enough positive charge to make the atom electrically neutral~\cite{Thomson:1904}. He then searched for configurations that were stable and had normal modes of vibration corresponding to the known frequencies of the spectral lines. One difficulty with all such models was that electrostatic forces alone cannot produce stable equilibrium. The nuclear model proposed by Rutherford pictures the atom as a heavy, positively-charged nucleus, around which much lighter, negatively-charged electrons circulate, much like planets in the Solar system~\cite{Rutherford:1911}. This model is however completely untenable from the standpoint of classical electromagnetic theory, for an accelerating electron (circular motion represents an acceleration) should radiate away its energy. In fact, a hydrogen atom should exist for no longer than $5 \times 10^{-11}~{\rm s}$, time enough for the electron's death spiral into the nucleus. Bohr sought to avoid such an atomic catastrophe by proposing that certain orbits of the electron around the nucleus could be exempted from classical electrodynamics and remain stable~\cite{Bohr:1913a,Bohr:1913b}. The Bohr model was quantitatively successful for the hydrogen atom, as we shall now show.

\begin{figure}[tbp] \postscript{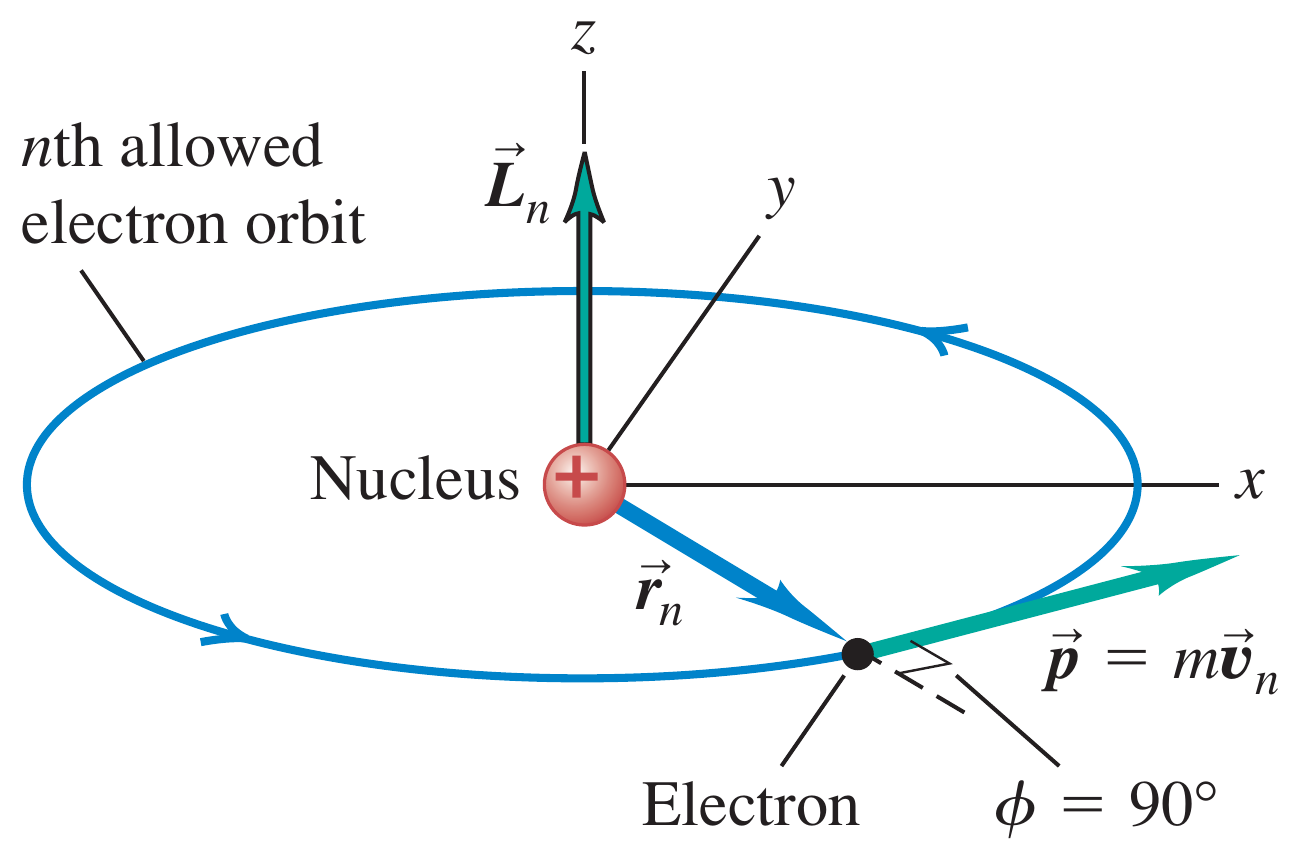}{0.9} \caption{Angular momentum of an electron in a circular orbit around an atomic nucleus. Angular momentum $\vec L_n$ of orbiting electron is perpendicular to plane of orbit and has magnitude $L = m v_n r_n$~\cite{Young:2012}.}
\label{fig:bohr-atom}
\end{figure}

In applications to atomic phenomena we prefer to use the Gaussian unit system.  We recall that the attraction between two opposite charges, such as the electron and proton, is given by Coulomb's law \begin{equation} \vec F = \frac{e^2}{r^2} \, \hat \imath_r \, , \label{Bohr3} \end{equation} where $\hat \imath_r$ denotes the unit radial vector~\cite{Coulomb}.  Since the Coulomb attraction is a central force (dependent only on $r$), the potential energy is related by \begin{equation} |\vec F| = -\frac{dV(r)}{dr} \, .  \label{Bohr4} \end{equation} We find therefore, for the mutual potential energy of a proton and electron, \begin{equation} V(r) = - \frac{e^2}{r} \, .  \label{Bohr5} \end{equation} Bohr considered an electron in a circular orbit of radius $r$ around the proton. To remain in this orbit, the electron must be experiencing a centripetal acceleration \begin{equation} a = v^2/r \label{Bohr6} \end{equation} where $v$ is the speed of the electron.  Using (\ref{Bohr4}) and (\ref{Bohr6}) in Newton's second law, we find \begin{equation} \frac{e^2}{r} = \frac{m_e v^2}{r} \label{Bohr7} \end{equation} where $m_e$ is the mass of the electron. For simplicity, we assume that the proton mass is infinite so that the proton's position remains fixed (actually $m_p \approx 1836m_e$). The energy of the hydrogen atom is the sum of the kinetic and potential energies: \begin{equation} E = K + V = \frac{1}{2} m_e v^2 - \frac{e^2}{r} \label{Bohr8} \end{equation} Using (\ref{Bohr7}), we see that \begin{equation} K = - \frac{1}{2} V \quad {\rm and} \quad E = \frac{1}{2} V = - K \, .  \label{Bohr9} \end{equation} This is the form of the virial theorem for a force law varying as $r^{-2}$. Note that the energy of a bound atom is negative, since it is lower than the energy of the separated electron and proton, which is taken to be zero.  For further progress, we need some restriction on the possible values of $r$ or $v$. This is where we can introduce the quantization of angular momentum $\vec L = \vec r \times \vec p$. Since $\vec p$ is perpendicular to $\vec r$ (see Fig.~\ref{fig:bohr-atom}), we can write simply \begin{equation} L = rp = mvr \, .  \end{equation} Using (\ref{Bohr9}), we find also that \begin{equation} r = \frac{L^2}{me^2} \end{equation} We introduce angular momentum quantization, writing \begin{equation} L = n \hslash, \quad {\rm with} \quad n = 1,2,\cdots \end{equation} excluding $n=0$, since the electron would then not be in a circular orbit.  The allowed orbital radii are then given by \begin{equation} r_n = n^2 a_0 \label{Bohr13} \end{equation} where \begin{equation} a_0 \equiv \frac{\hslash^2}{m_e e^2} = 5.29 \times 10^{-11}~{\rm m} \simeq 0.529~\AA \,, \end{equation} which is known as the Bohr radius. The corresponding energy is \begin{equation} E_n = - \frac{e^2}{2 a_0 n^2} = - \frac{m_e e^4}{2 \hslash^2 n^2}, \quad n = 1,2\cdots \, .  \label{Bohr15} \end{equation}
 Balmer-Rydberg-Ritz formula  (\ref{Bohr1}) can now be deduced from the Bohr model. We have
\begin{equation}
\frac{hc}{\lambda} = E_{n_2} - E_{n_1} = \frac{2 \pi^2 m_ee^4}{h^2} \left(\frac{1}{n_1^2} - \frac{1}{n_2^2} \right)
\end{equation}
and the Rydbeg constant can be identified as
\begin{equation}
{\cal R} = \frac{2 \pi m_e e^4}{h^3 c} \approx 1.09737 \times 10^7~{\rm m^{-1}} \, .
\end{equation}
The slight discrepency with the experimental value for hydrogen is due to the finite proton mass. 
 
The Bohr model can be readily extended to hydrogen-like ions, systems in which a single electron orbits a nucleus of arbitrary atomic number $Z$. Thus $Z=1$ for hydrogen, $Z=2$ for He$^+$, $Z=3$ for Li$^{++}$, and so on. The Coulomb potential (\ref{Bohr5}) generalizes to
\begin{equation}
V(r) = - \frac{Ze^2}{r} \,,
\end{equation}
the radius of the orbit (\ref{Bohr13}) becomes
\begin{equation}
r_n = \frac{n^2 a_0}{Z} \,,
\end{equation}
and the energy (\ref{Bohr15}) becomes 
\begin{equation}
E_n = - \frac{Z^2 e^2}{2a_0 n^2} \, .
\end{equation}

\begin{figure}[tbp] \postscript{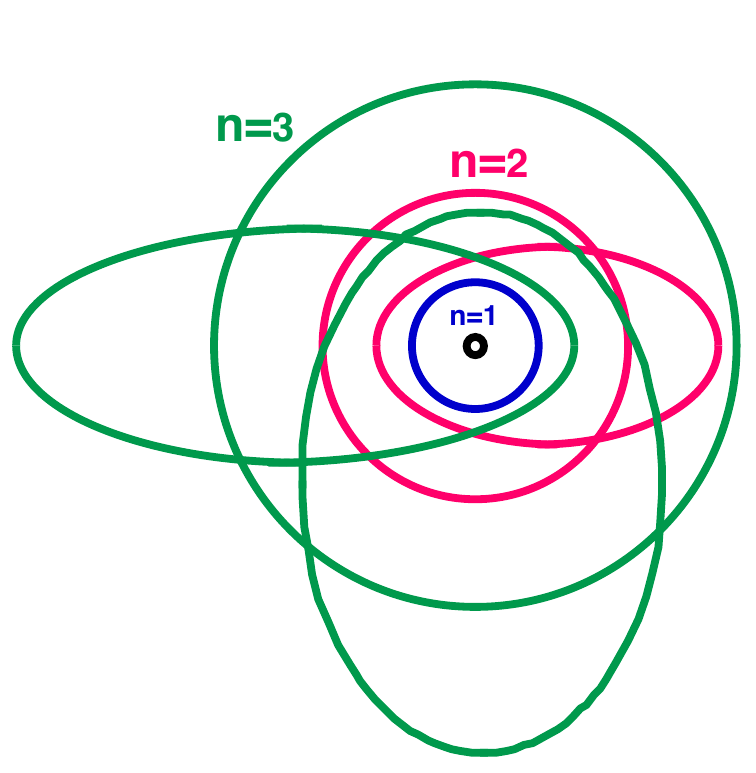}{0.85} 
\caption{Bohr-Sommerfeld-Wilson orbits for n = 1,2,3 (not to scale).}
\label{fig:Bohr-Sommerfeld}
\end{figure}

Wilson~\cite{Wilson} and Sommerfeld~\cite{Sommerfeld} generalized Bohr's formula for the allowed orbits to 
\begin{equation}
\oint p \ dr = nh, \quad {\rm with} \quad n = 1,2,\cdots .
\end{equation}
The Sommerfeld-Wilson quantum condition (23) reduce to Bohr's results for circular orbits, but allow, in addition, elliptical orbits along which the momentum $p$ is variable. According to Kepler's first law of planetary motion, the orbits of planets are ellipses with the Sun at one focus. Some of these generalized orbits are drawn schematically in Fig.~\ref{fig:Bohr-Sommerfeld}. The lowest energy state $n = 1$ is still a circular orbit. However, $n = 2$ allows an elliptical orbit in addition to the circular one; $n = 3$ has three possible orbits, and so on. The energy still depends on $n$ alone, so that the elliptical orbits represent degenerate states. 

The Bohr model was a major step in the development of quantum mechanics. It introduced the quantization of atomic energy levels (allowing transition between states $n_1$ and $n_2$ via absorption or emission of photons \cite{Einstein:1917zz}) and gave quantitative agreement with the atomic hydrogen spectrum. With the Sommerfeld-Wilson generalization, it accounted as well for the degeneracy of hydrogen energy levels. The Bohr model was able to sidestep the atomic catastrophe. However, as we will discuss next,  the assumption of well-defined electron orbits around a nucleus is completely contrary to the basic premises of quantum mechanics. Another flaw in the Bohr picture is that the angular momenta are all too large by one unit, for example, the ground state actually has zero orbital angular momentum (rather than $\hslash$).\\

{\bf EXERCISE 9.9}~According to general principles of classical electrodynamics, accelerated charged particles always radiate electromagnetic waves. This is the basic rule upon which all radiation sources are based. At the end of the last century Larmor calculated the total power radiated by an accelerated non-relativistic electron ($v \ll c$)~\cite{Larmor}. His well known result is (in Gaussian units)
\begin{equation}
P = \frac{2}{3} \frac{q^2 a^2}{c^3} \,,
\label{larmor257}
\end{equation}
where $a$ is the acceleration and $q$ is the charge; see Appendix~\ref{appC}.  In the Rutherford model of the hydrogen atom's ground state, the electron moves in a circular orbit of radius $a_0 = 0.529~\AA$ around the proton, which is assumed to be rigidly fixed in space. Since the electron is accelerating, a classical analysis suggests that it will continuously radiate energy, and therefore the radius of the orbit would shrink with time. Assuming that the electron is always in a nearly circular orbit and that the rate of radiation of energy is sufficiently well approximated by classical, nonrelativistic elec- trodynamics, how long is the fall time of the electron, i.e., the time for the electron to spiral into the origin?

\subsection{Wave-particle duality and uncertainty principle}

In view of the particle properties for light waves -- photons -- de Broglie ventured to consider the reverse phenomenon, he proposed to assign wave properties to matter,  which we will formulate here in the following way: To every particle with mass $m$ and momentum $\vec p$ we associate 
\begin{equation}
\lambda = h/ |\vec p| \,,
\label{deBroglie}
\end{equation}
called the de Broglie wavelength of the particle~\cite{deBroglie:1923,deBroglie:1970}. The above statement can be easily understood when assigning energy and momentum to matter in (reversed) analogy to photons
\begin{equation}
E = \hslash \omega \quad {\rm and} \quad |\vec p| = \hslash |\vec k| = h/\lambda \, .
\label{deBroglie-rel}
\end{equation}
The definitive evidence for the wave nature of light was deduced from the double-slit experiment discussed in Sec.~III~\cite{Anchordoqui:2015xca}. In principle, it should be possible to do double-slit experiments with particles and thereby directly observe their wavelike behavior. However, the technological difficulties of producing double slits for particles are formidable, and such experiments did not become possible until long after the time de Broglie hypothesized the wave-particle duality. The first double-slit experiment with electrons was done in 1961~\cite{Jonsson:1961}.

\begin{figure*}[tbp] \postscript{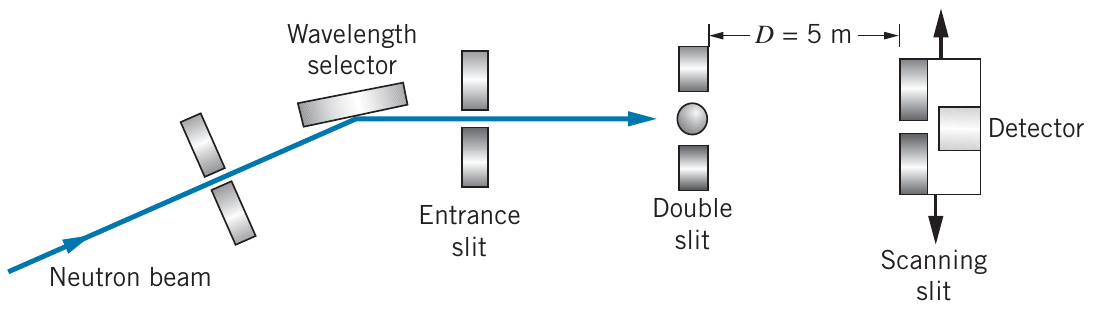}{0.95} \caption{Double-slit apparatus for neutrons.Thermal neutrons from a reactor are incident on a prisma which together with the entrance slit select the energy of the neutrons. After passing through the double slit, the neutrons are counted by the scanning slit assembly, which moves laterally~\cite{Krane}.}
\label{neutron-experiment}
\end{figure*}

Perhaps the best way to crystallize our ideas about the wave-particle duality is to consider the ``simple'' double-slit experiment for neutrons sketched in Fig.~\ref{neutron-experiment}.  
A parallel beam of neutrons falls on a double slit, which has individual openings much smaller than the center-to-center distance between the two slits $d$.  At a distance from the slits $D \gg d$ is an neutron detector capable of detecting individual neutrons. It is important to note that the detector always registers discrete particles localized in space and time. This can be achieved if the neutron source is weak enough. If the detector collects neutrons at different positions for a long enough time, a typical wave interference pattern for the counts per minute or probability of arrival of electrons is found. If one imagines a single neutron to produce in-phase ``wavelets'' at the slits, standard wave theory can be used to find the angular separation $\theta$ of the central probability maximum from its neighboring minimum.  As we discuss  now in detail, the similarity of the observed intensity pattern with the double-slit interference pattern for light is striking.

\begin{figure}[tbp] \postscript{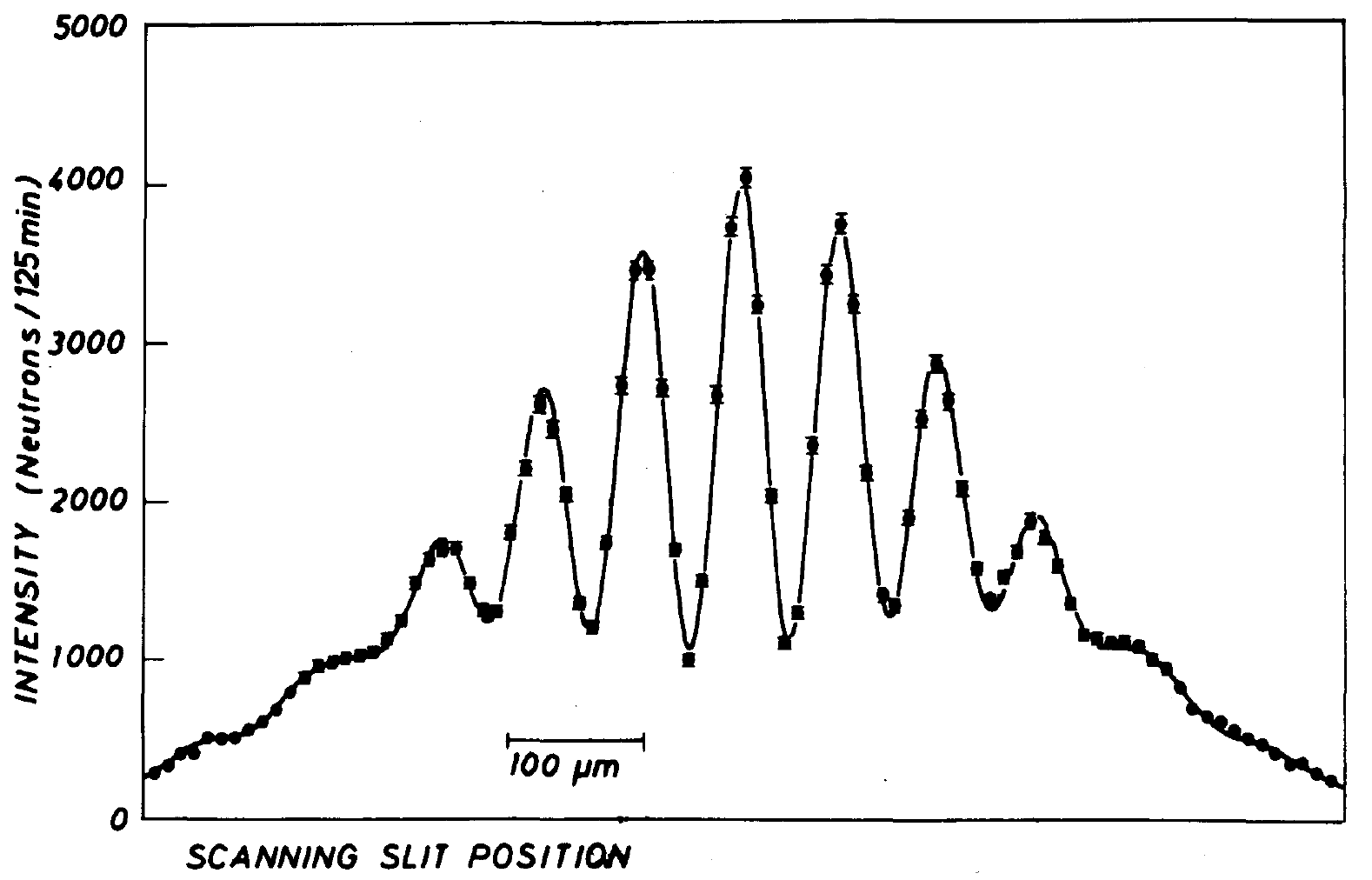}{0.9} \caption{Measured neutron distribution after inter at a double slit where a boron wire was used to define the two individual slits. The boron wire was opaque for the neutrons used in the experiment. The solid line represents the first-principles theoretical calculation.~\cite{Gahler}.}
\label{boron}
\end{figure}

In a nuclear reactor, the neutrons set free by the fission process have energies around 2~MeV, which corresponds to speeds around $20000~{\rm km/s}$. These neutrons are slowed down to a room-temperature ``thermal'' energy by collisions with the moderator nuclei. Their energy distribution is thus Maxwellian with a temperature around 300~K. Beams of neutrons with larger wavelengths are available at the Institute Laue-Langevin (ILL) high-flux reactor as emerging from a cold source. The ILL cold source neutron moderator is liquid deuterium at temperature $\approx 25~{\rm K}$  (average $K \approx kT \approx 2.5 \times 10^{-2}~{\rm eV}$). The neutron beam  from the source eneters a prism refraction monochromator system. Because of the dispersive properties of the prism medium (quartz glass), for neutrons the radiation emerging behind the prism is fanned out exhibiting a correlation between wavelength and direction very much analogous to the raninbow-colored light radiation field emerging from a glass prism. In contrast to light, the neutron refractive index is smaller than unity for most materials. As a consequence the prism orientation is reversed as compared to the customary light case. The 20-$\mu$m-wide optical bench entrance slit then selects neutrons of a specific wavelength out of the radiation field emerging from the prism. The selected neutron beam, with kinetic energy $2.4 \times 10^{-4}~{\rm eV}$ and de Broglie wavelength 1.85~nm, passed through a gap of diameter $148~\mu {\rm m}$ in a material that absorbs virtually all of the neutrons incident on it. In the center of the gap was a boron wire (also highly absorptive for neutrons) of diameter $104~\mu {\rm m}$. The neutrons could pass on either side of the wire through slits of width $22~\mu {\rm m}$.  (Actually, due to difficulties in mounting the boron wire exactly centered, the two resulting slits were slightly unequal, 21.5~$\mu$n and 22.3~$\mu$m.) The intensity of neutrons that pass through this double slit is observed by sliding another slit across the beam and measuring the intensity of neutrons passing through this ``scanning slit.''  In Fig.~\ref{boron} we show the resulting pattern of intensity maxima and minima, which leaves no doubt that interference is occurring and that the neutrons have a corresponding wave nature. The wavelength can be deduced from the slit separation using (26) to obtain the spacing between adjacent maxima, $y_{n+1} - y_n$. Estimating the spacing from Fig.~\ref{boron} to be about $75~\mu {\rm m}$, it follows that
\begin{equation}
\lambda = \frac{d \ (y_{n+1} - y_n)}{D} \simeq 1.89~{\rm nm} \, ,
\end{equation}
where we have taken $d \simeq 126~\mu{\rm m}$. This result agrees very well with the de Broglie wavelength selected for the neutron beam.

\begin{figure*}[tbp] \postscript{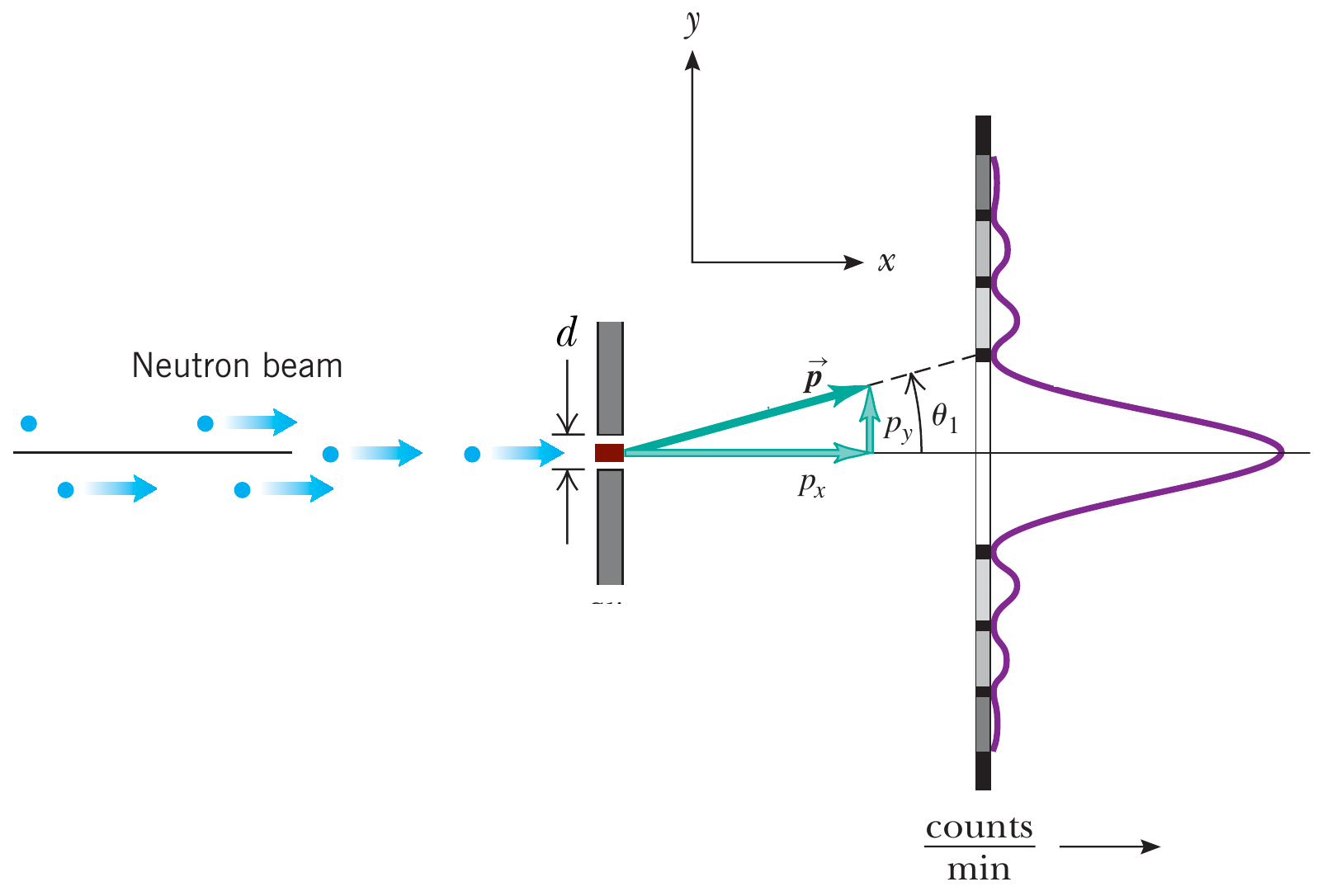}{0.9} \caption{Interpreting double-slit experiment in terms of neutron momentum~\cite{Young:2012}.}
\label{uncertainty-principle}
\end{figure*}

Furthermore, as first emphasized by Heisenberg, there are fundamental {\it uncertainties} in both the position and the momentum of an individual neutron, and these uncertainties are related inseparably~\cite{Heisenberg:1927zz}. To clarify this point, we adopt the reasoning introduced by Feynman~\cite{Feynman}.  We use $\theta_1$  to denote the angle between the central maximum and the first minimum, see Fig.~\ref{uncertainty-principle}. Using (26) with $m = 1$ , we find that 
\begin{equation}
\sin \theta_1 = \lambda/d \, .
\label{UMI312}
\end{equation}
A neutron that strikes the screen at the outer edge of the central maximum, at angle $\theta_1$, must have a component of momentum $p_y$ in the $y$-direction, as well as a component $p_x$ in the $x$-direction, despite the fact that initially the beam was directed along the $x$-axis. From the geometry of the situation the two components are related by $p_y/p_x = \tan \theta_1$. Since $\theta_1$ is small, we may use the approximation $\tan \theta_1 = \theta_1$, and 
\begin{equation}
p_y = p_x \, \theta_1 \, . 
\label{UMI313}
\end{equation} 
Substituting (\ref{UMI312}) into (\ref{UMI313}) gives
\begin{equation}
p_y = p_x \ \lambda/d \, .
\label{UMI314}
\end{equation}
Now, (\ref{UMI314}) says that for neutrons that strike the detector within the central maximum (that is, at angles between $-\lambda/d$ and $\lambda/d$), the $y$-component of momentum is spread out over a range from $-p_x \lambda/d$ to $+p_x \lambda/d$. Next, we consider all the neutrons that pass through the slits and strike the screen. Again, they may hit above or below the center of the pattern, so their component $p_y$ may be positive or negative. However, the symmetry of the interference pattern shows us the average value $\langle p_y \rangle = 0$. There will be an {\it uncertainty} $\Delta p_y$ in the $y$-component of momentum at least as great as $p_x \lambda/d$. That is,
\begin{equation}
\Delta p_y \geq p_x \ \lambda/d \, .
\label{UMI15}
\end{equation}
The narrower the separation between slits $d$, the broader is the interference pattern and the greater is the uncertainty in the $y$-component of momentum $p_y$.  The neutron wavelength $\lambda$ is related to the momentum $p_x$ by  (\ref{deBroglie}), which we can rewrite as $\lambda = h/p_x$. Substitutying this expression  into (\ref{UMI15}) and simplifying, we find
\begin{equation}
\Delta p_y \geq p_x \frac{h}{p_x d} = \frac{h}{d}
\label{UMI16}
\end{equation}
What does (\ref{UMI16}) mean? The 
center-to-center distance between the two slits $d \equiv \Delta y$ represents an uncertainty in the $y$-component of the position of a neutron as it passes through the double-slit gap. We don't know exactly where in the gap each neutron passes through. So both the $y$-position and the $y$-component of momentum have uncertainties, and the two uncertainties are related by (\ref{UMI16}), or equivalently
\begin{equation}
\Delta p_y \ \Delta y \geq h \, .
\label{dos66}
\end{equation}
We can reduce the momentum uncertainty $\Delta p_y$ only by reducing the width of the interference pattern. To do this, we have to increase the gap width $d$, which increases the position uncertainty $\Delta y$. Conversely, when we decrease the position uncertainty by narrowing the doubl-slit gap,  the interference  pattern broadens and the corresponding momentum uncertainty increases.\\

{\bf EXERCISE 9.10}~The neutrons produced in a reactor are known as thermal
neutrons, because their kinetic energies have been reduced
(by collisions) until $K = \frac{3}{2} kT$, where $T$ is room temperature (about 293~K). {\it (i)}~What is the kinetic energy of such neutrons? {\it (ii)}~What is their de Broglie wavelength? 

\section{Schr\"odinger Equation}
\label{sec:sch}

In the Fall of 1925 Schr\"odinger was invited by Debye to give a talk at a seminar in Zurich on de Broglie's thesis. During the discussion that followed, Debye commented that he thought this approach to wave-particle duality to be somewhat ``childish.'' After all, said Debye, ``to deal properly with waves one had to have a wave equation...''
Perhaps stimulated by this comment, Schr\"odinger left for holiday in the Swiss Alps just before Christmas 1925, and when he returned on 9 January 1926, he had discovered wave mechanics and the equation that governs the evolution of de Broglie waves~\cite{Schrodinger:1926a,Schrodinger:1926b,Schrodinger:1926c,Schrodinger:1926zz,Schrodinger:1926d}.

For non-relativistic quantum physics the basic equation to be solved is the Schr\"odinger equation. Like Newton's laws, the Schr\"odinger equation must be written down for a given situation of a quantum particle moving under the influence of some external forces, although it turns out to be easier to frame this in terms of potential energies instead of forces. However, unlike Newton's laws, the Schr\"odinger equation does not give the trajectory of a particle, but rather the wave function of the quantum system, which carries information about the wave nature of the particle, which allows us to only discuss the probability of finding the particle in different regions of space at a given moment in time. In this section, we introduce the Schr\"odinger equation, obtain solutions in a few situations, and learn how to interpret these solutions.

\subsection{Motivation and derivation}
\label{sMOTI}

It is not possible to derive the Schr\"odinger equation in any rigorous fashion from classical physics. However, it had to come from somewhere, and it is indeed possible to ``derive'' the Schr\"odinger equation using somewhat less rigorous means. If we first start by considering a particle with mass $m$, momentum $p_x$ moving in one dimension in a potential $V(x)$, we can express the total energy as
\begin{equation}
E = \frac{p_x^2}{2m} + V(x) \, ,
\label{Scho1}
\end{equation}
Multiplying both sides of (\ref{Scho1}) by the wave function $\psi (x, t)$ should not change the equality
\begin{equation}
E \ \psi(x,t) = \left[\frac{p_x^2}{2m} + V(x) \right] \ \psi(x,t) \, .
\label{Scho2}
\end{equation} 
Now, recall the de Broglie relations between energy and frequency as well as momentum and wave vector (\ref{deBroglie-rel}).
Assume the wave function in (\ref{Scho2}) is a plane wave traveling in the $x$
direction with a well defined energy and momentum, that is,
\begin{equation}
\psi (x,t) = A_0 e^{i(k_x x - \omega t)} \,,
\label{Scho5}
\end{equation}
where $p_x= \hslash k_x$ and $E = \hslash \omega$. If we now combine  (\ref{Scho2}) and (\ref{Scho5}), using de Broglie's relations, we obtain  
\begin{equation}
\hslash \omega A_0 e^{i(k_x x - \omega t) } = E A_0 e^{i(k_x x - \omega t)} 
\label{Scho6}
\end{equation}
and
\begin{eqnarray}
\left[\frac{\hslash^2 k_x^2}{2m} + V(x) \right]  A_0  e^{i(k_x x - \omega t)}  & = &  \left[\frac{p_x^2}{2m} + V(x) \right] \nonumber \\
& \times & A_0 e^{i(k_x x - \omega t)} \, .
\label{Scho7}
\end{eqnarray}
{}From (\ref{Scho6}) we see that for the equality to hold, the product of energy times the wave function $E\psi (x, t)$ must be equal to the first derivative of the wave function with respect to time multiplied by $i \hslash$, that is
\begin{equation}
E \psi (x,t) = i \hslash \frac{\partial }{\partial t} \psi (x,t) \, .
\label{Scho8}
\end{equation}
Similarly, by examining (\ref{Scho7}), we see that the product of momentum times the wave function $p_x \psi (x, t)$ must be equal to the first derivative of the wave function with respect to position $x$ multiplied by $-i \hslash$, that is
\begin{equation}
p_x \psi(x,t) = - i \hslash \frac{\partial}{\partial x} \psi (x,t) \, .
\label{Scho9}
\end{equation}
Combining (\ref{Scho2}), (\ref{Scho8}) and (\ref{Scho9}), we arrive at the time-dependent Schr\"odinger equation
\begin{equation}
i \hslash \frac{\partial}{\partial t} \psi(x,t) = \left[- \frac{\hslash^2}{2m} \frac{\partial^2}{\partial x^2} + V(x) \right] \psi (x,t) \, .
\label{Scho10}
\end{equation} 
This is a second-order linear differential equation. The term on the left-hand side of (\ref{Scho10}) represents the total energy of the particle. The first term on the right-hand side represents the kinetic energy of the particle, while the second term represents the potential energy of the particle. The time-dependent Schr\"odinger equation has three important properties: {\it (i)}~it is consistent with energy conservation; {\it (ii)}~it is linear and singular value, which implies that solutions can be constructed by superposition of two or more independent solutions; {\it (iii)}~the free-particle solution, $V(x) = 0$, is consistent with a single de Broglie wave.\footnote{An {\it eigenvalue} and {\it eigenvector} of a square matrix $A$ are a scalar $\lambda$ and a nonzero vector $\vec x$ so that $A \vec x = \lambda \vec x$.  A singular value and pair of singular vectors of a square or rectangular matrix $A$ are a nonnegative scalar $\sigma$ and two nonzero vectors $\vec u$ and $\vec v$ so that $A \vec v = \sigma \vec u$ and $A^\dagger \vec u = \sigma \vec v$. The superscript on $A^\dagger$ stands for Hermitian adjoint and denotes the complex conjugate transpose of a complex matrix. If the matrix is real, then $A^T$ denotes the same matrix.}

If the potential energy is independent of time, as we have written above, we can separate (\ref{Scho10}) into a time-independent form using the mathematical technique known as separation of variables. Here, we assume that our wave function can be written as a product of a temporal and spatial function
\begin{equation}
\psi(x,t) = \psi(x)  \ \chi(t) \, .
\label{Scho11}
\end{equation}
Substituting (\ref{Scho11}) into (\ref{Scho10}) we find 
\begin{equation}
i \hslash \ \psi(x) \ \dot \chi (t) = - \frac{\hslash^2}{2m} \chi (t) \ \left[ \psi"(x) + V(x) \ \psi(x) \right]
\end{equation}   
and so 
\begin{equation}
i \hslash \frac{\dot \chi (t)}{\chi(t)} = - \frac{\hslash^2}{2m}  \ \left[  \frac{\psi"(x)}{\psi(x)} + V(x) \right]
\end{equation}   
Since the left-hand side does not depend on $x$ while the right-hand side does not depend on $t$ it follows that
\begin{equation}
i \hslash \frac{\partial}{\partial t} \chi(t) = E \chi(t) = \hslash \omega \chi(t)
\label{Scho12}
\end{equation}
and
\begin{equation}
\left[-\frac{\hslash^2}{2m} \frac{\partial^2}{\partial x^2} + V(x) \right] \psi(x) = E \psi(x) \, ,
\label{Scho13}
\end{equation}
where $E$ is the separation constant. (\ref{Scho13}) is called the time-independent Schr\"odinger equation. The solution to (\ref{Scho12}) can be easily verified to be an oscillating complex exponential
\begin{equation}
\chi(t) = e^{-iEt/\hslash} = e^{-i\omega t} \, .
\end{equation}

The next steps involve solving (\ref{Scho13}) for a given potential energy $V(x)$. The techniques involved in solving this equation are similar regardless of the functional form of the potential and can thus be summarized in a set of steps. We assume that the potential $V(x)$ is known and we wish to determine the wave function $\psi (x)$ and its corresponding energy $E$ for that potential. This differential equation problem is known as an eigenvalue problem. There are only particular values of $E$ that satisfy the differential equation, which are called eigenvalues. We will not go into the general theory of solving such equations, but simply go through a few examples. However, before moving on to that, we note three further properties of the solutions of (\ref{Scho13}). {\it (i)~Continuity:} The solutions $\psi(x)$ to (\ref{Scho13}) and its first derivative $\psi'(x)$ must be continuous for all values of $x$ (the latter holds for finite potential $V(x)$).
{\it (ii)~Normalizable:} The solutions $\psi(x)$ to (\ref{Scho13}) must be square integrable, i.e. the integral of the modulus squared of the wave function over all space must be a finite constant so that the wave function can be normalized to give $\int |\psi(x)|^2 \, dx = 1$. {\it (iii)~Linearity:} Given two independent solutions $\psi_1(x)$ and $\psi_2(x)$, we can construct other solutions by taking an appropriate superposition of these $\psi (x) = \alpha_1 \, \psi_1(x)+ \alpha_2 \, \psi_2(x)$, where $\alpha_i \in \mathbb{C}$, satisfying $|\alpha_1|^2 + |\alpha_2|^2 = 1$ to ensure normalization.

\subsection{Expectation value, observables, and operators}

In quantum mechanics, a probability amplitude $\psi$ is a complex function  used to describe the behaviour of systems. The modulus squared of this quantity represents a probability or probability density. Probability amplitudes provide a relationship between the wave function (or, more generally, a quantum state vector) of a system and the results of observations of that system, a link first proposed by Born~\cite{Born:1926}.  Indeed, the wave function gives the probability density (probability per unit length in one dimension)
\begin{equation}
P(x) \, dx = |\psi(x)|^2 dx \, .
\end{equation}
This interpretation helps us understand the continuity constraint on the wave function. We do not want the probability of the particle to be zero at
a point $x$ and jump to a non-zero value infinitessimally close by. 
From this interpretation, we see that we can calculate the probability to find a particle between two points $x_1$ and $x_2$ from the wave function $\psi(x)$, 
\begin{equation}
P(x_1 < x < x_2) = \int_{x_1}^{x_2} |\psi (x)|^2  \, dx \, .
\end{equation}
Related to this is the concept of normalization of the wave function. We require that the particle must be found somewhere in space, and thus the probability to find the particle between $(-\infty,+\infty)$ should be equal to one, i.e.
\begin{equation}
\int_{-\infty}^{+\infty} |\psi (x)|^2 \, dx = 1 \, .
\end{equation}
\vspace{0.25cm}

{\bf EXERCISE 10.1}~Starting with Schr\"odinger equation (\ref{Scho10}) derive the one-dimensional {\it continuity equation}
\begin{equation}
\frac{\partial \rho}{\partial t} + \frac{\partial j}{\partial x} = 0 \,,
\label{exercise101}
\end{equation}
where
\begin{equation}
\rho \equiv |\psi (x,t)|^2 
\end{equation}
and
\begin{equation}
j \equiv - \frac{i\hslash}{2m} \left[\psi^* \frac{\partial \psi}{\partial x} - \frac{\partial \psi^*}{\partial x} \psi \right] \, .
\end{equation}
What does this equation imply in terms of conservation of probability?\\

Since we can no longer speak with certainty about the position of the particle, we can no longer guarantee the outcome of a single measurement of any physical quantity that depends on position. However, we can calculate the most probable outcome for a single measurement (also known as the expectation value), which is equivalent to the average outcome for many measurements. For example, suppose we wish to determine the expected location of a particle by measuring its $x$ coordinate. Performing a large number of measurements, we find the value $x_1$ a certain number of times $n_1$, $x_2$ a number of times $n_2$, etcetera, and in the usual way, we can calculate the average position
\begin{equation}
\langle x \rangle = \frac{n_1 x_1 + n_2 x_2 + \cdots}{n_1 + n_2 + \cdots} = \frac{\sum_i n_i x_i}{\sum_i n_i} \, .
\end{equation}
Here we use the notation $\langle x \rangle$ to represent the average value of the quantity within brackets. The number of times $n_i$ that we measure each position $x_i$ is proportional to the probability $P(x_i) \, dx$ to find the particle in the interval $dx$ at $x_i$. Making this substitution and changing sums to integrals, we have
\begin{equation}
\langle x \rangle = \frac{\int_{-\infty}^{+\infty} P(x) \, x \, dx }{\int_{-\infty}^{+\infty} P(x) \, dx} \ ,
\label{speed19}
\end{equation}
and thus
\begin{equation}
\langle x \rangle = \int_{-\infty}^{+\infty} x |\psi(x)|^2 \, dx \,,
\label{speed20}
\end{equation}
where in the last step we assume that the wave function is normalized so that the integral in the denominator of (\ref{speed19}) is equal to one. The expectation value of any function of $x$ can be found in a similar way, by replacing $x$ with $f(x)$ in (\ref{speed20})
\begin{equation}
\langle f(x) \rangle = \int_{-\infty}^{+\infty} f(x) |\psi (x)|^2 \ dx \, .
\label{speed21}
\end{equation}

As a common application of (\ref{speed21}) we can calculate the variance, denote by $\Delta x^2$ 
(or equivalently the standard deviation, which is the square root of the variance $\Delta x = \sqrt{\Delta x^2}$) in the position of a particle. The variance in position is given by
\begin{eqnarray}
\Delta x^2 & = & \langle (x - \langle x \rangle)^2 \rangle = \langle (x^2 - 2 x \langle x \rangle + \langle x \rangle^2) \rangle \nonumber \\  &= & \langle x^2 \rangle - \langle x \rangle^2 \,,
\end{eqnarray}
where we used the fact that $\langle x \langle x\rangle \rangle = \langle x \rangle^2$. Hence, we see that we can express the variance as
\begin{equation}
\Delta x^2 = \int_{-\infty}^{+\infty} x^2 |\psi (x)|^2 \, dx - \left(\int_{-\infty}^{+\infty} x |\psi (x)|^2 \, dx \right)^2 \, . \nonumber
\end{equation}

In the {\it parlons} of mathematics, square integrable functions (such as wave-functions) are said to form  an infinite dimensional vector space, much like the familiar $n$-dimensional vector spaces. In the Dirac notation, a state vector or wave-function, $\psi$, is represented as a ``ket'', $|\psi \rangle$. Just as we can express any $n$-dimensional vector in terms of the basis vectors,  so we can expand any wave function as a superposition of basis state vectors,
\begin{equation}
|\psi \rangle = \lambda_1 |\psi_1 \rangle + \lambda_2 |\psi_2 \rangle + \cdots \, .
\end{equation}
Alongside the ket, we can define the ``bra'', $\langle \psi|$. Together, the bra and ket
define the {\it scalar product}
\begin{equation}
\langle \phi |\psi \rangle \equiv \int_{-\infty}^{+\infty} dx \ \phi^*(x) \ \psi (x) \,,
\end{equation}
from which follows the identity, $\langle \phi|\psi \rangle^* = \langle \psi | \phi \rangle$. As with an $n$-dimensional vector,  the magnitude of the scalar product is limited by the magnitude of the vectors
\begin{equation}
\langle \psi | \phi \rangle \leq \sqrt{\langle \psi | \psi \rangle \langle \phi | \phi \rangle} \,,
\end{equation}
a relation known as the Schwartz inequality.

An operator $\hat A$ is a ``mathematical object'' that maps one state vector, $|\psi \rangle$, into another, $|\phi \rangle$, i.e. $\hat A|\psi \rangle = |\phi \rangle$. If $\hat A |\psi \rangle = a | \psi \rangle$, 
with $a$ real, then $|\psi \rangle$ is said to be an eigenstate (or eigenfunction) of $\hat A$ with eigenvalue $a$.  An observable is any particle property that can be measured. The position,  momentum, and energy of a particle are observables.\footnote{By contrast, the wavefunction $\psi$ although clearly indispensable to the quantum description, is not directly measurable and so is not an observable.} In quantum mechanics,  
for any observable $A$, there is an operator $\hat A$ which
acts on the wave function so that, if a system is in a state described by $|\psi \rangle$, the expectation value of $A$ is
\begin{equation}
\langle A \rangle = \langle \psi |\hat A | \psi \rangle = \int_{-\infty}^{+ \infty} dx \psi^* (x) \hat A \psi (x) \, .
\label{speedB37}
\end{equation}
An operator $\hat A$ is said to be {\it linear} if $\hat A(c \psi(x)) = c \hat A \psi (x)$ and $\hat A (\psi(x) + \phi(x)) = \hat A \psi(x) + \hat A \phi(x)$, where $\psi(x)$ and $\phi(x)$ are any two functions on the domain of $\hat A$ and $c$ is a complex number.  It is straightforward to check by explicit calculation that the operators $\hat x$ and $\hat p = - i\hslash \, \partial/\partial x$ are linear operators.
The operator $A^\dagger$ is called the {\it hermitian conjugate} (or adjoint) of $\hat A$ if
\begin{equation}
\int_{-\infty}^{+\infty} (\hat A^\dagger \phi)^* \psi dx = \int_{-\infty}^{+\infty} \phi^* \hat A \psi \, dx \,;
\label{operation1}
\end{equation}
in bra-ket notation (\ref{operation1}) becomes $\langle A^\dagger \phi| \psi \rangle = \langle \phi | A \psi \rangle$.
Its easy to show that $(c \hat A)^\dagger = c^* \hat A^\dagger$ and $(\hat A + \hat B)^\dagger = \hat A^\dagger  + \hat B^\dagger$ just from the properties of the dot product.\\ 

{\bf EXERCISE 10.2} Using  (\ref{operation1})  show that $(\hat A^\dagger)^\dagger = \hat A$ and $(\hat A \hat B)^\dagger = \hat B^\dagger \hat A ^\dagger$.\\

The operator $\hat A$ is called hermitian if $\hat A^\dagger = \hat A$, i.e.
\begin{equation}
\int_{-\infty}^{+\infty} (\hat A \phi)^* \psi dx = \int_{-\infty}^{+\infty} \phi^* \hat A \psi \, dx \,;
\label{operation2}
\end{equation}
in bra-ket notation (\ref{operation2}) becomes $\langle A \phi| \psi \rangle = \langle \phi | A \psi \rangle$. \\

{\bf EXERCISE 10.3}~Convince yourself that $\hat x$ and $\hat p = -i \hslash \partial /\partial x$ are hermitian operators. \\

{\bf EXERCISE 10.4}~A physical variable must have real expectation values (and eigenvalues). By computing the complex conjugate of the expectation value of a physical variable  
  show that  every operator corresponding to an observable is hermitian.\\

Note that operators are associative but not (in general) commutative,
\begin{equation}
\hat A \hat B |\psi \rangle = \hat A (\hat B | \psi \rangle ) = (\hat A \hat B) |\psi \rangle \neq \hat B \hat A |\psi \rangle \, .
\end{equation}
For example, take $\hat A = \hat x$ and $\hat B = \hat p$, then we have 
\begin{equation}
(\hat x \hat p - \hat p \hat x) \psi (x) = - i \hslash \left\{x \frac{\partial \psi}{\partial x} - \frac{\partial}{\partial x} [x \psi (x) ]\right \}
\end{equation}
and hence by the product rule of differentiation:
\begin{equation}
(\hat x \hat p - \hat p \hat x) \psi (x) = i \hslash \psi (x) \, .
\end{equation}
and since this must hold for any differentiable function $\psi (x)$, we can write this as an operator equation:
\begin{equation}
\hat x \hat p - \hat p  \hat x = i \hslash
\end{equation}
Therefore, we have shown that the operator product of $\hat x$ and $\hat p$ is non-commuting.
Because combinations of operators of the form $\hat A \hat B - \hat B \hat A$ do frequently arise in quantum mechanical  calculations, it is customary to use a short-hand notation: $[\hat A, \hat B] \equiv \hat A \hat B - \hat B \hat A$ and this is called the commutator of $\hat A$ and $\hat B$ (in that order!). If $[\hat A, \hat B] \neq 0$, then one says that $\hat A$ and $\hat B$ do not commute,
if $[\hat A, \hat B] =0$, then $\hat A$ and $\hat B$ are said to commute with each other. An operator equation of the form of \mbox{$[\hat A, \hat B] =$ something} is called a commutation relation; e.g. $[\hat x , \hat p] = i \hslash$.\\

{\bf EXERCISE 10.5}~Let $\hat A$ and $\hat B$ be Hermitian operators {\it (i)}~Show that $\hat A \hat B = [\hat A, \hat B]/2 + \{\hat A, \hat B \}/2$ where $\{\hat A, \hat B\} = \hat A \hat B + \hat B \hat A$ is the anticommutator. {\it (ii)}~Show that the anticommutator is Hermitian and the commutator is anti- Hermitian (that is, $[\hat A, \hat B]^\dagger = -[\hat A, \hat B]$). {\it (iv)}~We know that expectation values of Hermitian operators are real. What can you say about the expectation value of an anti-Hermitian operator?\\

{\bf EXERCISE 10.6}~We have seen that  the Heisenberg uncertainty principle~\cite{Heisenberg:1927zz}  captures the difference between classical and quantum states, and sets a limit on the precision of incompatible quantum measurements. It has been introduced in the early days of quantum mechanics, but its form has evolved with the understanding and formulation of quantum physics throughout the years. In particular,  Robertson~\cite{Robertson:1929zz} and  Schr\"odinger~\cite{Schrodinger:1930ty} presented a lower bound for the product of the dispersion of two non-commuting observables. This lower bound is more general than the one given in (\ref{dos66}). Let $A$ and $B$ be Hermitian operators. Define the ``uncertainty'' in $A$ by the square root of the mean square deviation from the mean: 
\begin{equation}
\Delta A = \sqrt{\langle (\hat A - \langle A \rangle)^2}
\end{equation}
Show that
\begin{equation}
\Delta A \Delta B \geq \tfrac{1}{2} |\langle [\hat A, \hat B]\rangle | \, .
\label{speedHP}
\end{equation}
Recall that the commutator of $\hat x$ and $\hat p_x$ is $i \hslash$, so from (\ref{speedHP}) we have  $\Delta x \,  \Delta p_x \geq \hslash/2$. [{\it Hint}: Use the Schwarz inequality and the results from exercise 10.1.]\\

{\bf EXERCISE 10.7}~Show that $\Delta E \, \Delta t \geq \hslash/2$, where $\Delta t$ is the shortest time, during which the average value of a certain quantity is changed by an amount equal to the standard deviation of this quantity~\cite{Mandelstam}.\\

{\bf EXERCISE 10.8}~It’s often the case that we want to find the “component” of a function “parallel” to another function. We just take the dot product with the second function, but then we also need to multiply by the second function. A handy notation is $|\psi \rangle \langle \psi |$. This {\it projector operator} projects onto $\psi$. Operating on $|\phi\rangle$, we get $|\psi\rangle \langle \psi |\phi \rangle$, which is what we want.
 Remember $\langle \psi |\phi \rangle$ is just a number and $|\psi \rangle$ is the vector. Similarly,
operating on $\langle \phi|$ we get $\langle \phi |\psi \rangle \langle \psi|$ which is the desired expression for the adjoint vector. Suppose you have a complete set of orthonormal basis vectors $|\psi_n\rangle$. What is a compact expression for transforming an arbitrary vector $|\phi \rangle$ into this basis set? (This is much easier to write down than to ask!)

\subsection{Free particle solution}

A ``free'' particle refers to a particle that has no external forces acting upon it, in other words the potential energy is constant $V_0$. The state of such a free particle is represented by its wave function $\psi(x)$. Starting with (\ref{Scho13}), and proposing a solution of the form (this is known as an educated-guess)
\begin{equation}
\psi (x) = A e^{ikx}
\end{equation}
we find four possible solutions of the Schr\"odinger equation that satisfy
\begin{eqnarray}
\frac{2m}{\hslash^2} (E - V_0) \psi (x) & = & - \frac{\partial^2}{\partial x^2} \psi (x) \nonumber \\
& = & k^2 \psi (x) \, .
\end{eqnarray}
The values $\pm k$ can take on real or imaginary values depending on the particle's energy and the potential
\begin{equation}
k = \pm \frac{1}{\hslash} \sqrt{2m (E - V_0)}, \quad (E > V_0),
\end{equation}
with corresponding wave functions of the form
\begin{equation}
\psi (x) = A e^{ikx} + B e^{-ikx} 
\end{equation}
which represent travelling waves solutions, and
\begin{equation}
i \kappa = \pm i \frac{1}{\hslash} \sqrt{2m (V_0 - E)}, \quad (E < V_0),
\end{equation}
with corresponding wave functions of the form
\begin{equation}
\psi (x) = A e^{ \kappa x} + Be^{- \kappa x}, 
\end{equation}
which represent exponentially decaying solutions. The allowed energies are given by
\begin{equation}
E = \frac{\hslash^2 k^2}{2m} + V_0 \, .
\end{equation}

The case in which $E > V_0$ is classically allowed, whereas the situation in which $E < V_0$ is classically forbidden. To understand this, imagine our particle rolling on a potential surface described by $V(x)$. If it has total energy $E$, it can only exist in a region of space in which $V(x) < E$, and once $V(x) \geq E$, the particle must turnaround (this is the classical turning point). However, in quantum physics, the particle has a non-zero probability to be found in this classically forbidden region. We will see how this manifests itself in another section to allow quantum tunneling, in which a particle can penetrate a barrier and emerge on the other side of the barrier.

For the traveling wave solutions, consider the time evolution of the probability density, $P (x, t)$, given by
\begin{eqnarray}
P (x, t) & = &  \psi^*(x, t) \psi (x, t) = \psi^*(x) e^{i\omega t} \psi (x) e^{-i\omega t} \nonumber \\ & = & \psi^*(x) \psi (x) \, .
\end{eqnarray}
This is independent of time. If we consider a particle traveling in only one
direction, say the $+x$ direction, then the probability density is
\begin{equation}
P (x, t) = \psi^*(x) \psi (x) = A^* e^{-ikx} A e^{ikx} = A^*A, 
\end{equation}
which is independent of position. This implies that the particle is equally likely to be anywhere in space. It is completely delocalized. For a superposition of both positive and negative going waves, we have
\begin{eqnarray}
P(x,t) & = & \left(A e^{ikx} + B^{-ikx} \right)^* (A e^{ikx} + B e^{-ikx}) \nonumber \\
& = & A^* A + B^* B + 2 \Re \{A^* B e^{-2ikx} + B^* A e^{2ikx} \} \,, \nonumber
\end{eqnarray}
where $\Re(z)$ gives the real value of $z$. For real-valued coefficients $A$ and $B$, this simplifies to
\begin{equation}
P (x, t) = A^2 + B^2 + 2AB \cos(2kx) \, . 
\end{equation}
This is the equation for a standing wave.

\subsection{Step potential}
\label{Sec54}

In this section we examine the behavior of a particle initially traveling in a region of space of constant potential suddenly moves into a region of different, but also constant potential. At $x = 0$, we have the transition between potential $V_1$ and $V_2$. The potential can thus be expressed as a piecewise function
\begin{equation}
V(x) = \left\{\begin{array}{cc} V_1 & ~~~{\rm for} \, x<0 \\
V_2 & ~~~{\rm for} \, x \geq 0 \end{array} \right. \ .
\end{equation}
The particle has fixed energy $E$. There are two situations of interest, first when $E>V_1$ and $E>V_2$ and second when $E>V_1$ and $E<V_2$.

\subsubsection{ Case1: $E>V_1$  and $E>V_2$}

Here we break the problem down into two sections and consider a piece- wise wave function defined on either side of the step. For region 1, $x < 0$, 
the wave function is given by
\begin{equation}
\psi_1 (x) = A e^{ik_1 x} + B e^{-ik_1x} \,,
\end{equation}
where $A$ and $B$ are coefficients to be determined, and the wave vector for region 1 is
\begin{equation}
k_1 = \sqrt{2m (E- V_1)}/\hslash \, .
\end{equation}
Similarly, in region 2 the wave function is given by
\begin{equation}
\psi_2 (x) = C e^{ik_2 x} + D e^{-ik_2 x} \,,
\end{equation}
where $C$ and $D$ are coefficients to be determined, and the wave vector for region 2 is
\begin{equation}
k_2 = \sqrt{2m (E-V_2)}/\hslash \, .
\end{equation}
The ratio of wave vectors is thus given by
\begin{equation}
\frac{k_2}{k_1} = \sqrt{\frac{1-V_2/E}{1- V_1/E}} \, .
\end{equation}
We can set $D = 0$ since we assume the particle initially comes from the $-x$ direction. The $A$ coefficient corresponds to the incident wave, while the $B$ is related to the reflected wave. Now, to determine the remaining coefficients, we use the continuity of the wave function and its first derivative at the origin to give
\begin{equation}
\psi_1 (0) = \psi _2(0) \Rightarrow A+B =C \,,
\end{equation} 
and
\begin{equation}
\psi'(0) = \psi'_2 (0) \Rightarrow ik_1 (A-B) = ik_2 C \,,
\end{equation}
where the prime implies the first derivative with respect to $x$. Combining
these and eliminating $C$, we find the ratio of $B/A$
\begin{equation}
\frac{B}{A} = \frac{k_1 - k_2}{k_1 + k_2} = \frac{1- k_2/k_1}{1 + k_2/k_1} \, ,
\end{equation}
which is the reflection coefficient of the barrier (the ``coefficient'' corresponds to the ratio of amplitudes). The reflectivity of the barrier, corresponding to the ratio of probabilities, or flux, is thus given by
\begin{equation}
R = \left|\frac{B}{A} \right|^2 = \left|\frac{1 - k_2/k_1}{1 + k_2/k_1}\right|^2 \, .
\label{speed44}
\end{equation}
Due to the conservation of particle number (or probability depending on how you want to think about the wave function), the transmissivity is simply given by
\begin{equation}
T = 1 - R = 1 - \left|\frac{1-k_2/k_1}{1+k_2/k_1} \right|^2 \, .
\end{equation}
Note that in going from region 1 to region 2, the de Broglie wavelength of the particle with energy $E$ changes and becomes longer for an increased potential step. The specific values of $A$ and $B$ are typically determined by considering the normalization of the wave function. However, since plane wave solutions are infinite in extent, they are not normalizable.

\subsubsection{Case2: $E>V_1$ and $E<V_2$}

In this case, we follow a similar approach to above. However, in region 2, we now have decaying solutions since $E < V_2$. Again we have the following wave function in region 1
\begin{equation}
\psi_1(x) = A e^{ik_1 x} + B e^{-ik_1 x} \,,
\end{equation}
where $A$ and $B$ are coefficients to be determined, and the wave vector for
region 1 is
\begin{equation}
k_1 = \sqrt{2m (E- V_1)}/\hslash \, .
\end{equation}
In region 2 the wave function is now given by exponentials
\begin{equation}
\psi_2 (x) = C e^{\kappa_2 x} + D e^{-\kappa_2 x} \,,
\end{equation}
where $C$ and $D$ are coefficients to be determined, and $\kappa_2$ is given by
\begin{equation}
\kappa_2 = \sqrt{2m(V_2 - E)} / \hslash \, .
\end{equation}
We can set $C = 0$ since we cannot have the probability amplitude growing infinitely large as $x \to \infty$. The $A$ coefficient corresponds to the incident wave, while the $B$ is related to the reflected wave. Now, to determine the remaining coefficients, we use the continuity of the wave function and its first derivative at the origin to give
\begin{equation}
\psi_1(0) = \psi_2(0) \Rightarrow A+B = D \,,
\end{equation}
and
\begin{equation}
\psi'_1 (0) = \psi'_2(0) \Rightarrow ik_1 (A-B) = -\kappa_2 D \, .
\end{equation}
Combining these and eliminating $D$, we find the ratio of $B/A$
\begin{equation}
\frac{B}{A} = \frac{k_1 - i \kappa_2}{k_1 + i \kappa_2} \,,
\label{speed52}
\end{equation}
which is the reflection coefficient of the barrier. The reflectivity of the barrier is thus given by
\begin{equation}
R = \left|\frac{B}{A} \right|^2 = \left(\frac{k_1 - i \kappa_2}{k_1 + i \kappa_2} \right) \left(\frac{k_1 + i \kappa_2}{k_1 - i \kappa_2} \right) = 1 \, .
\end{equation}
Thus, we see that even though the particle has non-zero probability to penetrate into the classically forbidden region, it will always be reflected (eventually). We could have also obtained the same result in (\ref{speed52}) from (\ref{speed44}) by allowing $k_2 = -i\kappa$. By allowing $k_1$ and $k_2$ to take on complex values, we do not have to consider multiple cases.

Note that the depth at which the particle penetrates into the classically forbidden region is given by the distance from $x = 0$ at which the probability drops by $1/e$,
\begin{equation}
P(\Delta x) = e^{-2 \kappa_2 \Delta x} = e^{-1} \,,
\end{equation}
which gives
\begin{equation}
\Delta x = \frac{1}{2 \kappa_2} = \frac{1}{2} \frac{\hslash}{\sqrt{2m (V_2 - E)}} \, .
\label{speed55}
\end{equation}

\subsection{Potential barrier and tunneling}

Another useful example is the scattering of a particle from a potential barrier of width $L$ and height $V_0$. The functional form of the potential can be expressed as a piecewise function
\begin{equation}
V(x) = \left\{\begin{array}{cl} V_0 & ~~~{\rm for} \, -L/2 <x < L/2 \\
0 & ~~~{\rm otherwise} 
\end{array}
\right. . 
\end{equation}
Again, there are two types of behavior that occur. The first, in which the particle energy is greater than the barrier $E > V_0$, in which the particle will have some reflection and some transmission as expected classically. The second situation, in which the particle energy is less than the barrier, is of more interest. Here, the classical prediction would be that the particle should be reflected completely and there is no transmission. There should be zero probability to find the particle on the right hand side of the barrier. However, due to the wave nature of quantum systems, there is a finite, non-zero probability to find the particle on the right-hand side of the barrier, and it will continue to propagate to $+ \infty$. This simple model is the precursor to discussing the decay of atomic nuclei as well as other quantum tunneling effects such as those associated with a scanning electron microscope.

On the left-hand and right-hand sides of the barrier, regions 1 and 3, the
wave vector is given by
\begin{equation}
k = \sqrt{2 m E}/\hslash \,,
\end{equation}
while within the barrier, the decay is governed by
\begin{equation}
\kappa = \sqrt{2m (V_0 - E)}/\hslash \, .
\end{equation}
Hence, we can  express the wave function in each region in terms of either plane waves or decaying exponentials with unknown coefficients to be determined
\begin{eqnarray}
\psi_1 (x) & = & A e^{ikx} + B e^{-ikx} \,, \\
\psi_2(x) & = & C e^{\kappa x} + D e^{-\kappa x} \,, \\
\psi_3(x) & = & F e^{ikx} + G e^{ikx} \, .
\end{eqnarray}
Assuming that the particle initially starts on the left-hand side of the barrier (i.e. $x \leq -L/2$), then we can set $G = 0$, since there is no way to obtain a solution on the right-hand side propagating in the negative $x$ direction. Next we apply the boundary conditions on the wave function at the edges of the barrier, which leads to the following four equations
\begin{eqnarray}
e^{-ikL/2} + \frac{B}{A} e^{ikL/2} &= & \frac{C}{A} e^{-\kappa L/2} + \frac{D}{A} e^{\kappa L/2} \nonumber \\
ik \left( e^{-ikL/2} - \frac{B}{A} e^{ikL/2} \right) & = & \kappa \left(\frac{C}{A} e^{-\kappa L/2} - \frac{D}{A} e^{\kappa L/2} \right), \nonumber \\
ik \left(\frac{F}{A} e^{ikL/2}\right) &= &\kappa \left(\frac{C}{A} e^{\kappa L/2} - \frac{D}{A} e^{-\kappa L/2} \right), \nonumber \\
\frac{F}{A} e^{ikL/2} & = & \frac{C}{A} e^{\kappa L/2} + \frac{D}{A} e^{-\kappa L/2} \,,  
\label{speed65}
\end{eqnarray}
where we have divided through by $A$, since we want to solve for the transmission and reflection coefficients given by the ratios of $F/A$ and $B/A$ respectively. By combining the last two equations in (\ref{speed65}), we can solve for $C/A$ and $D/A$ in terms of $F/A$. For example, $1/\kappa$ times the 3rd equation added to the 4th  gives
\begin{equation}
\frac{C}{A} = \frac{F}{A} \frac{e^{ikL/2} e^{-\kappa L/2}}{ 2\kappa} (\kappa + ik) \,,
\end{equation}
and subtracting $1/\kappa$ times the 3rd equation from 4th gives
\begin{equation}
\frac{D}{A} = \frac{F}{A} \frac{e^{ikL/2} e^{\kappa L/2}}{2 \kappa} ( \kappa - ik) \, .
\end{equation}
We can now solve for the transmission and reflection coefficients by combining the first two equations in (\ref{speed65}) and substituting in for $C/A$ and $D/A$ to give
\begin{eqnarray}
\frac{F}{A} & = & \frac{4 i k \kappa e^{-ikL}}{(\kappa + ik)^2 e^{-\kappa L} - (\kappa - ik)^2 e^{\kappa L}} \, \nonumber \\
\frac{B}{A} & = & \frac{F}{A} \frac{\kappa^2 + k^2}{2 i k \kappa} \cosh (\kappa L) \, .
\end{eqnarray}
The transmittance $T = |F/A|^2$ is thus given by
\begin{eqnarray}
T \!\! \! & = & \!\!\! \left|\frac{F}{A} \right|^2 \nonumber \\
& = & \!\! \!\frac{8 k^2 \kappa^2} {(k^2 + \kappa^2)^2 \cosh (2 \kappa L) - (k^4 - \kappa^4 - 6 k^2 \kappa^2)} .
\end{eqnarray}
In the limit of weak transmission, i.e. $\kappa L \ll 1$, the transmittance can be approximated by setting $\cosh(2 \kappa L) \approx e^{2 \kappa L/2}$, and dropping the smaller terms in the denominator, leading to
\begin{equation}
T \approx \frac{16 k^2 \kappa^2}{(k^2 + \kappa^2)^2} e^{-2 \kappa L} \, .
\end{equation}

In the last few sections we found the allowed solutions to the Schr\"odinger equation for a particle of mass $m$ traveling in free space, and through different potentials -- a step and a barrier. We found that the allowed energy and momentum the particle may take on have a continuous range of values, which is in line with our classical notions of energy and momentum for a particle. A key signature of non-classical behavior is the non-zero probability to find the particle in a region of space where it is classically forbidden to go, in which its total energy is less than the potential in that region, i.e. $E < V_0$. This quantum penetration is due to the wave nature of quantum objects and can be shown to be consistent with the uncertainty principle, in which the particle is allowed to penetrate the potential for a finite period of time $\Delta t$, which is on the order of $\hslash \Delta E$ where $\Delta E = V_0 -E$ is the difference between the total particle energy and the potential energy of the barrier. The depth into the barrier can be approximated by multiplying  this allowed time by the ``particle velocity'' which we approximate as $v = p/m = \sqrt{2m (V_0 - E)}/m$. The reason to take $p = \sqrt{􏰅2m(V_0 - E)}$ as the value of the momentum is that $K = V_0 - E$ is the maximum value of kinetic energy the particle could have and still not propagate in the potential barrier region, it is also the only relevant energy quantity by dimensional analysis. This gives us a penetration depth from the uncertainty principle of
\begin{eqnarray}
\Delta x & = &  \frac{1}{2} v \Delta t = \frac{1}{2} \frac{\hslash}{V_0- E} \frac{\sqrt{2m(V_0 -E)}}{m} \nonumber \\
& = & \frac{\hslash}{\sqrt{2m (V_0 - E)}} \,,
\end{eqnarray}
where the factor of $1/2$ arises from the fact that the particle must travel into the forbidden region and return in the time $\Delta t$. This is consistent with the result we obtained in (\ref{speed55}) (to within a factor of two), and thus in good agreement.

\subsection{Particle in a box}

The final example that we will look at is that of a particle confined to an infinite potential well. Here a particle of mass $m$ is trapped (or bound) to the well that has a width $L$, and finite potential $V_0$ inside the infinite walls. We want to determine what the allowed energies and wave functions that the particle can have. We choose to center the well (or potential ``box'') on the origin of our coordinate axes, so that the walls of the well are located at $x = -L/2$ and $x = +L/2$. The potential can thus be written as
\begin{equation}
V(x) = \left\{ \begin{array}{cl} \infty & ~~~{\rm for} \, x < L/2 \\
V_0 & ~~~{\rm for} \, -L/2 \leq x \leq L/2 \\
\infty & ~~~{\rm for} \, x > L/2 
\end{array} \right. .\end{equation}
We again break up the problem into different regions of space as dictated by the boundaries of the potential. In regions 1 and 3 outside the box where the potential is infinite, we must have
\begin{equation}
\psi (x) = 0 , \quad x<-L/2 \wedge x > L/2 \,,
\end{equation}
so that the Schr\"odinger equation can be satisfied (otherwise the term $V(x)\psi (x)$ would be infinite, which makes no sense). Inside the box, we have a 1-dimensional plane wave again, similar to the situation of the finite potential in Sec.~\ref{Sec54},
\begin{equation}
\psi (x) = A e^{ikx} + B e^{-ikx}, \quad -L/2 \leq x \leq L/2 \,,
\end{equation}
where the magnitude of the wave vector comes from the Schr\"odinger equation, in which the energy is
\begin{equation}
E = \frac{\hslash^2 k^2}{2m} + V_0 \,,􏰔
\end{equation}
￼which gives the square of the wave vector
\begin{equation}
k^2 = \frac{2m (E - V_0)}{\hslash^2} \,,
\end{equation}
while the expansion coefficients $A$ and $B$, and the allowed energy values $E$
are still to be determined. These unknown values of $A,\, B$ and $E$ are set by the initial conditions of the system and the boundary conditions on the wave function respectively. The boundary conditions for the wave function at the left-hand side of the box, $x = -L/2$ implies that
\begin{equation}
\psi (-L/2) = A e^{-ikL/2} + B e^{ikL/2} = 0 \,,
\label{speed78}
\end{equation}
and similarly on the right hand side of the box $x = L/2$
\begin{equation}
\psi (L/2) = A e^{ikL/2} + B e^{-ikL/2} = 0 \,,
\label{speed79}
\end{equation}
Adding (\ref{speed78}) to (\ref{speed79}) gives
\begin{equation}
2 (A+B ) \cos (kL/2) = 0 
\label{speed80}
\end{equation}
while subtracting (\ref{speed78}) from (\ref{speed79}) gives
\begin{equation}
2 i (A-B ) \sin (kL/2) = 0 
\label{speed81}
\end{equation}
Both conditions in (\ref{speed80}) and (\ref{speed81}) must be met. There are two cases we can consider. First, when $A = B$ (\ref{speed81}) is met, and to satisfy (\ref{speed80}), we see that the wave vector can only take on discrete values
\begin{equation}
k = \frac{2 \pi n_1}{L} + \frac{\pi}{L} \, ,
\label{speed83}
\end{equation}
where $n_1 = 0, 1, 2, 3, \cdots$. The second case is when $A = -B$ in which (\ref{speed80}) is met, and to satisfy  (\ref{speed81}) the wave vector can only take on discrete values
\begin{equation}
k = \frac{2 \pi n_2}{L} \,,
\label{speed82}
\end{equation}
where $n_2 = 1,2,3, \cdots$. We can consolidate these conditions and rewrite the solutions by noting that both (\ref{speed82}) and (\ref{speed83}) are satisfied by
\begin{equation}
k = \frac{\pi n}{L} \,,
\end{equation}
where $n = 1,2,3 \cdots$ and the solution to the time-independent Schr\"odinger equation is
\begin{eqnarray}
\psi_n (x) & = & A \left\{ \begin{array}{cl} \cos (n\pi x/L) & ~~~{\rm for} \, n \, {\rm odd} \\
\sin (n \pi x/L) & ~~~{\rm for} \, n \, {\rm even} \end{array} \right. \nonumber \\
& = & A \sin \left[ \frac{n \pi}{L} \left ( x + \frac{L}{2} \right) \right] \, . 
\end{eqnarray} 
Note that this implies that not only is the wave vector quantized, but also the particle momentum and energy
\begin{equation}
p = \hslash k = \hslash \pi n/L \,,
\end{equation}
and
\begin{equation}
E = V_0 + \frac{\hslash^2 k^2}{2 m} = V_0 + \frac{\hslash^2 \pi^2 n^2}{2mL^2} \, .
\end{equation}
The amplitude $A$ can be found by considering the normalization condition of the wave function
\begin{eqnarray}
\int_{-\infty}^{+\infty} |\psi_n (x) |^2 dx & = & \int_{-L/2}^{+L/2} \left|A \sin \left[\frac{n \pi}{L} \left( x + \frac{L}{2} \right) \right] \right|^2 dx \nonumber \\
& = & |A|^2 \frac{L}{2} \,,
\label{speed88}
\end{eqnarray}
where we have made use of the fact that the wave function is zero outside
the box, and the following definite integral
\begin{equation}
\int_{-L/2}^{+L/2} \left|\sin \left[\frac{n \pi}{L} \left( x + \frac{L}{2} \right) \right] \right|^2 dx = \frac{L}{2} \, .
\end{equation}
For the normalization integral in (\ref{speed88}) to equal unity, we see that the normalization amplitude must be
\begin{equation}
A = \sqrt{\frac{2}{L}} \,,
\label{speed90}
\end{equation}
since we require $|A|^2L/2 = 1$. The solutions for the time independent Schr\"odinger equation can thus be expressed as
\begin{equation}
\psi_n (x) = \sqrt{\frac{2}{L}} \, \sin \left[ \frac{n \pi}{L} \left ( x + \frac{L}{2} \right) \right] \ .
\end{equation}
Note that the normalization in  (\ref{speed90}) can be met for a range of complex amplitudes
\begin{equation}
A = e^{i \phi} \sqrt{\frac{2}{L}} \,, 
\end{equation}
in which the phase $\phi$ is arbitrary. This implies that the outcome of a measurement about the particle position, which is proportional to $|\psi(x)|^2$ is invariant under a {\it global} phase factor. However, as we will see shortly, if there are two or more amplitudes that contribute to a measurement outcome, it is important to keep track of the relative phase between amplitudes, as in the double slit experiment, where the fringe pattern arises from the interference of two probability amplitudes with different phases.

Each solution $\psi_n (x)$, labelled by the quantum number $n$, has quantized wave number, momentum, and energy. The wave function $\psi_n(x)$ is often referred to as an energy eigenstate with corresponding eigenvalue (or eigen-energy $E_n$), because it satisfies the eigenvalue problem
\begin{equation}
\hat H \psi_n(x)  = E_n \psi_n (x) \, 
\end{equation}
where we have introduced the Hamiltonian operator,
\begin{equation}
\hat H = \left[ - \frac{\hslash^2}{2m} \frac{\partial^2}{\partial x^2} + V(x) \right]\,,
\end{equation}
which in this case is the sum of kinetic energy and potential energy operators.
The wave vector and momentum scale linearly with the quantum number ($k_n = k_1n$ and $p_n = p_1n$), while the energy scales quadratically ($E_n = E_1n^2$). The ground state, labelled by $n = 1$ has the lowest allowed energy, followed by the first excited state $n = 2$, which has four times the energy. 

 We often speak of the state of the system, which implies specification of the wave function at some initial time, although this is not always explicitly written. Furthermore, the solutions are orthogonal to one another, that is, they have zero overlap across the entire box
\begin{equation}
\int_{-L/2}^{+L/2} \psi_m^* (x) \, \psi_n(x) \, dx  = \delta_{mn}
\label{speed95}
\end{equation}
where $\delta_{mn}$ is the Kronecker delta, which is one for $m = n$ and zero otherwise.

\subsection{Superposition and time dependence}

The time dependence of quantum states is governed by the time dependent Schr\"odinger equation  (\ref{Scho10}). We can separate out the time and spatial degrees of freedom if the potential energy is time-independent ($V(x)$ is independent of time) by using the method of separation of variables, as described in Sec.~\ref{sMOTI}. In this case the temporal evolution of the wave function is described by an oscillating exponential phase factor
\begin{equation}
\psi_n (x,t) = \psi_n e^{-i E_nt/\hslash} \,,
\end{equation}
where $\psi_n(x)$ is the solution to the time independent Schr\"odinger equation with quantum number $n$ and $E_n$ is its corresponding energy eigenvalue. The states, $\psi_n(x)$ are called energy eigenstates and have special properties in terms of their temporal evolution. Owing to the linearity of the Schr\"odinger equation, a general solution for the ``particle in a box'' can be expressed as a sum of different solutions
\begin{equation}
\Psi (x,t) = \sum_{n=1}^{\infty} c_n \psi_n (x,t) \,,
\end{equation}
where the amplitudes $c_n$ that weight the superposition can be complex, and must obey the normalization condition
\begin{equation}
\sum_{n=1}^\infty |c_n|^2 = 1 \, .
\end{equation}
This follows from considering the normalization condition, and the orthogonality of the states as in (\ref{speed95}). The set of amplitudes $\{c_n\}$, are determined by the specific initial conditions of the problem. The modulus squared of each coefficient gives the probability to find the particle in that state
\begin{equation}
P_n = |c_n|^2 \, .
\end{equation}
For example, we can consider a particle initially prepared in the symmetric superposition of the ground and first excited states, i.e. an equally weighted superposition of the ground and first excited states, at time $t = 0$
\begin{equation}
\Psi^{(+)} (x, t=0) = \frac{1}{\sqrt{2}} \left[\psi_1 (x) + \psi_2 (x) \right]\, .
\end{equation}
The probability to find the particle in state 1 or 2 is 1/2. The state will then evolve in time, with each amplitude having a different time-dependent phase
\begin{eqnarray}
\Psi^{(+)} (x, t) & = & \frac{1}{\sqrt{2}} \left[\psi_1 (x) e^{-i\omega_1 t} + \psi_2 (x) e^{-i\omega_2 t} \right] \nonumber \\
& = & e^{-i\omega_1 t} \frac{1}{\sqrt{2}}  \left[\psi_1 (x)  + \psi_2 (x) e^{-i \Delta \omega t} \right]
\label{speed103}
\end{eqnarray}
where we have introduced the angular frequency associated with the energy
eigenvalues $\omega_n = E_n/\hslash$ and the frequency difference $\Delta \omega = \omega_2 - \omega_1$. The probability to find the particle in state 1, is given by the modulus squared of the corresponding amplitude coefficient, which for the state in  (\ref{speed103}) is just 1/2. The same holds true for the probability to find the particle in state 2. However, the probability to find the particle in the initial superposition state is not time independent.\\

{\bf EXERCISE 10.9} Consider the bound state problem $E<V_0$ in the asymmetric infinite well potential 
\begin{equation}
V(x) = \left\{ \begin{array}{cl}
\infty & ~~~ x\leq 0 \\
0 & ~~~ 0 < x < L \\
V_0 & ~~~ L < x < 2L \\
\infty & ~~~ x > 2L
\end{array} \right. \, .
\end{equation}
{\it (i)}~Solve the time-dependent Schr\"odinger equations  in regions I $(0 < x < L)$ and II $(L <x < 2L)$ and impose appropiate boundary conditions. {\it (ii)}~Use the results in  (i) to derive an equation in terms of $E$, $V_0$, $L$ whose solution will determine the possible energy eigenvalues $E$. {\it (iii)}~Duplicate the procedure to determine an equation in terms of $E$, $V_0$, $L$ whose solution will resolve the possible energy eigenvalues for  $E>V_0$. \\

{\bf EXERCISE 10.10}~Consider the square well of width a which is infinite on the left side and finite on the right,
\begin{equation}
V(x) = \left\{\begin{array}{cl} \infty & ~~~~ x<0 \\
0  & ~~~~ 0<x<L \\
V_0 & ~~~~ x>L \end{array} \right. . \end{equation} 
{\it (i)} Apply appropriate boundary conditions and derive the transcendental equation that determines the energy eigenvalues for the bound states. {\it (ii)}~Show that if the parameter $2 m L^2 V_0/\hslash^2$ is smaller than a critical value, there are no bound states. What is this critical value?\\

{\bf EXERCISE 10.11}~Consider a ``downstep'' potential, which drops at $x = 0$ as one goes from left to right.
\begin{equation}
V(x) = \left\{ \begin{array}{cl} 0 & ~~~~ x<0 \\
-V_0 & ~~~~ x>0 
\end{array} \right. .
\end{equation}
A particle of mass $m$ and kinetic energy $E>0$ approaches the abrupt potential drop.  {\it (i)}~Derive the reflection and transmission coefficients in terms of $E$ and $V_0$. {\it (ii)}~When a free neutron enters a nucleus, it experiences a sudden drop in potental energy, from $V=0$ outside to around $-12~{\rm MeV}$ inside. Suppose a neutron, emitted with kinetic energy $4~{\rm MeV}$ by a fission event, strikes such a nucleus. What is the probability it will be absorbed, thereby initiating another fission? [{\it Hint}: The transmission coefficient expresses the probability of transmission through the surface.]

\subsection{Particle in a central potential}
\label{Lisforam}

We begin by recalling that a prescription for obtaining the (three-dimensional) Schr\"odinger equation for a free particle of mass $m$ is to substitute into the classical energy momentum relation
\begin{equation}
E = \frac{|\vec p\,|^2}{2m}
\end{equation}
the differential operators
\begin{equation}
E \to i \hslash \frac{\partial}{\partial t} \quad {\rm and} \quad \vec p \to - i \hslash \vec \nabla \, .
\label{operadores}
\end{equation}
The resulting operator equation is understood to act on a complex wave function $\psi (\vec x,t)$. That is
\begin{equation}
- \frac{\hslash^2}{2m} \nabla^2 \psi = i \hslash \frac{\partial}{\partial t} \psi \, ,
\label{HM-Sch}
\end{equation}
where we interpret $\rho = |\psi|^2$ as the probability density ($|\psi|^2 d^3x$ gives the probability of finding the particle in a volume element $d^3x$.)

We are often concerned with moving particles, for example the collision of one particle with another. We therefore need to be able to calculate the density flux of a beam of particles $\vec \jmath$. Now, from the conservation of probability, the rate of decrease of the number of particles in a given volume is equal to the total flux of particles out of that volume, i.e
\begin{equation}
- \frac{\partial}{\partial t} \int_V \rho \ dV = \int_S \vec \jmath \cdot \hat n \ dS = \int_V \vec \nabla \cdot \vec \jmath \ dV \,,
\end{equation}
where the last equality is Gauss' theorem and $\hat n$ is a unit vector along the outward normal to the element $dS$ of the surface $S$ enclosing the volume $V$. The probability and the flux densities are therefore related by the continuity equation
\begin{equation}
\frac{\partial \rho}{\partial t} + \vec \nabla \cdot \vec \jmath =0 \, . 
\label{HM-continuity}
\end{equation}
To determine the flux, we first form $\partial \rho/\partial t$ by substracting the wave equation (\ref{HM-Sch}), multiplied by $-i\psi^*$ from the complex conjugate equation multiplied by $-i \psi$. We then obtain
\begin{equation}
\frac{\partial \rho}{\partial t} - \frac{\hslash}{2m} (\psi^* \nabla^2 \psi - \psi \nabla^2 \psi^*) = 0 \, .
\end{equation}
Comparing this with (\ref{HM-continuity}), we identify the probability flux density as
\begin{equation}
\vec \jmath = - \frac{i \hslash}{2m} (\psi^* \nabla \psi - \psi \nabla \psi^*) \, .
\end{equation}
For example, a solution of (\ref{HM-Sch}),
\begin{equation}
\psi = N e^{i \vec p \cdot \vec x - i Et} \,,
\end{equation}
which describes a free particle of energy $E$ and momentum $\vec p$, has
$\rho = |N|^2$ and $\vec \jmath = |N^2|\,  \vec p/m$.\\

{\bf EXERCISE 10.12}~Calculate the eigenfunctions and energy
levels for a free particle, enclosed in a box with edges of lengths
$a$, $b$, and $c$. [Hint: The presence of the box (because of
continuity) requires the wave function to vanish at the edges.]\\

\begin{widetext}

We now consider the time-independent Schr\"odinger equation for a central potential 
\begin{equation}
V(\vec r) = V(|\vec r \,|) = V(r) \,.
\end{equation}
Since the potential depends only on the distance from the origin, the hamiltonian is spherically symmetric.
Therefore, instead of using cartesian coordinates $\vec x = \{x, y, z\}$, it is convenient to use spherical coordinates $\vec x = \{r, \vartheta, \varphi\}$ defined by
\begin{equation}
\left\{\begin{array}{l} x = r \sin \vartheta \cos \varphi \\
y = r \sin \vartheta \sin \varphi\\
z = r \cos \vartheta \end{array} \right\} \Leftrightarrow \left\{ \begin{array}{l} r = \sqrt{x^2 + y^2 +z^2}  \\
\vartheta = \arctan \left(z/\sqrt{x^2 + y^2} \right) \\
\varphi = \arctan (y/x) \end{array} \right\} \, .
\label{polaco1}
\end{equation}
First, we express the Laplacian $\nabla^2$ in spherical coordinates. It follows, after some tedious analysis, that
\begin{equation}
\nabla^2  =  \frac{1}{r^2} \frac{\partial}{\partial r} \left(r^2 \frac{\partial}{\partial r} \right) + \frac{1}{r^2 \sin \vartheta} \frac{\partial}{\partial \vartheta} \left(\sin \vartheta \frac{\partial}{\partial \theta} \right)   +  \frac{1}{r^2 \sin^2 \vartheta} \frac{\partial^2}{\partial \varphi^2} \, .
\end{equation}
To look for solutions, we use again the separation of variable methods, writing $\psi(\vec x) = \psi(r, \vartheta, \varphi) = R(r)Y (\vartheta, \varphi)$:
\begin{equation}
-\frac{\hslash^2}{2m} \left[\frac{Y}{r^2} \frac{d}{d r} \left(r^2 \frac{d R}{d r} \right) + \frac{R}{r^2 \sin \vartheta } \frac{\partial}{\partial \vartheta} \left( \sin \vartheta \frac{\partial Y}{\partial \vartheta} \right)+ \frac{R}{r^2 \sin^2 \vartheta } \frac{\partial ^2Y}{\partial \varphi^2} \right] + V(r) RY = ERY \, .
\end{equation}
We then divide by $RY/r^2$ and rearrange the terms as
\begin{equation}
-\frac{\hslash^2}{2m} \left[\frac{1}{R} \frac{d}{d r} \left(r^2 \frac{d R}{d r} \right)\right]+ r^2 (V-E) = \frac{\hslash^2}{2m Y} \left[\frac{1}{\sin \vartheta} \frac{\partial}{\partial \vartheta} \left(\sin \vartheta \frac{\partial Y}{\partial \vartheta} \right) + \frac{1}{\sin^2 \vartheta} \frac{\partial^2 Y}{\partial \varphi^2} \right] \, .
\end{equation} 
Each side is a function of $r$ only and $\vartheta,\varphi$, so they must be independently equal to a constant $\varkappa$ that we set (for
reasons to be seen later) equal to 
\begin{equation}
\varkappa = - \frac{\hslash^2}{2m} l (l+1) \, .
\end{equation}
We obtain two equations:
\begin{equation}
\frac{1}{R} \frac{d}{dr} \left(r^2 \frac{dR}{dr} \right) - \frac{2mr^2}{\hslash^2} (V-E) = l (l+1) 
\end{equation}
and
\begin{equation}
\frac{1}{\sin \vartheta} \frac{\partial}{\partial \vartheta} \left(\sin \vartheta \frac{\partial Y}{\partial \vartheta} \right) + \frac{1}{\sin^2 \vartheta} \frac{\partial^2 Y}{\partial \varphi^2} = - l(l+1) Y \, .
\end{equation}
This last equation is the angular equation. Notice that it can be considered an eigenvalue equation for the operator
\begin{equation}
\hat {\cal O} \equiv \frac{1}{\sin \vartheta} \frac{\partial}{\partial \vartheta} \left( \sin \vartheta \frac{\partial}{\partial \vartheta}\right) + \frac{1}{\sin^2 \vartheta} \frac{\partial^2}{\partial \varphi^2} \, .
\end{equation}
What is the meaning of this operator?

We take one step back and look at the angular momentum operator. From its classical form $\vec L  = \vec r \times \vec p$  we can define the quantum mechanical operator:
\begin{equation}
\hat{\vec L} = \hat{\vec r} \times \hat{\vec p} = - i \hslash \ \hat{\vec r} \times \hat{\vec \nabla} \, .
\end{equation}
In cartesian coordinates this reads
\begin{eqnarray}
\hat L_x & = & \hat y \hat p_z - \hat p_y \hat z = - i \hslash \left(y \frac{\partial}{\partial z} - \frac{\partial}{\partial y} z \right), \nonumber \\
\hat L_y & = & \hat z \hat p_x - \hat p_z \hat x = - i \hslash \left(z \frac{\partial}{\partial x} - \frac{\partial}{\partial z} x \right), \nonumber \\
\hat L_z & = & \hat x \hat p_y - \hat p_x \hat y = - i \hslash \left(x \frac{\partial}{\partial y} - \frac{\partial}{\partial x} y \right) \, .
\end{eqnarray}
Some very important properties of this vector operator regard its commutator. Consider for example
\begin{equation}
[\hat L_x , \hat L_y] = [\hat y \hat p_z - \hat p_y \hat z, \hat z \hat p_x - \hat p_z \hat x] 
 =  [\hat y \hat p_z,\hat z \hat p_x] - [\hat p_y \hat z, \hat z \hat p_x] - [\hat y \hat p_z, \hat p_z \hat x] + [\hat p_y \hat z, \hat p_z \hat x]. 
\end{equation} 
Now remember that $[x_i,x_j] = [p_i,p_j] = 0$ and $[x_i,p_j] = i \hslash \delta_{ij}$. Also $[AB,C] = A[B,C] + [A,C]B$. This simplifies
matters a lot
\begin{equation}
[\hat L_x , \hat L_y]  =  \hat y [ \hat p_z, \hat z] \hat p_x - \cancel{[\hat p_y \hat z, \hat z \hat p_x]} - \cancel{[\hat y p_z, \hat p_z \hat x]} + \hat p_y [\hat z ,\hat p_z] \hat x 
 =  i \hslash (\hat x \hat p_y - \hat y \hat p_x) = i \slash \hat L_z \, .
\end{equation}
By performing a cyclic permutation of the indexes, we can show that this holds in general:
\begin{equation}
[\hat L_i, \hat L_j] = i \hslash \ \varepsilon_{ijk} \ \hat L_k \, ,
\label{gugupa1}
\end{equation}
where 
\begin{equation}
\varepsilon_{ijk} = \left\{ \begin{array}{rl} 1 & ~~~{\rm if} \ (i, j, k) \ {\rm is \ an \ even \ permutation \ of} \ (1,2,3); \ {\rm that \ is} \ (1,2,3), (2,3,1), (3,1,2)\,, \\
 -1&  ~~~{\rm if} \ (i, j, k) \  {\rm is \ an \ odd\ permutation \ of} \ (1,2,3);\  {\rm that \ is} \ (3,2,1), (1,3,2), (2,1,3) \,,\\
0 &  ~~~{\rm if \ any \ index \ is \ repeated}; \ {\rm that \ is \ if} \ i = j\ {\rm or} \ j=k\ {\rm or} \  k = 1 \end{array} \right. 
\end{equation}
 is the Levi-Civita tensor. (The following relations are useful:
$\varepsilon_{ikl} \varepsilon_{imn} = \delta_{km} \delta_{ln} - \delta_{kn} \delta_{lm}$, $\varepsilon_{ikl} \varepsilon_{ikm} = 2 \delta_{lm}$, and $\varepsilon_{ikl} \varepsilon_{ikl} = 6$.) Since the different components of the angular momentum do not commute, they do not possess common eigenvalues and there is an uncertainty relation for them. If for example we know with absolute precision the angular momentum along the $z$ direction, we cannot have any knowledge of the components along $x$ and $y$.
From the uncertainty relations, 
\begin{equation}
\Delta L_x \Delta L_z \geq \frac{\hslash}{2} |\langle L_y \rangle| \quad 
{\rm and }
\quad 
\Delta L_y \Delta L_z \geq \frac{\hslash}{2} |\langle L_x \rangle |, 
\end{equation}
which derive from (\ref{speedHP}) and (\ref{gugupa1}), we have that if $\Delta L_z = 0$ (perfect knowledge) then we have a complete uncertainty in $L_x$ and $L_y$.\\

{\bf EXERCISE 10.13}~Consider the squared length of the angular momentum vector $\hat L^2 = \hat L^2_x +\hat L_y^2+L_z^2$. Show that  $[\hat L^2 ,\hat L_i]=0$, for $i = {x, y, z}.$\\ 

All in all, we can always know the length of the angular momentum plus one of its components. For example, choosing the $z$-component, we can represent the angular momentum as a cone, of length $\langle L \rangle$, projection on the $z$-axis ($L_z$) and with complete uncertainty of its projection along $x$ and $y$, see Fig.~\ref{fig:LzL2}.

\begin{figure}[tbp] \postscript{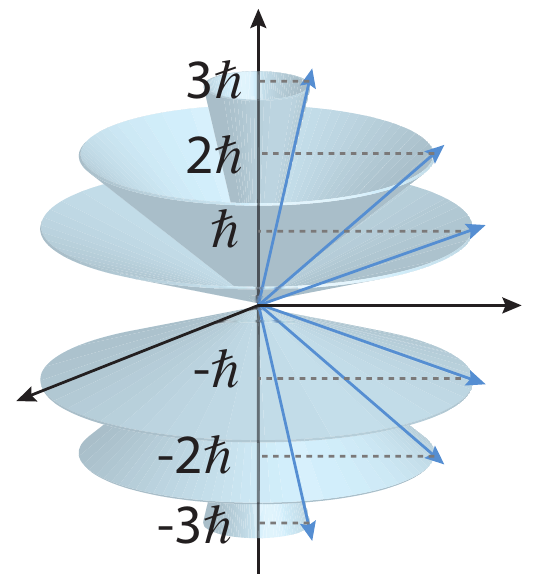}{0.6} \caption{Graphical representation of the angular momentum, with fixed $L_z$ and $L^2$, but complete uncertainty in $L_x$ and $L_y$.}
\label{fig:LzL2}
\end{figure}

We now express the angular momentum using spherical coordinates. This simplifies particularly how the azimuthal angular momentum $\hat L_z$ is expressed.
It is also convenient to define a new set of orthogonal unit vectors appropriate to the spherical coordinate system (\ref{polaco1}). These are
\begin{eqnarray}
\hat e_r & = & \frac{\vec r}{r} = \cos \varphi \sin \vartheta \hat e_x + \sin \varphi \sin \vartheta \hat e_y + \cos \vartheta \hat e_z \nonumber \\
\hat e_\theta & = & \frac{\partial \hat e_r}{\partial \theta} = \cos \varphi \cos \vartheta  \hat e_x + \sin \varphi \cos \vartheta \hat e_y - \sin \vartheta \hat e_z \nonumber \\
\hat e_\phi & = & \frac{1}{\sin \vartheta} \frac{\partial \hat e_r}{\partial \varphi} = - \sin \varphi \hat e_x + \cos \varphi \hat e_y \, .
\end{eqnarray}
We can now rewrite the gradient operator by using the chain rule for partial
differentiation as
\begin{eqnarray}
\vec \nabla & = & \hat e_x \frac{\partial}{\partial x} + \hat e_y \frac{\partial}{\partial y} + \hat e_z \frac{\partial}{\partial z} \nonumber \\
& = & \hat e_x \left(\frac{\partial r}{\partial x} \frac{\partial}{\partial r} + \frac{\partial \vartheta}{\partial x} \frac{\partial}{\partial \vartheta} + \frac{\partial \varphi}{\partial x} \frac{\partial }{\partial \varphi} \right) 
 +  \hat e_y \left(\frac{\partial r}{\partial y} \frac{\partial}{\partial r} + \frac{\partial \vartheta}{\partial y} \frac{\partial}{\partial \vartheta} + \frac{\partial \varphi}{\partial y} \frac{\partial }{\partial \varphi} \right) 
+  \hat e_z \left(\frac{\partial r}{\partial z} \frac{\partial}{\partial r} + \frac{\partial \vartheta}{\partial z} \frac{\partial}{\partial \vartheta} + \frac{\partial \varphi}{\partial z} \frac{\partial }{\partial \varphi} \right) .
\end{eqnarray}
The various partial derivatives can be evaluated using (\ref{polaco1}) to give
\begin{eqnarray}
\frac{\partial r}{\partial x} & = &  \frac{x}{r} = \cos \varphi \sin \vartheta \,,  \quad \quad \quad \quad
\frac{\partial r}{\partial y}  =  \frac{y}{r} = \sin \varphi \sin \vartheta \,, \quad \quad \quad \quad
\frac{\partial r}{\partial z}  =  \frac{z}{r} = \cos \vartheta \,, \nonumber \\
\frac{\partial \vartheta}{\partial x} & = & \frac{1}{r} \cos \varphi \cos \vartheta \,,  \quad \quad \quad \quad
~~~\frac{\partial \vartheta}{\partial y}  =  \frac{1}{r} \sin \varphi \cos \vartheta \,,  \quad \quad \quad \quad
~~~\frac{\partial \vartheta}{\partial z}  =  - \frac{1}{r} \sin \vartheta\,, \nonumber \\
\frac{\partial \varphi}{\partial x} & = &  - \frac{\sin \varphi}{ r \sin \vartheta} \,, \quad \quad \quad \quad
~~~~~~~~\frac{\partial \varphi}{\partial y}  =   \frac{\cos \varphi}{ r \sin \vartheta} \,,  \quad \quad \quad \quad
~~~~~~~~~~\frac{\partial \varphi}{\partial z}  =  0 \, .
\end{eqnarray}
These along with the definitions of the spherical unit vectors can be used to write the gradient as
\begin{equation}
\vec \nabla = \hat e_r \frac{\partial}{\partial r} + \frac{\hat e_\vartheta}{r} \frac{\partial}{\partial \vartheta} + \frac{\hat e_\varphi}{r \sin \vartheta} \frac{\partial}{\partial \varphi} \, .
\end{equation}
We can now write the angular momentum operator as
\begin{eqnarray}
\hat {\vec L} & = & \vec r \times \frac{\hslash }{i} \vec \nabla = \frac{\hslash}{i} r \hat e_r \times \left(\hat e_r \frac{\partial }{\partial r} + \frac{\hat e_\vartheta}{r} \frac{\partial}{\partial \vartheta} + \frac{\hat e_\varphi}{r \sin \vartheta} \frac{\partial}{\partial \varphi} \right) \nonumber \\
& = & \frac{\hslash}{i} r \left(\frac{\hat e_\varphi}{r} \frac{\partial}{\partial \vartheta} - \frac{\hat e_\vartheta}{r \sin \vartheta } \frac{\partial }{\partial \varphi} \right) = \frac{\hslash}{i} \left(\hat e_\varphi \frac{\partial}{\partial \theta} - \frac{\hat e_\vartheta}{\sin \vartheta }\frac{\partial}{\partial \varphi} \right) \nonumber \\
& = & \frac{\hslash}{i} \left[( - \sin \varphi \hat e_x + \cos \varphi \hat e_y) \frac{\partial}{\partial \vartheta} - \frac{\cos \varphi \cos \vartheta \hat e_x + \sin \varphi \cos \vartheta \hat e_y - \sin \vartheta \hat e_z}{\sin \vartheta } \frac{\partial}{\partial \varphi} \right] \nonumber \\
 & = & \frac{\hslash}{i} \left[ \hat e_x \left( - \sin \varphi \frac{\partial}{\partial \vartheta} - \frac{\cos \varphi \cos \vartheta}{\sin \vartheta} \frac{\partial }{\partial \varphi} \right)+ \hat e_y \left( \cos \varphi \frac{\partial}{\partial \vartheta} - \frac{\sin \varphi \cos \vartheta}{\sin \vartheta} \frac{\partial}{\partial \varphi} \right)  + \hat e_z \frac{\partial }{\partial \varphi} \right] \, .
\end{eqnarray}
From this we can extract the cartesian components of the angular momentum vector in terms of spherical coordinates yielding
\begin{eqnarray}
\hat L_x & =&  i \hslash \left(\sin \varphi \frac{\partial}{\partial \theta} + \cot \varphi \cos \varphi \frac{\partial}{\partial \varphi} \right), \nonumber \\
\hat L_y & = & -i \hslash \left(\cos \varphi \frac{\partial}{\partial \vartheta} - \cot \vartheta \sin \varphi \frac{\partial}{\partial \varphi} \right), \nonumber \\
\hat L_z & = & - \hslash \frac{\partial}{\partial \varphi} \, .
\end{eqnarray}
The spherical representation of the components of the angular momentum operator can be used to express the square of the angular momentum operator as
\begin{eqnarray}
\hat L^2 & = & - \hslash^2 \left[ \sin^2 \varphi \frac{\partial^2}{\partial \vartheta^2} \color{blue} + \sin \varphi \cos \varphi \frac{\partial \cot \vartheta}{\partial \vartheta} \frac{\partial}{\partial \vartheta} \color{red} + \frac{\sin \varphi \cos \varphi \cos \vartheta}{\sin \vartheta} \frac{\partial^2}{\partial \vartheta \partial \varphi} \right. \nonumber \\
& +  &  \frac{\cos^2 \varphi \cos \vartheta}{\sin \vartheta} \frac{\partial}{\partial \vartheta} \color{green} + \frac{\sin \varphi \cos \varphi \cos \vartheta}{\sin \vartheta} \frac{\partial^2}{\partial \varphi \partial \vartheta}  \color{magenta} - \frac{\sin \varphi \cos \varphi \cos^2 \vartheta}{\sin^2 \vartheta} \frac{\partial}{\partial \varphi} \nonumber \\
\nonumber \\
& + & \frac{\cos^2 \varphi \cos^2 \vartheta}{\sin^2 \vartheta} \frac{\partial^2}{\partial \varphi^2} \nonumber \\
& + & \cos^2 \varphi \frac{\partial^2}{\partial \vartheta^2} \color{blue} - \sin \varphi \cos \varphi \frac{\partial \cot \vartheta}{\partial \vartheta} \frac{\partial}{\partial \vartheta} \color{red}  - \frac{\sin \varphi \cos \varphi \cos \vartheta}{\sin \vartheta} \frac{\partial^2}{\partial \vartheta \partial \varphi} \nonumber \\
& + & \frac{\sin^2 \varphi \cos \vartheta}{\sin \vartheta} \frac{\partial}{\partial \vartheta} \color{green} - \frac{\sin \varphi \cos \varphi \cos \vartheta}{\sin \vartheta} \frac{\partial^2}{\partial \varphi \partial \vartheta}  \color{magenta} + \frac{\sin \varphi \cos \varphi \cos^2 \vartheta}{\sin^2 \vartheta} \frac{\partial}{\partial \varphi} \nonumber \\
& + & \left. \frac{\sin^2 \varphi \cos^2 \vartheta}{\sin^2 \vartheta} \frac{\partial^2}{\partial \varphi^2} + \frac{\partial^2}{\partial \varphi^2} \right] \, .
\end{eqnarray}
The colored terms cancel in pairs and the remaining terms can be simplified to give a ralation that should be familar
\begin{equation}
\hat L^2 = - \hslash^2 \left[\frac{1}{\sin \vartheta} \frac{\partial}{\partial \vartheta} \left( \sin \vartheta \frac{\partial}{\partial \vartheta}\right) + \frac{1}{\sin^2 \vartheta} \frac{\partial^2}{\partial \varphi^2} \right]\, .
\end{equation}
We can then write the eigenvalue equations for these two operators:
and
\begin{equation}
\hat L^2 Y(\vartheta, \varphi) = \hslash^2 l(l+1) Y(\vartheta, \varphi)
\quad \quad {\rm and} \quad \quad
\hat L_z Y (\vartheta, \varphi) = \hslash m Y(\vartheta,\varphi)
\end{equation}
where we already used the fact that they share common eigenfunctions. Then, by its very nature we can label these eigenfunctions by $l$ and $m$, i.e.  $Y_{l,m} (\vartheta, \varphi)$.\\

{\bf EXERCISE 10.14}~Show that the allowed values for $l$ and $m_z$ are integers such that $l = 0,1,2,\cdots$ and $m_z = −l, \cdots ,l - 1,l$. [{\it Hint:} This result can be inferred from the commutation relationship.]\\

\begin{figure}[tbp] \postscript{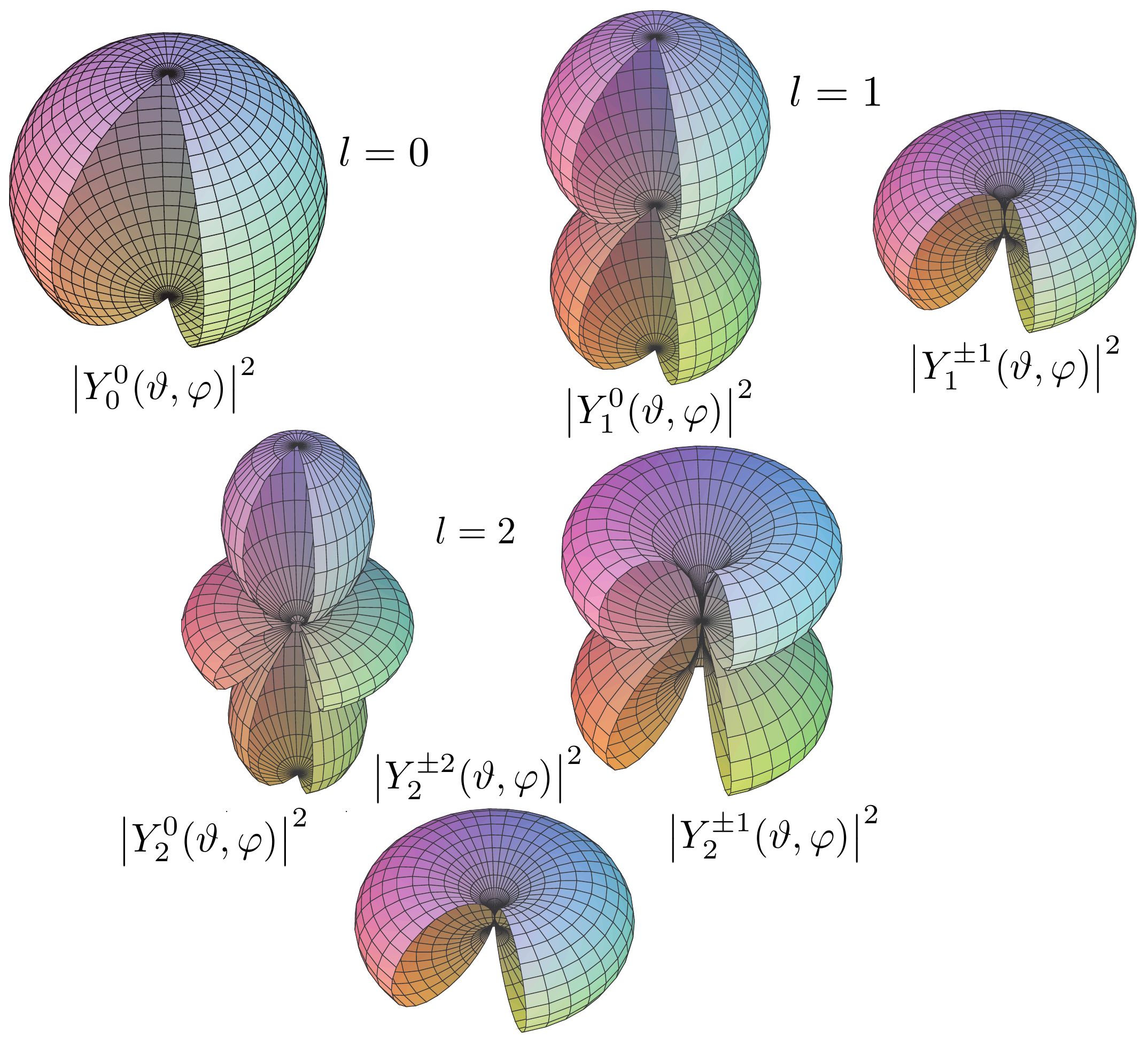}{0.8} \caption{Representations of $|Y_l^m|^2$ for different sets of quantum numbers. The $z$ axis is the vertical direction. The probability densities have rotational symmetry about the $z$ axis~\cite{Anchordoqui:2013}.}
\label{fig:Ylm}
\end{figure}

\begin{table*}
\caption{Associated Legendre polynomials. \label{LegP}}
\begin{tabular}{c|cccc}
\hline
\hline
\backslashbox{l}{m} & 0 & 1 & 2 & 3 \\
\hline
0 & $P_0^0 = 1$ & & & \\
1& $P_1^0 = \cos \vartheta$ & $P_1^1 \sin \vartheta$ &  & \\
2 & $P_2^0 = (3 \cos^2 \vartheta -1)/2$ & $P_2^1 = 3 \cos \vartheta \sin \vartheta$ & $P_2^2 = 3 \sin^2 \vartheta$ & \\
3& ~~~$P_3^0 (5 \cos^3 \vartheta - 3 \cos \vartheta)/2$~~~ & ~~~$P_3^1 = 3 (5 \cos^2 \vartheta -1)/2 \, \sin \vartheta$~~~ & ~~~$P_3^2 = 15 \cos \vartheta \sin^2 \vartheta$~~~ & ~~~$P_3^3 = 15 \sin^3 \vartheta$~~~ \\
\hline \hline
\end{tabular}
\end{table*}

We now go back to the Schr\"odinger equation in spherical coordinates and we consider the angular and radial equation separately to find the energy eigenvalues and eigenfunctions. The angular equation was found to be
\begin{equation}
\frac{1}{\sin \vartheta} \frac{\partial}{\partial \vartheta} \left(\sin \vartheta \frac{\partial Y_l^m (\vartheta, \varphi)}{\partial \vartheta} \right) + \frac{1}{\sin^2 \vartheta } \frac{\partial^2 Y_l^m (\vartheta, \varphi)}{\partial \varphi^2} = -l (l+1) Y_l^m (\vartheta, \varphi) \, .
\end{equation}
Note that this equation does not depend at all on the potential, thus it will be common to all problems with a central potential.

We can solve the equation by using again separation of variables: $Y ( \vartheta, \varphi) = \Theta (\vartheta )\Phi(\varphi)$. By multiplying both sides of the equation by $\sin^2 \vartheta/Y (\vartheta, \varphi)$ we obtain:
\begin{equation}
\frac{1}{\Theta (\vartheta)} \left[ \sin \vartheta \frac{d}{d \vartheta} \left(\sin \vartheta \frac{d\Theta}{d \vartheta} \right) \right] + l (l+1) \sin^2 \vartheta = - \frac{1}{\Phi (\varphi)} \frac{d^2\Phi}{d\varphi^2} \, .
\end{equation}
\end{widetext}
As usual we separate the two equations in the different variables and introduce a constant $m^2$:
\begin{equation}
\frac{d^2 \Phi}{d \varphi^2} = - m^2 \Phi (\varphi) 
\end{equation}
and 
\begin{equation}
\sin \vartheta \frac{d}{d\vartheta} \left(\sin \vartheta \frac{d\Theta}{d \vartheta}\right) = [m^2 - l(l+1) \sin^2 \vartheta ] \Theta (\vartheta) \, .
\end{equation}
The first equation is easily solved to give $\Phi(\varphi) = e^{im\varphi}$ with $m = 0, \pm1, \pm 2, \cdots$ since we need to impose the periodicity
of $\Phi$, such that $\Phi(\varphi+ 2\pi) = \Phi(\varphi)$.
The solutions to the second equation are associated Legendre polynomials: $\Theta(\vartheta) = A P_l^m (\cos\vartheta)$, the first few of 
which are in Table~\ref{LegP}. Note that, as previously found when solving for the eigenvalues of the angular momentum, we have that $m = -l, -l + 1, . . . , l$, with $l = 0, 1, \cdots$.

The normalized angular eigenfunctions are then spherical harmonic functions, given by the normalized associated Legendre polynomial times the solution to the equation in $\varphi$, 
\begin{equation}
Y_l^m (\vartheta, \varphi) = \sqrt{\frac{(2l+1)}{4 \pi} \frac{(l-m)!}{(l+m)!}} \ P_l^m (\cos \vartheta) e^{im\varphi} \, .
\label{spherical-harmonics}
\end{equation}
As we expect from eigenfunctions, the Spherical Harmonics are orthogonal:
\begin{equation}
\int_{S^2} {Y_l^m}^*(\vartheta,\varphi) \, Y_{l'}^{m'} d\Omega = \delta_{ll'} \delta_{mm'} \,,
\end{equation}
where $\int_{S^2} d \Omega = \int_0^\pi \int_0^{2\pi} \sin \vartheta d \vartheta d \varphi$ denotes the integral over the spherical surface.
For this phase convention,
$Y_{lm}^*(\vartheta, \varphi) = (-1)^m \, Y_{l-m} (\vartheta, \varphi)$, and hence for
$l=0,1,2$ we obtain
\begin{eqnarray}
Y_{0}^0 (\vartheta,\varphi) & = & \frac{1}{\sqrt{4\pi}} \nonumber \\
Y_{1}^{-1} (\vartheta, \varphi) & = & \sqrt{\frac{3}{8\pi}}\, \sin \vartheta \,
e^{i\varphi} \nonumber \\
Y_{1}^0(\vartheta, \varphi) & = & \sqrt{\frac{3}{4 \pi}} \ \cos \vartheta
\nonumber \\
Y_{1}^{1} (\vartheta, \varphi) & = & - \sqrt{\frac{3}{8\pi}} \, \sin \vartheta \,
e^{i\varphi}  \nonumber
\end{eqnarray}
\begin{eqnarray}
Y_{2}^{-2} (\vartheta, \varphi) & = &  \sqrt{\frac{15}{32 \pi}} \, \sin^2 \vartheta \,
  e^{- i 2 \varphi} \nonumber \\
Y_{2}^{-1} (\vartheta, \varphi) & = & \sqrt{\frac{15}{8\pi}} \, \sin \vartheta \, \cos
  \vartheta \, e^{-i \varphi} \nonumber \\
Y_{2}^{0} (\vartheta, \varphi) & = & \sqrt{\frac{5}{16 \pi}} (-1 + 3 \cos^2 \vartheta)
    \nonumber \\
Y_{2}^{1} (\vartheta, \varphi) & = &- \sqrt{\frac{15}{8\pi}} \, \sin \vartheta \, \cos
  \vartheta \, e^{i \varphi} \nonumber \\
Y_{2}^{2} (\vartheta, \varphi) & = &  \sqrt{\frac{15}{32 \pi}} \, \sin^2 \vartheta \,
  e^{ i 2 \varphi}  \, ; \nonumber
\end{eqnarray}
the surfaces $r= |Y_{lm} (\vartheta, \varphi)|$ for these spherical
harmonics are shown in Fig.~\ref{fig:Ylm}. (For details, see e.g.,~\cite{Anchordoqui:2013}.)

We now turn to the radial equation:
\begin{equation}
\frac{d}{dr} \left(r^2 \frac{dR(r)}{dr} \right) - \frac{2m r^2}{\hslash^2} (V-E) = l (l+1) R(r) \, .
\end{equation}
To simplify the solution, we introduce a different function $u(r) = rR(r)$. Then the equation reduces to:
\begin{equation}
- \frac{\hslash^2}{2m} \frac{d^2 u}{dr^2} + \left[ V + \frac{\hslash}{2m} \frac{l(l+1)}{r^2} \right] u(r) = E u(r) \, .
\label{chachoC}
\end{equation}
If we define an effective potential 
\begin{equation}
V'(r) = V (r) + \frac{\hslash^2}{2m} \frac{l (l+1)}{r^2} \, ,
\end{equation}
(\ref{chachoC}) is very similar to the one-dimensional Schr\"odinger equation. 
The second term in this effective potential is called the {\it centrifugal} term. 
Solutions can be found for some forms of the potential $V (r)$, by first calculating the equation solutions $u_{n,l}(r)$, then finding $R_{n,l}(r) = u_{n,l}(r)/r$ and finally the wavefunction
\begin{equation}
\psi_{n,l,m}(r, \vartheta, \varphi) = R_{n,l}(r) Y_l^m(\vartheta, \varphi) \, .
\end{equation}
Note that we need 3 quantum numbers $(n, l, m)$ to define the eigenfunctions of the hamiltonian in three dimensions. For example, we can have a simple spherical well: $V(r)=0$ for $r<r_0$ and $V(r)=V_0$ otherwise. In the case of $l=0$, this is the same equation as for the square well in one dimension. Note, however, that since the boundary conditions need to be such that $R(r)$ is finite for all $r$, we need to impose that $u(r = 0) = 0$, hence only the odd solutions are acceptable.  For $l > 0$, we can find solutions in terms of Bessel functions~\cite{Anchordoqui:2013}. Two other important examples of potential are: the harmonic oscillator potential 
\begin{equation}
V(r) = V_0 \frac{r}{r_0^2} - V_0 \,, 
\end{equation}
which is an approximation for any potential close to its minimum, and the Coulomb potential 
\begin{equation}
V(r) = - \frac{e^2}{4\pi \epsilon_0} \frac{1}{r} \,,
\end{equation}
which describes the atomic potential and in particular the Hydrogen atom.

\section{Stern-Gerlach experiment}
\label{Sec:Stern-Gerlach}

Within the framework of classical mechanics one can show that an electron in a circular orbit has an angular momentum $\vec L = m_e r^2 \vec \omega $ and an associated magnetic moment 
\begin{equation}
\vec \mu = - \frac{e}{2m_e} \vec L \,,
\label{SG5}
\end{equation}
where $m_e$ and $e$ are, respectively, the mass and charge of the electron, and $r$ and $\vec \omega$ are the
radius and angular velocity of the orbital motion. In a magnetic field $\vec B$ the atom will be
acted on by a torque 
\begin{equation}
\vec \tau = \vec \mu \times \vec B \,,
\label{torque-equation}
\end{equation}
which causes $\vec L$ to precess about the direction of $\vec B$ with some fixed value of the projection $\mu_z = |\vec \mu| \ \cos \theta$ of its magnetic moment along the direction of the field; see Appendix~\ref{appD}. The atom will also have a potential energy $- \vec \mu \cdot \vec B$, and if the field is inhomogeneous such that at a certain point it is in the $z$ direction and varies strongly with $z$, then the atom will be acted on by a force 
\begin{equation}
F_z = - \vec \nabla (- \vec \mu \cdot \vec B) = \mu_z \frac{\partial B_z}{\partial z} \,,
\end{equation}
which may have any of a continuous set of values from $−|\vec \mu| \, \partial B_z/\partial z$ to $+|\vec \mu| \, \partial B_z/\partial z$. One would then expect a monoenergetic beam of  atoms, initially randomly oriented and passing through an inhomogeneous magnetic field, to be deflected in the $+z$ and $-z$ directions with a distribution of deflection angles that has a maximum value at zero deflection and decreases monotonically in either direction. This is not what is observed experimentally.  

In 1921, Stern and Gerlach generated a beam of neutral silver atoms by evaporating silver from an oven~\cite{Stern:1921,Gerlach:1922ur,Gerlach:1922dv,Gerlach:1922zz}. The process was performed in a vacuum so that the silver atoms moved without scattering. The atoms were collimated by slits and sent through a region with a large non-uniform magnetic field.  The silver atoms were thus deflected and allowed to strike a cold metallic plate (hereafter refer to as the detection screen). After about 8-10 hours the number of condensed silver atoms was large enough to show a visible trace on the screen. The trace showed two marks implying that the silver atoms had two possible components of $\mu_z$.

From the classical relation (\ref{SG5}) we see that the quantum-mechanical operator corresponding to the observable $\mu_L$ (the magnetic moment due to the orbital motion) must be
\begin{equation}
\hat {\vec \mu}_L = - \frac{e}{2m_e} \ \hat {\vec L} = - \frac{e}{2m_e} \ \hat {\vec r} \times \hat {\vec p} \, .
\label{SG435}
\end{equation}
The operator relation (\ref{SG435}) suggests that the magnetic moment $\vec \mu_L$ due to the orbital motion is quantized in the same way as the orbital angular momentum. This means that the size and one of the components can have discrete values simultaneously, i.e.
\begin{equation}
|{\vec \mu}_L| =  \frac{e}{2m_e} |\vec {L}| = \mu_B \sqrt{l (l+1)}, 
\end{equation}
with $l = 0,1,2,\cdots$, while for example the $z$ component can take the values
\begin{equation}
(\mu_L)_z = - \frac{e}{2m_e}  L_z = - m \mu_B, 
\end{equation}
with $m = 0,\pm1,\pm2, \cdots, \pm l$. Here the quantity,
\begin{equation}
\mu_B \equiv \frac{e \hslash}{2m_e} \quad \quad (1 \, {\rm Bohr \, magneton})
\end{equation}
is the natural unit for the magnetic moment for the electron, just as $\hslash$ is the natural unit for angular momenta, $1~\mu_B = 5.788 \times 10^{-5}~{\rm eV/T(esla)}$~\cite{Gerlach:1924}.
We note that the quantization of $L_z$ and $(\mu_L)_z$ corresponds to a so-called ``space quantization,'' that is, quantized values of the angle between the vector $\mu_L$ and the $z$ axis.
If the total angular momentum of the atom is ``integral'' (given by an integral angular-momentum quantum number $l$), we should then expect to find $2l + 1$ discrete deflections, that is, an odd number of pictures of the slit on the detection screen. In plain English, Schr\"odinger equation predicts an odd number of possible states, also in contradiction with experiment.

The explanation for the even number of images observed in the Stern-Gerlach experiment was developed by Uhlenbeck and Goudsmit~\cite{Uhlenbeck:1925,Uhlenbeck:1926}, and independently by Pauli~\cite{Pauli:1925,Pauli:1927}. It is most easily
understood in the hydrogen case: Even with $l = 0$ the electron in the hydrogen atom has a magnetic moment $\mu_S$, which causes the deflection of the orbit of each atom. This magnetic moment is connected with an intrinsic angular momentum of the electron, the so-called intrinsic spin angular momentum $S$, (spin for short). As all other angular momenta, the spin can be characterized by an angular-momentum quantum number which is usually denoted by $s$, so that $|\vec S|$ = $\hslash \sqrt{s (s+1)}$ and such that the $z$ component can take the values
\begin{equation}
S_z = m_s \hslash \,,
\end{equation}
where $m_s = -s, -s+1, -s+2, \cdots, + s$. This is analogous to $m = L_z/ \hslash$ taking the values $-l, -l + 1, -l + 2,\cdots, +l$ for a given orbital angular-momentum quantum number $l$. From the general discussion of angular momenta in Sec.~\ref{Lisforam}, it follows that a spin quantum number must in general take one of the values $s = 0, \frac{1}{2}, 1, \frac{3}{2}, 2, \cdots$. The number of $m_s$ values is $2s+1$, in analogy with the $2l + 1$ values of $L_z$.

\begin{figure}[tbp] \postscript{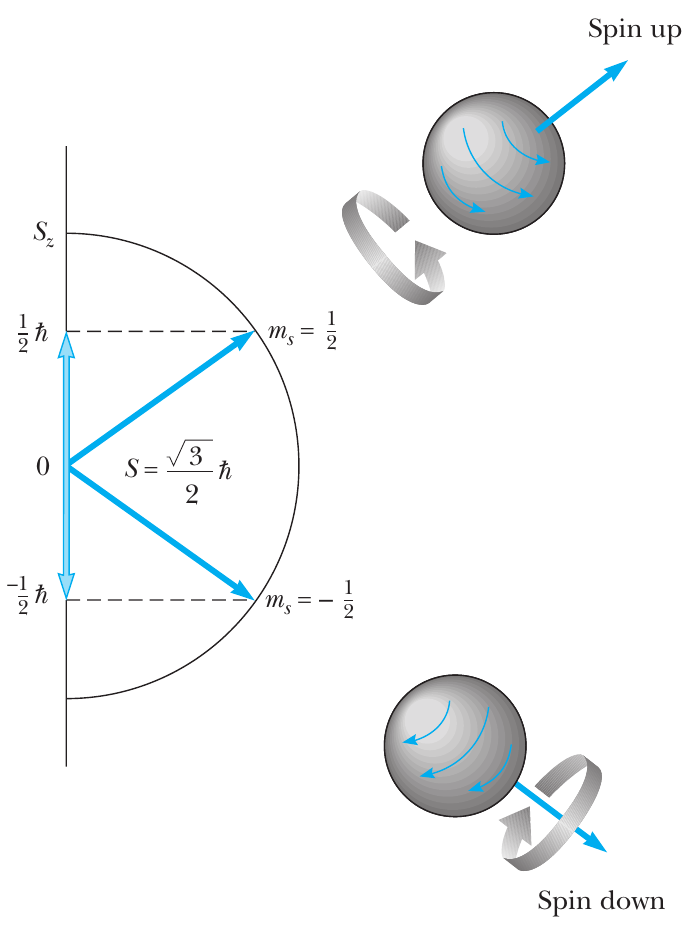}{0.9} \caption{Two
allowed orientations of the spin
vector $\vec S$ for a spin-$\frac{1}{2}$ particle, such as the electron~\cite{Serway:2005}.}
\label{Spin}
\end{figure}

The fact that we observe two discrete deflections in the Stern-Gerlach experiment then
leads to the interpretation that in this case $2s + 1$ is equal to 2; that is, the spin quantum number $s$ of the electron is equal to $\frac{1}{2}$. This corresponds to $|\vec S| = \hslash \sqrt{s(s+1)} = \hslash \sqrt{3/4} = 0.866\hslash$. The two possible values of the magnetic quantum number of the electron spin then are $m_s = \pm \frac{1}{2}$, corresponding to $S_z = \pm \frac{1}{2} \hslash$. These two spin states are commonly denoted by spin up and spin down, see Fig.~\ref{Spin}.

Experiment show that the intrinsic magnetic moment connected with the spin is
\begin{equation}
\vec \mu_S = - \frac{e}{2m_e} g_e \vec S \,,
\end{equation}
where the factors in front of $\vec S$ is the product of the gyromagnetic ratio we found for the orbital motion, and $g_e$ which is a dimensionless factor known as the gyromagnetic factor of the electron. It is conventional to express $g_e$ in terms of the   so called ``anomalous magnetic moment'',  
\begin{equation}
a_e = (g_e -2)/2, 
\end{equation}
which is one of most accurately determined quantities experimentally. Recent measurements of the anomalous
magnetic moment of the electron reached the fabulous
relative precision of 0.7 parts per billion~\cite{Odom:2006zz,Hanneke:2008tm},
\begin{equation}
a_e^{\rm exp} =11596521807.3(2.8) \times 10^{-13} \, .
\end{equation}

Pauli realized that, for intrinsic spin, only a matrix representation
is possible~\cite{Pauli:1926ma}.\footnote{Quantum mechanics was initially developed in the language of matrices by Heisenberg~\cite{Heisenberg:1925} and only later in the language of waves by Schr\"odinger. Due to the greater familiarity of physicists with waves than with matrices, Schr\"odinger's formulation quickly became very popular.}  The electron (ignoring the spatial dependence of the wave function) can be in two states, called {\it spin up} and {\it spin down}, which we denote by
\be
\textrm{spin up} \quad \Leftrightarrow   \left(\begin{array}{c}
    1 \\ 
    0 \\ 
  \end{array}\right) \quad \textrm{} \quad \textrm{spin down} \quad \Leftrightarrow   \left(\begin{array}{c}
    0 \\ 
    1 \\ 
  \end{array}\right) \, .
\ee
The $\hat S_z$ spin operator is defined by
\be
\hat S_z = \frac{\hslash}{2}\left(\begin{array}{cc}
    1 &0\\ 
    0 &-1\\ 
  \end{array}\right) \, ,
\ee
and acts on the spin up and spin down states by ordinary matrix multiplication. Denoting the spin up and spin down state by $|\uparrow\rangle $ and $|\downarrow\rangle $, it follows that
\begin{eqnarray}
\hat S_z |\uparrow\rangle &  = & \frac{\hslash}{2}\left(\begin{array}{cc}
    1 &0\\ 
    0 &-1\\ 
  \end{array}\right) 
 \left(\begin{array}{c}
    1 \\ 
    0 \\ 
  \end{array}\right)
=\frac{\hslash}{2}  \left(\begin{array}{c}
    1 \\ 
    0 \\ 
  \end{array}\right) \nonumber \\
& = & \frac{\hslash}{2} |\uparrow\rangle 
\label{spinup}
\end{eqnarray}
and similarly
\begin{eqnarray}
\hat S_z |\downarrow\rangle & = & \frac{\hslash}{2}\left(\begin{array}{cc}
    1 &0\\ 
    0 &-1\\ 
  \end{array}\right) 
 \left(\begin{array}{c}
    0 \\ 
    1 \\ 
  \end{array}\right)
=-\frac{\hslash}{2}  \left(\begin{array}{c}
    0 \\ 
    1 \\ 
  \end{array}\right) \nonumber \\
& = & -\frac{\hslash}{2} |\downarrow\rangle \, .
\label{spindown}
\end{eqnarray}
From (\ref{spinup}) and (\ref{spindown}) we conclude that the spin up state $|\!\uparrow\rangle $ has eigenvalue $\hslash/2$ under the operator $\hat S_z$, and the spin down state $|\!\downarrow\rangle $ has eigenvalue $-\hslash/2$. This corresponds precisely to the two spots observed on the detection screen. When the electron is in the up state, it is deflected upwards by the Stern-Gerlach apparatus, whereas when it is in the down state, it will hit the spot lower on the screen.

According to the rules of quantum mechanics, all linear combinations of $|\uparrow \rangle$ and $|\downarrow \rangle$ should also be physical states. Therefore, any state of the form
\be
\alpha \left(\begin{array}{c}
    1 \\ 
    0 \\ 
  \end{array}\right) + \beta \left(\begin{array}{c}
    0 \\ 
    1 \\ 
  \end{array}\right) = 
\left(\begin{array}{c}
    \alpha \\ 
    \beta \\ 
  \end{array}\right)
\ee
is a viable physical state, for any complex numbers $\alpha$ and $\beta$. (Requiring normalization yields  $|\alpha|^2 + |\beta|^2 =1$.)Now, what happens to such an state when it enters the Stern-Gerlach apparatus? We have already seen that the Schr\"{o}dinger equation (for whatever system of hamiltonian) is linear. Thus, if a state is the sum of two parts, each part will evolve in its own way -- as if the other was not there. From this we infer that the $\alpha|\!\uparrow\rangle$ component will travel upwards, while the $\beta|\downarrow\rangle$ will fly downwards. As a consequence the wave function will {\it split} in two parts. The total amplitudes of the two parts of the wave function are given by $|\alpha|^2$ and $|\beta|^2$. We know that the measuring device (detecting screen) will then report impact on the upper spot with probability $|\alpha|^2$ , and impact on the lower spot with probability $|\beta|^2$. Note that the electron can be: {\it (i)}~in the up state; {\it (ii)}~in the down state; {\it (iii)}~in a superposition.
Nevertheless, for any of these possibilities there will only be one place at which detection occurs, so it \textit{always} looks as if the electron was strictly up or down, even if the state entering the apparatus was a mixture~\cite{Gaasbeek}. 

The preceeding discussion clarifies the $z$-component of the spin, but what about the other components, $x$ and $y$? Interestingly, Stern-Gerlach apparatuses are like Lego pieces: you can put them together in many different ways, and have lots of fun. Consider drilling a hole in the detecting screen, at the spin up spot, and behind it placing another Stern-Gerlach apparatus. The outcome is not so surprising: the screen after the second apparatus will only detect spin up. Indeed, the hole served as some kind of  filter letting through only the $|\uparrow\rangle $ states. Therefore, at a second splitting, there will only be particles going upwards, as confirmed by experiment. Now we can do something more radical. Keep the first magnet in the same position, but twist the second apparatus over an angle of $90^\circ$, so that the $\vec B$-field is in the $x$-direction. In this way it selects states of the $\hat S_x$ operator, the $x$-component of their intrinsic magnetic moment. The outcome is not so surprising, as we again see two spots on the final screen: one with magnetic moment $S_x = \frac{\hslash}{2}$, and one spot for $S_x = - \frac{\hslash}{2}$. As expected, the state $|\uparrow\rangle$ of definite $S_z$ does not have a definite $S_x$. It is a superposition of an spin up and spin down when viewed in the $x$-direction. A similar result holds when studying $S_y$. 

Both $S_x$ and $S_y$ are, like $S_z$,  physical quantities which can be measured. They have to correspond to \textit{operators} acting on the spin state. So, just like $\hat S_z$ they have to be represented by a two-by-two matrix acting on the states $(^\alpha_\beta)$. Without loss of generality, we can write the $\hat S_x$ and $\hat S_y$ opertaors in terms of complex numbers $a_x, \cdots, d_y$,  
\be
\hat S_x = \left(\begin{array}{cc}
    a_x &b_x\\ 
    c_x & d_x \\ 
 \end{array}\right) \quad \textrm{and } \quad 
\hat S_y =\left(\begin{array}{cc}
    a_y & b_y \\ 
    c_y & d_y \\ 
 \end{array}\right) \, .
\label{operameesta}
\ee
How can we guess the form of $S_x$ and $S_y$? Because the operators $S_x$ and $S_y$ are observables, they have to be hermitian. Hence the corresponding matrices need to be so too, and  (\ref{operameesta}) becomes
\be
\left(\begin{array}{cc}
    a & b \\ 
    c & d \\ 
 \end{array}\right) = \left(\begin{array}{cc}
    a^* & c^* \\ 
    b^* & d^* \\ 
 \end{array}\right).
\ee
This is a very restricting property. It is easily seen that (up to linear combinations) there are only 4 hermitian 2-by-2 matrices: the identity $\mathds{1}$ and the three Pauli matrices $\sigma_1$, $\sigma_2$, and $\sigma_3$, 
\begin{eqnarray}
\mathds{1} & = \left(\begin{array}{cc}
    1 & 0 \\ 
    0 & 1 \\ 
 \end{array}\right), 
\quad \quad &
\sigma_1 =
\left(\begin{array}{cc}
    0 & 1 \\ 
    1 & 0 \\ 
 \end{array}\right), \nonumber \\
\sigma_2 & = \left(\begin{array}{cc}
    0 & -i\\ 
    i & 0 \\ 
 \end{array}\right), \quad \quad
&\sigma_3 = \left(\begin{array}{cc}
    1 & 0 \\ 
    0 & -1 \\ 
 \end{array}\right). 
\end{eqnarray}
Note that $\sigma_3$ is $S_z$ modulo the factor $\hslash/2$ which has been stripped of. A fond guesser might then conjecture that
\be
\hat S_x =\frac{\hslash}{2}\left(\begin{array}{cc}
    0 & 1 \\ 
    1 & 0 \\ 
 \end{array}\right)\quad \textrm{and} \quad
\hat S_y =\frac{\hslash}{2}\left(\begin{array}{cc}
    0 & -i \\ 
    i & 0 \\ 
 \end{array}\right) \,,
\ee
which is actually a good guess. To convince you a bit, we will use the above guess to explain the outcome of the  ``twisted'' Stern-Gerlach experiment. On the first place, 
\be
\hat S_x \left(\begin{array}{c}
    1  \\ 
    0  \\ 
 \end{array}\right)=\frac{\hslash}{2} \left(\begin{array}{cc}
    0 & 1 \\ 
    1 & 0 \\ 
 \end{array}\right) \left(\begin{array}{c}
    1 \\ 
    0  \\ 
 \end{array}\right) = \frac{\hslash}{2}\left(\begin{array}{c}
    0 \\ 
    1  \\ 
 \end{array}\right)
\ee
so clearly $|\uparrow\rangle$ is \textit{not} and state with definite $S_x$. However, it can be broken in two parts
\be
\left(\begin{array}{c}
    1  \\ 
    0  \\ 
 \end{array}\right) = 
\left(\begin{array}{c}
    1/2  \\ 
    1/2  \\ 
 \end{array}\right) +
\left(\begin{array}{c}
    1/2  \\ 
    -1/2 \\ 
 \end{array}\right)
\ee
And these parts \textit{are} eigenstates of $S_x$:
\bea
\hat S_x \left(\begin{array}{c}
    1/2  \\ 
    1/2  \\ 
 \end{array}\right)&=&
\frac{\hslash}{2}\left(\begin{array}{c}
    1/2 \\ 
 1/2  \\ 
 \end{array}\right) \, , \nonumber\\
\hat S_x \left(\begin{array}{c}
    \phantom{-}1/2  \\ 
    -1/2  \\ 
 \end{array}\right)&=&
-\frac{\hslash}{2}\left(\begin{array}{c}
    \phantom{-} 1/2 \\ 
    -1/2  \\ 
 \end{array}\right) \, .
\eea
This explains the outcome of the ``twisted'' Stern-Gerlach experiment. The first part selects the state to be in $|\uparrow \rangle$. This is an eigenstate of $\hat S_z$, but not of $\hat S_x$. Hence, when passing through the second $\vec B$-field, the state gets split again.  The part $(^{1/2}_{1/2})$ had positive $S_x$, hence feels a force $F_x > 0$ and goes to positive $x$. The part $(^{\phantom{-}1/2}_{-1/2})$ has negative $S_x$, so it is subject to $F_x < 0$ and goes the other way. Similar reasonings hold when considering $\hat S_z$ and $\hat S_y$ or $\hat S_x$ and $\hat S_y$: if an electron is in a definite state with respect to one of the three operators, it will be a superposition of the up and down states along any other direction, i.e. $[\hat S_i, \hat S_j] = i \hslash \, \epsilon_{ijk} \hat S_k$. \\

{\bf EXERCISE 11.1}~{\it(i)}~Show that $\{\mathds 1, \sigma_1, \sigma_2, \sigma_3 \}$
are linearly independent.  {\it (ii)}~Prove that $\{\mathds 1, \sigma_1, \sigma_2,
\sigma_3 \}$ form a basis in $2\times 2$ matrix space, by showing that
any arbitrary matrix $$\mathbb M= \left(\begin{array}{cc} m_{11} &
    m_{12} \\ m_{21} & m_{22} \end{array} \right)$$ can be written on
the form $\mathbb M = a_0 \mathds 1 + \vec a \, .\, \vec \sigma$, where
$a_0 = \frac{1}{2} {\rm Tr}~(M)$, $\vec a = \frac{1}{2} {\rm
  Tr}~(\mathbb M \vec \sigma),$ and $\vec \sigma = (\sigma_1, \sigma_2,
\sigma_3)$ is the Pauli vector.

\section{Klein-Gordon Equation}

Unless otherwise stated hereafter we work with natural
(particle physicist's) Heaviside-Lorentz units $\hslash = c = 1$. In
natural units the quantities energy, momentum, mass, (length)$^{-1}$,
and (time)$^{-1}$ all have the same dimension.\\

{\bf EXERCISE 12.1}~In a unit system where $\hbar = c = 1$ show that:
{\it (i)} $1~{\rm kg} = 5.62 \times 10^{26}~{\rm GeV};$ {\it (ii)}~$1~{\rm GeV}^{-2} = 0.389~{\rm mb};$
{\it (iii)}~$1~{\rm m} = 5.068 \times 10^{15}~{\rm GeV}^{-1};$
{\it (iv)}~$1~{\rm s} = 1.52 \times 10^{24}~{\rm GeV}^{-1}$.
{\it (v)}~The Compton wavelength for an electron is given by $\lambda_c = m_e^{-1},$ calculate the numerical value;
{\it (vi)}~the Bohr radius of a hydrogen atom is given by $r_B = (\alpha m_e)^{-1},$ calculate the numerical value;
{\it (vii)}~the velocity of an electron in the lowest Bohr orbit is $\alpha$, calculate the numerical value. Show that:
{\it (viii)}~due to the fact that the electromagnetic interaction is relatively weak, we can use the non-relativistic Schr\"odinger equation to describe the hydrogen atom;
{\it (ix)}~the energy scale where quantum gravity effects become important  is $M_{\rm Pl} \sim 10^{19}~{\rm GeV}.$ Estimate the length scale at which this happens.
[{\it Hint:} $\hbar c = 197.3~{\rm MeV \, fm}$, $c = 3 \times 10^8~{\rm m/s}$, $\alpha = 1/137,$  $m_e = 0.511~{\rm MeV},$
 $G = 6.67 \times 10^{-11}~{\rm J \, m/kg}^2$.]\\

The wave equation (\ref{HM-Sch}) violates Lorentz invariance and is not suitable for a particle moving relativistically. 
It is tempting to repeat the steps of Sec.~\ref{sec:sch}  but starting from the relativistic energy momentum relation (110)~\cite{Anchordoqui:2015xca}. Making the operator substitution (\ref{operadores}), we obtain the hyperbolic wave 
equation for the quantum mechanical description of a relativistic free particle
\begin{equation}
- \frac{\partial^2 \psi}{\partial t^2} + \nabla^2 \psi = m^2 \psi \,,
\label{KG1}
\end{equation}
which is known as the Klein-Gordon equation~\cite{Klein:1926tv,Gordon:1926}. Introducing the covariant form of (\ref{operadores}), 
$
p^\mu \to i \partial^\mu$,
with
\begin{equation}
\partial^\mu = \left(\frac{\partial}{\partial t}, - \vec \nabla \right)
\quad
{\rm and}
\quad
\partial_\mu = \left(\frac{\partial}{\partial t}, \vec \nabla \right) \,,
\end{equation}
we can form the invariant (D'Alembertian) operator
$\Box^2 \equiv \partial_\mu \partial^\mu$ 
and rewrite (\ref{KG1}) as
\begin{equation}
\partial_\mu \partial^\mu \psi + m^2 \psi \equiv (\Box^2 + m^2) \psi = 0 \, .
\label{Klein-Gordon}
\end{equation}
Recall  $\psi (\vec{x}, t)$ is a scalar, complex-valued 
wave function.  Multiplying (\ref{KG1}) by $-i\psi^*$ and the
complex conjugate equation by $-i\psi$, and subtracting, leads the continuity
equation 
\begin{widetext}
\begin{equation}
\partial_ t \ \underbrace{[ i (\psi^*\, \partial_t \psi - \psi \, \partial_t \psi^*)]}_{\rho} + \vec \nabla . \underbrace{[ -i (\psi^* \, \vec \nabla \psi - \psi \, \bm \nabla \psi^*)]}_{\vec \jmath} = 0 \, . 
\label{continuity33}
\end{equation}
\end{widetext}

{\bf EXERCISE 12.2}~Convince yourself that the continutiy equation follows from (\ref{KG1}).\\

Considering the motion a free particle of energy $E$ and momentum $\vec p$,
described by Klein-Gordon solution, 
\begin{equation}
\psi = N \, e^{i (\vec p . \vec x - Et)} \, \, ,
\label{KG-solution}
\end{equation}
 from (\ref{continuity33}) we find
$\rho =2 \, E \, |N|^2$ and $\vec \jmath = 2 \, \vec p \, |N|^2.$
We note that the probability density $\rho$ is the timelike component
of a 4-vector 
\begin{equation}
\rho \propto E = \pm (\vec p^{\phantom{1}\! 2} + m^2)^{1/2} \, . 
\end{equation}
Thus, in addition to the acceptable $E>0$ solutions, we have negative
energy solutions which have associated a negative probability
density. We cannot simply discard the negative energy solutions as we
have to work with a complete set of states, and this set inevitably
includes the unwanted states.

Pauli and Weisskopf gave a natural interpretation to positive and
negative probability densities by inserting the charge $e$ into
(\ref{continuity33}),
\begin{equation}
 j^\mu = - i \, e \, (\psi^*\,\, \partial^\mu \psi - \psi \,\, 
\partial^\mu \psi^*) \, ,
\label{Pauli-Weisskopf}
\end{equation}
and interpreting it as the electromagnetic charge-current
density~\cite{Pauli:1934}.   With this in mind, $j^0$ represents a charge
density, not a probability density, and so the fact that it can be
negative is no longer objectable. In some sense, which we will make
clear in a moment, the $E<0$ solutions may then be regarded as $E>0$
solutions for particles of opposite charge (antiparticles).

The prescription for handling negative energy configurations was put
forward by St\"uckelberg~\cite{Stuckelberg} and by
Feynman~\cite{Feynman:1948km,Feynman:1948fi,Feynman:1949hz,Feynman:1949zx}.  Expressed most simply, the idea is that a negative energy
solution describes a particle which propagates backwards in time or,
equivalently, a positive energy {\em antiparticle} propagating forward in
time. 

Consider a spin-0 particle of energy $E$, 3-momentum $\vec p$, and
charge $e$, generally referred to as the ``spinless electron.'' From
(\ref{KG-solution}) and (\ref{Pauli-Weisskopf}), we know that the
electromagnetic 4-vector current is
\begin{equation}
j^\mu (e^-)
= - 2 e |N|^2 (E,\ \vec p) \, . 
\end{equation}
Now, taking its antiparticle $e^+$ 
of the same ($E, \vec p$), because its charge is $+e$, we obtain
\begin{eqnarray}
j^\mu (e^+) & = & 
=+ 2 e
|N|^2 (E, \vec p) \nonumber \\
 &= & - 2 e |N|^2 (-E, -p) \, , 
\label{simetriaC}
\end{eqnarray}
which is exactly the same as the current of the original particle
with $-E, -\vec p$. Hence, as far as a system is concerned, the
emission of an antiparticle with energy $E$ is the same as 
the absorption
of a particle of energy $-E$. 
In other words, negative-energy particle solutions going backward in
time describe positive-energy antiparticle solutions going forward in
time. Of course the reason why this identification can be made is
simply because 
\begin{equation}
e^{-i(-E)(-t)} = e^{-iEt} \, .
\end{equation}

A point worth noting at this juncture: No
  spinless quark or lepton has ever been observed in an
  experiment. Spinless hadrons exist ({\it e.g.} the $\pi$-meson), but
  they are complicated composite structures of spin-$\frac{1}{2}$
  quarks and spin-1 gluons. The spin-zero leptons, that is, leptons
  satisfying the Klein-Gordon equation, are completely fictitious
  objects. We have ignored the complications due to the spin of the
  electrons, leaving complete developments of Dirac equation~\cite{Dirac:1928hu}  to specialized textbooks;
  see e.g.~\cite{Bjorken:1965zztop}.

\section{The Standard Model}

The ``Standard Model'' is our most modern attempt to answer
two simple questions that have been perplexing (wo)mankind
throughout the epochs: What is the Universe made of? Why is our world
the way it is?

\subsection{Quarks, leptons, and gauge bosons}

  If we look deep inside a rock, we can see that it is
made up of only a few types of elementary ``point-like'' particles.  The elementary-particle model accepted today views quarks
and leptons as the basic constituents of ordinary matter.  By pointlike, we understand that quarks and leptons show no evidence
of internal structure at the current limit of our resolution, which is
about $r \sim 2 \times 10^{-20}~{\rm m}$.

\begin{figure}[tbp] \postscript{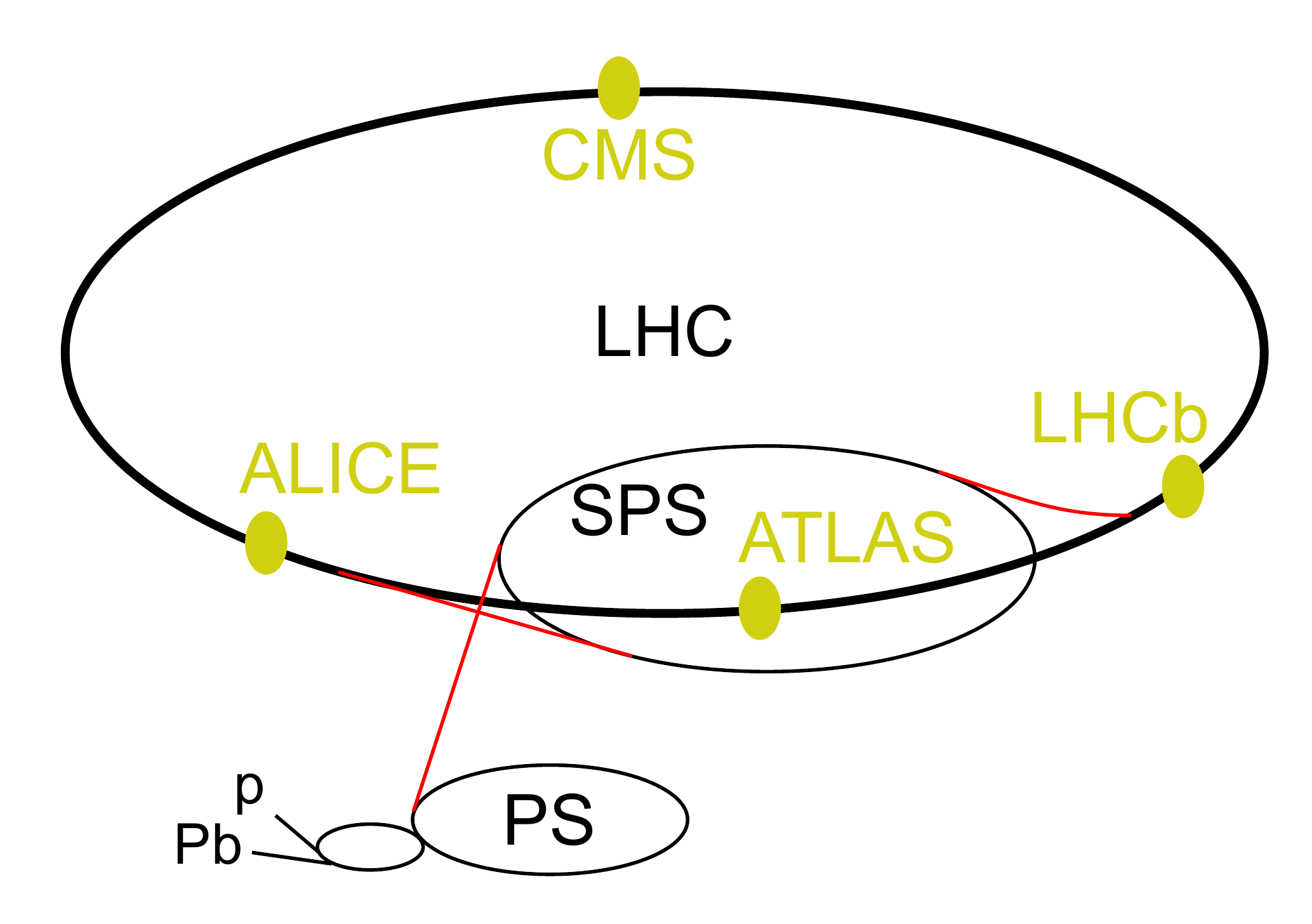}{0.9} \caption{The four large detectors at the LHC. ATLAS and CMS are designed to cover the full solid angle -- a goal that is achieved using a cylindrical configuration with a central barrel and end-caps on both sides that also detect particles traversing the detector under a shallow angle with respect to the beamline (i.e. in the large pseudo-rapidity region). LHCb and ALICE are conceptualized for studies at well-defined solid angles. Both consist of a main detector section and an extended forward arm in one direction. Within the chosen solid angle of an experiment -- be it the full solid angle or only a fraction of it -- no particle should escape detection, except for weakly interacting particles (such as e.g. neutrinos) that do not interact on the length scale of a typical detector. Alongside  measuring cross sections at central rapidity in $pp$ colissions, ALICE is optimized to study heavy-ion (Pb-Pb nuclei) collisions at a CM energy of 2.76~{\rm TeV} per nucleon pair. The resulting temperature and energy density are expected to be high enough to produce quark-gluon plasma, a state of matter wherein quarks and gluons are free. Similar conditions are thought to have existed a fraction of the second after the Big Bang before quarks and gluons bound together to form hadrons.}
\label{LHC}
\end{figure}

The colossal microscope attaining such an increidble resolution is the Large Hadron Collider (or LHC), a machine for collisions of ultrarelativistic protons~\cite{Evans:2008zzb}.  A schematic representation of the acceleration process is visible in Fig.~\ref{LHC}. At the beginning Hydrogen atoms are injected into the source chamber of the linear accelerator, CERN's Linac2, where the electrons are srtipped off to become packets of protons, which are subsequently accelerated by an electric field. The beam of protons is further accelerated at the booster, the Proton Synchrotron (PS), and the Super Proton Synchrotron (SPS) up to energies of 1.4~GeV, 25~GeV, and 450~GeV, respectively. The packets of protons are then launched into the orbit of the gigantic Large Hadron Collider that has a circumference of 27~km. There are two vacumm pipes containing proton beams travelling in opposite directions. The conunter rotating beams cross over in the four detectors (ATLAS~\cite{Aad:2008zzm}, CMS~\cite{Chatrchyan:2008aa}, LHCb~\cite{Alves:2008zz}, ALICE~\cite{Aamodt:2008zz}) caverns where they can be made to collide. It is the debris of this collisions that are tracked in the detectors. Remarkably, 70\% of the energy carried into the collision by the protons emerges perpendicular to the incident beams. At a given transverse energy $E_\perp$, we may roughly estimate the resolution as $r \approx \hslash c/E_\perp \approx 2 \times 10^{-19}~{\rm TeV} \, {\rm
  m}/E_\perp$. For CM collisions of 13~TeV, we obtain a
resolution of $2 \times 10^{-20}~{\rm m}$~\cite{ATLAS:2015nsi,Khachatryan:2015dcf}. (Throughout LHC8 refers to the run at $\sqrt{s} = 8~{\rm TeV}$ and LHC14 to the future run at $\sqrt{s} = 14~{\rm TeV}$.)

Now, an understanding of how the world is put together requires a theory of how quarks and leptons interact with one another. Equivalently, it requires a theory of the basic forces of nature. Four such forces have been identified. Two of the forces, gravitation and electromagnetism, have an unlimited range; largely for this reason they are familiar to everyone. They can be felt directly as agencies that pull or push. The remaining forces, which are called simply the weak force and the strong force, cannot be perceived directly because their influence extends only over a short range, no larger than the radius of an atomic nucleus. The strong force binds together the quarks inside the protons, the neutrons, and various other particles generically called hadrons. Indirectly, it also binds protons and neutrons into atomic nuclei. The weak force is mainly responsible for the decay of certain particles.  Its best-known effect is to transmute a down quark into an up quark, which in turn causes a neutron to become a proton plus an electron and a neutrino.

\begin{table}[t]
\caption{Relative force  strength for protons in  a nucleus.} 
\begin{tabular}{cc}
\hline
\hline
~~~~~~~~~~~Force~~~~~~~~~~~~&~~~~~~~~~~~~Relative Strength~~~~~~~~~~~~\\
\hline
Strong  & 1  \\
Electromagnetic & $10^{-2}$  \\
Weak  & $10^{-6}$ \\
Gravitational & $10^{-38}$  \\
\hline
\hline
\end{tabular}
\label{fstrength}
\end{table}

The forces can be characterized on the basis of the following four criteria: the types of particles that experience the force,  the relative strength of the force, the previously alluded range over which the force is effective, and the nature of the particles that mediate the force.  The electromagnetic force is carried by the photon, the strong force is mediated by gluons, the $W$ and $Z$ bosons transmit the weak force, and the quantum of the gravitational force is called the graviton.  A comparison of the (approximate) relative force strengths for two protons inside a nucleus is given in Table~\ref{fstrength}. Though gravity is the most obvious force in daily life, on a nuclear scale it is the weakest of the four forces and its effect at the particle level can nearly always be ignored.

The quarks are fractionally charged spin-$\frac{1}{2}$ strongly interacting
objects which are known to form the composites collectively called
hadrons~\cite{GellMann:1961ky,Ne'eman:1961cd,GellMann:1964nj}:
\begin{widetext}
\begin{equation}
\left\{ 
\begin{tabular}{llll} 
 $q \bar q$  &~~~~(quark + antiquark)~~~~&~~~~mesons~~~~& 
{\rm integral}\ {\rm spin} $\to$ {\rm Bose-Eisntein}\ {\rm statistics}~\cite{Bose:1924s,Einstein:1925s} \\
 $qqq$  & ~~~~(three quarks)~~~~ & ~~~~baryons~~~~ & 
   half-integral spin $\to$ Fermi-Dirac statistics~\cite{Fermi:1926s,Dirac:1926s} 
\end{tabular}
\right. . 
\label{GellM}
\end{equation}
\end{widetext}
There are six different types of quarks, known as flavors, forming three generations; their properties are given in Table~\ref{table:ql}.  (Antiquarks have opposite signs of electric charge.)

Quarks are fermions with spin-$\frac{1}{2}$ and therefore should obey the Pauli exclusion principle~\cite{Pauli}; see Appendix~\ref{appE}. Yet for three particular baryons ($\Delta^{++} = uuu$, $\Delta^-= ddd,$ and $\Omega^- = sss$), all three quarks would have the same quantum numbers, and at least two quarks have their spin in the same direction because there are only two choices, spin up $(\uparrow)$ or spin down $(\downarrow).$ This would seem to violate the exclusion principle!

Not long after the quark theory was proposed, it was suggested that
quarks possess another ``charge'' which enables them to interact
strongly with one another. This ``charge'' is a three-fold degree of
freedom which has come to be known as color~\cite{Fritzsch:1973pi}, and so the field theory
has taken on the name of quantum chromodynamics, or QCD.
Each quark flavor can have three colors usually designated red, green,
and blue. The antiquarks are colored antired, antigreen, and antiblue.
Baryons are made up of three quarks, one with each color. Mesons
consist of a quark-antiquark pair of a particular color and its
anticolor. Both baryons and mesons are thus colorless or white.
Because the color is different for each quark, it serves to
distinguish them and allows the exclusion principle to hold.  Even
though quark color was originally an {\it ad hoc} idea, it soon became
the central feature of the theory determining the force binding quarks
together in a hadron.

One may wonder what would happen if we try to see a single quark with
color by reaching deep inside a hadron. Quarks are so tightly bound to
other quarks that extracting one would require a tremendous amount of
energy, so much that it would be sufficient to create more quarks.
Indeed, such experiments are done at modern particle colliders and all
we get is not an isolated quark, but more hadrons (quark-antiquark
pairs or triplets). This property of quarks, that they are always
bound in groups that are colorless, is called confinement.  Moreover,
the color force has the interesting property that, as two quarks
approach each other very closely (or equivalently have high energy),
the force between them becomes small. This aspect is referred to as
asymptotic freedom~\cite{Gross:1973id,Politzer:1973fx}.
 When probed at small distances compared to the size of a hadron
(i.e., about 1~fm = $10^{-15}$~m) the ``bare'' masses of the quarks are those  given in Table~\ref{table:ql}. 
However, the effective quark masses in composite hadrons are
significantly larger; namely, $0.3$~GeV, $0.3$~GeV, $0.5$~GeV,
$1.5$~GeV and $4.9$~GeV, for $u$, $d$, $s$, $c$, and $b$;
respectively. The lightest flavors are generally stable and are very
common in the universe as they are the constituents of protons ($uud$)
and neutrons ($ddu$). More massive quarks are unstable and rapidly
decay; these can only be produced as quark-pairs under high energy
conditions, such as in particle accelerators and in cosmic rays.

Leptons are fractionally spin-$\frac{1}{2}$ particles which do
not strongly interact. There are six types of leptons, known as flavours, forming three generations (see Table~\ref{table:ql}). Each charged flavor has an associated neutrino.  

One important aspect of on-going research is the attempt to find a unified basis for the different forces.  For example, the weak and electromagnetic forces are indeed two different manifestations of a single, more fundamental {\em electroweak} interaction~\cite{Glashow:1961tr,Weinberg:1967tq,Salam:1968rm}. The electroweak theory has had many notable successes, culminating in the discovery of the predicted Higgs boson~\cite{Higgs:1964pj,Englert:1964et} by the ATLAS~\cite{ATLAS:2012ae} and CMS~\cite{Chatrchyan:2012tx} collaborations.\\

\begin{table*}
\caption{The three generations of quarks and leptons in the Standard Model. \label{table:ql}}
\begin{tabular}{ccccccc}
\hline
\hline 
~~~~~~~~~~~~~~~~~~~~~~~~~~~~~~~~~& ~~~~Fermion~~~~ & ~~~~Short-hand~~~~ & ~~~~Generation~~~~ & ~~~~Charge~~~~ & ~~~~Mass~~~~ & ~~~~Spin~~~~ \\
\multirow{7}{4em}{Quarks} & up & $u$ & I & & $2.3^{+0.7}_{-0.5}~{\rm MeV}$ & \\
& charm & $c$ & II & $+\frac{2}{3}$ & $1.275 \pm 0.025~{\rm GeV}$ & $\frac{1}{2}$ \\
& top & $t$ & III & & $173.21\pm 0.51~{\rm GeV}$ & \\
\cline{4-7}
& down & $d$ & I &  & $4.8^{+0.5}_{-0.3}~{\rm MeV}$ & \\
& strange & $s$& II & $-\frac{1}{3}$ & $95^\pm5~{\rm MeV}$ & $\frac{1}{2}$ \\
& bottom & $b$& III &  & $4.18\pm 0.03~{\rm GeV}$ & \\
\hline
\multirow{7}{4em}{Leptons} & electron neutrino & $\nu_e$ & I & & $<2~{\rm eV}$ 95\% CL & \\
& muon neutrino & $\nu_\mu$ & II & 0 & $<0.19~{\rm MeV}$ 90\% CL & $\frac{1}{2}$ \\
& tau neutrino & $\nu_\tau$ & III & & $< 18.2~{\rm MeV}$ 95\%CL \\
\cline{4-7}
& electron &$e$ & I & & 0.511~{\rm MeV} & \\
& muon & $\mu$ & II & $-1$ & $105.7~{\rm MeV}$ & $\frac{1}{2}$ \\
& tau & $\tau$ & III & & 1.777~{\rm GeV} \\
\hline
\hline
\end{tabular}
\end{table*}

\begin{table*}
\caption{The four force carriers in the Standard Model. \label{tablegb}}
\begin{tabular}{cccccc}
\hline \hline
~~~~~~~~~Force~~~~~~~~~ & ~~~~~~~~~Boson~~~~~~~~~ & ~~~~~~~~~Short-hand~~~~~~~~~ & ~~~~~~~~~Charge~~~~~~~~~ & ~~~~~~~~~Mass~~~~~~~~~ & ~~~~~~~~~Spin~~~~~~~~~ \\
\hline
Electromagnetic & photon & $\gamma$ & 0 & 0 & 1\\
Weak & $W$ & $W^\pm$ & $\pm 1$ & $80.385 \pm 0.015~{\rm GeV}$ & 1\\
Weak & $Z$ & $Z^0$ & 0 & $91.1876 \pm 0.0021~{\rm GeV}$ & 1 \\
Strong & gluon & $g$ & 0 & 0 & 1 \\
\hline
\hline
\end{tabular}
\end{table*}

{\bf EXERCISE 13.1}~The LHC beam is not continuous. Instead, it has
a bunch structure. Bunches of particles will collide every 25 ns
at $\sqrt{s} = 14~{\rm TeV}$ with a luminosity (number of
particles per second per square centimeter) of $10^{34}~{\rm
cm}^{-2}\, {\rm s}^{-1}$. The total inelastic proton-proton cross
section at $14~{\rm TeV}$ is 70~mb. {\em (i)}~Compute the number
of interactions occurring per second as well as every time two
bunches collide. These are called minimum bias event and are of
almost no interest. {\em (ii)}~The Higgs particle is produced at the LHC mainly through gluon fusion ($gg \to H$)
which has a cross section of about 40~pb. One
of the Higgs discovery channels involves searching for the Higgs
decaying to two photons which has a probability (branching ratio)
of $\sim 0.2 \times 10^{-2}$. How many bias events are expected to
be produced for every Higgs event observed via the two photon
channel if one ignores detector
effects?\\

The main properties of the four force carriers are summarized in Table~\ref{tablegb}. In the next sections we will discuss the properties of the forces in more detail, but the style coverage will remain qualitatively. The discussion will follow closely the heuristic approach of~\cite{'tHooft:1992rg,Weinberg:1974uk,'tHooft:1980us,Quigg:1985ai,Veltman:1986br}. For a rigurous introduction to gauge theories of the strong and elecrtoweak interactions, see e.g.~\cite{Halzen:1984mc,Barger:1987nn,Quigg:1983gw,Anchordoqui:2009eg}.

\subsection{Gauge symmetries and field theories}

\begin{figure}[tbp] \postscript{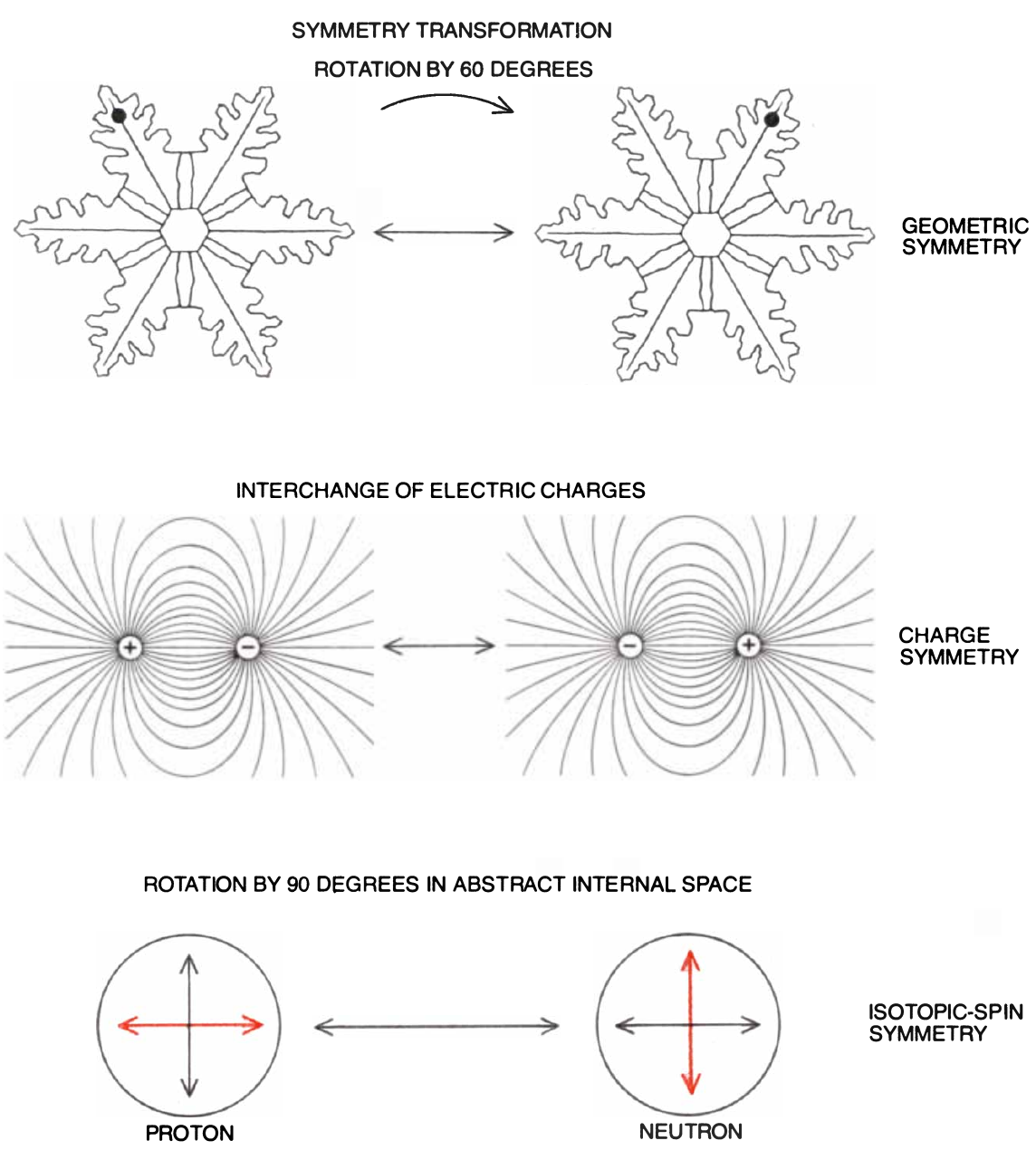}{0.9} \caption{Discrete and continuous symmetries in nature~\cite{'tHooft:1980us}.}
\label{fig:symmetries}
\end{figure}

We have seen that  since the time of Galileo and Newton symmetries and apparent symmetries in the laws of nature have played an important role in the development of physical theories. The most recognizable symmetries are spatial or geometric ones. For instance, in a snowflake the existence of a symmetrical pattern can be detected at first sight. The symmetry can be defined as an invariance in the pattern that is observed when some transformation is enforced to it. For the snowflake, the transformation is a rotation by $\pi/3$, or else by $1/6$ of a circle. In plain English, if the initial position is identified (say, the balck dot in the upper left branch of the snowflake shown in Fig.~\ref{fig:symmetries}) and the snowflake is then rotated by $\pi/3$ (or by any integer multiple of $\pi/3$), no change will be noted. The snowflake is invariant with respect to  rotations by $\pi/3$. Applying the same rule, a square is invariant with respect to  rotations by $\pi/2$ and a circle is said to have continuous symmetry because rotation by any angle leaves it unchanged. The invariance of the circle is known in the physics vernacular as $U(1)$ symmetry; see Appendix~\ref{appF}.

Despite the fact the notion of symmetry had its origin in geometry, it is general enough to encompass invariance with respect to transformations of other kinds. For example,  a nongeometric symmetry is the charge symmetry of electromagnetism shown in Fig.~\ref{fig:symmetries}. Assume a number of electrically charged particles have been arranged in some definite configuration and all the forces acting between pairs of particles have been measured. If the polarity of all the charges is then inverted,  the forces remain unchanged.

Another symmetry with nongeometric origin reveals the isotopic spin, a property of neutrons and protons. The essence of this symmetry manifests in the observation that the neutron and the proton are rather similar particles. They differ in mass by only $\approx 10\%$, and aside for their electric charge they are identical in all other properties. Then it looks like all neutrons and protons could be interchanged and the strong interactions would hardly be altered. Should the electromagnetic forces (which depend on electric charge) could somehow be turned off, the isotopic-spin symmetry would be exact; in the real world, however, it is only approximate.

Despite the fact the neutron and proton  seem to be distinct particles and it is hard to envissage a state of matter intermediate between them, it turns up that the symmetry with respect to isotopic spin is a continuous symmetry, like the symmetry of the circle rather than like that of a snowflake. We can provide a simplified explanation of why that is so. Contemplate that inside each particle there are imaginary black and red arrows crossing each other representing the proton and neutron component of the particle, respectively. As shown in Fig.~\ref{fig:symmetries}, if the proton arrow is pointing up (it makes no difference what direction is defined as up) then the particle is a proton, and  if the neutron arrow is up then it is a neutron. Intermediate positions correspond to quantum-mechanical superpositions of the two states, and the particle then looks sometimes like a proton and sometimes like a neutron. The symmetry transformation associated with isotopic spin rotates the internal indicators of all protons and neutrons everywhere in spacetime by the same amount and at the same time. If the rotation is exactly by $\pi/2$, every proton mutates to a neutron and every neutron mutates to a proton. The isotopic spin symmetry, to the extent it is exact, states that no effects of this transformation can be detected.

So far we have discussed symmetries which can be defined as global symmetries. Herein the word global means ``occurring everywhere at the same time.'' In our characterization of isotopic-spin symmetry this constraint has been made explicit: the internal rotation that transforms protons into neutrons and neutrons into protons is to be carried out everywhere in spacetime at once. Supplementary to global symmetries, which are almost always present in a physical theory, it is desirable to have a ``local'' symmetry, in which the convention can be decided independently at every point in space and at every instant of time. Even though in general ``local'' could appear as something of more modest scope than  ``global'', in fact the requirement of local symmetry places a far more stringent constraint on the structure of a theory. A global symmetry states that some law of physics remains invariant when the same transformation is applied everywhere at the same time. For a local symmetry to be observed the law of physics must retain its validity even when a different transformation takes place at each point in space and time.

The so-called ``gauge theories'' can be assembled with either a global or a local symmetry (or both), but it is the theories with local symmetry that hold particular interest here. To construct a theory invariant with respect to a local transformation something new must be added: {\it a force}. Before showing how this comes about, however, it is necessary to reexamine  how forces are described in theories of elementary-particle interactions.

Nowadays, the building blocks of particle theory include not only particles and forces, but also fields. We have already introduced the concept of a field  in Sec.~III. A field is a quantity defined at every point throughout some region of space and time. For example, the quantity might be temperature and the region might be the surface of the Earth. Then the field consists of temperature values for every point on the Earth's surface.

\begin{figure}[tbp] \postscript{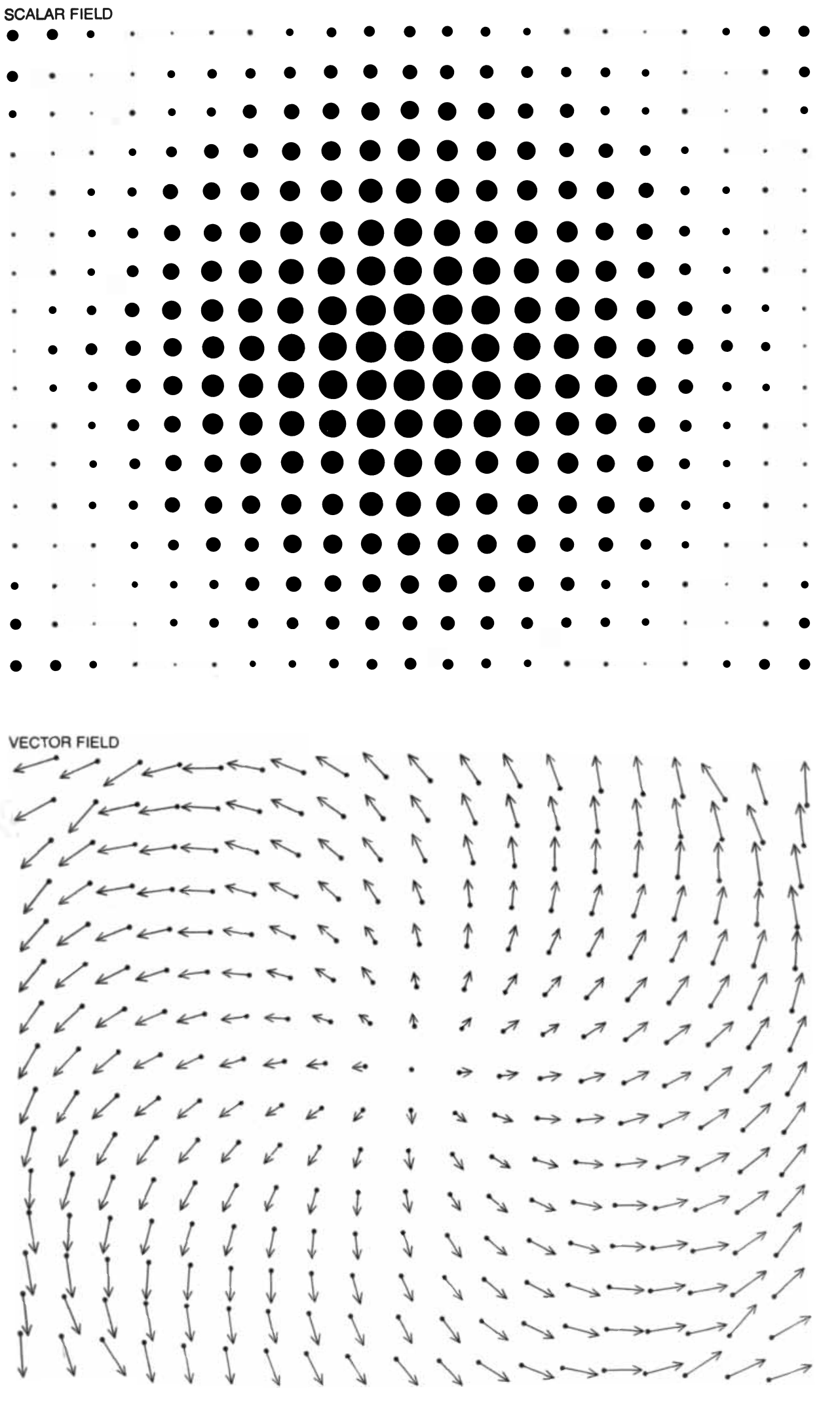}{0.9} \caption{
A scalar field (top) has only a magntude at each point; in the example shown here the magnitude is given by the area of the dots. A vector field (bottom) has both a magnitude and a direction and can be illustrated by drawing an arrow at each point. The number of components in the fields is reflected in the number of distinct orientations of the particle, which in turn depends on its spin angular momentum. A scalar field has just one component (its value can be given by a single number) and is represented by spin-0 particle with one spin state, or orientation. A vector field in 3-dimensional space has three components (a magnitude and two angles), and it corresponds to a spin-1 particle with three spin states~\cite{'tHooft:1980us}.}
\label{fig:fields}
\end{figure}

The temperature is a scalar quantity, because it can be represented by position along a line, or scale. The associated temperature field is a scalar field, in which each point has correspondence with  a single number, or magnitude. There are other types of field, such as the vector field, where at each point a vector, or arrow, is drawn. A familiar example of a vector field is the velocity field of a fluid; at each point throughout the volume of the fluid an arrow can be drawn to show the speed and direction of flow. Examples of vector and scalar fields are shown in Fig.~\ref{fig:fields}.

When studying electrically charged objects a field is a handy mechanism for expressing how the force of electromagnetism is conveyed from one point to another. All charged particles are expected to emanate an electromagnetic field; each particle then interacts with the sum of all the fields rather than directly with the other particles.

We have seen that in quantum systems the particles themselves  can be represented  as fields. For example, the  electron can be treated as a packet of waves with some finite extension in space. Conversely, it is often convenient to represent a quantum mechanical field as if it were a particle. The interaction of two particles through their interpenetrating fields can be analyzed as a two particles exchange a third particle, which is dubbed the quantum of the field. As an illustration, if two electrons, each surrounded by an electromagnetic field, approach near each other and bounce apart, we may conjecture they have exchanged a quantum of the electromagnetic field, i.e. a photon.

The exchanged quantum has only a fugacious existence. Shortly after it has been emitted it must be reabsorbed, either by the same particle or by another one, within a finite time. It cannot keep going forever, and it cannot be detected in an experiment. This kind of entities are called virtual particles. The larger their energy, the shorter their existence. In effect a virtual particle borrows or embezzles an amount of energy, but it must repay its debt before the shortage can be noticed.
The range of an interaction is linked to the mass of the exchanged quantum. If the quantum field is heavy, more energy must be borrowed in order to support its existence, and the debt must be repaid sooner lest the discrepancy be discovered. The length a virtual particle can travel before it must be re-absorbed is thus reduced and so the associated  force has a short range. For the particular case in which the exchanged quantum is massless, the range is infinite.

The number of components in a field coincides with the number of quantum-mechanical states of the field quantum. The number of possible states is in turn associated to the particle's intrinsic spin angular momentum, which can only take  discrete values, i.e. the measured magnitude  of the spin in fundamental units is always an integer or a half integer. Furthermore, it is not only the magnitude of the spin that is quantized but also its direction or orientation. (As we have seen in Sec.~\ref{Sec:Stern-Gerlach}, the spin can be defined by a vector parallel to the spin axis, and the projections, or components, of this vector along any direction in space must have values that are integers or half integers.) The number of possible orientations (a.k.a. spin states)  equals  twice the magnitude of the spin, plus one. Hence, a particle with a spin-$\frac{1}{2}$, such as the electron, has two spin states. The spin can be oriented  parallel to the particle's direction of motion or antiparallel to it.  Using the arrow notation introduced in Sec.~\ref{Sec:Stern-Gerlach} the spin orientation becomes up $|\uparrow\rangle$ or  down $|\downarrow\rangle$. A spin-1 particle has three orientations: parallel, antiparallel, and transverse. A spin-0 particle has no spin axis, because all orientations are equivalent;  it is said to have just one spin state. A scalar field that has only one component, its magnitude, must be characterized by a field quantum that has also one component: a spin-0 particle. Such a particle is then called a scalar particle. Likewise, a three-component vector field requires a spin-1 field quantum with three spin states: a vector particle. The electromagnetic field is a vector field, and the photon, in conformity with these specifications, has a spin-1. Actually, since the photon is massless and propagates at $c$ it carries a property not shared by particles with  finite mass: the transverse spin state does not exist. Even though in some formal interpretation the photon has three spin states, in practice only two of the spin states can be detected.

Electrodynamics describes how light and matter interact. It is the simplest field theory where full agreement between quantum mechanics and special relativity is achieved. We discuss this next.

\subsection{Electromagnetic interaction}

We have already noted that electromagnetism is a local field theory of fundamental interactions. The foundation of Maxwell's theory is the proposition that an electric charge is surrounded by an electric field $\vec E$ stretching to infinity, and that the movement of an electric charge gives rise to a magnetic field $\vec B$ also of infinite extent. Both $\vec E$ and $\vec B$ are vector quantities, which are defined at each point in space by a magnitude and a direction. In electromagnetism the value of $\vec E$ at any given point is determined ultimately by the charge distribution around that point. However, in general it is convenient to define a potential $V$ that is also determined by the charge distribution: the greater the density of charges in a region, the higher its potential. The electric field between two points is then given by the voltage difference between them.

The nature of the electromagnetic symmetry is evident in the following Gedanken-experiment. Assume a system of electric charges is arranged in a laboratory and we measure $\vec E$ and its properties. If the charges are stationary there can be no $\vec B$-field. In such experimental footing a global symmetry is readily noted. The symmetry transformation consists in raising the entire laboratory to a high voltage, or in other words to a higher $V$. If the measurements are then repeated, no change in $\vec E$ will be detected. The reason is simple,  $\vec E$ is determined only by differences in electric potential $\Delta V$, not by the absolute value of the potential.  This property carries a symmetry: $\vec E$ is invariant with respect to the addition or subtraction of an arbitrary overall potential. Note that this is a global symmetry, because the result of the experiment remains constant only if $V$ is changed everywhere at the same time. If $V$ were raised in one region and not in another, any experiment that crossed the boundary would be affected by the potential difference. 

A complete theory of electromagnetic fields must contain not only static arrays of charges but also moving charges. To consolidate this, the global symmetry of the theory must be converted into a local symmetry. If the electric field were the only one acting between charged particles, it would not have a local symmetry. Indeed if the charges are in motion not  only $\vec E$ but also $\vec B$ are at play. It is the effects of $\vec B$ that restore the local symmetry. In the same way $\vec E$ depends ultimately on the distribution of charges but can conveniently be derived from $V$, so the $\vec B$-field is generated by the motion of the charges but is more easily described as resulting from a magnetic potential $\vec A$. It is in this system of potential fields that local transformations can be carried out leaving all the original electric and magnetic fields unaltered. The system of dual, interconnected fields has an exact local symmetry even though $\vec E$ alone does not. Any local change in $V$ can be combined with a compensating change in $\vec A$ in such a way that $\vec E$ and $\vec B$ are invariant.
￼

Maxwell's theory of electromagnetism is a classical field theory, but a related symmetry can be demonstrated in the quantum theory of electromagnetic interactions. We have seen that in quantum mechanics massive particles must be characterized by a wave or a field. In particular, this convention can be adopted for electrons. In their quantum description a change in $V$ yields a change in the phase of the electron wave.
The electron is spin-$\frac{1}{2}$ particle and therefore it has two spin states (parallel and antiparallel). This implies that the associated field should have two components. Each component is characterized by a complex number. The electron field is a moving packet of waves, which are oscillations in the amplitudes of the real and the imaginary components of the field. At this stage, it is worthwhile to point out  that this field is not the electric field emanating from the electron but instead is a massive field. It would exist even if the electron had no electric charge. What the field defines is the probability of finding the electron in a specified spin state at a given point in space and at a given instant in time. The probability is given by the sum of the squares of the real and the imaginary parts of the field.

We have seen that in the absence of electromagnetic fields the frequency of the oscillations in the electron field is proportional to the energy of the electron $E = \hslash \omega$, and the wavelength of the oscillations is proportional to the momentum $\lambda = h/p$. In order to completely  define the oscillations an additional quantity must be determined: the phase. Recall that the phase measures the displacement of the wave from some arbitrary reference point and is generally expressed by the angle $\phi$. If at some point the real part of the oscillation, say, has its maximum positive amplitude, then at that point $\phi =0$. On the hand, where the real part falls to zero $\phi = \pi/2$, and where it reaches its negative maximum $\phi = \pi$. In general the imaginary part of the amplitude is out of phase by $\pi/2$ with the real part, so that whenever one part has a maximal value the other part is zero.

It is evident that the only way to determine $\phi$ is to disentangle the contributions of the real and the imaginary parts of the amplitude, which is a {\it mission impossible}. The sum of the squares of the real and the imaginary parts can be measured, but there is no way of resolving at any given point or at any moment how much of the total derives from the real part and how much from the imaginary part. As a matter of fact, an exact symmetry of the theory demands that the two contributions are indistinguishable. Phase differences  of the field at two points or at two moments (of course) can be measured, but never the absolute phase.
The conclusion that the phase of an electron wave is inaccessible to measurement has a corollary: {\it the phase cannot have an effect on the outcome of any conceivable experiment}. If it did so, such an experiment could be used to determine the phase. Hence the electron field exhibits a symmetry with respect to arbitrary changes of phase. In summary any phase angle $\phi$ can be added to or subtracted from the electron field and the results of all possible experiments will remain invariant.

\begin{figure}[tbp] \postscript{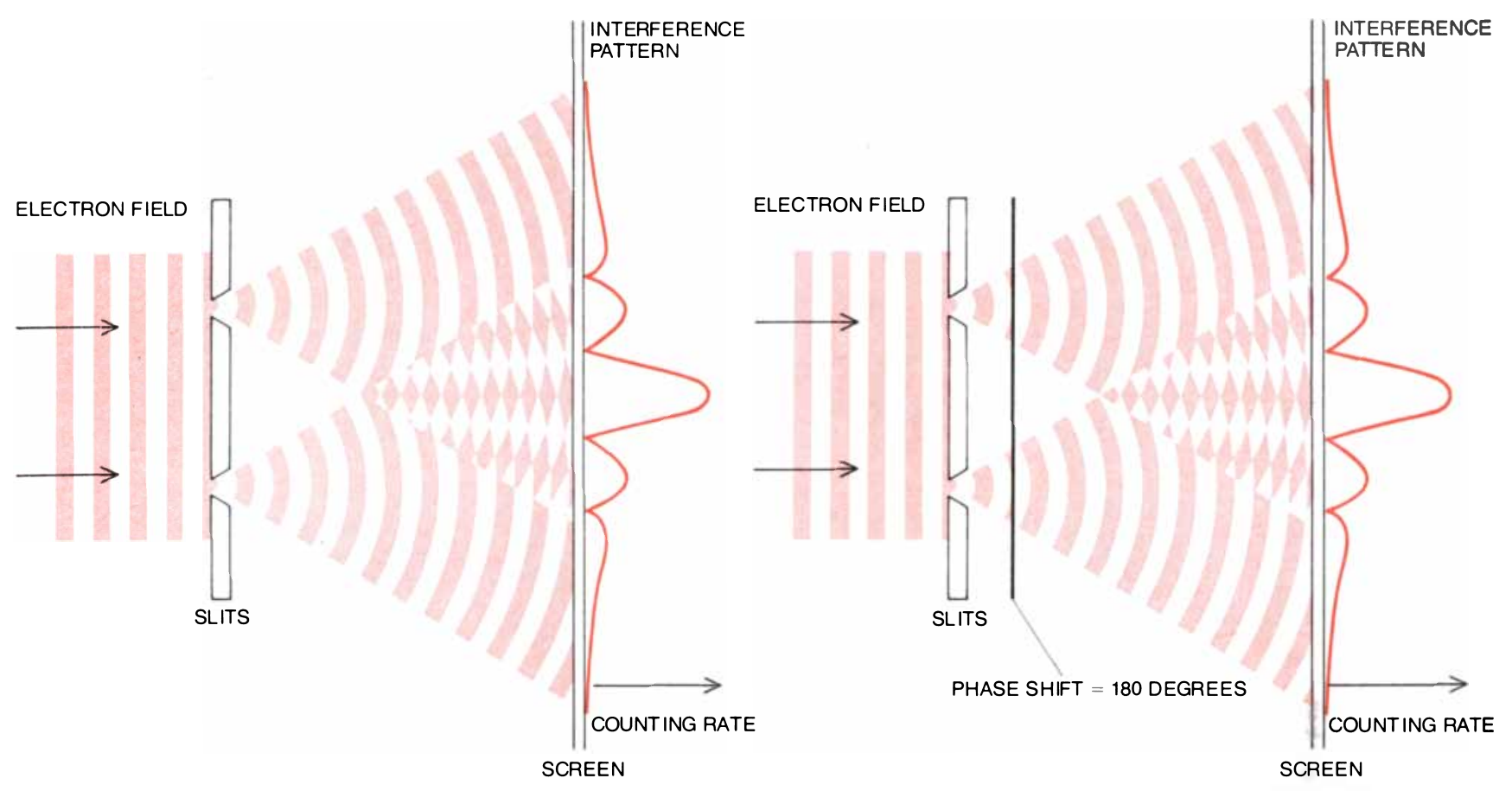}{0.99} \caption{Global gauge symmetry of electromagnetism~\cite{'tHooft:1980us}.}
\label{fig:global-symmetry}
\end{figure}

We have dropped enough clues and hints that a really attentive reader might guess this symmetry principle can be best illustrated by considering  a double-slit  experiment of electrons. Consider again a beam of massive   particles (electrons in this case) passing through two narrow slits in a screen, as shown in Fig.~\ref{fig:global-symmetry}. The number of particles reaching a second screen located at a distance $D \gg d$  is counted. As expected, the distribution of electrons across the surface of the second screen forms an interference pattern of alternating peaks and valleys.

\begin{figure}[tbp] \postscript{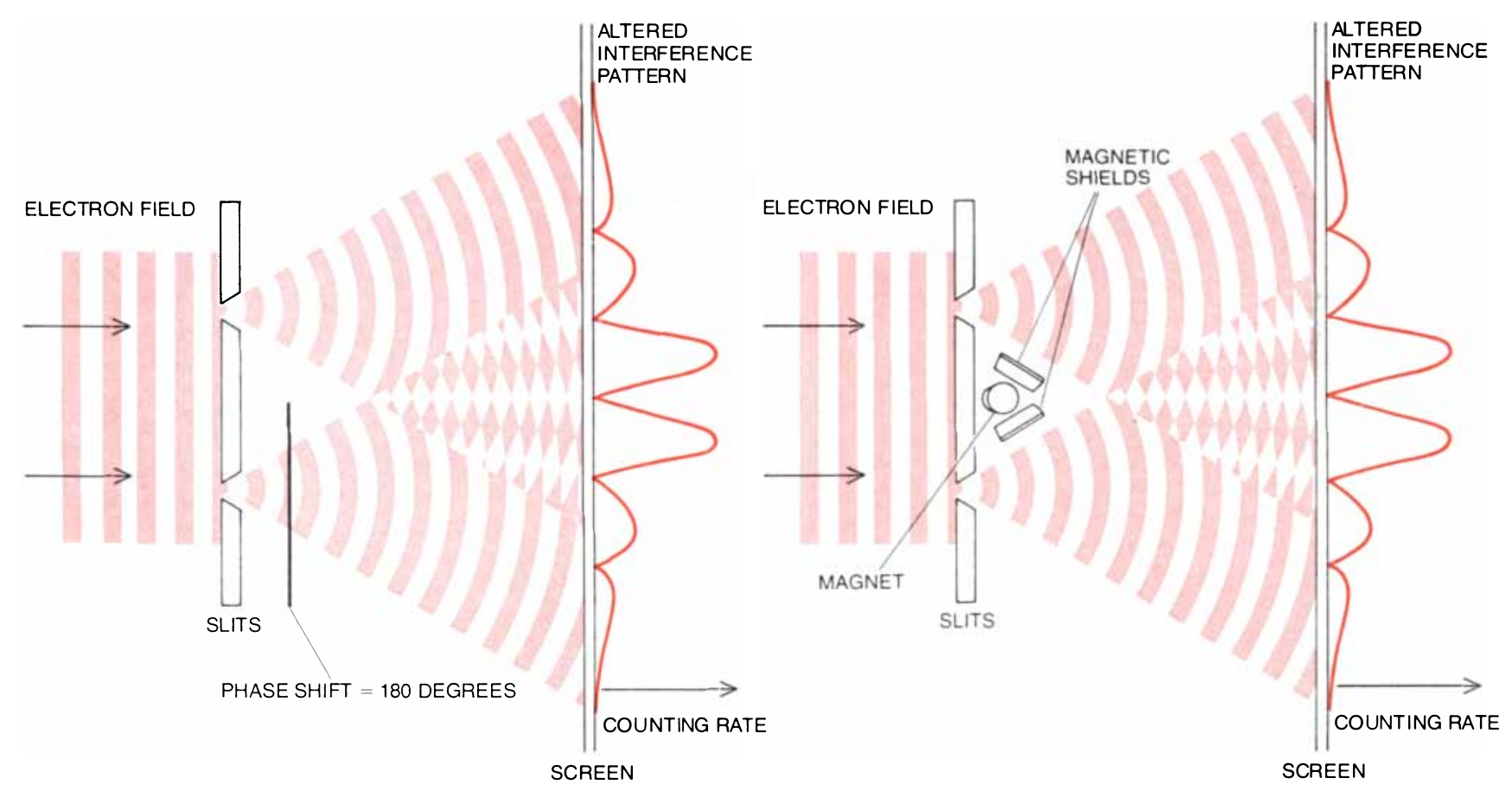}{0.99} \caption{Local gauge symmetry of electromagnetism~\cite{'tHooft:1980us}.}
\label{fig:local-symmetry}
\end{figure}

The quantum-mechanical interpretation of this experiment is that the electron wave splits into two segments on striking the first screen and the two waves then interfere with each other. Again, if the waves are in phase the interference is constructive and many electrons are counted at the second screen. However,  if the waves are out of phase a destructive interference reduces the count. Undoubtedly, it is only the difference in phase $\Delta \phi$ that regulates the pattern formed. In other words, if we shift  both waves  phases  by the same amount, the phase difference at each point would be unaffected and the same pattern of constructive and destructive interference would be observed. This particular type of symmetries, in which  the phase of a quantum field can be adjusted at will, are the so-called ``gauge symmetries.'' Despite the fact the absolute value of the phase is irrelevant to the outcome of experiments, in constructing a theory of electrons it is still necessary to specify such a phase. The choice of a particular value is known as a gauge-fixing condition.

The preceding electron symmetry  is a global symmetry. Note that  the phase of the field must be shifted in the same way everywhere in space at the same time. It is easily seen that a theory of electron fields alone, with no other forms of matter or radiation, is not invariant with respect to a corresponding local gauge transformation. Consider again the double-slit electron experiment. An initial experiment is carried out as before and the interference pattern is recorded. Then the experiment is repeated, but one of the  slits is fitted with the electron-optical equivalent of a half-wave plate, a device that shifts the phase of a wave by $\pi$. When the waves emanating from the two slits now interfere, the phase difference between them will be altered by $\pi$. As a result wherever the interference was constructive in the first experiment it will now be destructive, and vice versa. The positions of all the peaks and depressions on the observed interference pattern will be interchanged; see Fig.~\ref{fig:local-symmetry}.

Pretend  we want to make the theory consistent with a local gauge symmetry, assuming e.g. another field could be added that would compensate for the changes in electron phase. The new field would of course have to accomplish more than restore the defects in this one experiment. It would have to preserve the invariance of all observable quantities when the phase of the electron field is altered in any way from point to point in space and from time to time. Strictly speaking, the phase shift must be allowed to vary as an arbitrary function of position and time.

At first glance it may look improbable to find a field that can accommodate these specifications. However, on a second look we can figure out that the required field is a vector, corresponding to a field quantum with a spin of one unit. We can also immediately realize that the field must have infinite range, as there is no limit to the distance over which the phases of the electron fields might have to be reconciled. The need for infinite range entails that the field quantum must be massless. These are the characteristics of a familiar field: the electromagnetic field, whose quantum is the photon. How does the electromagnetic field protect the gauge invariance of the electron field?  Recall that the effect of the electromagnetic field is to transmit forces between charged particles. These forces can modify the state of motion of the particles. Of particular interest here the forces can change the phase. When an electron absorbs or emits a photon, the phase of the electron field is shifted. We have already seen that $\vec E$ itself exhibits an exact local symmetry; by describing the two fields together the symmetry can be extended to both of them.

The connection between the two fields relies in the interaction of $\vec E$ with the charge of the electron. Because of this interaction the propagation of an electron wave through the $\vec E$ field can be described properly only if $V$ is specified. Likewise, to describe an electron in a magnetic field $\vec A$ must be specified. Once these two potentials are assigned definite values the phase of the electron wave becomes fixed everywhere. However, the local symmetry of electromagnetism allows $V$ to take any arbitrary value, which can be chosen independently at every point in space and at every instant of time. For this reason the phase of the electron matter field can also take on any value at any point, but the phase will always be consistent with the convention adopted for $V$ and $\vec A$. This local structure can be reproduced in  the double-slit experiment because  the effects of an arbitrary shift in the phase of the electron wave can be mimicked by applying an electromagnetic field. As an illustration, the change in the observed interference pattern caused by interposing a half-wave plate in front of one slit could be generated instead by placing the slits between the poles of a magnet; see Fig.~\ref{fig:local-symmetry}. By observing the resulting pattern it would be impossible to discern which procedure had been followed. Since the gauge-fixing condition for $V$ and $A$ can be chosen locally, so can the phase of the electron field. The theory that results from combining electron massive fields with electromagnetic fields is known as quantum electrodynamics (QED).

We have seen that quantum mecahnics predicts probabilities of events. These probabilities must not be negative, and all the probabilities taken together must add up to one. In addition, energies must be assigned positive values but should not be infinite. It is not immediately apparent whether QED could satisfy all of these conditions. One challenge  repeatedly shows up in any attempt to calculate  the probability that one
particle will scatter off another. This is the case even in the simplest electromagnetic interaction, such as the interaction between two electrons. The likeliest sequence of events in  Bhabha scattering is that one electron emits a single virtual photon and the other electron absorbs it. Many more complicated exchanges are also possible, however; indeed, their number is infinite. For example, the electrons could interact by exchanging two photons, or three, and so on. The total probability of the interaction is determined by the sum of the contributions of all the events. This sum can be most easily carried out using Feynman diagrams, i.e.  by drawing  diagrams of the events in one spatial dimension and one time dimension, which are shorthand representations of a well-defined mathematical procedure for tabulating all of these contributions~\cite{Feynman:1948km,Feynman:1948fi,Feynman:1949hz,Feynman:1949zx}. A notably troublesome class of diagrams are those that include ``loops,'' such as the loop in spacetime that is formed when a virtual photon is emitted and later reabsorbed by the same electron. As was shown above, the maximum energy of a virtual particle is limited only by the time needed for it to reach its destination. When a virtual photon is emitted and reabsorbed by the same particle, the distance covered and the time required can be reduced to zero, and so the maximum energy can be infinite. For this reason some diagrams with loops make an infinite contribution to the strength of the interaction.
These infinities would spoil even the description of an isolated electron: because the electron can emit and reabsorb virtual particles it has infinite mass and infinite charge. The cure for this plague of infinities is the procedure called renormalization. Roughly speaking, it works by finding one negative infinity for each positive infinity, so that in the sum of all the possible contributions the infinities cancel and a finite residue could be obtained. The finite residue is the theory's prediction. It is uniquely determined by the requirement that all interaction probabilities come out finite and positive. The rationale of this procedure can be summarized as follows. When a measurement is made on an electron, what is actually measured is not the mass or the charge of the pointlike particle with which the theory begins but the properties of the electron together with its enveloping cloud of virtual particles. Only the net mass and charge, the measurable quantities, are required to be finite at all stages of the calculation. The properties of the pointlike object, which are called the ``bare'' mass and the ``bare'' charge, are not well defined.

The symmetry properties of QED are unquestionably appealing~\cite{Schwinger:1948yk,Schwinger:1948yj,Tomonaga:1946zz,Feynman:1948ur,Dyson:1949bp,Dyson:1949ha,Feynman:1950ir}. Moreover,
QED has yielded results that are in agreement with experiment to an accuracy of about one part in a billion~\cite{Schwinger:1948iu}, which makes the theory the most accurate physical theory ever devised. It is the model for theories of the other fundamental forces and the standard by which such theories are judged.\\

{\bf EXERCISE 13.2}~Consider that the fields $\vec E$ and $\vec B$ are the components of a second rank tensor,
\begin{equation}
F^{\mu \nu} = \left(\begin{array}{cccc}
0 & - E_x & - E_y & - E_z \\
  & & &  \\
E_x & 0 & -B_z & B_y \\
  & & &  \\
E_y & B_z & 0 & -B_x \\
  & & & \\
E_z &  -B_y  & B_x & 0 \\
\end{array} \right)\,,
\label{fmunu}
\end{equation}
and that $j^\mu = (\rho, \vec \jmath\, )$ is a four-vector. Maxwell's
equations of classical electrodynamics are, in vacuo,
\begin{equation}
\overset{\rightharpoonup}{\nabla} \times \overset{\rightharpoonup}{E} +  \frac{\partial \overset{\rightharpoonup}{B}}{\partial t} = 0 \, ,
\label{max1}
\end{equation}
\begin{equation}
\overset{\rightharpoonup}{\nabla}  \cdot \overset{\rightharpoonup}{E} =  \rho \,,
\label{max2}
\end{equation}
\begin{equation}
\overset{\rightharpoonup}{\nabla} \times \overset{\rightharpoonup}{B} -
\frac{\partial \overset{\rightharpoonup}{E}}{\partial t} = 
\overset{\rightharpoonup}{\jmath}, \label{max3}
\end{equation}
\begin{equation}
\overset{\rightharpoonup}{\nabla} \, \cdot\, \overset{\rightharpoonup}{B} = 0 \, ,
\label{max4}
\end{equation}
where we are using Heaviside-Lorentz rationalized units. {\it (i)}~Show that in terms of the antisymmetric field strength tensor
$F^{\mu\nu}$ Maxwell's equations can be written in the
compact forms
\begin{eqnarray}
\partial_\mu F^{\mu \nu} & = &  j^\nu\nonumber \\
\partial^{\alpha} F^{\beta \gamma} + \partial^{\beta} F^{\gamma \alpha}+\partial^{\gamma} F^{\alpha \beta} & = & 0 \, ,
\label{compact}
\end{eqnarray}
while the current conservation
$$\partial_\nu j^\nu = \frac{\partial \rho}{\partial t} + \overset{\rightharpoonup}{\nabla}
\cdot  \overset{\rightharpoonup}{\jmath}  = 0 \,,$$
follows as a natural compatibility. {\it (ii)}~Show that if the electric and magnetic fields can be
written in terms of a four-vector potential $A^\mu = (V,
\overset{\rightharpoonup}{A})$ as
\begin{equation}
\overset{\rightharpoonup}{E} = - \frac{\partial
\overset{\rightharpoonup}{A}}{\partial t} -
\overset{\rightharpoonup}{\nabla} V, \qquad
\overset{\rightharpoonup}{B} = \overset{\rightharpoonup}{\nabla}
\times \overset{\rightharpoonup}{A} \, , \label{maxrelation}
\end{equation}
then the homogeneous Maxwell's equations (\ref{max1}) and
(\ref{max4}) are trivially satisfied, while the inhomogeneous
ones, (\ref{max2}) and (\ref{max3}), follow from the covariant
equation
\begin{equation} \Box^2 A^\mu -
\partial^\mu (\partial_\nu  A^\nu) =  j^\mu \, .
\label{maxcovariant}
\end{equation}
The tensor $F^{\mu\nu}$ can be written now as $F^{\mu\nu} = \partial^\mu A^\nu
- \partial^\nu A^\mu$. [{\it Hint:} note that $\overset{\rightharpoonup}{\nabla} \times
(\overset{\rightharpoonup}{\nabla} \times \overset{\rightharpoonup}{A}) =
- \nabla^2 \overset{\rightharpoonup}{A} +
\overset{\rightharpoonup}{\nabla} (\overset{\rightharpoonup}{\nabla} \cdot
\overset{\rightharpoonup}{A})]$. ({\it iii})~Verify that Maxwell's equations remain invariant under arbitrary Lorentz
boosts.

\subsection{Yang-Mills bonanza}

In early 1954, in an attempt to explain the strong interaction,  Yang and Mills  generalized the symmetry concepts associated with QED~\cite{Yang:1954ek}. The symmetry at play in the Yang-Mills theory is isotopic-spin symmetry, the guideline stating that the strong interactions of matter remain invariant (or nearly so) when the identities of neutrons and protons are interchanged. In the global symmetry any rotation of the internal arrows that pinpoint the isotopic-spin state must be made simultaneously everywhere. Postulating a local symmetry allows the orientation of the arrows to fluctuate independently from place to place and from time to time. In essence, rotations of the arrows can depend on any arbitrary function of position and time. This freedom to adopt different conventions for the identity of a nuclear particle in different places constitutes a local gauge symmetry. As in other examples where a global symmetry is converted into a local one, the invariance can be maintained only if something more is added to the theory. Because the Yang-Mills theory is more complicated than earlier gauge theories it turns out that a large number of degrees of freedom must be added. By and on itself, when isotopic-spin rotations are made arbitrarily from point to point in spacetime, the laws of physics remain invariant only if six new fields are introduced. They are all vector fields which have infinite range.

\begin{figure}[tbp] \postscript{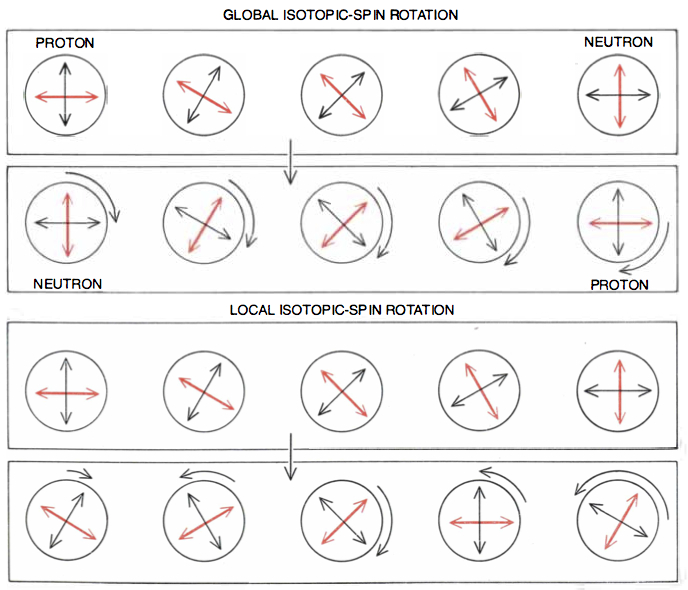}{0.9} \caption{Isotopic-spin symmetry~\cite{'tHooft:1980us}.}
\label{fig:isotopic-spin-rotation}
\end{figure}

The Yang-Mills fields are constructed using electromagnetism as a prototype, and actually two of them can be identified with the ordinary electric and magnetic fields. In plain English, they describe the field of the photon. The remaining Yang-Mills fields can also be taken in pairs and interpreted as electric and magnetic fields, but the ``photons'' they describe differ in a central respect from the well known properties of the photon: they
are still massless spin-1 particles, but they carry an electric charge. One ``photon'' is  positive and the other one negative. The wicked effects of ``charged photons'' become most apparent when a local symmetry transformation is applied more than once to the same particle. We have seen that in QED the symmetry operation is a local change in the phase of the electron field, each such phase shift being accompanied by an interaction with the electromagnetic field. It is simple to picture an electron undergoing two phase shifts in succession, say by  absorbing a
photon and later emitting one. Intuition guess that if the sequence of the phase shifts were reversed, so that first a photon was emitted and later one was absorbed, the end result would be the same. This is indeed the case. An unlimited series of phase shifts can be made, and the final result will be simply the algebraic sum of all the shifts no matter what their sequence. The symmetry of electromagnetism can be described by phase rotations in a circle, i.e. a $U(1)$ symmetry.

\begin{figure*}[tbp] \postscript{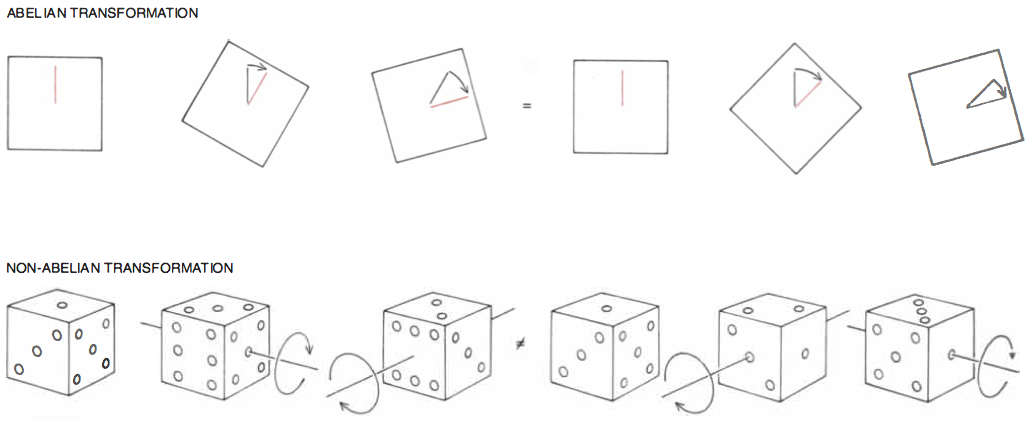}{0.9} \caption{Effects of repeated Abelian and non-Abelian transformations~\cite{'tHooft:1980us}.}
\label{fig:abelian-transformation}
\end{figure*}

In the Yang-Mills theory, where the symmetry operation is a local rotation of the isotopic-spin arrow, the result of multiple transformations is indeed different. Assume a hadron is subjected to a gauge transformation ${\cal A}$, followed shortly by a second transformation ${\cal B}$; at the end of the sequence the isotopic spin arrow is found in the orientation that corresponds to a proton. Now assume the same transformation were applied to the same hadron but in reverse sequence: ${\cal B}$ followed by ${\cal A}$. Generically, the expected final state will not necessarily be the same, e.g.
the particle may be a neutron instead of a proton. The net effect of the two transformations depends explicitly on the sequence in which they are applied. Because of this contrast QED is called an Abelian theory and the Yang-Mills theory is called a non-Abelian one. The terms are mooch from group theory. Abelian groups are made up of transformations that, when they are applied one after another, have the commutative property; non-Abelian groups are not commutative. A day-to-day example is the group of rotations. All possible rotations of a 2-dimensional object are commutative, and so the group of such rotations is Abelian. As an illustration, rotations by $\pi/3$ and $-\pi/2$ degrees yield a net rotation of $-\pi/6$ degrees no matter which is applied first. For a three-dimensional object free to rotate about three axes the commutative law does not hold, and the group of three-dimensional rotations is non-Abelian; an illustration is exhibited in Fig.~\ref{fig:abelian-transformation}.

The Yang-Mills theory has proved to be of paramount importance in the development of the Standard Model, but as it was originally formulated it does not describe the real world. A first objection to the theory is that isotopic-spin symmetry is exact, with the outcome that protons and neutrons are indistinguishable. This prediction is of course excluded by experiment. Even more troubling is the prediction of electrically charged ``photons.'' The photon is necessarily massless because it must have an infinite range. The existence of any electrically charged particle lighter than the electron would modify the world beyond recognition. Obviously, no such particle has been observed.  One possible pathway to contest these problems is to artificially endow the charged field quanta with a mass greater than zero. Note that by imposing a mass on the quanta of the charged fields we can confine them to a finite range. If the mass is large enough, the range can be made as small as is desired. If the long-range effects are removed then the existence of the charged fields can be easily reconciled with experimental observations. Furthermore, by selection the neutral Yang-Mills field as the only real long-range we can automatically distinguish protons from neutrons. Since the long-range field is simply the electromagnetic field, the proton and the neutron can be distinguished by their differing interactions with it, or expressly by their differing electric charges. After introducing this adjustment the local symmetry of the Yang-Mills theory would no longer be exact but approximate, since rotation of the isotopic-spin arrow would now have observable consequences. That is not a radical objection: approximate symmetries are usually commonplace in nature, e.g. the bilateral symmetry of the human body is only approximate.  What is more, at distance scales much smaller than the range of the massive components of the Yang-Mills field, the local symmetry becomes better and better. Therefore, in principle the microscopic structure of the theory could remain locally symmetric, but not its predictions of macroscopic, observable events. The Yang-Mills theory begun as a model of the strong interactions, but as we discuss below, the real interest in it centers on applications to the weak interactions.

\subsection{Parity violation}

\begin{figure}[tbp] \postscript{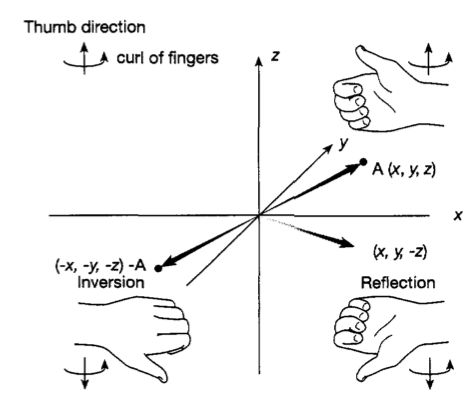}{0.9} \caption{Parity transformation.}
\label{fig:parity}
\end{figure}

Parity says something about the symmetry of a system.  The parity transformation inverts an object completely (back to front, side to side, and top to bottom) by the simultaneous flip in the sign of all three spatial coordinates $(x,y, z )$ to $(-x, -y, -z)$, i.e. by a point reflection. As an  illustration,  in Fig.~\ref{fig:parity}, we show a parity transformation, which  is equivalent to a reflection (in the $x-y$ plane) together with a rotation through $180^\circ$ (around the $z$ axis).  Note that during the transformation a right-handed system changes to a left-handed system. As a consequence, a point reflection can also be thought of as a test for chirality of a physical phenomenon, in that a parity inversion transforms a phenomenon into its mirror image. 

In classical physics, space inversion is merely a geometrical transformation, a rule to map each point in space to its image by inversion through a chosen origin. The time is left unaffected. A particle trajectory, for example, would be mapped onto another possible trajectory.  Wigner showed that in quantum mechanics, parity is more than a transformation, it is a physical observable whose value can be experimentally measured~\cite{Wigner}. In order to gain an understanding of parity in quantum systems, let us examine the symmetry properties of the wavefunction $\psi(\vec x,t)$ under space inversion. We define a new operator, $\hat {\rm P}$ which transforms $\psi(\vec x,t)$ into $\psi (-\vec x, t)$, i.e. 
\begin{equation}
\hat  {\rm P} \, \psi (\vec x,t) = \psi (-\vec x,t) \, . 
\end{equation}
If the system is symmetric  ($\psi (\vec x, t) = \psi (-\vec x, t)$) or antisymmetric ($\psi (\vec x, t) = -\psi (\vec x, t)$) then it obeys a simple eigenvalue problem when operated on by 
\begin{equation}
 \hat {\rm P} \, \psi (\vec x, t) = {\rm P} \ \psi (\vec x,t), 
\end{equation}
where ${\rm P} \in \{-1,1\}$ is the parity of the system.  The possible results of measurements are then  $\pm 1$, and the corresponding quantum states are said to posses even or odd partity, respectively.

Parity has three  important properties which we should state before we proceed: {\it (i)}~Parity is a multiplicative quantum number. This means that the quantity that is conserved in parity conserving interactions is not the sum of the parities of the constituent particles $\sum_j {\rm P}_j$, but is instead the product of these constituent parities $\prod_j {\rm P}_j$. This is actually true for any operator $\hat {\cal O}$ which, when applied twice, does nothing: $\hat {\cal O}^2 = \mathds{1}$. The parity operator, which performs a mirror reflection $\vec x  \to -\vec x$, if applied twice, does nothing to the original system, and therefore fits this rule. {\it (ii)}~Besides the parity connected to its spatial state, a particle $a$ has an intrinsic parity ${\rm P}_a$. The total parity of the particle is the product of the intrinsic and spatial parities. If the overall wavefunction of a particle (or system of particles) contains spherical harmonics then we must take this into account to get the total parity of the particle (or system of particles). For a wavefunction containing spherical harmonics, it is easily seen that the spatial parity is given by:
\begin{equation}
\hat {\rm P}  \, \psi (r, \theta, \phi)  = (-1)^l \, \psi(r, \theta,\phi)\, .
\label{parityYlm}
\end{equation}
The total parity of a state consisting of particles $a$ and $b$ is then $(-1)^L {\rm P}_a {\rm P}_b$ where $L$ is their relative orbital momentum, and ${\rm P}_a$ and ${\rm P}_b$ are the intrinsic parity of the two particles
 {\it (iii)}~Vector quantities change sign under a parity transformation, e.g. $\vec p \to - \vec p$. However, axial vectors (a.k.a. pseudovectors)  like the angular momentum, does not switch sign: $\vec L = \vec r \times \vec p \to \vec L = -\vec r \times -\vec p$. In this direction, it is interesting to see what happens if we take the scalar product of  $\vec S$ (which is an axial vector) and the unit vector in the direction of the momentum $\hat p$ (which is a vector), i.e. $\lambda = \vec S \cdot \hat p$. Usually scalars do not change sign under reflection but, surprisingly, we can clearly see that, since $\hat p$ changes sign but $\vec S $ does not, $\lambda \to  -\lambda$. This is not a usual scalar, it is actually a pseudoscalar. This particular pseudoscalar, the projection of the spin onto the direction of a particle's momentum is called the helicity of the particle. For example, for a spin-$\frac{1}{2}$ particle, the eigenvalues of the helicity operator are 
\begin{equation}
\lambda = \left\{\begin{array}{cc}
+\frac{1}{2} \ {\rm positive \, helicity}, & \bm{\longrightarrow \!\!\!\!\!\!\!\!\! \Rightarrow} \\
-\frac{1}{2} \ {\rm negative \, helicity}, & \bm{\longrightarrow \!\!\!\!\!\!\!\!\!\! \Leftarrow} \\
\end{array}
\right .
\end{equation} 
Left handed particles are defined as having a negative helicity while right handed particles are defined as having a positive helicity. \\

{\bf EXERCISE 13.3} Convince yourself that wavefunctions containing spherical harmonics satisfy (\ref{parityYlm}), i.e., $\hat {\rm P} \left[R(r) \, Y_m^l (\theta, \phi) \right] = (-1)^l R(r) \, Y_m^l (\theta, \phi)$.\\

The parity of the neutron, the proton, and the electron is conventionally taken to be equal to $+1$. Let us assume that parity is conserved in the strong interaction. No experimental indication is reported so far to sustain the contrary. We can determine the intrinsic parity of the pion by studying pion capture by a deuteron, $\pi^- d \to nn$. The pion is known to have spin-0, the deuteron spin-1, and the neutron spin-$\frac{1}{2}$. The internal parity of the deuteron is $+1$. The pion is capture by the deuteron from the ground state, implying $l=0$ in the initial state. So the total angular momentum quantum number of the initial state is $j=1$. The parity of the initial state is $(-1)^0 {\rm P}_\pi P_d = (-1)^0 {\rm P}_\pi {\rm P}_d = {\rm P}_\pi$. The parity of the final state is ${\rm P}_n {\rm P}_n (-1)^l = (-1)^l$. Because neutrons are identical fermions, the only allowed state has spin oriented in opposite directions. To conserve angular $j =1$ which requires $l  = 1$, and so ${\rm P}_\pi = -1$.

It is the role of parity in quantum mechanics that was shown by Wigner to be the explanation for Laporte's selection rule in atomic spectroscopy:  atomic states undergoing photon absorption or emission always end up, after the transition, in a final state of opposite parity~\cite{Laporte}. We can use the properties defined above to show that this rule is simply a statement of the law of conservation of parity. 
The intrinsic parities of both the proton  and the electron are $+1$;  actually, by convention all fermions have even parity. However, the electron state also has a spatial parity of $(-1)^l$, where $l$ is the orbital angular momentum. Thus, the total parity of the hydrogen atom is ${\rm P}_{{\rm H}_i} = (+1)(+1)(-1)^l = (-1)^l$. Using Laporte's rule, we know that the final state of an atom which has undergone photon absorption or emission will have a parity opposite to its original state. Thus, the wavefunction of the hydrogen atom will have a parity of ${\rm P}_{{\rm H}_f} = (-1)  {\rm P}_{{\rm H}_i} = (-1)^{l+1}$ after the transition. Including the intrinsic parity of the photon, which is defined as $-1$, we see that the total parity of the system in its final state is ${\rm P}_{{\rm tot}_f} = {\rm P}_{{\rm H}_f} {\rm P}_\gamma = (-1)^{l+2} = (-1)^l = {\rm P}_{{\rm H}_i}$. 
 Thus we see that Laporte's rule (the atom's wavefunction must change parity after photon emission or absorption) is nothing more than a statement of the conservation of parity in electromagnetic interactions.

In the early 1950's there was an ``odd'' experimental observation: two particles with identical mass, spin, charge, lifetime, etcetera decayed (weakly) into states of opposite parity:
\begin{eqnarray}
\theta^+ \to  \pi^+ \pi^0 ~~~~~~~~~~ & ~~~({\rm P}_\theta = + 1) \phantom{\, .} \nonumber \\
\tau^+ \to  \left\{\begin{array}{c} \pi^+ \pi^0 \pi^0 \\ \pi^+ \pi^+ \pi^- \\
\end{array} \right\} & ~~~({\rm P}_\tau = -1) \, .
\end{eqnarray}
There are two radical hypotheses that can explain the $\theta$-$\tau$ puzzle: {\it (i)}~there are two particles with identical properties except for parity; {\it (ii)} parity is not conserved in the weak interaction.  Lee and Yang soon realized that if parity is violated by the weak force then reflection symmetry would also be broken, implying that the average value of some pseudoscalars would not be zero in weak interactions~\cite{Lee:1956qn}.  \\

{\bf EXERCISE 13.4} Assuming parity conservation is an exact symmetry of nature and that the quarks and antiquarks of the mesons do not have any relative orbital angular momentum, determine the parities of $\tau^+$ and $\theta^+$~\cite{Lesov:2009pb}. [{\it Hint:} The mesons all have well-defined parity values which turn out to be $-1$, i.e. they are parity eigenstates $\hat {\rm P} |\pi^+ \rangle = - | \pi^+ \rangle$; along with being spin 0.]\\

\begin{figure}[tbp] \postscript{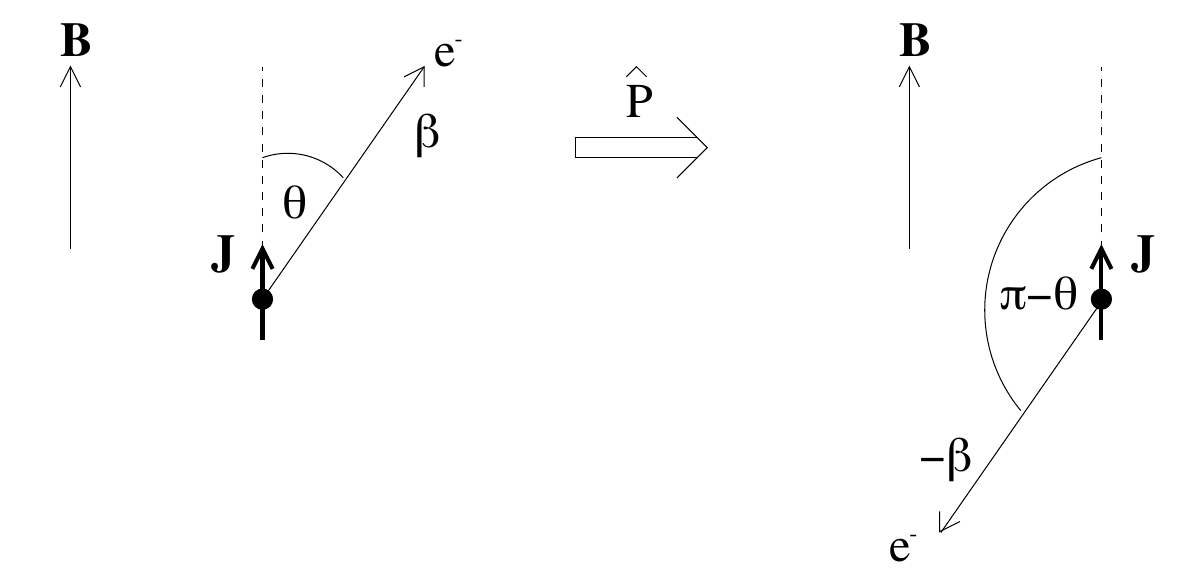}{0.9} \caption{The Wu experiment~\cite{Wu:1957my}.}
\label{fig:Wu}
\end{figure}

The classical experiment which actually established that parity is not a good symmetry of nature operates with the archetype of weak processes; namely, the $\beta$-decay of $^{60}{\rm Co} \to\, ^{60}{\rm Ni} \, e^- \, \overline \nu_e$~\cite{Lee:1956qn,Wu:1957my}. This is a weak interaction as it involves a neutrino, which has no electromagnetic or strong charge. The cobalt nuclei were cooled to temperatures around 0.01~K and placed in a magnetic field. At such low temperatures, the magnetic moments of the nuclei cause the nuclear spins to align themselves with the magnetic field direction, so the nuclei are all polarised along the field direction.  The experiment then simply consisted of measuring the angular distribution of the emitted electrons. The pseudoscalar to be measured in this case is the projection of the electron's momentum on the spin $\vec J$ of the cobalt nucleus. Consider a system where $\vec J$ is lined up on the $+z$ axis with a mirror at the $x-y$ plane as shown in Fig.~\ref{fig:Wu}. In the reflection of the system $\vec J$ still points in the same direction as it had before since it is a pseudovector. On the other hand, the momentum of an electron $\vec p_e$ which had been at an angle of $\theta$ relative to the $+z$ axis is at an angle $\pi - \theta$ in the reflection, because momentum is a vector.  Accordingly, the correlation $\vec J \cdot \hat p_e$ is manifestly parity violating. If the process obeys reflection symmetry then electron will be just as likely to be emitted at an angle $\theta$ as they would be at $\pi -\theta$. Does this make sense with regard to the magnetic field? Yes; since $\vec B = \vec \nabla \times \vec A$, then $\vec B$ is itself an axial vector like $\vec J$. Hence, the nuclear spin is still aligned along the magnetic field, as this alignment is due to the electromagnetic force which conserves parity.  The angular distribution of the emitted electrons is found to be $1 - \beta \cos \theta$, where $\beta = v/c$ is the electron velocity, meaning more electrons were emitted in a direction opposite the magnetic field than along it (see Appendix~\ref{appG}). In the parity inverted experiment, the electron comes off at $\pi - \theta$ so if parity is a symmetry, then the rate for electrons at $\theta$ must equal the rate at $\pi - \theta$. However, a distribution of $1 - \beta \cos \theta$ clearly does not have the same rate, since $ 1 - \beta \cos (\pi - \theta) = 1 + \beta \cos \theta \neq 1 - \beta \cos \theta$. Hence, this simple observation is all that is required to conclude that parity is not a symmetry of the weak interactions.  One thing to note is the velocity dependence. For the electrons emitted with $v \to 0$, the distribution becomes isotropic, which clearly then is parity symmetric. In contrast, for electrons with $v \to c$, the maximum parity violation occurs.

Another symmetry that we might expect nature to respect is the charged conjugation C, i.e. swapping the sign of all charges.  For example, if we suddenly swapped the charge of every proton and electron in the universe, we would expect nothing to change since the force between them would be the same. Electromagnetism, gravity and the strong interaction all obey the C-symmetry. However, it is easily seen that weak interactions violate the C-symmetry.

\subsection{Electroweak interaction}

Experiments in the late 1950's established that (charged-current) weak interactions are left-handed, and motivated the construction of a manifestly parity-violating theory of the weak interactions. The electroweak theory takes three crucial clues from experiment: {\it (i)} the existence of left-handed weak-isospin doublets,
\begin{equation}
\left(\begin{array}{c}
\nu_e\\
e \end{array} \right) \quad \left(\begin{array}{c}
\nu_\mu\\
\mu \end{array} \right) 
\quad 
\left(\begin{array}{c}
\nu_\tau\\
\tau \end{array} \right) 
\end{equation}
and
\begin{equation}
\left(\begin{array}{c}
u\\
d \end{array} \right) \quad \left(\begin{array}{c}
c\\
s \end{array} \right) 
\quad 
\left(\begin{array}{c}
t\\
b \end{array} \right) \,;
\end{equation}
{\it (ii)}~the universal strength of the weak interactions; {\it (iii)}~the idealization that neutrinos are massless. To incorporate electromagnetism into a theory of the weak interactions, we add to the $SU(2)_L$ isospin family symmetry (suggested by the first two experimental clues) a $U(1)_Y$ weak-hypercharge phase symmetry (see Appendix~\ref{appF}).  The electroweak theory then implies two sets of gauge fields a weak isovector, $\vec W$, and a weak isoscalar $B$~\cite{Glashow:1961tr,Weinberg:1967tq,Salam:1968rm}. From the fermionic point of view, the electroweak gauge symmetry $SU(2)_L \otimes U(1)_Y$ is chiral: the left-handed and the right-handed fermions form different types of multiplets. Specifically, all the left-handed quarks and leptons form $SU(2)$ doublets,  whereas the right-handed quarks and leptons are $SU(2)$ singlets. As a consequence, the weak interactions do not respect the parity or the charge-conjugation symmetries. 
Note that the particle-antiparticle conjugation (\ref{simetriaC})  does not alter the chirality of particles. Therefore, a left-handed neutrino would be taken by charge conjugation into a left-handed antineutrino, which does not interact in the Standard Model. This property is what is meant by the ``maximal violation'' of C-symmetry in the weak interaction.  We define the electric charge quantum number as $Q = T^3 + Y$, where the weak-isospin and hypercharge quantum numbers of
leptons and quarks are given in Table~\ref{TYQ}~\cite{Halzen:1984mc}.

\begin{table}
\caption{Weak-isospin and hypercharge quantum numbers.}
\begin{tabular}{ccccc|ccccc}
  \hline
  \hline
  ~~Lepton &  $T$ & $\phantom{-}T^3$ & $\phantom{-}Q$ & $\phantom{-}Y$ ~~~~~&~~~~~ Quark & 
  $T$ & $\phantom{-}T^3$ & $\phantom{-}Q$ & $\phantom{-}Y$~~ \\[1mm]
  \hline
  ~~$\nu_e$ & $\frac{1}{2}$ & $\phantom{-}\frac{1}{2}$ & $\phantom{-}0$ & 
  $-\frac{1}{2}$ ~~~~~&~~~~~ $u_L$ 
  & $\frac{1}{2}$ & $\phantom{-}\frac{1}{2}$ & $\phantom{-}\frac{2}{3}$ & 
  $\phantom{-}\frac{1}{6}$~~ \\[1mm]
  ~~$e^-_L$ & $\frac{1}{2}$ & $-\frac{1}{2}$ & $-1$ & $-\frac{1}{2}$ 
~~~~~&~~~~~ $d_L$ 
  & $\frac{1}{2}$ & $-\frac{1}{2}$ & $-\frac{1}{3}$ & 
  $\phantom{-}\frac{1}{6}$~~ \\[1mm]
  & & & & &~~~~~ $u_R$ & 0 & $\phantom{-}0$ & $\phantom{-}\frac{2}{3}$ & 
  $\phantom{-}\frac{2}{3}$~~ \\
  ~~$e^-_R$ & 0 & $\phantom{-}0$ & $-1$ & $-1$ ~~~~~&~~~~~ $d_R$ & 0 
& $\phantom{-}0$ & $-\frac{1}{3}$ & 
  $-\frac{1}{3}$~~ \\[1mm]
  \hline 
  \hline
\end{tabular}
\label{TYQ}
\end{table}

\begin{figure*}[tbp] \postscript{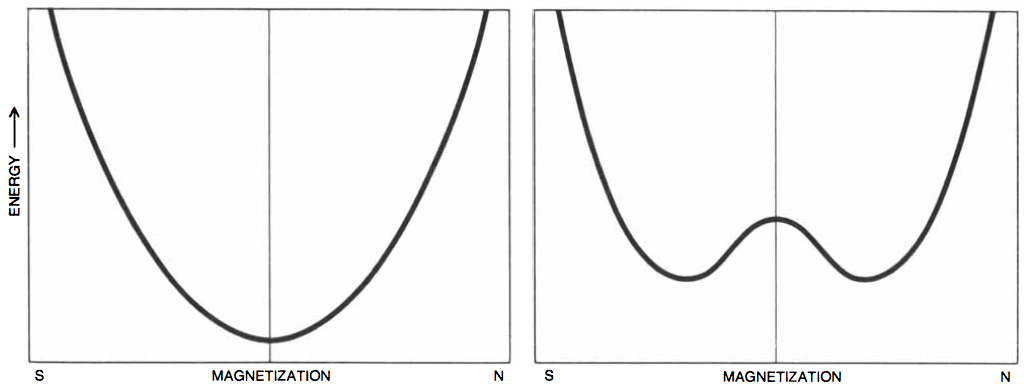}{0.9} \caption{Free energy as a function of the  magnetization for a bar magnet at high temperature (left) or at low temperature (right).  The magnet naturally seeks a state of minimum free energy. At high temperature this is a state of zero magnetization, a state that exhibits perfect symmetry between north and south. At low temperature the equilibrium state shifts to one of nonzero magnetization, which can be eitber north or south, even though the free energy curve is still perfectly symmetrical between north and south. In this case we say that the symmetry is spontaneously broken~\cite{Weinberg:1974uk}.}
\label{fig:magnetization}
\end{figure*}

A central question of the Standard Model is why the electroweak forces are asymmetrical: electromagnetism is long-ranged, whereas the weak nuclear force is short-ranged (less than $10^{-17}$~{\rm m}).  According to Heinsenberg's uncertainty principle, this limited range implies that the force particles must have a mass approaching $\sim$ GeV energies. The answer is that both these forces are actually symmetrical, but their symmetry is hidden, or ``broken.''  A simple analogy is provided by the familiar phenomenon of ferromagnetism.  The equations governing the electrons and iron nuclei in a bar of iron obey rotational symmetry, so that the free energy of the bar is the same whether one end is made the north pole by magnetization or the south.
Strictly speaking, the free energy of a ferromagnet is related to its magnetization $M$ by
\begin{equation}
G = \alpha M^2 + \beta M^4 \, .
\end{equation}
At high temperatures the curve of energy versus magnetization has a simple $U$ shape ($\alpha >0$ and $\beta >0$) that has the same rotational symmetry as the underlying equations, see Fig.~\ref{fig:magnetization}. The equilibrium state, the state of lowest energy at the bottom of the $U$, is also a state of zero magnetization, which shares this symmetry. On the other hand, when the temperature is lowered, the lowest point on the $U$-shaped curve humps upward so that the curve resembles a $W$ with rounded corners ($\alpha < 0$ and $\beta >0$), see Fig.~\ref{fig:magnetization}. The curve still has the same rotational symmetry as the underlying equations, but now the equilibrium state has a definite nonzero magnetization, which can be either north or south but which in either case no longer exhibits the rotational symmetry of the equations. We say in such cases that the symmetry is spontaneously broken.  The spontaneous symmetry breaking then describes systems where the equations of motion obey certain symmetries, but the lowest energy solutions do not exhibit that symmetry.

\begin{figure}[tbp] \postscript{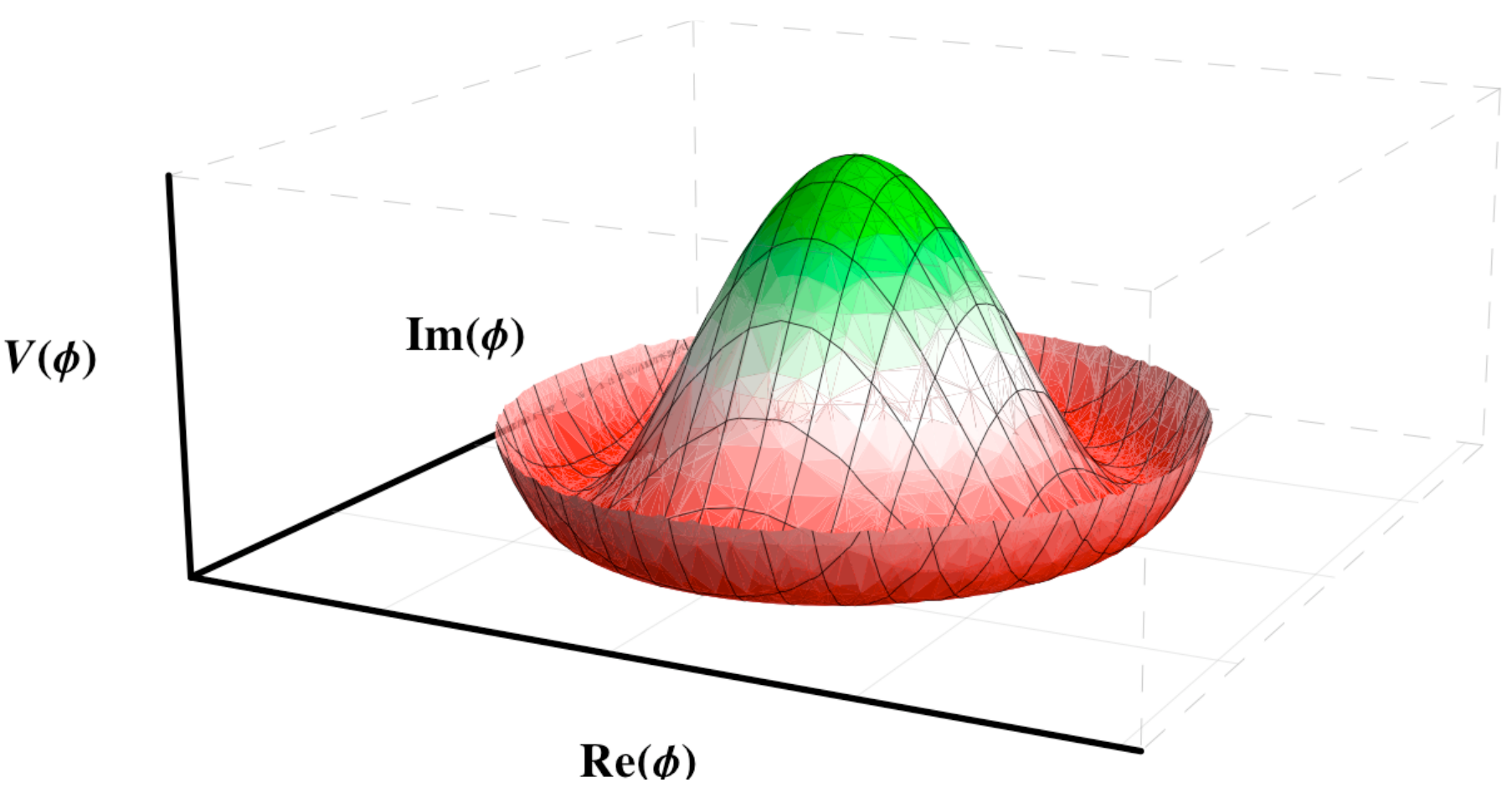}{0.9} \caption{Mexican hat potential, $V(\phi) =  \mu^2 (\phi^* \phi) + |\lambda| (\phi^* \phi)^2$, 
 of complex scalar field $\phi$~\cite{Quevedo:2010ui}.}
\label{fig:mexican-hat}
\end{figure}

To approach our goal of generating a mass for the gauge bosons we first note that the Mexican hat potential of Fig.~\ref{fig:mexican-hat} has a circle of mininma. The minima of this system are thus degenerate; there are multiple states with the same vacuum energy. The different orientations in the complex plane define different states. The orientation of these states is comparable to the direction of alignment of the spins in the ferromagnet. Note that the potential is symmetric under rotations in $\phi$ space, but the ground states are asymmetric. Applying the $U(1)$ transformation to any of the vacuum states will rotate it to a different orientation that describes a different physical state. 

 In the electroweak theory the Higgs field brings about the symmetry breaking~\cite{Higgs:1964pj,Englert:1964et}.  The process is analogous to that of the ferromagnet, but it requires more dimensions than we can comfortably draw. The idea is that there are originally four Higgses which are all parts of a single ``complex Higgs doublet'' \begin{equation} \phi \equiv \left(\begin{array}{c} \phi^+ \\ \phi^0 \end{array} \right) \,, \label{Hdoublet} \end{equation} with weak hypercharge $Y_\phi = +1/2$. The scalar potential is given by \begin{equation} V(\phi) = \mu^2 (\phi^\dagger \phi) + |\lambda| (\phi^\dagger \phi)^2 \, .  \end{equation} The electroweak symmetry is spontaneously broken if the parameter $\mu^2 < 0$. In the unified theory, where electroweak symmetry is unbroken, these four Higgses can be rotated into one another and the physics does not change. However, when the system rolls into the bottom of the Mexican hat  one of the Higgses obtains a vacuum expectation value (VEV) \begin{equation} \langle \phi \rangle_0 = \left(\begin{array}{c} 0 \\ v/\sqrt{2} \end{array} \right) \ , \label{vacuum} \end{equation}
 while the others do not, where $v = \sqrt{-\mu^2/|\lambda}|$. Performing a
``rotation'' then moves the VEV from one Higgs to the others and the
symmetry is broken as the four Higgses are no longer being treated
equally; see Appendix~\ref{appH}. The electroweak symmetry makes all the electroweak force
particles massless. The broken symmetry gives masses to the gauge
bosons, thereby restricting their range.  The $W_1$ and $W_2$ combine
into the $W^+$ and $W^-$ by eating the charged Higgses. A similar story goes through for the $W_3$ and $B$, which combine and eat the neutral Higgs to form the massive $Z$ boson. Meanwhile, the photon is the leftover combination of the $W_3$ and $B$. 
There are no more Higgses to eat, so the photon remains massless. The apparent extra degree of freedom is actually spurious, because it corresponds only to the freedom to make a gauge transformation. Such a transformation leads to the massive scalar $H$, usually refer to as the Higgs boson.

Fermion masses arise from Yukawa interactions, which couple the right-handed fermion singlets to the left-handed fermion doublets and the Higgs field~\cite{Yukawa:1935xg}. In the process of spontaneous symmetry breaking these interactions lead to charged fermion masses, but leave the neutrinos massless.
 
The electroweak symmetry is more abstract than the symmetry
of the ferromagnet. It means the freedom to decide which leptons are
electrons and which are neutrinos or how to label up and down
quarks. In the symmetrical case, the lepton-naming convention is set independently at each
point in space. What one person calls an electron, another might call
some mixture of electron and neutrino, and it would make no difference
to their predictions. In the broken symmetry, the convention is fixed
everywhere. What one person calls an electron, all do.

\subsection{Strong interaction}

The development  of a successful gauge theory of the strong interaction, which is unique to hadrons, cannot not be undertaken until an inherent property about the hadrons is understood: they are not elementary particles. A model of hadrons as composite objects was proposed independently by Gell-Mann and Ne'eman~\cite{GellMann:1961ky,Ne'eman:1961cd,GellMann:1964nj}. In this model hadrons are composed of point-like particles named quarks. A hadron can be made up of quarks according to either of archetypes in (\ref{GellM}). Binding together three quarks leads to a baryon, a class of hadrons that includes the neutron and the  proton.  Combining one quark and one antiquark makes a meson, a class typified by the pions. Every known hadron can be accounted for as one of these allowed composites of quarks. The keystone of any theory of the strong interactions is to explain the peculiar rules for building hadrons out of quarks. The structure of a meson is not so hard to account for: since the meson is made  out of a quark and an antiquark, it is merely necessary to assume that the quarks carry some property analogous to electric charge. The binding of a quark and an antiquark would then be explained on the principle that opposite charges attract, just as they do in electromagnetism. However, the structure of the baryons is a far profound enigma. To describe how three quarks can produce a bound state we must assume that three like charges attract. 

The theory that describes the strong force, QCD, lays down exactly these interactions. The analogue of electric charge is a property called color. The term color refers to the fact that the rules for producing hadrons can be expressed succinctly by requiring all allowed combinations of quarks to be white, or colorless. The quarks are assigned the primary colors red, green and blue, whereas the antiquarks have the complementary anticolors cyan, magenta and yellow. Each of the quark flavors comes in all three colors, so that the introduction of the color charge triples the number of distinguishable quarks.

There are two prescriptions of producing the white color with the available quark pigments: by mixing all three primary colors or by mixing one primary color with its complementary anticolor. The baryons are made according to the first prescription: the three quarks in a baryon are required to have different colors. In a meson a color is always accompanied by its complementary anticolor.

\begin{figure*}[tbp] \postscript{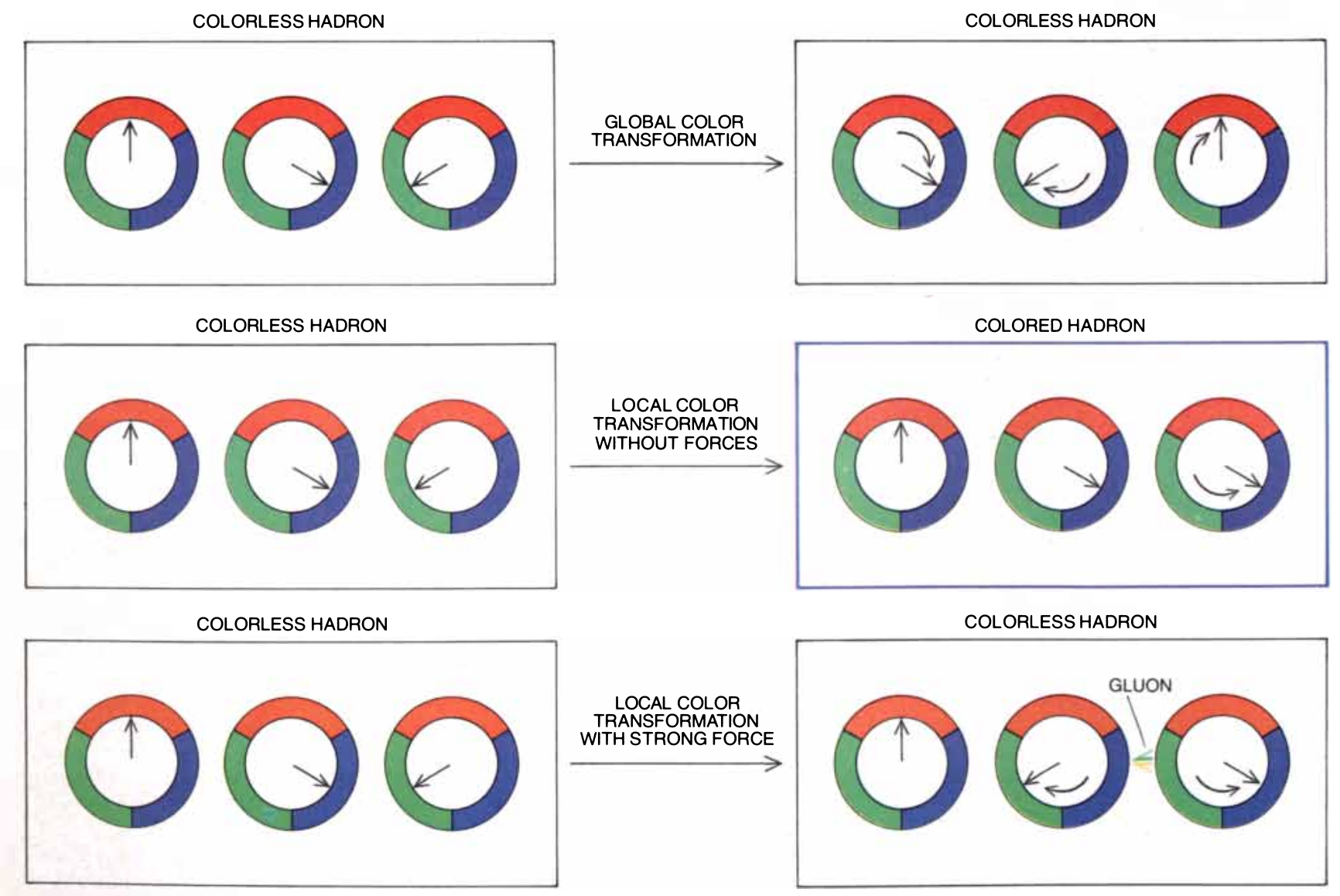}{0.9} \caption{Color symmetry~\cite{'tHooft:1980us}.}
\label{fig:color}
\end{figure*}

QCD a non-Abelian gauge theory. The gauge symmetry is an invariance with respect to local transformations of quark color. It is easy to visualize a global color symmetry. The quark colors, like the isotopic-spin states of hadrons, could be indicated by the orientation of an arrow in some imaginary internal space. Successive rotations of $\frac{1}{3}$ of a circle would change a quark from red to green to blue and back to red again. Thus, in a baryon there would be three arrows with one arrow set to each of the three colors. By definition a global symmetry transformation must affect all three arrows in the same way and at the same time. For example, as shown in Fig.~\ref{fig:color} all three arrows might rotate clockwise a third of a turn. As a result of such a transformation all three quarks would change color, but all observable properties of the hadron would remain as before. In particular there would still be one quark of each color, and so the baryon would remain colorless.

QCD requires that the color invariance be retained even when the symmetry transformation is a local one. In the absence of forces or interactions the invariance is of course lost. Hence a local transformation can change the color of one quark but leave the other quarks unaltered, which would give the hadron a net color. As in QED, we restore the invariance with respect to local symmetry operations by introducing  new fields. In QCD the fields needed are analogous to the electromagnetic field but are much more complicated; they have eight times as many components as the electromagnetic field has. It is these fields that govern the strong force.

The quanta of the color fields are called gluons (as they glue the quarks together). There are eight of them, they are all massless, and have a spin-1; i.e. they are massless vector bosons like the photon. In addition,  the gluons are electrically neutral like the photon, but they are not color-neutral. Each gluon carries one color and one anticolor. There are nine possible combinations of a color and an anticolor, but one of them is equivalent to white and is excluded, leaving eight definite gluon fields.

\begin{figure}[tbp] \postscript{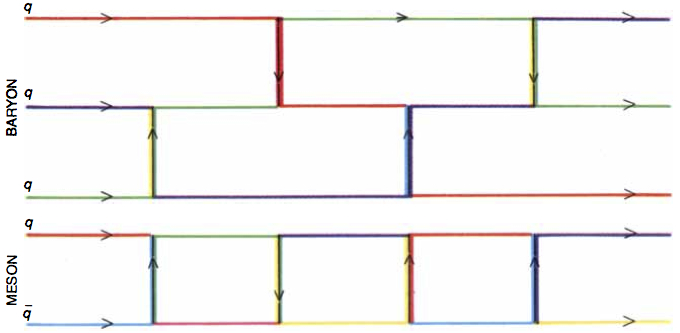}{0.9} \caption{Gluon exchange~\cite{'tHooft:1980us}.}
\label{fig:gexchange}
\end{figure}

The gluons preserve local color symmetry as foolows: a quark is allowed to change its color, and it can do so independently of all other quarks, but every color transformation must be accompanied by the emission of a gluon, just as an electron can shift its phase only by emitting a photon. The gluon, propagating at the speed of light, is subsequently absorbed by another quark, which will have its color shifted in exactly the way needed to compensate for the original change. An example is shown in Fig.~\ref{fig:color}: a green quark changes its color to blue and in the process emits a gluon that bears the colors green and antiblue. The gluon is then absorbed by a blue quark, and in the ensuing reaction the blue of the quark and the antiblue of the gluon annihilate each other, leaving the second quark with a net color of green. As a consequence, in the final state as in the initial state there is one green quark and one blue quark. Because of the continual arbitration of the gluons there can be no net change in the color of a hadron, even though the quark colors vary freely from point to point, see Fig.~\ref{fig:gexchange}. All hadrons remain white, and the strong force is nothing more than the system of interactions needed to maintain that condition.

Notwithstanding the complex structure of the gluon fields, QED and QCD are remarkably similar in form. Most notably, we have seen that  the gluon and the photon  are identical in their spin and in their lack of mass and electric charge. It is hence intruiguing  that quarks interactions are very different from those of electrons.
Both electrons and quarks form bound states: electrons form atoms and quarks form hadrons.  Electrons, however, are also observed as independent particles; a small quantity of energy suffices to isolate an electron by ionizing an atom. An isolated quark has never been detected. It seems to be impossible to ionize a hadron, no matter how much energy is supplied. The quarks are evidently bound so tightly that they cannot be pried apart; paradoxically, however, probes of the internal structure of hadrons show quarks moving freely, as if they were not bound at all.  Gluons too have not been seen directly in experiments. 

\begin{figure}[tbp] \postscript{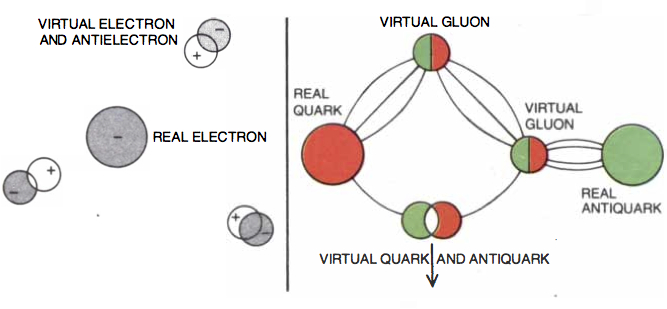}{0.9} \caption{Polarization of the vacuum~\cite{'tHooft:1980us}.}
\label{fig:pola}
\end{figure}

The resolution to this conundrum was developed not by modifying the color fields but by examining their properties in greater detail. We have seen that an isolated electron is 
surrounded by a cloud of virtual $e^+e^-$ pairs, which it constantly emits and reabsorbs; see Fig.~\ref{fig:pola}.  It is the charged virtual particles in this cloud that under ordinary conditions camouflage the ``infinite'' negative bare charge of the electron. In the vicinity of the bare charge the $e^+ e^-$ pairs become slightly polarized: the virtual positrons, under the attractive influence of the bare charge, stay closer to it on the average than the virtual electrons, which are repelled. As a result the bare charge is partially neutralized; what we observe at long range is the difference between the bare charge and the screening charge of the virtual positrons. Only when a probe approaches to within less than about $10^{-10}~{\rm cm}$ do the unscreened effects of the bare charge become significant.

It is feasible to guess the same process would operate among color charges, and indeed it does. A red quark is surrounded by pairs of quarks and antiquarks, and the antired charges in the cloud are attracted to the central quark and tend to screen its charge. However, in  QCD there is a competing effect which is not present in QED. Whereas the photon carries no electric charge and therefore has no direct influence on the screening of electrons, gluons do bear an $SU(3)_C$ color charge. This distinction expresses the fact that QED is a $U(1)$ Abelian theory and QCD is an $SU(3)$ non-Abelian one (see Appendix~\ref{appF}). Virtual gluon pairs also encompass a cloud around a colored quark (see Fig.~\ref{fig:pola}), but it turns out that the gluons tend to enhance the color charge rather than attenuate it. This can be understood visualizing that the red component of a gluon is actually attracted to a red quark and hence could add its charge to the total effective charge. If there are no more than 16 flavors of quark (and as of today only 6 have been observed), the ``antiscreening'' by gluons is the dominant component.  As a result of this``antiscreening'' the effective color charge of a quark grows larger at long range than it is close by. A distant quark reacts to the combined fields of the central quark and the reinforcing gluon charges; at close range, once the gluon cloud has been penetrated, only the smaller bare charge is effective. The quarks inside the hadron thus move as if they were connected by rubber bands: at very close range, where the bands are feeble, the quarks move almost independently, but at a  large distances, where the bands are stretched tense, the quarks are tightly bound~\cite{Gross:1973id,Politzer:1973fx}. Therefore, quarks and gluons could be permanently confined in hadrons. By its very nature the color symmetry is an exact symmetry and the colors of particles are completely indistinguishable. The theory is a pure gauge theory of the kind first proposed by Yang and Mills. The gauge fields are inherently long-range and fundamentally are similar to the photon field. However, the constraints on those fields from quatum mechanics are so strong that the observed interactions are quite unlike those of QED and even lead to the imprisonment of an entire class of particles.\\

{\bf EXERCISE 13.5}~A proton or a neutron can sometimes ``violate'' conservation of energy by emitting and then reabsorbing a $\pi$ meson, which has a mass of $135~{\rm MeV}$. This is possible as long as the $\pi$ meson is reabsorbed within a short enough time  consistent with the uncertainty principle. {\it (i)}~Consider $p \to p \pi$. By what amount $\Delta E$ is energy conservation violated? (Ignore any kinetic energies.) {\it (ii)}~For how long a time can the $\pi$ meson exist? {\it (iii)}~Assuming the $\pi$ meson to travel at very nearly the speed of light, how far from the proton can it go? (This procedure, gives us an estimate of the range of the nuclear force, because protons and neutrons are held together in the nucleus by exchanging $\pi$ mesons.)

\section{Multidiscipline Approach to the UV Completion of the Standard Model}

The conspicuously well-known accomplishments of the $SU(3)_C \otimes SU(2)_L \otimes U(1)_Y$ Standard Model (SM) of strong and electroweak forces can be considered as the apotheosis of the gauge symmetry principle to describe particle interactions. Most spectacularly, the recent discovery~\cite{ATLAS:2012ae,Chatrchyan:2012tx} of a new boson with scalar quantum numbers and couplings compatible with those of the SM Higgs has possibly plugged the final remaining experimental hole in the SM, cementing the theory further.

However, the saga of the SM is still exhilarating because it leaves all questions of consequence unanswered. The most evident of unanswered questions is the huge disparity between the strength of gravity and of the SM forces. Even if one abandons this hierarchy problem, which does not conflict with any experimental measurement, the SM has many other (perhaps more basic) shortcomings. Roughly speaking, the SM is incapable of explaining some well established observational results.
Among the most notable of these are neutrino masses and the presence of a large non-baryonic dark matter (DM) component of the energy density in the universe.
In addition, luminosity distance measurements of Type Ia supernovae strongly imply the presence of some unknown form of energy density, related to otherwise empty space, which appears to dominate the recent gravitational dynamics of the universe and yields a stage of cosmic acceleration. We still have no solid clues as to the nature of such dark energy (or perhaps more accurately dark pressure). In this section we give a brief description of the most characteristic features of these fundamental physical phenomena that the SM does not adequately explain.

\subsection{Neutrino oscillations}

At present, convincing experimental evidence exists for (time dependent) oscillatory transitions $\nu_\alpha \rightleftharpoons \nu_\beta$ between the different neutrino flavors. The simplest and most direct interpretation of the atmospheric data is that of muon neutrino oscillations~\cite{Fukuda:1998tw,Fukuda:1998mi,Fukuda:2000np,Ashie:2005ik}.
The evidence of atmospheric $\nu_\mu$ disappearing is now at $> 15
\sigma$, most likely converting to $\nu_\tau$.  The angular
distribution of contained events shows that for neutrino energy $E_\nu \sim 1~{\rm
  GeV},$ the deficit comes mainly from $L_{\rm atm} \sim 10^2 - 10^4~{\rm km}.$ These results have been confirmed by the KEK-to-Kamioka (K2K) experiment which observes the disappearance of accelerator $\nu_\mu$'s at a distance of 250~km and finds a distortion of their energy spectrum with a CL of $2.5-4\sigma$~\cite{Ahn:2001cq,Ahn:2002up,Ahn:2004te}.  Data collected by the Sudbury Neutrino Observatory (SNO) in conjuction with data from Super-Kamiokande (SK) show that solar $\nu_e's$ convert to $\nu_{\mu}$ or $\nu_\tau$ with CL of more than 7$\sigma$~\cite{Fukuda:1998fd,Fukuda:1998rq,Fukuda:2001nk,Fukuda:2002pe,Ahmed:2003kj}. The KamLAND Collaboration has measured the flux of $\overline \nu_e$ from distant reactors and find that $\overline{\nu}_e$'s disappear over distances of about 180~km~\cite{Araki:2004mb}. All these data suggest that the neutrino eigenstates that travel through space are not the flavor states that we measured through the weak force, but rather mass eigenstates.\footnote{Contrariwise, charged leptons are states of definite mass and hence cannot undergo oscillations~\cite{Pakvasa:1981ci,Akhmedov:2007fk}.}  This is the first compelling experimental evidence for new physics beyond the SM.

\begin{figure*}[tbp] 
\begin{minipage}[t]{0.46\textwidth}
\postscript{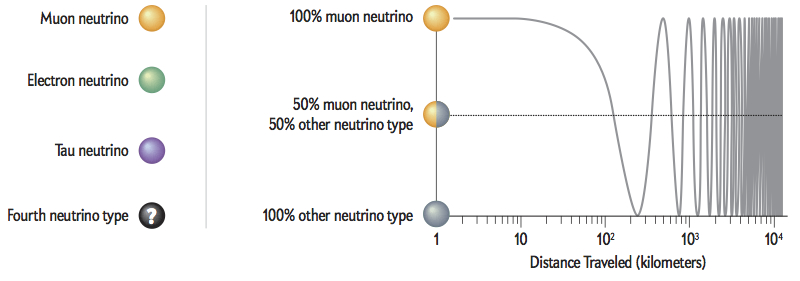}{0.99} 
\end{minipage}
\hfill
\begin{minipage}[t]{0.46\textwidth}
\postscript{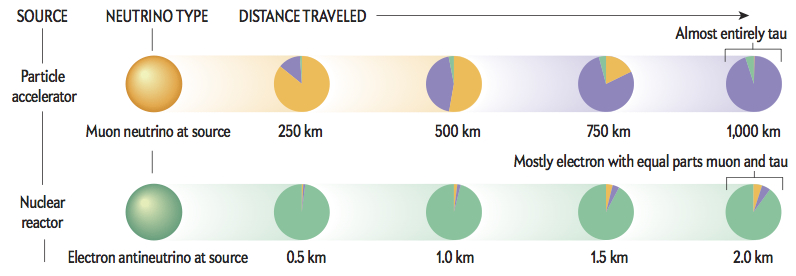}{0.99} 
\end{minipage}
\caption{The graph on the left illustrates the probability that a muon neutrino, having traversed a given distance, will switch flavors. In practice, oscillation distances depend on the neutrino's energy. The illustration on the right shows idealized source-to-detector oscillation patterns for experiments at particle accelerators and nuclear reactors~\cite{Hirsch:2013zca}.}
\label{fig:nu-osc}
\end{figure*}

The superposition of neutrino mass eigenstates produced in association
with the charged lepton of flavor $\alpha$ is the state we refer to as
the neutrino of flavor $\alpha$.  The flavor eigenstates and the mass
eigenstates are related by a unitary transformation $U$
(mass-to-flavor mixing matrix, fundamental to particle physics) 
\begin{eqnarray}
|\nu_\alpha\rangle = \sum_i U_{\alpha_i} | \nu_i \rangle \Leftrightarrow
  |\nu_i \rangle & = &  \sum_\alpha (U^\dagger)_{i\alpha} |\alpha \rangle \nonumber \\ & = &
  \sum_\alpha U_{\alpha i}^* |\nu_\alpha\rangle \,,
\label{state}
\end{eqnarray}
with $U^\dagger U = 1$, that is to say $\sum_i U_{\alpha i} U^*_{\beta i} =
\delta_{\alpha \beta}$ and $\sum_\alpha U_{\alpha i} U^*_{\alpha j} =
\delta_{ij}$~\cite{Pontecorvo:1957cp,Pontecorvo:1967fh,Maki:1962mu}. For antineutrinos we have to replace $U_{\alpha i}$ by
$U_{\alpha i}^*$, i.e.
\begin{equation}
|\overline \nu_\alpha \rangle = \sum_i U^*_{\alpha i} | \ \overline \nu_i
\rangle \, .
\end{equation}
A summary of the oscillation length for 3 neutrino species is exhibited in Fig.~\ref{fig:nu-osc}.\footnote{A sterile neutrino is a neutral lepton with no ordinary weak interactions except those induced by mixing~\cite{Abazajian:2012ys}. They are present in most extensions of the SM and in principle can have any mass. Throughout we only consider the three weakly-interacting neutrino species.}

For many years, the sparse data on the angle $\theta_{13}$ allowed
consistency with zero. However, in spring of 2012, the angle was
definitively measured to be nonzero (but still small on the scale of $\theta_{23}\sim 45^\circ$ and $\theta_{12}\sim 35^\circ$), 
$\theta_{13} \approx 9^\circ$~\cite{An:2012eh,Ahn:2012nd,An:2012bu}.  
At present, the low statistics  statistics limits our capacity to disentangle
neutrino flavors with sufficient precision to be sensitive to small $\theta_{13}$, 
granting us the privilege of certain amount of laziness in subsequent explanation. 
To simplify the  discussion, we will adopt maximal mixing for
atmospheric $\nu_\mu \leftrightharpoons \nu_\tau$ neutrinos (i.e. $\theta_{23} \sim 45^\circ$) along with a negligible
$|U_{e3}|^2 = \sin^2(\theta_{13})$.  The latter approximation allows
us to ignore CP violation and assume real matrix elements.  (The small
effects of nonzero $\theta_{13}$ have been investigated
in~\cite{Fu:2012zr}.)

With our simplifying assumptions in mind, one can define a mass basis as follows,
\begin{equation}
|\nu_1 \rangle = \sin \theta_\odot |\nu^\star\rangle +  \cos
\theta_\odot |\nu_e\rangle, 
\end{equation}
\begin{equation}
|\nu_2 \rangle =  \cos \theta_\odot |\nu^\star\rangle  -\sin \theta_\odot |\nu_e\rangle, 
\end{equation}
and
\begin{equation}
|\nu_3 \rangle = \frac{1}{\sqrt{2}} (|\nu_\mu \rangle + |\nu_\tau \rangle) \,\,,
\label{3rd}
\end{equation}
where $\theta_\odot \sim 34^\circ$ is the solar mixing angle and 
\begin{equation}
|\nu^\star\rangle = \frac{1}{\sqrt{2}} (|\nu_\mu\rangle - |\nu_\tau \rangle)
\label{orthogonal}
\end{equation}
is the eigenstate orthogonal to $|\nu_3 \rangle.$ Inversion of the neutrino mass-to-flavor 
mixing matrix leads leads to
\begin{equation}
|\nu_e \rangle = \cos \theta_\odot |\nu_1\rangle - \sin \theta_\odot |\nu_2 \rangle
\end{equation}
and
\begin{equation}
|\nu^\star \rangle = \sin \theta_\odot |\nu_1\rangle + \cos \theta_\odot |\nu_2 \rangle \,\,.
\end{equation}
Finally, by adding~(\ref{3rd}) and (\ref{orthogonal}) one obtains the $\nu_\mu$ flavor eigenstate,
\begin{equation}
|\nu_\mu \rangle = \frac{1}{\sqrt{2}} \left[ |\nu_3 \rangle + \sin \theta_\odot |\nu_1 \rangle + 
\cos \theta_\odot |\nu_2\rangle \right] \,\,,
\end{equation}
and by substracting these same equations the $\nu_\tau$ eigenstate.

The evolution in time of the $\nu_i$ component of a neutrino initially born as $\nu_\alpha$ in the rest 
frame of that 
component is described by  Schr\"odinger's equation,
\begin{equation}
|\nu_i (\tau_i) \rangle = e^{-i m_i \tau_i} |\nu_i (0)\rangle \,\,,
\end{equation}
where $m_i$ is the mass of $\nu_i$ and $\tau_i$ is the proper time. 
In the lab frame, the Lorentz invariant phase factor may be written as $e^{-i(E_i t - p_i L)}$, where $t,$ 
$L, $ $E_i,$ and $p_i,$ are respectively, the time, the position, the energy, and the momentum of $\nu_i$ 
in the lab frame. Since the neutrino is extremely 
relativistic $t \approx L$ and 
\begin{equation}
E_i  =  \sqrt{p^2 + m_i^2} = p \sqrt{1 +( m_i/p)^2}
\approx  p + m_i^2/2p \,,
\end{equation}
where we have used that $(1+x^2)^{1/2} = 1 + x^2/2 + \cdots$.  Hence, from 
(\ref{state}) it follows that the state vector of a neutrino born as $\nu_\alpha$ after propagation 
of distance $L$ becomes
\begin{equation}
|\nu_\alpha(L) \rangle \approx \sum_i U_{\alpha i} \, e^{-i (m_i^2/2E)\,\, L}\,\,|\nu_i \rangle \,\,,
\label{ritazza}
\end{equation}
where $E \approx p$ is the average energy of the various mass eigenstate components of the neutrino.
Using the unitarity of $U$ to invert (\ref{state}), from (\ref{ritazza}) one finds that 
\begin{equation}
|\nu_\alpha(L) \rangle \approx \sum_\beta \left[\sum_i U_{\alpha i} \, e^{-i (m_i^2/2E)\,\, L}\,\,U_{\beta i}\,\,\right]\,|\nu_\beta \rangle \,\,.
\end{equation}
In other words, 
the propagating mass eigenstates acquire relative 
phases giving rise to flavor oscillations. Thus, 
after traveling a distance $L$ an initial state 
$\nu_\alpha$ becomes a superposition of all flavors, with probability of transition to flavor $\beta,$ 
$|\langle \nu_\beta| \nu_\alpha(L) \rangle|^2$, given by~\cite{GonzalezGarcia:2002dz}
\begin{equation}
P(\nu_\alpha \to \nu_\beta) = \delta_{\alpha \beta} - 4 \sum_{i>j} U_{\alpha i}\, U_{\beta i}\, 
U_{\alpha j} \, U_{\beta j} \, \sin^2 \Delta_{ij}\,, \nonumber
\end{equation}
where $\Delta_{ij} \sim \delta m_{ij}^2 L/ 2E,$ and $\delta m_{ij}^2 = m_i^2 - m_j^2.$ 

For $\Delta_{ij} \gg 1$, the phases will be erased by uncertainties in $L$ and $E$. Consequently,
averaging over $\sin^2 \Delta_{ij}$ one finds
\begin{equation}
P(\nu_\alpha \to \nu_\beta) = \delta_{\alpha \beta} - 2 \sum_{i>j} U_{\alpha i}\, U_{\beta i}\, 
U_{\alpha j} \, U_{\beta j} \,. 
\label{paco}
\end{equation}
Now, using $2 \sum_{1>j} = \sum_{i,j} - \sum_{i=j},$ (\ref{paco}) can be re-written as
\begin{eqnarray}
P(\nu_\alpha \to \nu_\beta) & = & \delta_{\alpha \beta} -  \sum_{i,j} U_{\alpha i}\, U_{\beta i}\, 
U_{\alpha j} \, U_{\beta j} \nonumber \\ & + &  \sum_{i} U_{\alpha i}\, U_{\beta i}\, 
U_{\alpha i} \, U_{\beta i}  \\
 & = & \delta_{\alpha \beta} - \left( \sum_{i} U_{\alpha i}  U_{\beta i} \right)^2 + \sum_{i}   
U_{\alpha i}^2  U_{\beta i}^2\,. \nonumber 
\label{PP}
\end{eqnarray}
Since $\delta_{\alpha \beta}$ = $\delta_{\alpha \beta}^2,$ the first and second terms in (\ref{PP}) 
cancel each other, yielding
\begin{equation}
P(\nu_\alpha \to \nu_\beta) = \sum_{i} U_{\alpha i}^2 \,\,U_{\beta i}^2 \,\,. 
\label{octonautsF}
\end{equation}

{\bf EXERCISE 14.1}~It is helpful to envision the cosmic ray engines as machines where protons are accelerated and (possibly) permanently confined by the magnetic fields of the acceleration region. The production of neutrons and charged pions and subsequent decay produces both neutrinos and cosmic rays: the former via $\pi^+ \to \mu^+ \nu_\mu \to e^+ \nu_e \nu_\mu \overline \nu_\mu$ (and the conjugate process), the latter via neutron diffusion from the region of the confined protons. Consequently, the expectation for the relative fluxes of each neutrino flavor at production in the cosmic sources, $(\alpha_e : \alpha_\mu : \alpha_\tau)_S$, is nearly $(1 : 2 : 0)_S$. After neutrino oscillations decohere over the astronomical propagation distances the flavor conversion is properly described by the mean oscillation probability. Show that the flux of “pionic” cosmic neutrinos should arrive at Earth with democratic flavor ratios, $(\alpha_e, : \alpha_\mu, : \alpha_τ)_\oplus \approx (1 : 1 : 1)_\oplus$. {\it (ii)}~Show that the flux of antineutrinos, which originates via neutron $\beta$-decay, arrives at Earth with $\sim (3:1:1)_\oplus$~\cite{Anchordoqui:2003vc}.\\

The most economic way to get massive neutrinos would be to
introduce the right-handed neutrino states (having no gauge
interactions, these sterile states would be essentially undetectable)
and obtain a Dirac mass term through a Yukawa coupling~\cite{GonzalezGarcia:2002dz}.\\

{\bf EXERCISE 14.2}~On 23 February 1987, astronomers were startled by the observation of a new supernova in the Large Magellanic Cloud, a satellite galaxy of our Milky Way. However, the first observation of this supernova occurred several hours earlier by two neutrino-detection experiments~\cite{Hirata:1987hu,Bionta:1987qt}. The fact that the neutrinos all arrived within a few seconds of each other after traveling for more than 100,000 lightyears allows us to put tight constraints on the mass of the neutrino~\cite{Arnett:1987iz}. {\it (i)} Suppose light takes a time $t_0$ to reach us from the location of the supernova. How long
would it take a neutrino of energy $E$ and mass $m \ll E$ to reach the Earth (work to
lowest non-trivial order in $m$)? For future reference, the light travel time is approximately
$t_0 = 5.3 \times 10^{12}~{\rm s}.$ {\it (ii)} The observation times and neutrino energies for the first 5 neutrinos observed
by the Kamioka detector in Japan are given in Table~\ref{SN1987a}. The first neutrino is defined to have arrived at time 
$t_{\rm obs} = 0$. For any given neutrino, $t_{\rm obs}$ is the sum of its
emission time (compared to neutrino $\#1$) and its travel time (again, subtracting neutrino
$\#1$'s travel time). Assume that all the neutrinos have the same mass and plot their
emission times vs. the common value of $m^2c^4$ (in eV$^2$). From this plot, argue that, if all
these neutrinos were emitted within 4~s of each other (a conservative upper limit), the maximum neutrino mass is no
more than $13~{\rm eV}/c^2.$ [{\it Hint:} For simplicity, you can drop the error bars.] {\em (iii)} Consider only events $1$ and 
$3.$ If neutrino 1 was emitted no more than 1 second later than
event 3, what is the maximum neutrino mass? In this part, consider the error bars. \\

\begin{table}
\caption{Observation times and inferred neutrino energies in the Kamioka experiment. The uncertainties on observation times are small enough to be neglected.}
\begin{center}
\begin{tabular}{ccc}
\hline \hline
Event No. & Time $t_{\rm obs}$ [s] & Energy $E$ [MeV] \\
\hline
1 & 0 ({\rm def}) & $21.3 \pm 2.9$ \\
2 & 0.107 & $14.8 \pm 3.2$ \\
3 & 0.303 & $8.9 \pm 2.0$ \\
4 & 0.324 & $10.6 \pm 2.7$ \\
5 & 0.507 & $14.4 \pm 2.9$ \\
\hline
\hline
\end{tabular}
\end{center}
\label{SN1987a}
\end{table}

\subsection{Dark matter}

\begin{figure}[tbp] \postscript{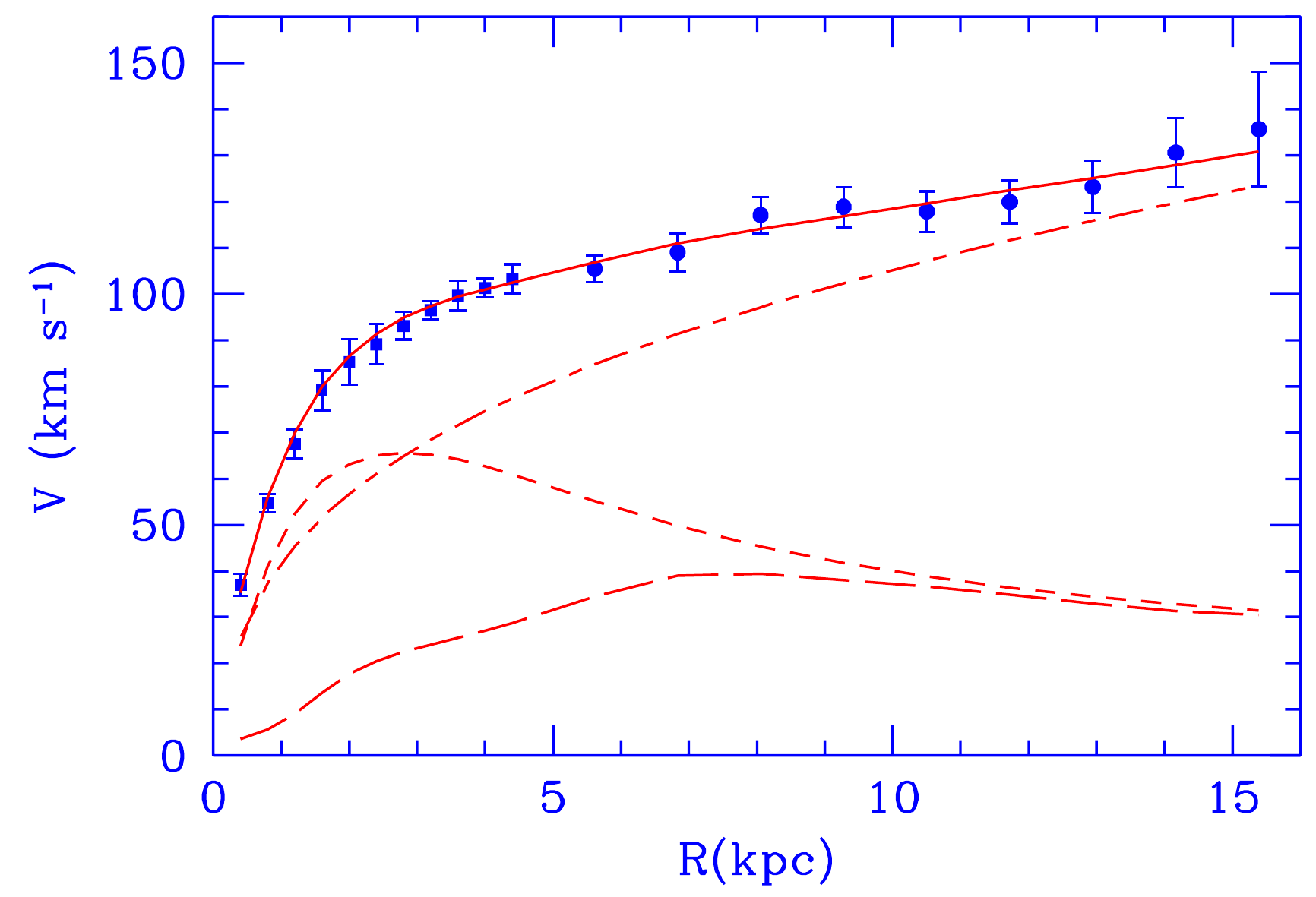}{0.9} \caption{M33 rotation curve (points) compared with the best fit model (continuous line). Also shown the halo contribution (dashed- dotted line), the stellar disk (short dashed line) and the gas contribution (long dashed line)~\cite{Corbelli:1999af}.}
\label{fig:M33}
\end{figure}

The evidence that DM is required to make sense of our Universe has been building for some time. In 1933 Zwicky found that the velocity dispersion of galaxies in the Coma cluster of galaxies was far too large to be supported by the luminous matter~\cite{Zwicky:1933gu}. In the 1970's, Rubin and collaborators measured the rotational velocities of stars in spiral galaxies and also found evidence for non-luminous matter~\cite{Rubin:1970zza,Rubin:1980zd,Rubin:1985ze}.  Spiral galaxies are flat rotating systems. The stars and gas in the disk are moving in nearly circular orbits, with the gravitational field of the galaxy providing the inward acceleration required for the circular motion.  Observed rotation curves usually exhibit a characteristic flat behavior at large distances, i.e. out towards, and even far beyond, the edge of the visible disks (see a typical example in Fig.~\ref{fig:M33}~\cite{Corbelli:1999af}). To a fair approximation, assuming Newtonian gravity, the rotational velocity $v(r)$ at radius $r$ is related to the total mass $M(r)$ within radius $r$ by the equation 
\begin{equation}
v^2(r) = \frac{GM(r)}{r} \,,
\end{equation}
 where $G$ is the gravitational constant. Here,  
\begin{equation}
M(r) = 4 \pi \int \rho(r) \ r^2 \ dr \, ,
\end{equation}
where $\rho(r)$ is the mass density profile, which should be falling $\propto 1/\sqrt{r}$ beyond the optical disc. The fact that $v(r)$ is  approximately constant implies the existence of a DM halo, with $M(r) \propto r$ and $\rho  \propto 1/r^2$.

\begin{figure}[tbp] \postscript{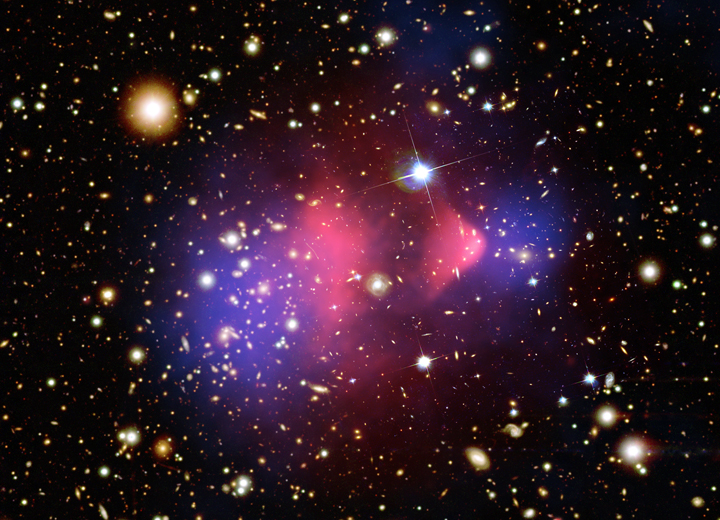}{0.9} \caption{The Bullet Cluster.}
\label{fig:BC}
\end{figure}

The most compelling evidence for DM is that
observed at the Bullet Cluster~\cite{Clowe:2006eq}. In Fig.~\ref{fig:BC} we show a composite image of
the Bullet Cluster (1E 0657-558) that shows the X-ray light detected
by Chandra in purple, (an image from Magellan and the Hubble space
telescope of) the optical light in white and orange, and the DM
map (drwan up using data on gravitational lensing from Magellan and
European Space Observatory telescopes at Paranal) in blue.  Galaxy
clusters contain not only the galaxies ($\sim 2\%$ of the mass), but
also intergalactic plasma ($\sim 10\%$ of the mass), and (assuming the
null hypothesis) DM ($\sim 88\%$ of the mass). Over time, the
gravitational attraction of all these parts naturally push all the
parts to be spatially coincident. If two galaxy clusters were to
collide/merge, we will observe each part of the cluster to behave
differently. Galaxies will behave as collisionless particles but the
plasma will experience ram pressure. Throughout the collision of two
clusters, the galaxies will then become separated from the
plasma. This is seen clearly in the Bullet Cluster, which is
undergoing a high-velocity (around 4500 km/s) merger, evident from the
spatial distribution of the hot, X-ray emitting gas.  The galaxies of
both concentrations are spatially separated from the (purple) X-ray
emitting plasma. The DM clump (blue), revealed by the
weak-lensing map, is coincident with the collisionless galaxies, but
lies ahead of the collisional gas.  As the two
clusters cross, the intergalactic plasma in each cluster interacts
with the plasma in the other cluster and slows down. However, the dark
matter in each cluster does not interact at all, passing right through
without disruption. This difference in interaction causes the DM to sail ahead of the hot plasma, separating each cluster into
two components: DM (and colissionless galaxies) in the lead
and the hot interstellar plasma lagging behind.
 
 The non-gravitational interactions of DM may be with any of the known particles or, 
for a complex hidden dynamical DM sector~\cite{Dienes:2011ja}, with other currently unknown
particles. A complete research program in DM therefore
requires a diverse set of experiments that together probe all possible
types of couplings~\cite{Feng:2010gw}. At a qualitative level,
the complementarity may be illustrated by the following observations
that follow from basic features of each approach. {\it (i)} With the information at hand one can ask the question:
``How often will a DM particle at Earth's location in the halo
interact with a particular nucleus, and what is the expected
distribution of recoil energies imparted to the nucleus?''
Direct detection experiments typically operate in deep
underground laboratories to reduce the background. The majority of
present experiments use one of two detector technologies: cryogenic
detectors, operating at temperatures below 100~mK, detect the heat
produced when a particle hits an atom in a crystal absorber such as
germanium. Noble liquid detectors detect the flash of scintillation
light produced by a particle collision in liquid xenon or argon.  {\it (ii)} DM may be detected indirectly when DM pair-annihilates somewhere, producing something, which is detected somehow. Searches for neutrinos are unique among indirect searches in that they are, given certain assumptions, probes of scattering cross sections, not annihilation cross sections, and so compete directly with the direct detection searches. The idea behind neutrino searches is the following: when DM particles pass through the Sun or the Earth, they may scatter and be slowed below escape velocity. Once captured, they then settle to the center, where their densities and annihilation rates are greatly enhanced. Although most of their annihilation products are immediately absorbed, neutrinos are not. Some of the resulting neutrinos then travel to the surface of the Earth, where they may convert to charged leptons through $\nu q \to l^+ q'$, and the charged leptons may be detected. Alternatively,
photons from DM annihilation at the Galactic center or in other galaxies can be seen by $\gamma$-ray telescopes.  {\it (iii)} Particle colliders, such as the LHC, could produce DM particles that escape the detector and be discovered as an excess of events with missing energy or momentum. {\it (iv)} The particle properties of DM are constrained through its impact on astrophysical observables. DM distributions and substructure in galaxies are unique probes of the ``warmth'' of DM and hidden DM properties, such as its self-interaction strength, and they measure the effects of DM properties on structure formation in the Universe. Examples include the self-interaction of DM particles affecting central DM densities in galaxies (inferred from rotation velocity or velocity dispersion measures), the mass of the DM particle affecting DM substructure in galaxies (inferred from strong lensing data), and the annihilation of DM in the early Universe affecting CMB fluctuations.

\subsection{Dark energy}

The expansion history of the cosmos can be determined
using as a ``standard candle'' any distinguishable class of
astronomical objects of known intrinsic brightness that can be
identified over a wide distance range. As the light from such beacons
travels to Earth through an expanding universe, the cosmic expansion
stretches not only the distances between galaxy clusters, but also the
very wavelengths of the photons en route. By the time the light
reaches us, the spectral wavelength $\lambda$ has thus been redshifted
by precisely the same incremental factor $z \equiv \delta
\lambda/\lambda$ by which the cosmos has been stretched in the time
interval since the light left its source. That time interval is the
speed of light times the object's distance from Earth, which can be
determined by comparing its apparent brightness to a nearby standard
of the same class of astrophysical objects.  The recorded redshift and
brightness of each such object thus provide a measurement of the
total integrated exansion of the universe since the time the light was
emitted. A collection of such measurements, over a sufficient range of
distances, would yield an entire historical record of the universe's
expansion.

\begin{figure}[tbp] \postscript{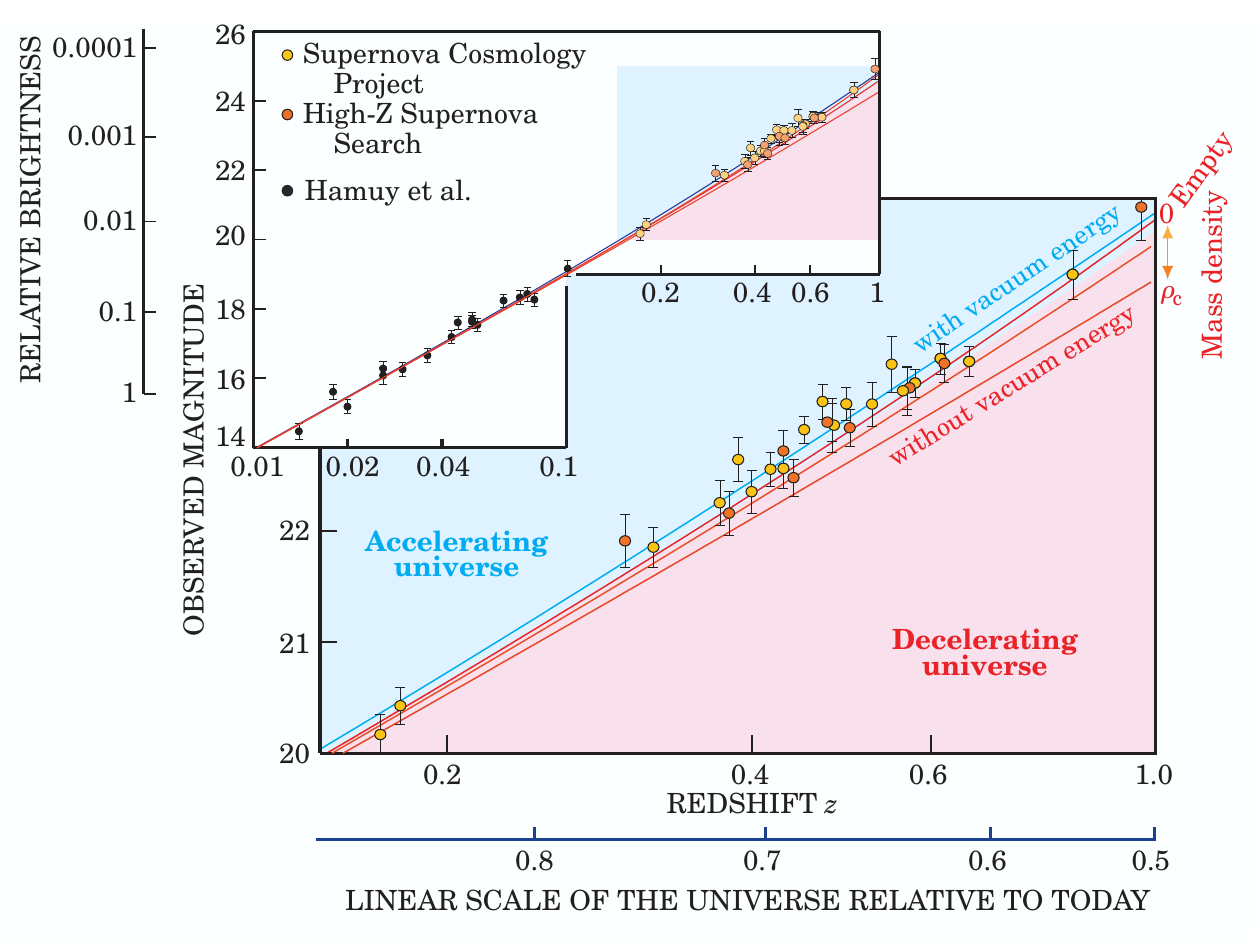}{0.9} \caption{Observed magnitude
versus redshift is plotted for well-measured distant~\cite{Riess:1998cb,Perlmutter:1998np} and (in the inset) nearby~\cite{Hamuy:1993,Hamuy:1995in} type Ia supernovae. For clarity, measurements at the same redshift are combined. At redshifts beyond $z = 0.1$ (distances greater than about 109 light-years), the cosmological predictions (indicated by the curves) begin to diverge, depending on the assumed cosmic densities of mass and vacuum energy. The red curves represent models with zero vacuum energy and mass densities ranging from the critical density $\rho_c$ down to zero (an empty cosmos). The best fit (blue line) assumes a mass density of about $\rho_c/3$ plus a vacuum energy density twice that large, implying an accelerating cosmic expansion~\cite{Perlmutter:2003}.}
\label{fig:SN}
\end{figure}

In Fig.~\ref{fig:SN}  we show the observed magnitude versus redshift
for well-measured distant and (in the inset) nearby type Ia
supernovae. The faintness (or distance) of the high-redshift supernovae
in Fig.~\ref{fig:SN}  comes as a dramatic surprise. In the simplest cosmological
models, the expansion history of the cosmos is determined entirely by
its mass density. The greater the density, the more the expansion is
slowed by gravity. Thus, in the past, a high-mass-density universe
would have been expanding much faster than it does today. So one
should not have to look far back in time to especially distant (faint)
supernovae to find a given integrated expansion
(redshift). Conversely, in a low-mass-density universe one would have
to look farther back. But there is a limit to how low the mean mass
density could be. After all, we are here, and the stars and galaxies
are here. All that mass surely puts a lower limit on how far-that is,
to what level of faintness we must look to find a given redshift.  However, the
high-redshift supernovae in Fig.~\ref{fig:SN} are fainter than would
be expected even for an empty cosmos. 

\begin{figure}[tbp] \postscript{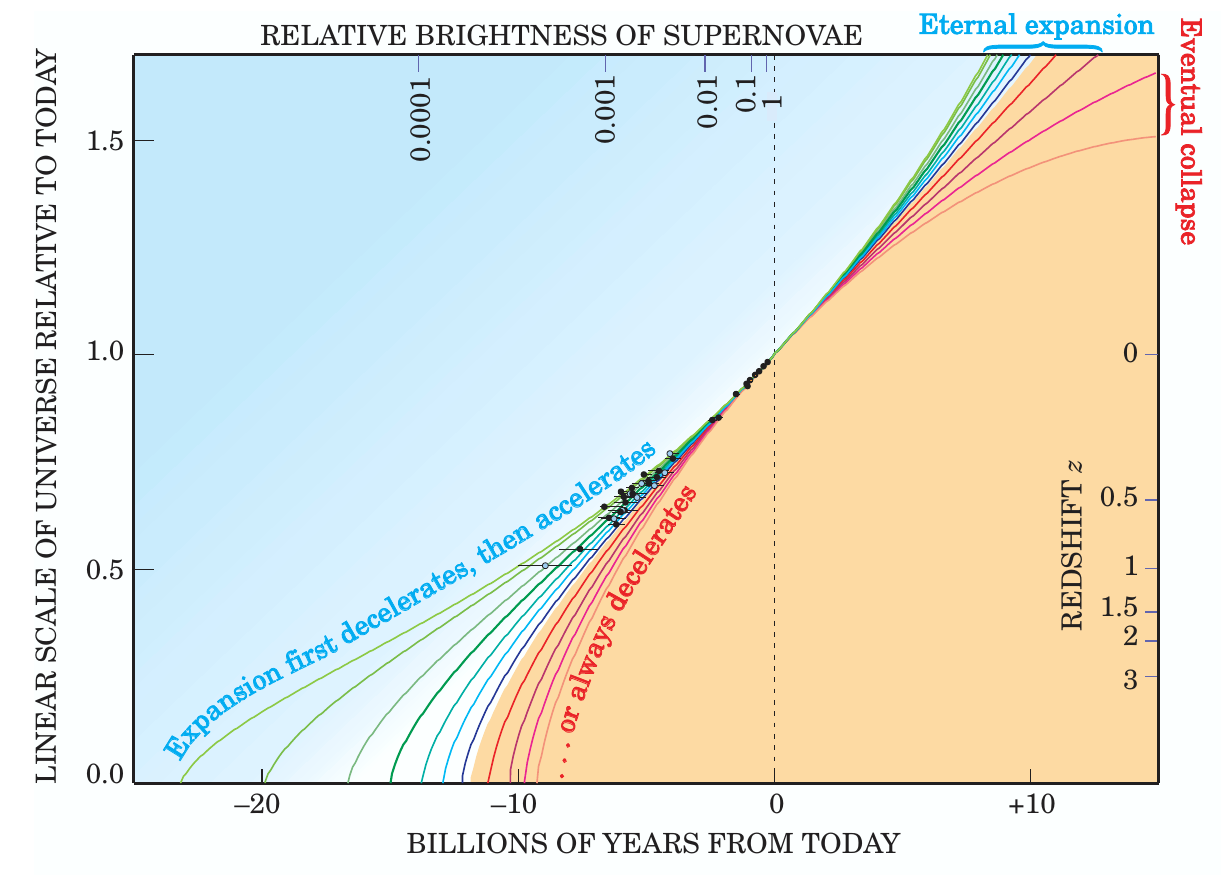}{0.9} \caption{The history of cosmic
expansion, as measured by the high-redshift supernovae (the black data points), assuming flat cosmic geometry. The scale factor $R$ of the universe is taken to be 1 at present, so it equals $1/(1 + z)$. The curves in the blue shaded region represent cosmological models in which the accelerating effect of vacuum energy eventually overcomes the decelerating effect of the mass density. These curves assume vacuum energy densities ranging from $0.95~\rho_c$ (top curve) down to $0.4~\rho_c$. In the yellow shaded region, the curves represent models in which the cosmic expansion is always decelerating due to high mass density. They assume mass densities ranging (left to right) from $0.8~\rho_c$ up to $1.4~\rho_c$. In fact, for the last two curves, the expansion eventually halts and reverses into a cosmic collapse~\cite{Perlmutter:2003}.}
\label{fig:SN2}
\end{figure}

If these data are correct, the obvious implication is that the simplest  model of cosmology must be too simple. The next to simplest model includes an expansionary term in the equation of motion driven by the cosmological constant $\Lambda$, which competes against gravitational collapse.  The best fit to the 1998 supernova data shown in Figs.~\ref{fig:SN} and \ref{fig:SN2} implies that, in the present epoch, the vacuum energy density $\rho_\Lambda$ is larger than the energy density attributable to mass $\rho_m$. Therefore, the cosmic expansion is now accelerating. If the universe has no large-scale curvature,
as the recent measurements of the CMB strongly indicate, we can say quantitatively that about 70\% of the total energy density is vacuum energy and 30\% is mass. In units of the critical density 
\begin{equation}
\rho_c = \frac{3H}{8 \pi G} \,,
\label{BJMwkee3}
\end{equation}
 one usually writes this result as $\Omega_\Lambda \equiv \rho_\Lambda/\rho_c \approx 0.7$ and $\Omega_m \equiv \rho_m/\rho_c \approx 0.3$~\cite{Weinberg:2008zzc}. Only about 4.6\% of the mass are atoms.  Hence, more than 95\% of the energy density in the universe is in a form that has never been directly detected in the laboratory. The actual density of atoms is equivalent to roughly 1 proton per $4~{\rm m}^3$.

The SM has no natural place for a vacuum energy density of the modest magnitude required by the astrophysical data. The simplest estimates would predict a vacuum energy $10^{120}$ times greater. So enormous $\Lambda$ would have engendered an acceleration so rapid that stars and galaxies could never have formed. Therefore it has long been assumed that there must be some underlying symmetry that precisely cancels the vacuum energy. Now, however, the supernova data appear to require that such a cancellation would have to leave a remainder of about one part in $10^{120}$. That degree of fine tuning is most unappealing.
The cosmological constant model requires yet another fine tuning. In the cosmic expansion, mass density becomes ever more dilute. Since the end of inflation, it has fallen by very many orders of magnitude. However, the vacuum energy density $\rho_\Lambda$, a property of empty space itself, stays constant. It seems a remarkable and implausible coincidence that the mass density, just in the present epoch, is within a factor of 2 of the vacuum energy density~\cite{Bahcall:1999xn}.
Given these two fine-tuning coincidences, it seems likely that the SM is missing some fundamental physics. \\

{\bf EXERCISE 14.3.}~{\it (i)} Write Gauss's law for a galaxy of mass $m$ inside a homogeneous and isotropic universe of mass density $\rho$, and show that
\begin{equation}
\frac{d^2 R/dt^2}{R} = -\frac{4 \pi}{3} G\rho,
\end{equation}
where $R$ is the distance of the galaxy to the center of the distribution. {\it (ii)}  Obtain an expression for $dR/dt$ from the fact that the total energy of the galaxy vanishes.
P.S. You have now derived Friedman's equations for a flat universe in the Newtonian limit.

\subsection{Across the Multiverse}

\begin{figure}[tbp] \postscript{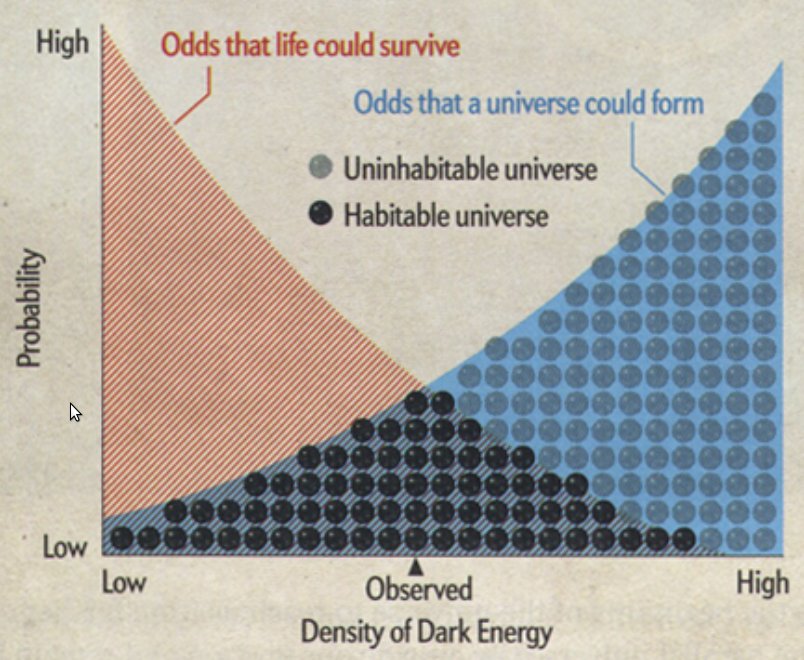}{0.9} \caption{The higher the value of $\Lambda$ the most probable it is for the universe to form, but the more hostile the universe would be to life~\cite{Ellis:2014}.}
\label{fig:multiverse}
\end{figure}

The extraordinary level of fine tuning required in generating a cosmological constant constant consistent with observations motivated a radiaclly different approach in deriving the laws of nature. The key insight of this new viewpoint is no longer symmetries and invariants, but rather statistics. A remarkable fact about our universe is that the physical constants has just the right values allow for complex structures, including living things. The so-called multiverse provides an explanation for this apparent coincidence: If all possible values occur in a large enough collection of universes, then viable ones for life will surely be found somewhere~\cite{Weinberg:1987dv,Weinberg:1988cp,Susskind:2003kw,Kachru:2003aw}. This reasoning has been applied, in particular, to explaining the density of dark energy that is speeding up the expansion of the universe today. Indeed, it is the only scientifically based option we have right now.  Inflation is the extremely rapid exponential expansion of the early universe by a factor of at least $10^{78}$ in volume, driven by a negative-pressure vacuum energy density.  Following the inflationary period, the universe continued to expand, but at a slower rate. As a direct consequence of this expansion, all of the observable universe originated in a small causally connected region. Inflation answers the classic conundrum of the Big Bang cosmology: why does the universe appear flat, homogeneous, and isotropic in accordance with the cosmological principle when one would expect, on the basis of the physics of the Big Bang, a highly curved, heterogeneous universe?  In theories of eternal inflation, the inflationary phase of the multiverse's expansion lasts forever in at least some regions of the multiuniverse. Because these regions expand exponentially rapidly, most of the volume of the multiverse at any given time is inflating. All models of eternal inflation produce an infinite multiverse. The process of eternal inflation, endows each universe with a random density of dark energy.  Relatively few universes have zero or a low dark energy density; most have higher values (blue area in Fig.~\ref{fig:multiverse}). But too much dark energy tears apart the complex structures needed to sustain life (red area in Fig.~\ref{fig:multiverse}). So most habitable universes should have a middling density of dark energy (peak of overlap region).

\section{Where do we stand and where are we going}

The LHC physics program has just started. The ATLAS and CMS experiments have reported the discovery of a new boson with a mass around 126~GeV consistent with the SM Higgs particle (within current experimental and theoretical uncertainties). This marks just the beginning of the physics exploration of the LHC and it opens a new chapter in the study of the mechanism of electroweak symmetry breaking. The physics accessible at the high-luminosity phase of the LHC14 extends well beyond that of the earlier LHC8 program.  Concurrent with LHC14 measurements will be many interesting missions to measure cosmological parameters, as well as dedicated experiments designed to detect cosmic rays, neutrinos, and DM particles. This new arsenal of data will certainly provide an ideal testing ground for the many interesting theories that extend the SM (see, e.g.~\cite {Quevedo:2010ui,Wells:2009kq,Quigg:2004is}), and, at the same time, a unique opportunity to view similar physics from various different points of view. Alternatively, it is possible that we may be guessing the future while holding too small a deck of cards and LHC14 will open a new world that we did not anticipate

\acknowledgments{L.A.A. is supported by U.S. National Science Foundation (NSF) CAREER Award PHY1053663 and by the National Aeronautics and Space Administration (NASA) Grant No. NNX13AH52G; he thanks the Center for Cosmology and Particle Physics at New York University for its hospitality. Any opinions, findings, and conclusions or recommendations expressed in this material are those of the author and do not necessarily reflect the views of the NSF or NASA.}

\onecolumngrid

\section*{Answers and Comments on the Exercises}

9.1~$\int_0^\infty P(E)  \, dE = \int_0^\infty e^{-E/(kT)}/(kT) \ dE  = \left. -(kT) e^{-E/(kT)}/(kT) \right|_0^\infty = 1$. To calculate $\int_0^\infty E \, P(E) \, dE$ we adopt the following change of variables, $u = E$ and $dv = e^{-E/(kT)} dE$, yielding $du = dE$ and $v = -kT e^{-E/(kT)}$. Then, $\int_0^\infty E e^{-E/(kT)} dE = uv - \int v du = \left. E (-kT) e^{-E/(kT)} \right|_0^\infty - \int_0^\infty (-kT) e^{-E/(kT)} dE = kT \int_0^\infty e^{-E/(kT) } dE = (kT)^2$. Thus
$\langle E \rangle = \int_0^\infty E \, P(E) \, dE = (kT)^2/(kT) = kT$. We now take $E_n = n h \nu = n h c/\lambda$ to compute (\ref{Planck-E}),
\begin{equation}
\langle E \rangle =  \frac{\sum_{n=0}^\infty \frac{nhc}{\lambda k T} e^{-nhc/(\lambda kT)}}{\frac{1}{kT} \sum_{n=0}^\infty e^{-nhc/(\lambda kT)}} = kT \frac{\sum_{n=0}^\infty n \alpha e^{-n \alpha}}{\sum_{n=0}^\infty e^{-n\alpha}} \,,
\label{france1}
\end{equation}
where $\alpha = hc/(\lambda kT)$. First we note that
\begin{equation}
-\alpha \frac{d}{d\alpha}\ln \left[ \sum_{n=0}^\infty e^{-n \alpha} \right] = \frac{\sum_{n=0}^\infty n \alpha e^{-n\alpha}}{\sum_{n=0}^\infty e^{-n\alpha}} \, ,
\label{france2}
\end{equation}
and so substituting (\ref{france2}) into (\ref{france1}) we have
\begin{equation}
\langle E \rangle = kT \left[-\alpha \frac{d}{d\alpha} \ln \left(\sum_{n=0}^\infty e^{-n\alpha} \right) \right] = - \frac{hc}{\lambda} \left[ \frac{d }{d\alpha} \ln \left(\sum_{n=0}^\infty e^{-n \alpha} \right) \right] = - \frac{hc}{\lambda} \left\{ \frac{d}{d \alpha} \ln \left[(1 - e^{-\alpha})^{-1} \right] \right\} \,,
\label{france3}
\end{equation}
where in the last equality we have used the sum of a geometric series $ \sum_{n=0}^\infty e^{-n\alpha} = (1 - e^{-\alpha})^{-1}$. Now, we calculate
\begin{equation}
\frac{d}{d \alpha} \ln (1 - e^{-\alpha})^{-1} = (-1) \frac{(1 - e^{-\alpha})^{-2} e^{-\alpha}  }{(1 - e^{-\alpha})^{-1}} = \frac{- e^{-\alpha}}{1- e^{-\alpha}} \left( \frac{e^\alpha}{e^{\alpha}}\right)  = - \frac{1}{(e^\alpha - 1)} \, .
\label{france4}
\end{equation}
Substituting (\ref{france4}) into (\ref{france3}) we obtain (\ref{Planck-E}). \\

9.2~In the frequency domain (\ref{Pumpkin1}) can be rewritten as
\begin{equation}
F(T) = \pi \int_0^\infty  \frac{2h \nu^3}{c^2} \frac{1}{e^{h \nu/(kT)} -1} d \nu \, .
\end{equation}
This is an integral over frequency alone. Change the variable to $x = h\nu /(kT)$ to obtain an integral of the form $\int[x^3 (e^x-1)^{-1}] dx$. Solutions of this type of integral include the Riemann zeta function, which is defined in Appendix~\ref{appB}. It is straightforward to show that the result of the integration can be written as $F(T) = \sigma T^4$, with $\sigma = 2 \pi^5 k^4/(15 c^2 h^3) = 5.67 \times 10^{-8}~{\rm W} \, {\rm m}^{-2} \, {\rm K}^{-4}$. \\

9.3~The first derivative of (\ref{yogabagabaP}) in the frequency domain reads
\begin{equation}
\frac{\partial}{\partial \nu} B_\nu(\nu,T) = \frac{2 h \nu^2}{c^2 \left[e^{h\nu/(kT)} - 1 \right]^2} \left\{3 \left[e^{h\nu/(kT)} -1 \right] - \frac{h \nu}{kT} e^{h\nu/(kT)} \right\} \, .
\end{equation}
This derivative is only equal to zero when the numerator is equal to zero. The corresponding denominator is larger than zero for $0 < \nu <\infty$. Demanding the numerator to vanish we obtain $3 (e^{x_\nu} - 1) = x_\nu e^{x_\nu}$, with $x_\nu = h\nu/(kT)$. This transcendental function can only be solved numerically. We find $x_\nu \simeq 2.82$, and in a further step $\nu_{\rm ext}/T = x_\nu k /h$, where $\nu_{\rm ext}$ is the frequency at which the extreme (either a minimum or a maximum) of the Planck function in the frequency domain occurs. It can simply be proved that for this extreme the second derivative fulfills the condition $\partial^2B_\nu/\partial \nu^2 <0$ so that the extreme corresponds to a maximum, $\nu_{\rm max}$. The form of Wien's displacement law in terms of maximum spectral emittance per unit wavelength is derived using similar methods, but starting with the form of Planck's law expressed in the wavelength domain. The effective result is to substitute 3 for 5 in the equation for the peak frequency, i.e. $5 (e^{x_\lambda} -1) = x_\lambda e^{x_\lambda}$, where $x_\lambda = hc/(\lambda kT)$. This solves with $x_\lambda = 4.96$, yielding $\lambda_{\rm max} T = ch/(x_\nu k) \simeq 2.90 \times 10^{-3}~{\rm m} \, {\rm K}$.\\

9.4~{\it (i)}~We can obtain a first estimate of the surface temperature of the Sun from the sensitivity of the human eye to light in the range $400-700~{\rm nm}$. Assuming that the evolution worked well, i.e. that the human eye uses optimal the light from the Sun, and that the atmosphere is for all frequencies in the visible range similarly transparent, we identify the maximum in Wien's law with the center of the frequency range visible for the human eye. Thus we set $\lambda_{\rm max,\odot} \approx 550~{\rm nm}$, and obtain $T_\odot \approx 5270~{\rm K}$ for the surface temperature of the Sun. {\it (ii)}~The bolometric luminosity $L$ of a star is given by the product of its surface $A = 4 \pi R^2$ and the radiation emitted per area $\sigma T^4$, i.e., $L = 4 \pi R^2 \sigma T^4$. The radiant flux is defined by $F = L/A$, so that  we recover the well known inverse-square law for the energy flux at the distance $r > R$ outside of the star, $F = L/(4\pi r^2)$. The validity of the inverse-square law $F(r) \propto r^{-2}$ relies on the assumptions that no radiation is absorbed and that relativistic effects can be neglected. The later condition requires, in particular, that the relative velocity of observer and source is small compared to the velocity of light. The energy flux received from the Sun at the distance of the Earth, $r_{\rm SE} = 1~{\rm AU}$, is equal to $F = 1365~{\rm W/m^2}$. The solar luminosity follows then as $L_\odot = 4 \pi d^2 F = 4 \times 10^{33}~{\rm erg} \, {\rm s}^{-1}$, and serves as a convenient unit in stellar astrophysics. The Stefan-Boltzmann law can then be used
to define, with $R_\odot \approx 7 \times 10^{10}~{\rm cm}$, the effective temperature of the Sun, $T_\odot \approx 5780~{\rm K}$. {\it (iii)}~The average temperature on the surface of Neptune is $T_{\rm N} = L_\odot/(4 \pi r_{\rm N}^2 \sigma) = 73~{\rm K}$.\\

9.5~The total energy density in the blackbody
radiation is 
\begin{equation}
 u = \int_0^\infty \frac{8 \pi h c}{\lambda^5}
\,d\lambda\, \frac{1}{e^{hc/\lambda k T} -1} \, .
\end{equation} 
Change the variable $x = hc/\lambda kT$ to obtain an integral of the form $\int[x^3 (e^x-1)^{-1}] dx$, which can be solved using the Riemann zeta function, see Appendix~\ref{appB}; we obatain  
\begin{equation}
 u = \frac{8 \pi^5 (kT)^4}{15 (hc)^3 }
= 7.56464 \times 10^{-15}\, (T/{\rm K})^4~{\rm erg}/{\rm cm}^3\,\,.  
\end{equation}
(Recall that $1~{\rm J} \equiv 10^7~{\rm erg} = 6.24 \times
10^{18}~{\rm eV}$.) We can easily interpret the Planck distribution in
terms of quanta of light or photons. Each photon has an energy $E_\gamma =
hc/\lambda.$ Hence the number $dn_\gamma$ of photons per unit volume in
blackbody radiation in a narrow range of wavelengths from $\lambda$ to
$\lambda + d \lambda$ is 
\begin{equation}
dn_\gamma = \frac{du}{hc/\lambda} = \frac{8
  \pi}{\lambda^4} \,d\lambda\, \frac{1}{e^{hc/\lambda k T} -1} \, .  
\end{equation}
Then the total number of photons per unit volume is 
\begin{equation}
 n_\gamma =
\int_0^\infty dn_\gamma = 8 \pi \left(\frac{kT}{hc}\right)^3 \int_0^\infty
\frac{x^2\,\, dx}{e^x - 1}, 
\end{equation}
where $x = hc/(\lambda kT).$ The integral cannot be expressed in terms of elementary functions, but $\int[x^2 (e^x-1)^{-1}] dx =\Gamma (3) \zeta (3) \approx 2.4$, see Appendix~\ref{appB}. Therefore, the number photon density is 
\begin{equation}
 n_\gamma = 60.42198 \left(\frac{kT}{hc}\right)^3 \nonumber = 20.28 \left(\frac{T}{{\rm
      K}}\right)^3~{\rm photons}\, {\rm cm}^{-3} \approx 400~{\rm
  photons}\, {\rm cm}^{-3}, 
\end{equation}
and the average photon energy is 
\begin{equation}
\langle E_\gamma \rangle = u/n_\gamma = 3.73 \times 10^{-16} \, (T/{\rm
  K})~{\rm erg} \,\,.  
\end{equation}
For a temperature of 2.726~K, the number density of CMB photons is $\approx 410~{\rm cm}^{-3}$ and the typical photon energy is $\approx 6 \times 10^{-4}~{\rm eV}$ in agreement with the values adopted in exercise~8.9. {\it (iii)}~Now, let's consider what happens
to blackbody radiation in an expanding universe. Suppose the size of
the universe changes by a factor $f$, for example, if it doubles in
size, then $f = 2.$ As predicted by the Doppler effect, the
wavelengths will change in proportion to the size of the universe to a
new value $\lambda' = f \lambda$. After the expansion, the energy
density $du'$ in the new wavelength range $\lambda'$ to $\lambda' +
d\lambda'$ is less than the original energy density $du$ in the old
wavelength range $\lambda + d\lambda$,
for two different reasons: 
$(1)$~since the volume of the universe has increased by a factor of
$f^3$, as long as no photons have been created or destroyed, the
number of photons
per unit volume has decreased by a factor of $1/f^3$;
$(2)$~the energy of each photon is inversely proportional to its
wavelength, and therefore is decreased by a factor of $1/f$. It
follows that the energy density is decreased by an overall factor
$1/f^3 \times 1/f = 1/f^4$:
\begin{equation}
du' = \frac{1}{f^4}\,\, du = \frac{8 \pi h c}{\lambda^{5} f^4} \,d\lambda \, \frac{1}{e^{hc/\lambda k T} -1}.
\end{equation}
If we rewrite the previous equation in terms of the new wavelengths
$\lambda'$, it becomes
\begin{equation}
du' = \frac{8 \pi h c}{\lambda^{'5}} \,d\lambda' \, \frac{1}{e^{hcf/\lambda' k T} -1},
\end{equation}
which is exactly the same as the old formula for $du$ in terms of
$\lambda$ and $d\lambda$, except that $T$ has been replaced by a new
temperature $T' = T/f$. Therefore, we conclude that freely expanding
blackbody radiation remains described by the Planck formula, but with
a temperature that drops in inverse proportion to the scale of
expansion. \\

9.6~{\it (i)}~According to classical theory, the energy in a light wave is spread out uniformly and continuously over the wavefront. Assuming that all absorption of light occurs in the top atomic layer of the metal, that each atom absorbs an equal amount of energy proportional to its cross sectional area, $A$, and that each atom somehow funnels this energy into one of its electrons, we find that each electron absorbs an energy $K$ in time $t$ given by $K = \epsilon IAt$
where $\epsilon$ is a fraction accounting for less than 100\% light absorption. Because the most energetic electrons are held in the metal by a surface energy barrier $\varphi$, these electrons will be emitted with $K_{\rm max}$ once they have absorbed enough energy to overcome the barrier. We can express this as $K_{\rm max} = \epsilon I A t - \varphi$.
Thus, classical theory predicts that for a fixed absorption period, $t$, at low light intensities when $\epsilon I A t < \varphi$, no electrons must be emitted. At higher intensities, when $\epsilon I A t> \varphi$ electrons should be emitted with higher kinetic energies the higher the light intensity. Therefore, classical predictions contradict experiment at both very low and very high light intensities. {\it (ii)}~According to classical theory, the intensity of a light wave is proportional to the square of the amplitude of the electric field, $|\vec E_0|^2$, and it is this electric field amplitude that increases with increasing intensity and imparts an increasing acceleration and kinetic energy to an electron. Substituting $I$ with a quantity proportional to $|\vec E_0|^2$ in part (i) shows that $K_{\rm max}$ should not depend at all on the frequency of the classical light wave, again contradicting the experimental results. {\it (iii)}~To estimate the time lag between the start of illumination and the emission of electrons, we assume that an electron must accumulate just enough light energy to overcome the work function. Setting $K_{\rm max} = 0$ gives $0 = \epsilon I A t - \varphi$ or $t = \varphi/ (\epsilon I A)$. Taking $\epsilon = 1$ and the cross sectional area of the atom $A = \pi r^2$, where $r = 1.0 \times 10^{-8}~{\rm cm}$ is a typical atomic radius, we have $t = 1.2 \times 10^7~{\rm s} \approx 130~{\rm days}$. Thus we see that the classical calculation of the time lag for photoemission does not agree with the experimental result, disagreeing by a factor of $10^{16}$!\\

9.7~At the threshold wavelength the photoelectrons have just enough energy to overcome the work function, so $K_{\rm max} = 0$. Hence we have $\varphi = hc/\lambda_0 = 3.808~{\rm eV}$. When 259.8~nm light is used, $eV_0 = hc/\lambda  - \varphi =0.964~{\rm eV}$, so $V_0 = 0.964~{\rm V}$.\\

9.8~{\it (i)}~Consider an incident photon of frequency $\nu_0$ which is scattered by a stationary electron to give a photon of frequency $\nu$ at an angle $\theta$ with respect to the original photon. Conservation of energy gives 
\begin{equation}
E_0 + mc^2 = E + E', 
\label{C-energy}
\end{equation}
while conservation of 3-momentum gives $\vec p_0 = \vec p + \vec p^{\,\prime}$ or $\vec p^{\, \prime} = \vec p_0 - \vec p$. Squaring this $\vec p^{\, \prime 2} = \vec p_o^{\, 2} + \vec p^{\, 2} - 2 \vec p_0 \cdot \vec p$ and using $E_0 = |\vec p_0 c|$, $E = |\vec pc|$, and $E^{\prime 2} = (\vec p^{\, \prime} c)^2 + m^2 c^4$, we get $E^{\prime 2} - m^2 c^4 = E_0^2 + E^2 - 2 E_0 E \cos \theta$. Now, from (\ref{C-energy}) the left-hand side of the previous relation equals $(E_0 - E)^2 + 2 mc^2 (E_0 - E)$, thus yielding 
\begin{equation}
2 mc^2 (E_0 - E) = 2 E_0 E - 2 E_0 E - 2 E_0 E \cos \theta \, .
\end{equation}
Using $E_0 = hc/\lambda_0$ and $E = h c/\lambda$ we obtain the well known formula for the change in wavelength as a function of angle
\begin{equation}
\lambda - \lambda_0 = \frac{h}{mc} (1 - \cos \theta) \,
\end{equation}
The quantity $h/(mc)$ is known as the Compton wavelength (in this case, of the electron). {\it (ii)}~Since $\lambda_c$ is very small, high energy radiation (X-rays) is needed to observe the effect. If we choose a wavelength of $7 \times 10^{-9}~{\rm cm}$ for the X-rays we estimate for a maximal scattering angle an effect of $\Delta \lambda/\lambda_0 = 2 \lambda_{\rm c}/\lambda_{\rm Xray} \approx 0.07.$\\

9.9~The dominant energy loss is from electric dipole radiation, which obeys the Larmor formula (\ref{larmor257}).  For an electron of charge $-e$ and  mass $m_e$ in an orbit of radius $r$ about a fixed nucleus of charge $+e$, the radial component of the nonrelativistic force law, $\vec F = m_e \vec a$, tells us that
$e^2/r^2 = m_e a_r \approx m_e \, v_\theta^2/r$,
in the adiabatic approximation that the orbit remains nearly circular at all times. In the same approximation, $a_\theta \ll a_r$, i.e., $a \approx a_r$, and hence,
\begin{equation}
\frac{dE}{dt} = - \frac{2 e^6}{3 r^4 m_e^2 c^3} = - \frac{2}{3} \frac{r_0^3}{r^4} m_e c^3 \,,
\label{prob3tres}
\end{equation}
where $r_0 = e^2/(m_ec^2) = 2.8 \times 10^{-15}~{\rm m}$ is the classical electron radius. The total nonrelativistic energy (kinetic plus potential) is
\begin{equation}
E = - \frac{e^2}{r} + \frac{1}{2} m_e v^2 = - \frac{e^2}{2r} = - \frac{r_0}{r} m_e c^2 \, .
\label{prob4cuatro}
\end{equation}
Equating the time derivative of  (\ref{prob4cuatro}) to (\ref{prob3tres}), we have
\begin{equation}
\frac{dE}{dt} = \frac{r_0}{2 r^2} \dot r m_e c^2 = - \frac{2}{3} \frac{r_0^3}{r^4} m_e c^3, \quad {\rm or \, equivalently} \quad
r^2 \dot r = \frac{1}{3} \frac{dr^3}{dt} = - \frac{4}{3} r_0^2 c, \quad {\rm yielding} \quad
r^3 = a_0^3 - 4 r_0^2 ct \,. 
\end{equation}
 The time to fall to the origin is then
\begin{equation}
t_{\rm fall} = \frac{a_0^3}{4 r_0^2 c} = 1.6 \times 10^{-11}~{\rm s} \, .
\end{equation}
This is of the order of magnitude of the lifetime of an excited hydrogen atom, whose
ground state, however, appears to have infinite lifetime.\\

9.10~{\it (i)} The average kinetic energy of the neutrons is $K = 3kT/2 = 0.0379~{\rm eV}$. {\it (ii)}~The neutrons are non-relativistic so the momentum is given by $p = \sqrt{2mK} =  \sqrt{2 mc^2 K}/c = 8.44 \times 10^3~{\rm eV}/c$, yielding $\lambda = hc/(pc) = 0.147~{\rm nm}$.\\

10.1~Perhaps the most direct solution is to begin with the continuity equation (\ref{exercise101}), substitute in the definitions of $\rho$ and $j$, and then prove the equality. First, calculate the partial derivatives:
\begin{equation}
\frac{\partial j}{\partial x} = - \frac{i \hslash}{2m} \left( \frac{\partial \psi^*}{\partial x} \frac{\partial \psi}{\partial x} + \psi^* \frac{\partial^2 \psi}{\partial x^2} - \frac{\partial^2 \psi^*}{\partial x^2} \psi - \frac{\partial \psi^*}{\partial x} \frac{\partial \psi}{\partial x} \right) = - \frac{i\hslash}{2m} \left( \psi^* \frac{\partial^2 \psi}{\partial x^2} - \frac{\partial^2 \psi^*}{\partial x^2} \psi \right) 
\end{equation}
and
\begin{equation}
\frac{\partial \rho}{\partial t} = \frac{\partial \psi^*}{\partial t} \psi + \psi^* \frac{\partial \psi}{\partial t} \, . 
\label{conti1}
\end{equation}
The connection between the time and space derivatives is given by rearranging the Schr\"odinger equation and its complex conjugate
\begin{equation}
\frac{\partial \psi}{\partial t} = \frac{i\hslash}{2m} \frac{\partial^2 \psi}{\partial x^2} - \frac{i}{\hslash} V \psi \quad {\rm and} \quad \frac{\partial \psi^*}{\partial t} = -\frac{i\hslash}{2m} \frac{\partial^2 \psi^*}{\partial x^2} + \frac{i}{\hslash} V \psi^* \, .
\label{conti2}
\end{equation} 
Substituting (\ref{conti2}) into (\ref{conti1})
\begin{equation}
\frac{\partial \rho}{\partial t} = - \frac{i\hslash}{2m} \frac{\partial^2 \psi^*}{\partial x^2} \psi + \frac{i}{\hslash} V \psi^* \psi + \frac{i \hslash}{2m} \psi^* \frac{\partial^2 \psi}{\partial x^2} - \frac{i}{h} V \psi^* \psi = - \frac{\partial j}{\partial x} \, .
\end{equation}
The continuity equation is equivalent to conservation of probability. One way to see this is to integrate the continuity equation over $x$, with the added restriction that $\psi$ and $\partial \psi/\partial x$ go to zero as $x \to \pm \infty$, 
\begin{equation}
0 = \int_{-\infty}^{+ \infty} dx \left(\frac{\partial \rho}{\partial t} + \frac{\partial j}{\partial x} \right) = \frac{\partial}{\partial t} \int_{-\infty}^{+\infty} d x |\psi|^2 + \int_{-\infty}^{+\infty} dx \frac{\partial j}{\partial x} = \left. \frac{\partial}{\partial t} \int_{-\infty}^{+\infty} d x |\psi|^2 + j(x) \right|_{-\infty}^{+\infty} =  \frac{\partial}{\partial t} \int_{-\infty}^{+\infty} d x |\psi|^2 \, .
\end{equation}
The last integral is the total probability (otherwise known as the normalization), and is shown to be constant with respect to time.\\

10.2~{\it (i)} From the indentity (\ref{operation1}) it follows that:
\begin{equation}
\langle (A^\dagger)^\dagger \phi | \psi \rangle = \langle \phi | A^\dagger \psi \rangle = \langle A^\dagger \psi| \phi \rangle^* = \langle \psi |\hat A \phi \rangle^* = (\langle A \phi | \psi \rangle^*)^* = \langle \hat A \phi |\psi \rangle \Rightarrow (\hat A^\dagger)^\dagger = \hat A 
\end{equation}
and
\begin{equation}
\langle \phi|\hat A \hat B \psi \rangle = \langle \hat A^\dagger \phi |B \psi \rangle = \langle \hat B^\dagger \hat A^\dagger \phi |\psi\rangle \Rightarrow (AB)^\dagger = B^\dagger A^\dagger \, . 
\end{equation}
\vspace{0.25cm}

10.3~{\it(i)}~The operator $\hat x$ is hermitian because 
\begin{equation}
\int_{-\infty}^{+\infty} (\hat x \phi)^* \, \psi \, dx = \int_{-\infty}^{+\infty} (x \phi(x))^* \, \psi(x) = \int_{-\infty}^{+\infty} \phi^*  x \, \psi \, dx = \int_{-\infty}^{+\infty} \psi^* \, \hat x \psi \, dx \, . 
\end{equation}
{\it (ii)} The operatror $\hat p$ is hermitian because
\begin{equation}
\int (\hat p \phi)^* \psi dx = \int_{-\infty}^{+\infty} \left(-i\hslash \frac{\partial \phi}{\partial x} \right)^* \psi \, dx = i \hslash \int_{-\infty}^{+\infty} \left(\frac{\partial \phi}{\partial x} \right)^* \psi \, dx 
\label{seqar}
\end{equation}
and after integration by parts, recognizing that the wave function tends to zero as $x \to infty$, the right-hand side of (\ref{seqar}) becomes
\begin{equation}
- i \hslash \int_{-\infty}^{+\infty} \phi^* \, \frac{\partial \psi}{\partial x} \, dx = \int_{-\infty}^{+\infty} \phi^* \, \hat p \psi dx \, .
\end{equation}
 
\vspace{0.25cm}

10.4~We may assert without proof that the expectation value of a physical observable is real, i.e. $\langle \psi |\hat A \psi \rangle = \langle \psi |\hat A \psi\rangle^*$. Now,
\begin{eqnarray}
\langle \psi |\hat A \psi\rangle^*  =  \left[\int_{-\infty}^{+\infty} \psi^*(x) \hat A \psi(x) dx\right]^*  =  \int_{-\infty}^{+\infty} \psi (x) [\hat A \psi(x)]^*  dx  =  \int_{-\infty}^{+\infty} [\hat A \psi(x)]^* \psi \, dx  = \langle \hat A \psi |\psi \rangle \, ,
\end{eqnarray} 
so from (\ref{operation2}) it follows that physical observables are represented by hermitian operators.\\

10.5~{\it (ii)}~$[\hat A, \hat B]/2 + \{\hat A, \hat B\}/2 = (\hat A \hat B - \hat B \hat A)/2 + (\hat A \hat B + \hat B \hat A)/2 = \hat A \hat B$. {\it (iii)}~$\{\hat A, \hat B \}^\dagger = (\hat A \hat B)^\dagger + (\hat B \hat A)^\dagger = \hat B^\dagger \hat A^\dagger + \hat A^\dagger \hat B^\dagger  = \{\hat A, \hat B\}$, so the anticommutator is hermitian.
$[\hat A, \hat B]^\dagger = (\hat A \hat B)^\dagger - (\hat B \hat A)^\dagger = \hat B^\dagger \hat A^\dagger - \hat A^\dagger \hat B^\dagger = - (\hat A \hat B - \hat B \hat A) = -[\hat A, \hat  B]$, so the commutator is anti-Hermitian. {\it (iv)}~An anti-hermitian operator is equal to the negative of its hermitian conjugate, that is $\hat A^\dagger = - \hat A$. In inner products this means $\langle \phi |\hat A \psi \rangle = \langle \hat A^\dagger \phi | \psi \rangle = - \langle \hat A \phi| \psi \rangle$. The expectation value of an anti-hermitian operator is: $\langle \psi | \hat A \psi \rangle = \langle \hat A^\dagger \psi | \psi \rangle = - \langle \hat A \psi | \psi \rangle = - \langle A \rangle^*$.  But $\langle \psi |\hat A \psi \rangle = \langle A \rangle$, so $\langle A \rangle = - \langle A \rangle^*$, which means the expectation value must be pure imaginary.\\

10.6~Define new Hermitian operators $\hat A' = \hat A - \langle \hat A \rangle$ and $\hat B' = \hat B - \langle \hat B \rangle$. Then, using the Schwarz's inequality  we obtain $\langle \hat A'{}^2 \rangle \langle \hat B'{}^2 \rangle \geq |\langle  \hat A' \hat B' \rangle |^2$, or $\Delta A \, \Delta B \geq |\langle \hat A' \hat  B'\rangle | = |\langle[\hat A',\hat B']\rangle /2 + \langle \{\hat A', \hat B'\} \rangle/2 \geq |\langle [\hat A', \hat B']\rangle |/2$. Since the expectation value of the commutator is imaginary and the anticommutator is
real, each makes a positive contribution to the absolute value, and the anticommutator can be dropped without changing the inequality in the last step. So, 
$\Delta A \, \Delta B \geq |\langle [\hat A', \hat B']\rangle |/2 = |\langle[\hat A, \hat B] - [\hat A, \langle \hat B \rangle] - [\langle \hat A \rangle , \hat B ] + [ \langle \hat A \rangle , \langle \hat B \rangle ]\rangle|/2 =|\langle [\hat A, \hat B] \rangle|/2$.
Note that 
$\langle \hat A \rangle$ and $\langle \hat B \rangle$ are just numbers, so they commute with the operators and the commutators involving them are 0.\\

10.7~For nonrelativistic quantum mechanics, it is not so surprising that time and space are treated differently, with position being an operator and {\it not time}. After all, this is also what happens in Newtonian mechanics: time is absolute, and part of the background, and all other observables are functions of time. This paradigm underlies the formulation of the fundamental problem of Newtonian physics: to determine how a system evolves in time. Time cannot be an observable because an observable is a function of what we consider the system's ``state'', but the state is considered a function of time in the first place (so time is the independent variable). In deriving the
time-energy uncertainty principle one should be careful in defining the meaning of the standard deviation $\Delta t$. It is well known that the total energy of an isolated quantum mechanical system in distinction to a classical one, does not, in general, have a definite constant value. Instead of this the probability to obtain in a measurement any specified value of the energy of the system remains constant in time. The energy can only be  determined exactly in the special case of a stationary state. But in this case, as easily seen, all dynamical variables or, more exactly, their distribution functions, remain constant in time. In other words, the {\it definiteness} of the total energy of the system entails the {\it constancy} with respect to the time of all dynamical variables. It can be concluded that there must exist a general connection between the dispersion of the total energy of the system and the time variation of coordinates, momenta, etcetera. The uncertainty relation with which are concern gives a quantitative formulation of this connection.
Let $A$ and and $B$ denote any two quantities and at the same time the corresponding Hermitian operators. From (\ref{speedHP}) we have
\begin{equation}
\Delta A \, \Delta B \geq \frac{1}{2} \langle AB - BA \rangle
\label{man3}
\end{equation}
where $\Delta A$ and $\Delta B$ are the standards of the quatities $A$ and $B$ and $\langle \cdot \rangle$ denotes as usual the quantum mechanical average. In addition, it is easily seen that
\begin{equation} 
\frac{\hslash}{2} \frac{\partial \langle B \rangle}{\partial t} = i (\langle H B - B H \rangle)
\label{man4}
\end{equation}
where $H$ is the Hamiltonian of the system not depending explicitely on time~\cite{Weinberg:2013qm}.
Putting in (\ref{man3}) $A \equiv H$ we obtain, with the help of (\ref{man4}) the desired uncertainty relation for energy, in the form of the follwoing inequality:
\begin{equation}
\Delta H \, \Delta B \geq \frac{\hslash}{2} \left| \frac{\partial \langle B \rangle}{\partial t} \right| \, .
\label{man5}
\end{equation}
This relation gives, thus, the connection between the standard $\Delta H$ of the total energy of an isolated system, the standard $\Delta B$ of some other dynamical quantity and the rate of change of the average value of this quantity. The relation (\ref{man5})
 can be put in a different form. The absolute value of an integral cannot exceed the integral of the absolute value of the integral. Thus, integrating (\ref{man5}) from $t$ to $t+\delta t$ and taking into account that $\Delta H$ is constant one gets
\begin{equation}
\Delta H \, \delta t \geq \frac{\hbar}{2} \frac{|\langle B_{t + \delta t} \rangle - \langle B_t \rangle|}{\Delta B} \,,
\label{man5a}
\end{equation}
where the denominator of the right-hand side denotes the average value of the standard $\Delta B$ during the time $\delta t$. Sometimes (especially in the case of a continuous spectrum of eigenvalues) it is convenient to refer the variations of the average value of a dynamical quantity to its standard. In such a case it is convenient to introduce a special notation $\Delta t$ for the shortest time, during which the average value of a certain quantity is changed by an amount equal to the standard of this quantity. $\Delta t$ can be called the standard time. With the help of this notation we can rewrite (\ref{man5a}) in the form of an uncertaity relation
\begin{equation}
 \Delta H \, \Delta t \equiv \Delta E \, \Delta t \geq \hbar/2 \, .
\end{equation}

\vspace{0.25cm}

10.8~We want to project $|\phi\rangle$ onto each basis vector (this gives the expansion coefficients) and then sum the coefficients times the basis vectors: $|\phi \rangle = \left( \sum_n  \psi_n\rangle \langle \psi_n | \right) |\phi \rangle$.\\

\begin{figure}[tbp] 
\begin{minipage}[t]{0.46\textwidth}
\postscript{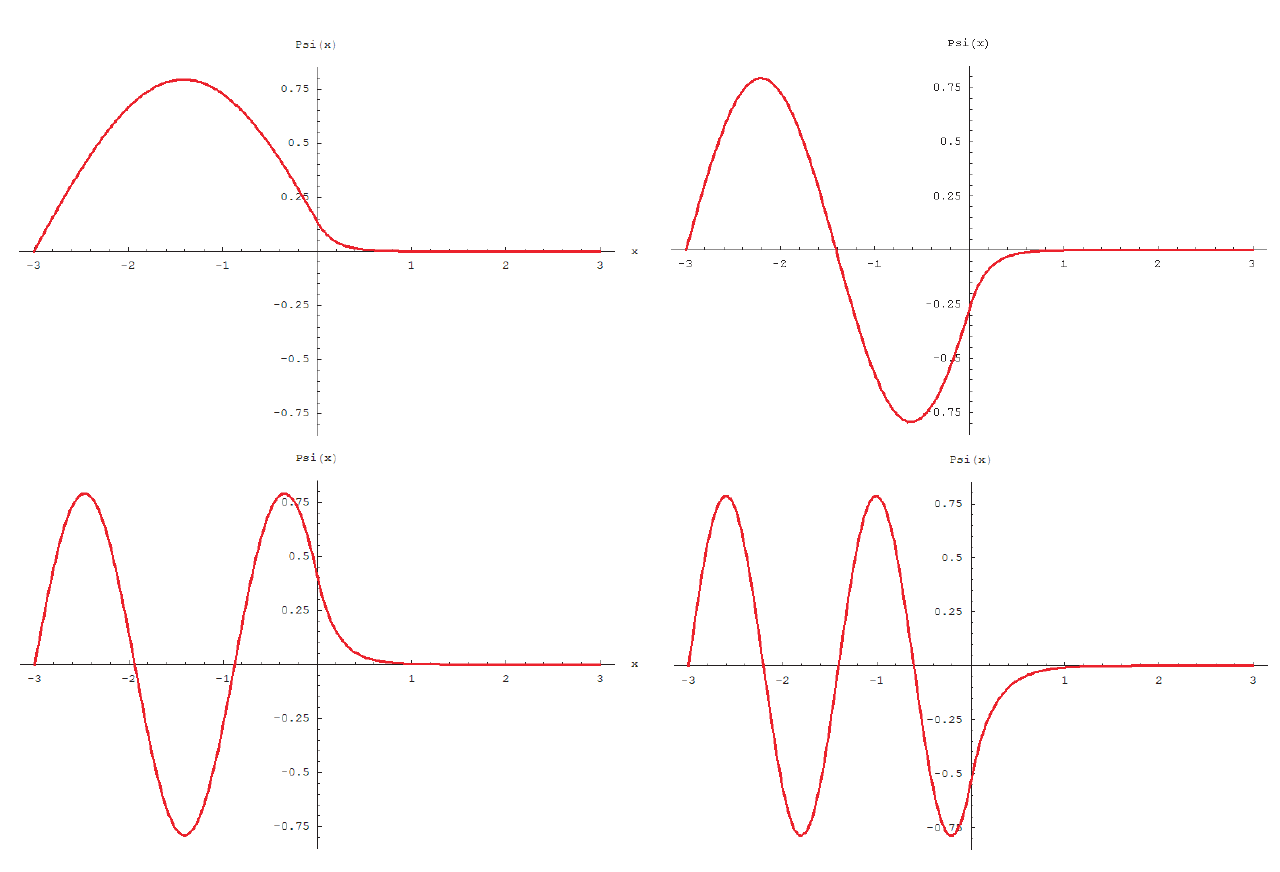}{0.99} 
\end{minipage}
\hfill
\begin{minipage}[t]{0.46\textwidth}
\postscript{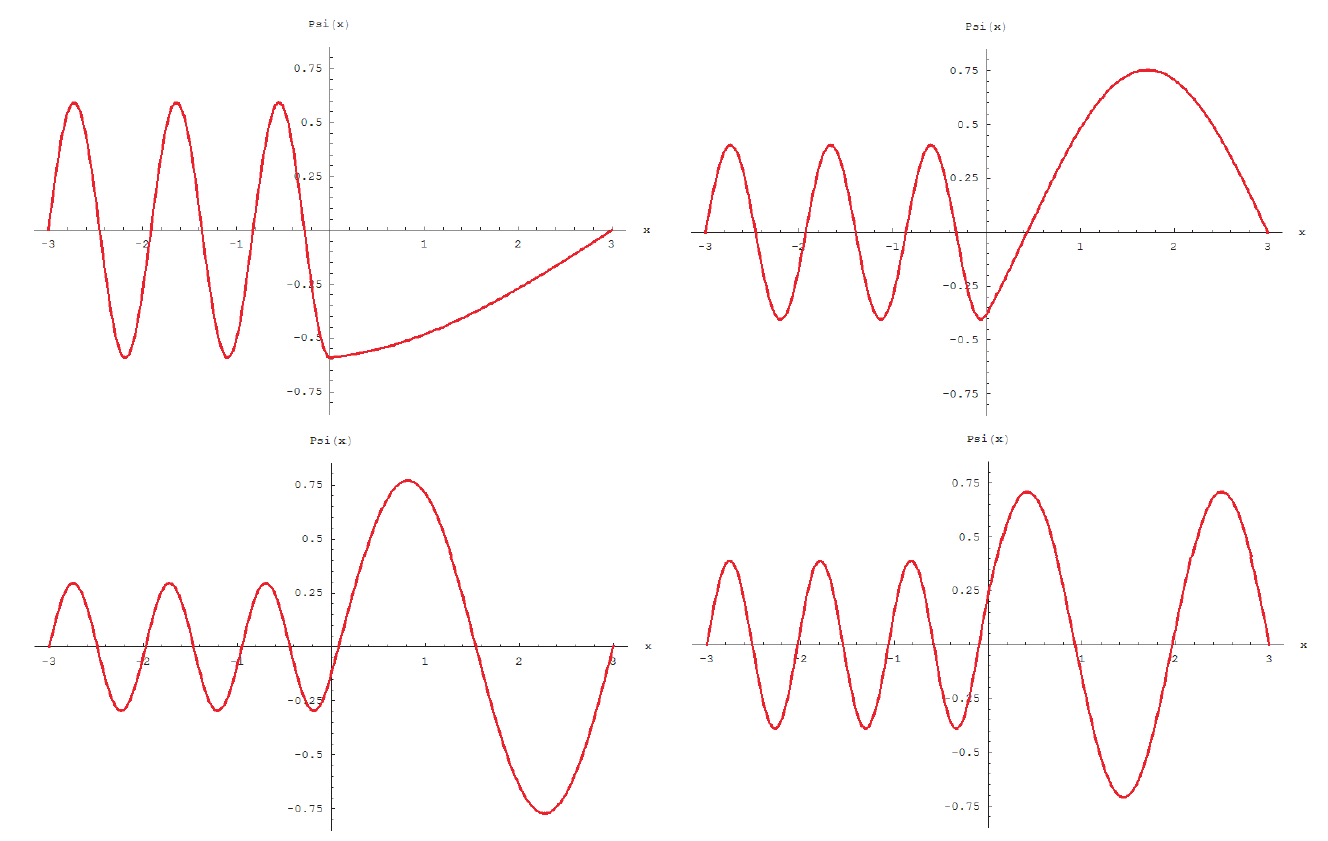}{0.99} 
\end{minipage}
\caption{The first four energy eigenstates for the asymmetric infinite square well with $E<V_0$ (left) and $E> V_0$ (right). In all of the images we have taken $\hslash = 2m =1$, $V_0 = 33$, and $L =3$~\cite{Gilbert:05}.}
\label{fig:Gilbert}
\end{figure}

10.9~For $E<V_0$, the solutions for zones I and II are 
\begin{equation}
\psi_I(x) = A \sin (k x) + B \cos (k x) \quad {\rm and} \quad \psi_{II}(x) = C e^{-\kappa x} + D e^{\kappa x},
\end{equation}
where $k = \sqrt{2mE}/\hslash$ and $\kappa = \sqrt{2m(V_0 -E)}/\hslash$.  Imposing the boundary conditions we have 
\begin{equation}
\psi_I(x=0) = 0 \Rightarrow B =0, 
\end{equation}
\begin{equation}
\psi_{II} (x =2L) = 0 \Rightarrow C e^{-2 \kappa L} + D e^{2 \kappa L} = 0 \Rightarrow  D = - C e^{-4 \kappa L} \,.
\end{equation}
Imposing continuity of the wave function at $x=L$ we obtain
\begin{equation}
\psi_I(x=L) = \psi_{II} (x=L) \Rightarrow A \sin k L = C e^{-\kappa L} + D e^{\kappa L} 
\label{arris1}
\end{equation}
and
\begin{equation}
 \psi'_I(x=L) = \psi'_{II} (x=L) \Rightarrow A k \cos k L = \kappa \left(-Ce^{-\kappa L} + D e^{\kappa L} \right) \, .
\label{arris2}
\end{equation}
Substituting the expression for $D$ in (\ref{arris1}) and (\ref{arris2}) we have
\begin{equation}
A \sin (k L) = C (e^{-\kappa L} - e^{-3 \kappa L})   \quad {\rm and} \quad A k \cos k L = -C \kappa (e^{-\kappa L} + e^{-3 \kappa L} ) \, . 
\end{equation}
Taking the ratio of these two expressions we get 
\begin{equation}
\kappa \tan (k L)   = -k \frac{1 - e^{-2 \kappa L}}{1 + e^{-2 \kappa L}} \,, \quad  {\rm or equivalently} \quad \kappa \tan (kL) = - k \tanh (2 \kappa L) \, .
 \end{equation}
{\it (iii)}~For $E>V_0$, the solutions for zones I and II are 
\begin{equation}
\psi_I(x) = A \sin (k_1 x)  \quad {\rm and} \quad \psi_{II}(x) = B \sin (k_2 x) \,.
\end{equation}
Imposing the continuity condition on $\psi(x)$ and $\psi'(x)$ at $x=L$ we obtain 
\begin{equation}
k_2 \tan (k_1 L) = - k_1 \tan (k_2 2L) \, .
\end{equation}
In Fig.~\ref{fig:Gilbert} we show the first four eigenstates for $E<0$ and $E>0$.\\

10.10~The wavefunction $\psi (x)$ for a particle with energy $E$ in a potential $V(x)$ satisfies the Schr\"odinger equation (\ref{Scho13}). Inside the well ($0 \leq x \leq L$), the particle is free. The wave function is
$\psi_{I}(x) = A \sin (kx)$, where $k = \sqrt{2mE}/\hslash$. Outside the well ($L < x < \infty$), the potential has constant value $V > E$. The wave function is $\psi_{II}(x) = B e^{-\kappa x}$, where $\kappa = \sqrt{2m(V_0-E)}/\hslash$. $\psi(x)$ and its derivative are continuous at $x = L$, then $A \sin kL = B e^{-\kappa L}$ and $A k \cos kL = - B \kappa e^{-\kappa L}$, from which 
\begin{equation}
k \cot kL = - \kappa \, . 
\label{trascenperrier}
\end{equation}
Now, since $\cot^2 \theta + 1 = \csc^2 \theta$, (\ref{trascenperrier}) can be rewritten as $k^2 (\csc^2 \theta - 1) = - \kappa^2$, where $\theta = kL$. After some algebra the trascendental equation can be rewritten as 
\begin{equation}
\theta \csc \theta = \pm a, 
\label{trascendentalperrier}
\end{equation}
where $a = \sqrt{2mV_0L^2}/h$. Note that (\ref{trascendentalperrier}) are equations for the allowed values of $k$. The equation with the positive sign yields values of $\theta$ in the second quadrant. The equation with the negative sign yields values of $\theta$ in the fourth quadrant.  Since $\sin \theta \leq \theta \, \forall \theta$, it follows from (\ref{trascendentalperrier}) that there are no bound states if $2mV_0L^2/\hslash^2 \leq 1$.\\

10.11~{\it (i)}~In region I, where $x<0$, Schr\"odinger equation is given by 
\begin{equation}
\frac{\partial^2 \psi_I}{\partial x^2} + \frac{2m E}{\hslash^2} \psi_I = 0 \Rightarrow \psi_{I} = A e^{ik_1 x} + B e^{-ik_1x},
\end{equation}
while in region II, where $x>0$, we have
\begin{equation}
\frac{\partial_{II}^2 \psi}{\partial x^2} + \frac{2m (E + V_0)}{\hslash^2} \psi_{II} = 0 \Rightarrow \psi_{II} = C e^{ik_2 x} \, .
\end{equation}
Demanding continuity of $\psi$ and $\psi'$ at $x-0$ we obtain
\begin{equation}
\psi_I (x=0) = \psi_{II} = 0 \Rightarrow A + B = C \,
\label{UFFFF1}
\end{equation}
and
\begin{equation}
\psi'_I(x=0) = \psi''_{II} (x =0) \Rightarrow i k_1 (A -B) = i k_2 C
\label{UFFFF2}
\end{equation}
Substituting (\ref{UFFFF1}) into (\ref{UFFFF2}) we get 
\begin{equation}
i k_1 (A-B) = i k_2 (A+B) \quad \quad {\rm or \, equivalently} \quad \quad A (k_1 - k_2) = B (k_1 + k_2) \, .
\end{equation}
This leads to $B/A = (k_1 - k_2)/ (k_1 + k_2)$. The reflection coeficcient is then
\begin{equation}
R = \left|\frac{B}{A} \right|^2 = \left( \frac{k_1 - k_2}{k_1 +k_2} \right)^2 = \left( \frac{\sqrt{E} - \sqrt{E+ V_0}}{\sqrt{E} + \sqrt{E+V_0} } \right)^2 \, ,
\end{equation}
whereas the transmissivity is given by
\begin{equation}
T = 1 -R = \frac{4 k_1 k_2}{(k_1 + k_2)^2 } = \frac{4 \sqrt{E (E+V_0)}}{(\sqrt{E+V_0} + \sqrt{E})^2} \, .
\end{equation}
{\it (ii)}~Taking $V_0 =12~{\rm MeV}$ and $E = 4~{\rm MeV}$ we obtain $T = 8/9$.\\

10.12~The time-independent Schr\"odinger equation of a particle of
mass $m$, which is constrained to remain within a finite region of
space (``a box'') is given by 
\begin{equation}
- \frac{\hslash^2}{2m} \ \nabla^2
\psi = E \psi.
\end{equation}
 Let $k^2 = 2mE /\hslash^2$, and note that it is
real. This equation can be solved with the help of the
separation of variables technique. Start out by trying a solution of
the following form $\psi (x,y,z) = X(x) \, Y(y), Z(z)$. Substitution
of this solution into the time-independent Schr\"odinger equation
yields: $YZ X'' + XZ Y'' + XY Z'' = - k^2 XYZ$. Then divide both sides
of the equation by $\psi$, obtain $X''/X + Y''/Y + Z''/Z = - k^2$, and
note that each of the three terms on the right-hand side is
independent of the others, because $x$, $y$, $z$ are independent
variables. In order for their sum to be equal to a constant, $-k^2$,
each of those terms must be independently equal to a constant, such
that the sum of all three constants is equal to $-k^2$. Denote those
three constants by $-k_x^2$, $-k_y^2$, $-k_z^2$, respectively, such
that Schr\"odinger equation now translates into three ordinary
differential equations: $X'' = -k_x^2 X$, $Y'' = -k_y^2 Y$, $Z'' = -
k_z^2 Z$. The solutions to these equations are: $X (x) = A \sin (k_x
x) + B \cos (k_x x)$, $Y(y) = C \sin (K_y y) + D \cos (k_y y)$, $Z (z)
= F \sin (k_z x) + G \cos (k_z x)$, where $A$, $B$, $C$, $D$, $F$, and
$G$ are (complex) undetermined parameters. Since the infinitely high
walls do not allow the particle to leave the box, the wave function is
zero at all times for $(x,y,z)<(0,0,0)$ and $(x,y,z) > (a,b,c)$, and
hence $\psi (0,0,0) =\psi (a,b,c) = 0$, because the wave function needs
to be continuous. Imposing $\psi (0,0,0) = 0$ implies $B=D = G =0$,
whereas applying the second boundary condition $\psi (a, b, c) = 0$
yields $k_x a = n \pi$, $k_y b = m \pi$, and $k_z c = l \pi$, with
$n,m,l \in \mathbb Z$. The particle is equally likely to be found
everywhere, $\int_0^a \int_0^b \int_0^c |\psi(x,y,z)|^2 dx dy dz
= 1$, and so $N = A C F$ can be determined from the requirement that
the wave function is normalized, i.e. 
\begin{equation}
|N|^2 \int_o^a \int_0^b \int_0^c
\sin^2 (n \pi x/a) \sin^2 (m \pi y/b) \sin^2(l \pi z/c) \, dx \, dy \, dz= \frac{1}{8}
|N|^2 a b c \Rightarrow |N| = \sqrt{\frac{8}{abc}} . 
\end{equation}
All in all, the
stationary states of a particle in a 3-dimensional box are given by
\begin{equation}
\psi_{nlm} (x,y,z) = \sqrt{\frac{8}{abc}} \sin (n \pi x/a) \, \sin (m
\pi y/b) \, \sin(l \pi z/c), 
\end{equation}
and the corresponding energy levels are
\begin{equation}
E_{n,m,l} = \frac{\hslash^2 \pi^2}{2m} (\frac{n^2}{a^2} +
\frac{m^2}{b^2} + \frac{l^2}{c^2}).
\end{equation}

\vspace{0.25cm}

10.13~Consider for instance the commutator $[\hat L^2,\hat L_z]$:
\begin{eqnarray}
[\hat L^2, \hat L_z] & = & [\hat L_x^2 + \hat L_y^2 + \hat L_z^2, \hat L_z]~~~{\rm from \ the \ definition \ of} \ \hat L^2 \nonumber \\
& = & [\hat L_x^2, \hat L_z] + [\hat L_y^2 , \hat L_z] + [\hat L_z^2, \hat L_z] \nonumber \\
& = & [\hat L_x^2, \hat L_z] + [\hat L_y^2, \hat L_z]~~~{\rm since}\ \hat L_z \ {\rm commutes \ with \ itself} \nonumber \\
& = & \hat L_x \hat L_x \hat L_z - \hat L_z \hat L_x \hat L_x + \hat L_y \hat L_y \hat L_z - \hat L_z \hat L_y \hat L_y \, .
\end{eqnarray}
We can use the commutation relation $[\hat L_z, \hat L_x] = i \hslash \hat L_y$ to rewrite the first term on the right-hand side as $\hat L_x \hat L_x \hat L_z = \hat L_x \hat L_z \hat L_x - i \hslash \hat L_x \hat L_y$, and the second term as $\hat L_z \hat L_x \hat L_x = \hat L_x \hat L_z \hat L_x + i \hslash \hat L_y \hat L_z$. In a similar way, we can use $[\hat L_y,\hat L_z] = i \hslash \hat L_x$ to rewrite the third term as $\hat L_y \hat L_y \hat L_z = \hat L_y \hat L_z \hat L_y + i \hslash \hat L_y \hat L_x$, and the fourth term $\hat L_z \hat L_y \hat L_y = \hat L_y \hat L_z \hat L_y - i \hslash \hat L_x \hat L_y$. Thus, on substituting in we find that
\begin{equation}
[\hat L^2, \hat L_z] = -i \hslash \hat L_x \hat L_y - i \hslash \hat L_y \hat L_x + i \hslash \hat L_y \hat L_x + i \hslash \hat L_x \hat L_y = 0 \, .
\end{equation} 
By performing a cyclic permutation of the indexes, we can show that this holds in general, i.e. $[\hat L^2, \hat L_i] =0$, for $i = x,y,z$.\\

10.14~Assume that the eigenvalues of $\hat L^2$ and $\hat L_z$ are unknown and denote them $\lambda$ and $\mu$. We introduce two new operators, the raising and lowering operators $\hat L_+ = \hat L_x + i \hat L_y$ and $\hat L_- = \hat L_x - i \hat L_y$. The commutator with $L_z$ is $[\hat L_z, \hat L_\pm] = \pm \hslash \hat L_\pm$ (while they of course commute with $L^2$). Now consider the function $f_\pm = \hat L_\pm f$, where $f$ is an eigenfunction of $\hat L^2$ and $\hat L_z$:
\begin{equation}
\hat L^2 f_\pm = \hat L_\pm  \hat L^2 f = \hat L_\pm \lambda f = \lambda f_\pm
\quad {\rm and} \quad \hat L_z f_\pm = [\hat L_z, \hat L_\pm] f + \hat L_\pm \hat L_z f = \pm \hslash \hat L_\pm f + \hat L_\pm \mu f = (\mu \pm \hslash ) f_\pm \, .
\end{equation}
Then $f_\pm = \hat L_\pm f$ is also an eigenfunction of $\hat L^2$ and $\hat L_z$. Moreover, we can keep finding eigenfunctions of $\hat L_z$ with higher and higher eigenvalues $\mu' = \mu + \hslash + \hslash + \cdots$, by applying the $\hat L_+$ operator (or lower and lower with $\hat L_-$), while the $\hat L^2$ eigenvalue is fixed. Of course there is a limit, since we want $\mu' \leq \lambda$. Then there is a maximum  eigenfunction such that $\hat L_+ f_M =0$ and we set the corresponding eigenvalue to $\hslash l_M$. Now note that we can write $\hat L^2$ instead of using $\hat L_{x,y}$ by using $\hat L_\pm$:
\begin{equation}
\hat L^2= \hat L_- \hat L_+ + \hat L_z^2 + \hslash \hat L_z \,. 
\end{equation}
Using this relationship on $f_M$ we find:
\begin{equation}
\hat L^2 f_M = \lambda f_M \Rightarrow (\hat L_-\hat L_+ + \hat L_z^2 + \hslash \hat L_z) f_M = [0 + \hslash^2 l_M^2 + \hslash ( \hslash l_M)] f_M \Rightarrow \lambda = \hslash^2 l_M (l_M +1) \, .
\end{equation}
In the same way, there is also a minimum eigenvalue $l_m$ and eigenfunction such that $\hat L_- f_m=0$ and we can find $\lambda = \hslash^2 l_m (l_m -1)$. Since $\lambda$ is always the same, we also have $l_m (l_m -1) = l_M (l_M +1)$, with solution $l_m = - l_M$ (the other solution would have $l_m > l_M$). Finally, we have found that the eigenvalues of $L_z$ are between $+\hslash l$ and $-\hslash l$ with integer increases, so that $l = -l+N$ giving $l = N/2$: that is, $l$ is either an integer or a half-integer. We thus set $\lambda = \hslash^2 l (l+1)$ and $\mu = \hslash m$, with $m= -l, -l+1, \cdots, l$.\\

11.1~{\it (i)}~First we show that the set $\{\mathds 1, \sigma_1,
\sigma_2, \sigma_3 \}$ is linearly independent. Suppose
\begin{equation}
\alpha \mathds 1 + \beta \sigma_1 + \zeta \sigma_2 + \xi \sigma_3 =
\left(\begin{array}{cc} \alpha + \xi & \beta - i \zeta \\
\beta + i \zeta & \alpha - \xi \end{array} \right) = \left(\begin{array}{cc}
0 & 0 \\  0 & 0 \end{array} \right) \, .
\end{equation}
Then $\alpha = - \xi$ and $\alpha = \xi \Leftrightarrow \alpha = \xi =
0$. Similarly, $\beta = - i \zeta$ and $\beta = i \zeta$, which
implies $\beta = \zeta = 0$. Now we show that the vectors $\{\mathds 1, \sigma_1,
\sigma_2, \sigma_3 \}$ span the $2 \times 2$ matrix space. Let
\begin{eqnarray}
\label{mathbbM}
\mathbb M & = & \left( \begin{array}{cc} m_{11} & m_{12} \\ m_{21} &
    m_{22} \end{array} \right)  \nonumber   \\
 & = & \frac{1}{2} (m_{11} + m_{22}) \left(\begin{array}{cc} 1 & 0 \\ 0
     & 1\end{array} \right) + \frac{1}{2} (m_{11} - m_{22})
 \left( \begin{array}{cc} 1 & 0 \\0 & -1 \end{array} \right)  + 
 \frac{1}{2} (m_{12} + m_{21}) \left(\begin{array}{cc} 0 & 1 \\ 1 &
   0 \end{array} \right) + \frac{i}{2} (m_{12} - m_{21}) \left(\begin{array}{cc} 0 & -i
   \\ i & 0 \end{array} \right) \nonumber \\
 & = & \frac{1}{2} (m_{11} + m_{22}) \mathds 1 + \frac{1}{2} (m_{12} + m_{21})
 \sigma_1 + \frac{i}{2} (m_{12} - m_{21}) \sigma_2   +  \frac{1}{2} (m_{11}
   - m_{22}) \sigma_3 \, .
\end{eqnarray}
Note that
\begin{equation}
\frac{1}{2} {\rm Tr}~\left[\mathbb M \right] = \frac{1}{2} (m_{11} + m_{22})
\end{equation}
and so the first term in (\ref{mathbbM}) can be written as
$\frac{1}{2} {\rm Tr}~\left[ \mathbb M\right] \mathds{1}$. Now,
\begin{eqnarray}
\frac{1}{2} {\rm Tr}~\left[\mathbb M \sigma_1 \right] & = & \frac{1}{2} {\rm Tr}
\left(\begin{array}{cc} m_{12} & m_{11} \\ m_{22} & m_{21} \end{array}
\right) = \frac{1}{2} (m_{12} + m_{21}) \nonumber \\
\frac{1}{2} {\rm Tr}~\left[\mathbb M \sigma_2 \right] & = & \frac{1}{2} {\rm Tr}
\left(\begin{array}{cc} i \ m_{12} & -i \ m_{11} \\ i \ m_{22} & - i \ m_{21} \end{array}
\right) = \frac{1}{2} (m_{12} - m_{21} )\nonumber \\
\frac{1}{2} {\rm Tr}~\left[\mathbb M \sigma_3 \right] & = & \frac{1}{2} {\rm Tr}
\left(\begin{array}{cc} m_{11} & -m_{12} \\ m_{21} & -m_{22} \end{array}
\right) = \frac{1}{2} (m_{11} - m_{22} )\nonumber \, .
\end{eqnarray}
We define, $ \mathbb M \vec \sigma = \left(\mathbb M \sigma_1, \mathbb
  M \sigma_2 , \mathbb M \sigma_3 \right)$ so that the last three
terms in (\ref{mathbbM}) can be written as $\frac{1}{2} {\rm
  Tr}~[\mathbb M \vec \sigma] \, \cdot \,  \vec \sigma$.  Therefore, any $2 \times
2$ matrix can be written as $\mathbb M = a_0 \mathds 1 + \vec{a}
\cdot \vec{\sigma}$, where $a_0 = \frac{1}{2} {\rm Tr}~[\mathbb M]$ and
$\vec a = \frac{1}{2} \, {\rm Tr}~[\mathbb M \vec \sigma]$.\\

12.1~{\it (i)}~Consider the equation $E=mc^2$.  If we use SI units and solve for $E$ with $m=1\mbox{ kg}$, we find
\begin{eqnarray}
E=(1\mbox{ kg})\cdot(3\times10^{8}\mbox{ m/s})^2 = 9\times10^{16}\mbox{ J}.
\end{eqnarray}
Converting from J to GeV using $1\mbox{ eV}=1.6\times10^{-19}\mbox{ J}$ and $10^{9}\mbox{ eV}=1\mbox{ GeV}$, we find
\begin{eqnarray}
E = 9\times10^{16}\mbox{ J}\left(\frac{1\mbox{ eV}}{1.6\times10^{-19}\mbox{ J}}\right)\left(\frac{1\mbox{ GeV}}{10^{9}\mbox{ eV}}\right)
 = \frac{9\times10^{16}}{(1.6\times10^{-19})\cdot(10^{9})}\mbox{ GeV}
= 5.6\times10^{26}\mbox{ GeV}.
\end{eqnarray}
Plugging in this value of $E$ into our original equation $E=mc^2$ and setting $c=1$, we can now solve for $1\mbox{ kg}$:
\begin{eqnarray}
(5.6\times10^{26}\mbox{ GeV})&=&(1\mbox{ kg})(1)^2
1\mbox{ kg} = 5.6\times10^{26}\mbox{ GeV}.
\end{eqnarray}
{\it (ii)}~We are given that $\hbar c=197.3\mbox{ MeVfm}$.  We can solve this equation for MeV$^{-1}$, set $\hbar=1$ and $c=1$, convert to GeV using $1\mbox{ GeV}=10^{3}\mbox{ MeV}$, and square both sides to get:
\begin{eqnarray}
\mbox{ MeV}^{-1}&=&\frac{197.3\mbox{ fm}}{\hbar c}\Rightarrow 
\frac{1}{\mbox{ MeV}}\left(\frac{10^{3}\mbox{ MeV}}{1\mbox{ GeV}}\right) = \frac{197.3\mbox{ fm}}{1}\\
1\mbox{ GeV}^{-1}&=&0.1973\mbox{ fm} \Rightarrow
1\mbox{ GeV}^{-2}=0.0389\mbox{ fm}^2
\end{eqnarray}
Converting from fm$^2$ to mb (note: b stands for barn, which is a measure of area) using $1\mbox{ m}=10^{15}\mbox{ fm}$, $1\mbox{ m}^2=10^{28}\mbox{ b}$, and $1\mbox{ b}=10^{3}\mbox{ mb}$, we find 
\begin{eqnarray}
1\mbox{ GeV}^{-2} = 0.0389\mbox{ fm}^2\left(\frac{1\mbox{ m}}{10^{15}\mbox{ fm}}\right)^2 \left(\frac{10^{28}\mbox{ b}}{1\mbox{ m}^2}\right)\left(\frac{10^{3}\mbox{ mb}}{1\mbox{ b}}\right) 
 = 0.389\mbox{ mb}.
\end{eqnarray}
{\it (iii)}~We are given that $\hbar c=197.3\mbox{ MeVfm}$.  We can solve this equation for fm, set $\hbar=1$ and $c=1$, convert from fm to m using $1\mbox{ m}=10^{15}\mbox{ fm}$, and convert from MeV to GeV using $1\mbox{ GeV}=10^{3}\mbox{ MeV}$ to get:
\begin{eqnarray}
\mbox{fm}&=&\frac{\hbar c}{197.3\mbox{ MeV}} \nonumber \\
\mbox{fm}\left(\frac{1\mbox{ m}}{10^{15}\mbox{ fm}}\right)&=&\frac{1}{197.3\mbox{ MeV}}\left(\frac{10^{3}\mbox{ MeV}}{1\mbox{ GeV}}\right) \nonumber\\
1\mbox{ m}&=&5.068\times10^{15}\mbox{ GeV}^{-1}.
\end{eqnarray}
{\it (iv)}~We know that $c=3\times10^8$ m/s in SI units.  Therefore, we can solve for 1 s by setting $c=1$ and plugging in that $1\mbox{ m}=5.068\times10^{15}\mbox{ GeV}^{-1}$:
\begin{eqnarray}
1\mbox{ s}&=&\frac{3\times10^8\mbox{ m}}{c}
=(3\times10^8)\cdot(5.068\times10^{15}\mbox{ GeV}^{-1})
=1.5\times10^{24}\mbox{ GeV}^{-1}.
\end{eqnarray}
{\it (v)}~Converting from MeV to GeV using that $1\mbox{ GeV}=10^{3}\mbox{ MeV}$, we can find the Compton wavelength to be
\begin{eqnarray}
\lambda_c = \frac{1}{0.511\mbox{ MeV}}\left(\frac{10^{3}\mbox{ MeV}}{1\mbox{ GeV}}\right) =1.957\times10^3\mbox{ GeV}^{-1}.
\end{eqnarray}
Using that $1\mbox{ m}=5.068\times10^{15}\mbox{ GeV}^{-1}$ and $1\mbox{ m}=10^{15}\mbox{ fm}$, we find that
\begin{eqnarray}
\lambda_c = 1.957\times10^3\mbox{ GeV}^{-1}
\left(\frac{1\mbox{ m}}{5.068\times10^{15}\mbox{ GeV}^{-1}}\right)\left(\frac{10^{15}\mbox{ fm}}{1\mbox{ m}}\right)
= 386\mbox{ fm}.
\end{eqnarray}
{\it (vi)} As you can see, $r_B=\lambda_c/\alpha$, where $\alpha=1/137$.  Therefore,
\begin{eqnarray}
r_B = \frac{386\mbox{ fm}}{1/137}
 = 5.29\times10^4\mbox{ fm}.
\end{eqnarray}
{\it (vii)}~In a unit system where $\hbar=1$ and $c=1$, $v=\alpha=1/137$.  To convert this into SI units where $c\ne1$, we must restore the $\hbar$ and $c$ variables.  Here, recognize that the 1 in the numerator must be $c$ in order to get the correct units for velocity.  Therefore,
\begin{eqnarray}
v= \frac{1}{137}
=\frac{c}{137}
=\frac{3\times10^8\mbox{ m/s}}{137}
=2.2\times10^6\mbox{ m/s}.
\end{eqnarray}
{\it (viii)} Since we just saw that
\begin{eqnarray}
v=\left(\frac{1}{137}\right)c
=0.007c \Rightarrow
v \ll c \,,
\end{eqnarray}
the non-relativistic Schr\"odinger equation can be used to describe the hydrogen atom.
{\it (ix)}~To estimate the length scale, let's find the associated Compton wavelength for $M_{\rm Pl}$ using that $1\mbox{ m}=5.068\times10^{15}\mbox{ GeV}^{-1}$ and $1\mbox{ m}=10^{15}\mbox{ fm}$:
\begin{eqnarray}
\lambda_{\rm Pl} \sim \frac{1}{10^{19}}\mbox{ GeV}^{-1}
\left(\frac{1\mbox{ m}}{5.068\times10^{15}\mbox{ GeV}^{-1}}\right)\left(\frac{10^{15}\mbox{ fm}}{1\mbox{ m}}\right)
\sim1.97\times10^{-20}\mbox{ fm}
\sim 10^{-20}\mbox{ fm}
\end{eqnarray}

\vspace{0.25cm}

12.2~Multiplying the Klein-Gordon equation by $-i\psi^*$ and the
complex conjugate equation by $-i\psi$, and subtracting it follows that
\begin{equation}
-i \psi^* \partial_\mu \partial^\mu \psi - i \psi^* m^2 \psi + i \psi \partial_\mu \partial^\mu \psi^* + i \psi m^2 \psi^* = - i \psi^* \partial_\mu \partial^\mu \psi + i \psi \partial_\mu \partial^\mu \psi^* = 0 \, .
\end{equation}
Note that the last equality holds because
\begin{equation}
\partial_\mu (\psi^* \partial^\mu \psi) = \partial_\mu \psi^* \partial^\mu \psi + \psi^* \partial_\mu \partial^\mu \psi \, .
\end{equation}

\vspace{0.25cm}

13.1~{\em (i)}~The number of minimum bias events per second can be computed as follows:
\begin{equation}
\frac{\Delta N}{\Delta t} = L \times \sigma
\end{equation}
where $L$ is the luminosity and $\sigma$ the cross section; this yields
\begin{equation}
\frac{\Delta N}{\Delta t} = 70 \times 10^{34}~{\rm mb} \, {\rm cm}^{-2}  \,{\rm s}^{-1} = 7 \times 10^{8}~{\rm s}^{-1} \, .
\end{equation}
Thus, we get 700 million interactions per second. Since the beams
collide every 25~ns, the number of events produced per bunch
crossing will be $\Delta N  = 17.5$. {\em (ii)}~To estimate the Higgs production rate we proceed in a
similar way, but since we are only looking for Higgs decaying into
two photons we need to multiply the Higgs production cross section
by the corresponding branching fraction,
\begin{eqnarray}
\frac{\Delta N}{\Delta t}  =  \sigma (pp \to H) \times {\rm BR} (H \to \gamma \gamma) \times L  =  8 \times 10^{-4}~{\rm s}^{-1} \, .
\end{eqnarray}
The ratio of the two rates is
\begin{equation}
R = \frac{7 \times 10^8}{8 \times 10^{-4}} \sim 10^{12} \, .
\end{equation}
This implies that LHC gets 1000
billion minimum bias events for every single Higgs event we
observe decaying into two photons. Clearly a very sophisticated
selection procedure must be employed to be able to select the
Higgs events from the
background.\\

13.2~{\it (i)} Let's show that inhomogeneous Maxwell's equations take the compact form
\begin{eqnarray}
    \partial_\mu F^{\mu \nu} = j^\nu \, .
\end{eqnarray}
This can be done just by separating temporal and spatial components and
looking at the matrix form (\ref{fmunu}) of the tensor $F^{\mu\nu}$.
Choosing $\nu = 0$, we find
\begin{eqnarray}\label{j0}
     j^0 = \rho = \partial_\mu F^{\mu 0} = \partial_0 F^{00} + \partial_1
F^{10} +\partial_2 F^{20} + \partial_3 F^{30}\, .
\end{eqnarray}
Substituting the matrix elements of $F^{\mu \nu}$ into \eqref{j0}
and identifying the partial derivatives, this becomes
\begin{eqnarray}
     \rho = \frac{ \partial E_x }{\partial x} + \frac{ \partial E_y
}{\partial y} + \frac{ \partial E_z }{\partial z} = \overset{\rightharpoonup}{\nabla} \cdot
\overset{\rightharpoonup}{E}
\end{eqnarray}
which is Gauss's law (\ref{max2}). Next, consider the spatial
components of $j^\nu$. For $\nu=1$, we have
\begin{eqnarray}\label{jx}
  j^1 & = & \partial_0 F^{01} + \partial_1 F^{11} + \partial_2
F^{21} + \partial_3 F^{31} \nonumber\\
  j_x & = & -\frac{\partial E_x}{\partial t} +
\frac{\partial B_z}{\partial y} - \frac{\partial B_y}{\partial z}
= -\frac{\partial E_x}{\partial t} +
(\overset{\rightharpoonup}{\nabla} \times
\overset{\rightharpoonup}{B})_x\, ,
\end{eqnarray}
and analogous relations can be obtained for $\nu=2$ and $\nu=3$. Thus we
get
\begin{eqnarray}
     \overset{\rightharpoonup}{\jmath} = -\partial_t \overset{\rightharpoonup}{E}
    + \overset{\rightharpoonup}{\nabla} \times \overset{\rightharpoonup}{B} \,
    ,
\end{eqnarray}
i.e.~(\ref{max3}). 

The homogeneous equations (\ref{max1}) and (\ref{max4}) follow from the
relation
\begin{equation}
\partial^{\alpha} F^{\beta \gamma} + \partial^{\beta} F^{\gamma \alpha}+\partial^{\gamma}
F^{\alpha \beta} = 0 \, .
\label{abg}
\end{equation}
Once again we consider here temporal and spatial components separately.
Owing to permutation symmetry it is enough to take into account four
cases: $(\alpha,\beta,\gamma) = (0,0,0),$ $(0,0,i),$ $(0,i,j)$ and
$(i,j,k)$. Moreover, taking into account that $F^{\mu\nu}$ is
antisymmetric, it is easily seen that in the first two cases the left hand
side of (\ref{abg}) trivially vanishes. For $(\alpha,\beta,\gamma) =
(0,i,j)$ we have
$$\partial_t F^{ij} + \partial^i F^{j0} + \partial^j
F^{0i} = 0 \, .$$
Here if $i=1$, $j=2$, by looking at (\ref{fmunu}) we get
$$\frac{\partial F^{12}}{\partial t} - \frac{\partial F^{20}}{\partial x}
 - \frac{\partial F^{01}}{\partial y}
 = -\frac{\partial B_z}{\partial t} - \frac{\partial E_y}{\partial x} +
\frac{\partial E_x}{\partial y} = -\left(\frac{\partial
\overset{\rightharpoonup}{B}}{\partial t} +
\overset{\rightharpoonup}{\nabla} \times
\overset{\rightharpoonup}{E}\right)_z = 0 \, .$$ Analogous
relations can be obtained for $i=2$, $j=3$ and $i=3$, $j=1$, while
the cases $i=j$ yield trivial identities. Thus we end up with~(\ref{max1}). Finally let us consider the case
$(\alpha,\beta,\gamma) = (i,j,k)$. From the permutation symmetry,
and noting that the sum in the left hand side of (\ref{abg}) is
zero if any two indices are equal, we just need to evaluate the
case $i=1$, $j=2$, $k=3$. One has
$$\partial^1 F^{23} + \partial^2 F^{31} + \partial^3
F^{12} = - \frac{\partial (-B_x)}{\partial x} - \frac{\partial (-B_y)}{\partial y}
 - \frac{\partial (-B_z)}{\partial z}
 = \overset{\rightharpoonup}{\nabla} \cdot
\overset{\rightharpoonup}{B} = 0 \, ,$$
i.e..~(\ref{max4}).

Now current conservation follows immediately from the fact that $F^{\mu
\nu}$ is antisymmetric: one has
\begin{eqnarray}
    \partial_\nu\partial_\mu F^{\mu \nu} =\partial_\nu( j^\nu) = 0 \, .
\end{eqnarray}

{\em (ii)} Let's verify that we can retrieve the homogeneous Maxwell's
equations from (\ref{maxrelation}). By substituting
(\ref{maxrelation}) into the left hand side of (\ref{max1}) we get
\begin{eqnarray}
\overset{\rightharpoonup}{\nabla } \times \overset{\rightharpoonup }{E} +  \partial_t\overset{\rightharpoonup
}{B} &=& \overset{\rightharpoonup }{\nabla } \times \left(-\partial _t\overset{\rightharpoonup }{A} - \overset{\rightharpoonup }{\nabla
}V  \right)
 +  \partial _t\left(\overset{\rightharpoonup }{\nabla } \times \overset{\rightharpoonup }{A}\right) \nonumber \\
& = & \overset{\rightharpoonup }{\nabla } \times \left(-\partial
_t\overset{\rightharpoonup }{A}\right) - \overset{\rightharpoonup
}{\nabla } \times \left(\overset{\rightharpoonup }{\nabla }V
\right) +  \partial _t\left(\overset{\rightharpoonup
}{\nabla } \times \overset{\rightharpoonup }{A}\right) \nonumber \\
& = & -\partial _t\left(\overset{\rightharpoonup }{\nabla } \times
\overset{\rightharpoonup }{A}\right) - \overset{\rightharpoonup
}{\nabla } \times \left(\overset{\rightharpoonup }{\nabla } V
\right)  +  \partial _t\left(\overset{\rightharpoonup
}{\nabla } \times \overset{\rightharpoonup }{A}\right) \nonumber \\
& = & - \overset{\rightharpoonup }{\nabla } \times
\left(\overset{\rightharpoonup }{\nabla
} V \right) \nonumber \\
& = & 0 \, ,
\end{eqnarray}
i.e.~(\ref{max1}) is satisfied. Now (\ref{max4}) can be verified
just by substituting $\overset{\rightharpoonup }{B}=\overset{\rightharpoonup
}{\nabla } \times \overset{\rightharpoonup }{A}$, since the divergence of
the curl is always zero:
\begin{eqnarray}
\overset{\rightharpoonup }{\nabla } \cdot
\overset{\rightharpoonup }{B}
 & = & \overset{\rightharpoonup }{\nabla } \cdot
 \left(\overset{\rightharpoonup }{\nabla } \times \overset{\rightharpoonup }{A}\right) \nonumber \\
& = & 0.
\end{eqnarray}

To obtain Eqs.~(\ref{max2}) and (\ref{max3}), we first note that
the d'Alembertian is $\square ^2 = \left(\partial _t^2 - \nabla
  ^2\right)$. Thus we can rewrite (\ref{maxcovariant}) as
\begin{equation}
\left(\partial _t^2 - \nabla ^2\right)A^{\mu } - \partial ^{\mu
}\left(\partial _{\nu }A^{\nu }\right) =  \mathit{j}^{\mu }\, .
\label{sprite}
\end{equation}
Let's look at the $\mu =0$ component of the left hand side of
(\ref{sprite}).  Recall, $A^0 = V$ and $\partial_\nu A^\nu =
\partial_t V + \overset{\rightharpoonup }{\nabla } \cdot
\overset{\rightharpoonup}{A}$. We get
\begin{eqnarray}
\left(\partial _t^2 - \nabla ^2\right) V - \partial
_t\left(\partial _{\nu }A^{\nu }\right) & = &
\partial _t^2V  - \nabla ^2 V  - \partial _t\left(\partial _t V  + \overset{\rightharpoonup }{\nabla } \cdot
 \overset{\rightharpoonup }{A}\right) \nonumber \\
 & = &
 \partial _t^2V  - \nabla ^2V  - \partial _t^2V  - \partial _t\left(\overset{\rightharpoonup }{\nabla
} \cdot  \overset{\rightharpoonup }{A}\right)\nonumber \\
 & = & - \partial _t\left(\overset{\rightharpoonup }{\nabla } \cdot  \overset{\rightharpoonup }{A}\right)\text{  }- \nabla ^2V
 \nonumber \\
 & = &\overset{\rightharpoonup }{\nabla } \cdot  \left(- \partial _t\overset{\rightharpoonup }{A}\right) + \overset{\rightharpoonup
}{\nabla } \cdot  \left(-\overset{\rightharpoonup }{\nabla } V \right) \nonumber       \\
& = &    \overset{\rightharpoonup }{\nabla } \cdot  \left(- \partial _t\overset{\rightharpoonup }{A} -\overset{\rightharpoonup }{\nabla
} V \right) \nonumber \\
&=& \overset{\rightharpoonup}{\nabla}\cdot \overset{\rightharpoonup}{E} \nonumber,
\end{eqnarray}
where we have used (\ref{maxrelation}) in the last line. Noting
that the $\mu = 0$ component of the right hand side of
(\ref{sprite}) is $\rho$, it is seen that we have retrieved
(\ref{max2}). Finally, let's look at the spatial components of
(\ref{sprite}), $\mu = 1,2,3$.  Once again using $\partial_\nu
A^\nu = \partial_t V + \overset{\rightharpoonup }{\nabla } \cdot
\overset{\rightharpoonup}{A}$, and recalling the $\mu = 1,2,3$
components of $A^\mu$ are the components of the vector potential
$\overset{\rightharpoonup}{A}$, we obtain
\begin{eqnarray}
\left(\partial _t^2 - \nabla ^2\right)\overset{\rightharpoonup }{A} + \overset{\rightharpoonup }{\nabla }\left(\partial _{\nu }A^{\nu }\right) & = &
\partial _t^2\overset{\rightharpoonup }{A} - \nabla ^2\overset{\rightharpoonup }{A} + \overset{\rightharpoonup }{\nabla
}\left(\partial _t V  + \overset{\rightharpoonup }{\nabla } \cdot  \overset{\rightharpoonup }{A}\right)\nonumber \\
& = &  -\nabla ^2\overset{\rightharpoonup }{A} +
\overset{\rightharpoonup }{\nabla }\left(\overset{\rightharpoonup
}{\nabla } \cdot  \overset{\rightharpoonup }{A} \right) + \partial
_t^2\overset{\rightharpoonup }{A} \text{  }+
\overset{\rightharpoonup }{\nabla }\left(\partial _t V \right)
\nonumber \, .
\end{eqnarray}
Using the triple cross product identity given in the hint,
$\overset{\rightharpoonup}{\nabla} \times
\left(\overset{\rightharpoonup}{\nabla}\times
\overset{\rightharpoonup}{A}\right) =
-\nabla^2\overset{\rightharpoonup}{A} +
\overset{\rightharpoonup}{\nabla}\left(\overset{\rightharpoonup}{\nabla}\cdot\overset{\rightharpoonup}{A}\right)$,
and rearranging the derivatives,
\begin{eqnarray}
\left(\partial _t^2 - \nabla ^2\right)\overset{\rightharpoonup }{A} + \overset{\rightharpoonup }{\nabla }\left(\partial _{\nu }A^{\nu }\right)& = &
\overset{\rightharpoonup }{\nabla } \times \left(\overset{\rightharpoonup }{\nabla } \times \overset{\rightharpoonup }{A}\right)
-\partial _t\left(- \partial _t\overset{\rightharpoonup }{A}\right)-\partial _t\left( -\overset{\rightharpoonup }{\nabla } V \right) \nonumber \\
 & = & \overset{\rightharpoonup }{\nabla } \times \left(\overset{\rightharpoonup }{\nabla } \times \overset{\rightharpoonup }{A}\right)
-\partial _t\left(- \partial _t\overset{\rightharpoonup }{A} -\overset{\rightharpoonup }{\nabla } V \right) \nonumber \\
& = & \overset{\rightharpoonup}{\nabla} \times \overset{\rightharpoonup}{B} - \partial_t \overset{\rightharpoonup}{E},
\label{sprite2}
\end{eqnarray}
where in the last line we have used (\ref{maxrelation}). We know that the
$\mu = 1,2,3$ components of the right hand side of (\ref{sprite}) are the
components of $\overset{\rightharpoonup }{\jmath}$, thus we get
(\ref{max3}). Therefore, we have shown that Maxwell's equations can be
retrieved from (\ref{maxrelation}) and (\ref{maxcovariant}).

Notice that (\ref{maxcovariant}) is equivalent to the compact expression
(\ref{compact}), written in terms of the four-vector potential. Indeed, we
have
\begin{equation}
\partial_\mu F^{\mu\nu} = \partial_\mu(\partial^\mu A^\nu - \partial^\nu
A^\mu) = \Box^2 A^\nu -
\partial^\nu (\partial_\mu  A^\mu)\, .
\end{equation}
Moreover, using the covariant notation it is easy to see that (\ref{abg}) ---and thus inhomogeneous Maxwell's equations--- follow
from $F^{\mu\nu} = \partial^\mu A^\nu - \partial^\nu A^\mu$. Indeed, one
has
\begin{equation}
\partial^{\alpha} F^{\beta \gamma} + \partial^{\beta} F^{\gamma \alpha}+\partial^{\gamma}
F^{\alpha \beta} =
\partial^{\alpha} \partial^\beta A^\gamma
- \partial^{\alpha} \partial^\gamma A^\beta
+ \partial^{\beta} \partial^\gamma A^\alpha
- \partial^{\beta} \partial^\alpha A^\gamma
+ \partial^{\gamma} \partial^\alpha A^\beta
- \partial^{\gamma} \partial^\beta A^\alpha = 0 \, .
\end{equation}

{\em (iii)} Given that Maxwell's Equations can be written in the covariant
form discussed above, they must remain invariant under arbitrary Lorentz
boosts. To show this explicitly, let us start with Maxwell's
equations in an unprimed frame,
\begin{eqnarray}
\label{MW_Eq1}
\partial_{\alpha}F^{\alpha\beta}&=& j^{\beta}\\
\label{MW_Eq2}
\partial_{\alpha}F_{\beta\gamma}+\partial_{\beta}F_{\gamma\alpha}+\partial_{\gamma}F_{\alpha\beta}&=&0,
\end{eqnarray}
apply an arbitrary Lorentz boost, and show that they have the same form in
the boosted, primed frame.

First, let us boost (\ref{MW_Eq1}):
\begin{equation}
{\Lambda^\gamma}_\beta \partial_\alpha F^{\alpha\beta} =  {\Lambda^\gamma}_\beta j^\beta\, .
\end{equation}
Using the properties of the delta function ${\delta^\alpha}_\beta$
(remember, it's just the identity matrix), we can rewrite this as
\begin{equation}
{\Lambda^\gamma}_\beta \partial_\alpha {\delta^\alpha}_\tau F^{\tau\beta}= {\Lambda^\gamma}_\beta j^\beta\, .
\end{equation}
Now from (76) in~\cite{Anchordoqui:2015xca}  (norm is preserved under Lorentz
transformations) we see that
${\delta^\alpha}_\beta={\Lambda_\gamma}^\alpha{\Lambda^\gamma}_\beta$. Thus we have
\begin{equation}
{\Lambda^\gamma}_\beta \partial_\alpha
{\Lambda_\delta}^\alpha{\Lambda^\delta}_\tau
F^{\tau\beta}={\Lambda^\gamma}_\beta j^\beta\, .
\end{equation}
By rearranging, we get
\begin{eqnarray}
({\Lambda_\delta}^\alpha\partial_\alpha)({\Lambda^\delta}_\tau{\Lambda^\gamma}_\beta F^{\tau\beta})
&=&({\Lambda^\gamma}_\beta j^\beta)\\
{\partial^\prime}_\delta {F^\prime}^{\delta\gamma}&=&{j^\prime}^\gamma\, ,
\end{eqnarray}
which means that (\ref{MW_Eq1}) remains invariant under arbitrary Lorentz boosts.

Finally, the invariance of (\ref{MW_Eq2}) is rather trivial. An arbitrary boost
leads to
\begin{eqnarray}
{\Lambda_\delta}^\alpha{\Lambda_\rho}^\beta {\Lambda_\sigma}^\gamma
(\partial_\alpha F_{\beta\gamma}+\partial_{\beta}F_{\gamma\alpha}+\partial_{\gamma}F_{\alpha\beta}) & = & 0
\nonumber \\
({\Lambda_\delta}^\alpha
 \partial_\alpha) ({\Lambda_\rho}^\beta {\Lambda_\sigma}^\gamma F_{\beta\gamma})
 +({\Lambda_\rho}^\beta\partial_{\beta}) ({\Lambda_\sigma}^\gamma{\Lambda_\delta}^\alpha
  F_{\gamma\alpha}) + ({\Lambda_\sigma}^\gamma \partial_{\gamma})
  ({\Lambda_\delta}^\alpha{\Lambda_\rho}^\beta F_{\alpha\beta}) & = & 0
\nonumber \\
{\partial^\prime}_\delta {F^\prime}_{\rho\sigma}
 + {\partial^\prime}_{\rho} {F^\prime}_{\sigma\delta} + {\partial^\prime}_{\sigma}
 {F^\prime}_{\delta\rho} & = & 0 \, .
\end{eqnarray}
Therefore, we have shown that Maxwell's equations remain invariant under arbitrary Lorentz boosts.\\

13.3~For any system bound by a central potential, $V(r)$, the wave function can be decomposed into radial and angular parts, with the angular parts described by spherical harmonics:
$\psi (r,\vartheta ,\varphi)=R (r)Y_l^m(\vartheta, \varphi)$. The spherical harmonics are given by (\ref{spherical-harmonics}).
The parity operation on spherical coordinates changes $r$ to $-r$, $\vartheta$ to $\pi - \vartheta$, and $\varphi$ to $\varphi + \pi$. Thus, $e^{im \varphi}$ goes to $e^{im \varphi + im \pi} = e^{im\pi} e^{im \varphi} = (-1)^m e^{im \phi}$  and $P_l^m (\cos \vartheta)$ goes to $P_l^m (\cos (\pi - \theta)) = (-1)^{l+m} P_l^m (\cos \vartheta)$.
Assembling this yields ${\rm P} Y_l^m = (-1)^l Y_l^m$.\\

13.4~The following decays had been observed: $\tau^+ \to \pi^- \pi^+ \pi^+$ and $\theta^+ \to \pi^+ \pi^0$. Let us find the parities of the $\tau^+$ and the $\theta^+$ from these decays. First off, we know that the pions have an intrinsic parity of $-1$. Now, we have to consider the spatial contribuition to parity in these decays. The $\tau$, the $\theta$, and the pions are all spin-0 particles. Thus, the total initial angular momentum in both of the above decays is $J = L+S = 0$, where $L = 0$ is the external angular momentum and $S = 0$ is the intrinsic spin. 
In the decay of the $\theta$, the orbital angular momentum of the two pions must be equal to zero in order to conserve the total angular momentum. In this case the spatial contribution to parity is $(-1)^l = (-1)^0 =
+1$. The total parity of the final state is then ${\rm P} = {\rm P}_{\pi^+} {\rm P}_\pi^0 {\rm P}_{\rm spatial} = (-1)(-1)(+1) = +1$. Thus, by parity conservation, ${\rm P} (\theta) = +1$. In the decay of the $\tau$, on the other hand, the situation is not so simple. The total orbital angular momentum has two components: The first is given by the angular momentum between the two $\pi^+$. The second, by the angular momentum of the remaining $\pi^-$ about the center of mass of the two $\pi^+$. This sum must be equal to zero in order to conserve total angular momentum. This implies that these two components of the total orbital angular momentum must have the same magnitude. In this case the spatial component of parity is given by the product of the parities given by the two components discussed above ${\rm P}_{\rm spatial} = (-1)^l (-1)^l  = +1$. So, the total parity of the final state of the $\tau$ decay, and thus of the $\tau$ itself, is ${\rm P}_\tau = {\rm P}_{\pi^+} {\rm P}_{\pi^+} {\rm P}_{\pi^-} {\rm P}_{\rm spatial} = (-1)^3(+1) = -1$. So the two particles have different parities.\\

13.5~{\it (i)}~$\Delta E \sim m_\pi = 135~{\rm MeV}$; {\it (ii)} $\Delta t \sim \hbar/(2 \Delta E) = 2.44 \times 10^{-24}~{\rm s}$; {\it (iii)}~$\Delta x = c \Delta t = 0.73~{\rm fm}$.\\ 

14.1~From (\ref{octonautsF}) it follows that the probabilities for flavor oscillation are 
\begin{equation}
P(\nu_\mu \to \nu_\mu) = P(\nu_\mu \to \nu_\tau)= \frac{1}{4}\,
(\cos^4 \theta_\odot + \sin^4 \theta_\odot + 1) \simeq 0.4 \,\,,
\end{equation}
\begin{equation}
P(\nu_\mu \to \nu_e) = P(\nu_e \to \nu_\mu) = P(\nu_e \to \nu_\tau) =
\sin^2 \theta_\odot \,\, \cos^2 \theta_\odot \simeq 0.2\,\,,
\end{equation} 
and
\begin{equation}
P(\nu_e \to \nu_e) = \cos^4 \theta_\odot + \sin^4 \theta_\odot \simeq 0.6\,\, .
\label{p3}
\end{equation}
{\it (i)}~For $(1:2:0)_S$,  the earth expectation for $\nu_\mu$  and 
$\nu_\tau$ is $2
\times 0.4 + 1 \times 0.2 =1$ and the $\nu_e$ expectation is $2 \times 0.2 + 1
\times 0.6 =1$, that is $(1:1:1)_{\oplus}$. {\it (ii)}~For  $n \to p e^- \bar \nu_e$, we have $(1:0:0)_S$, yielding $(3:1:1)_\oplus$.\\

\begin{figure*}[tbp]
\begin{center}
\includegraphics[width=4.75in]{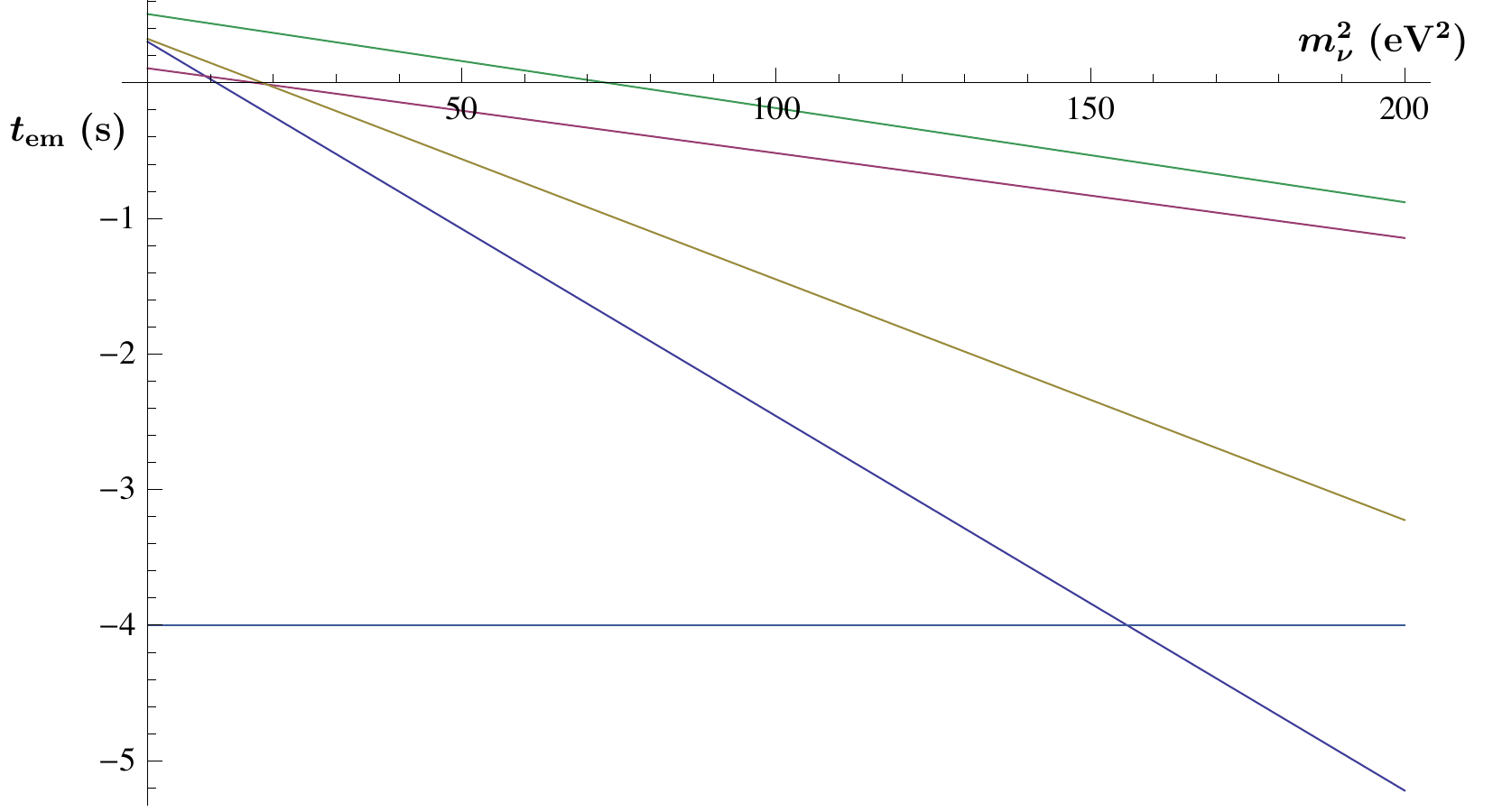}
\caption{\label{RRF}{\small Emission time as a function of the square neutrino mass. The $4\, \rm{s}$ emission 
window is indicated by the horizontal line.}}
\label{fig:sn1987a}
\end{center}
\end{figure*}

14.2~{\it (i)}~The travel time is determined by $v = dx/dt$.  Given
that SN1987A was a fixed distance $d$ away, $t_0 = d/c$. On the other hand, a massive neutrino
satisfies the relation $E = \gamma m_\nu c^2$, where $\gamma = (1-v^2/c^2)^{-1/2}$ is the Lorentz factor. Equating these two relations we see that
\begin{equation}
\frac{v}{c} = \sqrt{1 - \frac{m_\nu^2 c^4}{E^2} } \approx 1 - \frac{1}{2} \frac{m_\nu^2 c^4}{E^2} + \cdots \, .
\end{equation}
Hence, the neutrino travel time is
\begin{equation}
\Delta t = t_{\rm em} - t_{\rm obs} = \frac{d}{v} = \frac{c t_0}{v} \approx  t_0 \left(1 + \frac{1}{2} \frac{m_\nu^2 c^4}{E^2} + \cdots \right) \, .
\end{equation}
{\em (ii)} The emission time of each neutrino relative to the first one is then
\begin{equation}
t_{\rm em} = t_{\rm obs} - \frac{t_0}{2} \frac{m_\nu^2 c^4}{E^2} + \frac{t_0}{2} \frac{m_\nu^2 c^4}{E_1^2} \,,
\end{equation}
where $E_1$ is the energy of the first neutrino to arrive. As shown in Fig.~\ref{fig:sn1987a}, neutrino 3 would
need to have been emitted more than 4~s earlier than neutrino 1 if $m_\nu^2 > 156~{\rm eV}^2$.
This translates into  a limit on the neutrino mass to be no more than about $12.5~{\rm eV}$,
as desired. {\em (iii)} The larger the neutrino mass, the larger the difference in the speeds of the
two neutrinos, which makes it more difficult for the third neutrino to have been emitted
only 1~s before neutrino 1. However, the difference in speeds is less if the energy of neutrino 1, $E_1$, is at the
bottom of its range and the energy of neutrino 3, $E_3$, is at the top of its range. Therefore, 
we take $E_1 = 18.4~{\rm MeV}$ and $E_3 = 10.9~{\rm MeV}$. The largest possible mass is the
one that saturates the inequality
\begin{equation}
-1 \leq t_{\rm obs} - \frac{t_0 m_\nu^2 c^4}{2} \left( \frac{1}{E_3^2} - \frac{1}{E_1^2} \right) \,
\end{equation}
where $t_{\rm obs}$ corresponds to neutrino 3. The corresponding limit is $m_\nu < 9.5~{\rm eV}$.\\

14.3~{\it (i)}  Gauss's law states that the force due to gravity is only a function of the gravitational field inside the sphere of radius $R$.  The gravitational field outside the sphere of radius $R$ will result in a net zero force on the galaxy.  Using Newton's second law, $\vec F=m\vec a$, and noting that the net gravitational force will be in the radial direction, which means $a=d^2R/dt^2\equiv \ddot R$, we can say
\begin{equation}
m \ddot R = - \frac{G m M}{R^2} =  -\frac{G m}{R^2} \left(\frac{4}{3} \pi R^3 \rho\right),
\end{equation}
thus,
\begin{equation}
\ddot R = -\frac{ 4 \pi}{3} G \rho R 
\end{equation}
and so  
\begin{equation}
\frac{\ddot R}{R} =  -\frac{4 \pi}{3} G\rho,
\label{BJMwkee1}
\end{equation}
which is the desired result.  {\it (ii)}  The total energy of the galaxy is given by the sum of its kinetic and potential energy,
\begin{equation}
E_{\rm tot} = \frac{1}{2} m \dot R^2 - \frac{G m M}{R}.
\end{equation}
Using that the total energy is zero and $M = \frac{4}{3} \pi R^3 \rho$,
\begin{equation}
\frac{1}{2} m \dot R^2 - \frac{G m}{R} \left(\frac{4}{3} \pi R^3 \rho\right) = 0
\end{equation}
which implies
\begin{equation}
\frac{1}{2} \dot R^2 = \frac{4 \pi}{3} G \rho R^2
\end{equation}
yielding
\begin{equation}
\left(\frac{\dot R}{R}\right)^2 = \frac{ 8 \pi}{3} G \rho .
\label{BJMwkee2}
\end{equation}
This is Friedman's equation for a flat universe in the Newtonian limit.  For comparison, the full general relativistic Friedman's equation is~\cite{Weinberg:2008zzc} 
\begin{equation}
\left(\frac{ \dot R}{R}\right)^2 = \frac{ 8 \pi}{3} G \rho - \frac{k c^2}{R^2},
\end{equation}
where $k$ is the curvature constant, which is zero for a flat universe, and $R$ has the meaning of the scale factor in this equation. The critical density is defined as the density of a spatially flat universe, i.e. $k=0$. Identifying $H = \dot R/R$ as the Hubble parameter, from (\ref{BJMwkee2}) we obtain (\ref{BJMwkee3}). From (\ref{BJMwkee1}) it follows that in the abscence of vacuum energy the cosmic expansion is always decelerating.

\newpage

\twocolumngrid

\appendix

\setcounter{section}{1}
\section{Riemann zeta function}
\label{appB}

The Riemann zeta function $\zeta (z)$ is defined as
\begin{equation}
\zeta (z) \equiv \sum_{n=1}^{\infty} \frac{1}{n^z} \, .
\end{equation}
It is easily seen that, for ${\rm Re} (z) >1$,
\begin{equation}
\zeta(z) = \frac{1}{\Gamma (z)} \int_0^\infty \frac{x^{z-1}}{e^x -1} dx \,,
\label{zorro}
\end{equation}
where the gamma function is related to the factorial function as $\Gamma (z) \equiv (z-1)!$. This enables one to evaluate integrals of the type (\ref{zorro}). The relevant values of $\zeta (z)$ are tabulated in Table~\ref{Tzeta}.

\begin{table}
\caption{Riemann zeta function.\label{Tzeta}}
\begin{tabular}{ll}
\hline
\hline
$z$~~~~~~~~~~~~~~~~~~~~~~~~~~ &$\zeta (z)$~~~~~~~~~~~~~~~~~~~~~~~~~~\\
\hline
1  & diverges \\
2 & $1.644934 \cdots \equiv \pi^2/6$\\
3 & $1.202057 \cdots$ \\
4 & $1.082323 \cdots \equiv \pi^4/90$\\
\hline
\hline
\end{tabular}
\end{table}

\section{Radiation from an accelerated charge}
\label{appC}

The path that leads from Maxwell's equations to the solution relevant in the case of the radiation field generated by an accelerated non-relativistic charge is somewhat arduous. It can be followed in your preferred electrodynamics book (see .e.g.~\cite{Jackson:1998nia}), and will not be given here. The essential insights can, however, be understood from an argument of Thomson as presented in~\cite{Longair:1992ze}.  

\begin{figure}[tbp] \postscript{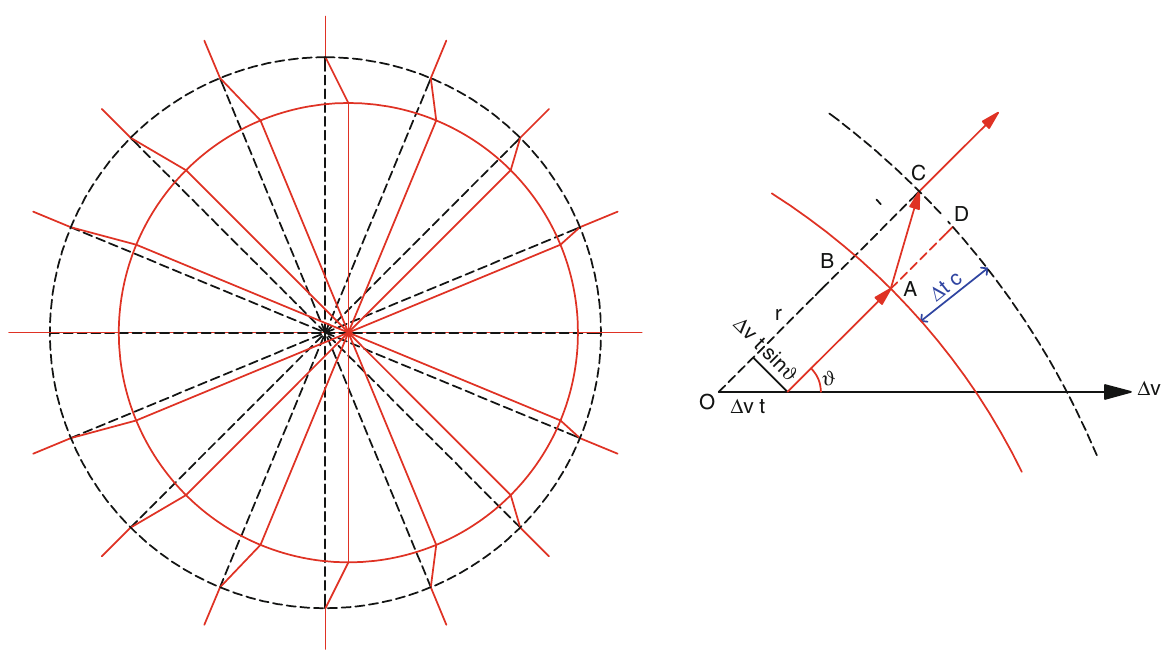}{0.95} \caption{ Schematical view of the electric field lines at time $t$ due to a charged particle accelerated to a velocity $\Delta v \ll c$ in a time interval $\Delta t$.}
\label{fig:Larmor1} 
\end{figure}

Consider a charge $q$ stationary at the origin $O$ of some inertial frame of reference $S$ at time $t = 0$.  The charge then suffers a small acceleration to velocity $\Delta v$ in the short time interval $\Delta t$. After a time $t$, we can distinguish between the field configuration inside and outside a sphere of radius $r = ct$ centred on the origin of $S$. Outside this sphere, the field lines do not yet know that the charge has moved away from the origin and so the field lines are radial, centred on $O$. Inside this sphere, the field lines are radial about the origin of the frame of reference centred on the moving charge. 
Between these two regions, there is a thin shell (perturbed zone) of thickness $c \Delta t$ in which the electric field lines are connected in a non radial way. 
In Fig.~\ref{fig:Larmor1} we draw the electric field lines that result from this arrangement at a time $t$.

There must be a component of the electric field in the $\hat \imath_\theta$ direction. This ``pulse'' of accelerated charged electromagnetic field is propagated away from the charge at the speed of light and is the energy loss of the accelerated charged particle.  The increment in velocity $\Delta v$ is very small, $\Delta v \ll c$, and therefore it can be assumed that the field lines are radial at $t = 0$ and also at time $t$ in the frame of reference $S$.  Consider a small cone of electric field lines at angle $\theta$ with respect to the acceleration vector of the charge at $t = 0$ and at some later time $t$ when the charge is moving at a constant velocity $\Delta v$. We join up electric field lines through the thin shell of thickness $c dt$ as shown in Fig.~\ref{fig:Larmor1}. The strength of the $E_\theta$ component of the field is given by number of field lines per unit area in the $\hat \imath_\theta$ direction. From the geometry of the diagram, you can read that the ratio of the tangential to the radial field line components in the perturbed zone is
\begin{equation}
\frac{E_\theta}{E_r} = \frac{\Delta v t \sin \theta}{c \Delta t} \, .
\end{equation}
Coulomb's law for the radial component $E_r$ a distance $r$ from a charge $q$ is (in Gaussian cgs units)
\begin{equation}
E_r = \frac{q}{r^2} \,,
\end{equation}
where $r= ct$.
We can therefore deduce the tangential field component and find
\begin{eqnarray}
E_\theta & = &  \frac{q (\Delta v/\Delta t) \sin \theta}{c^2 r} \, \nonumber \\
& = & \frac{q \dot v \sin \theta}{r c^2} \, .
\end{eqnarray} 
Note that  according to Coulomb's law the radial component of the field decreases as $r^{-2}$, but the field in the pulse decreases only as $r^{-1}$. This is a characteristics of the radiation field in the far zone. The only electrical field component that is relevant for radiation is that which is perpendicular to the direction of propagation, i.e. $E_\theta$ . It is the one we consider further here.

\begin{figure}[tbp] \postscript{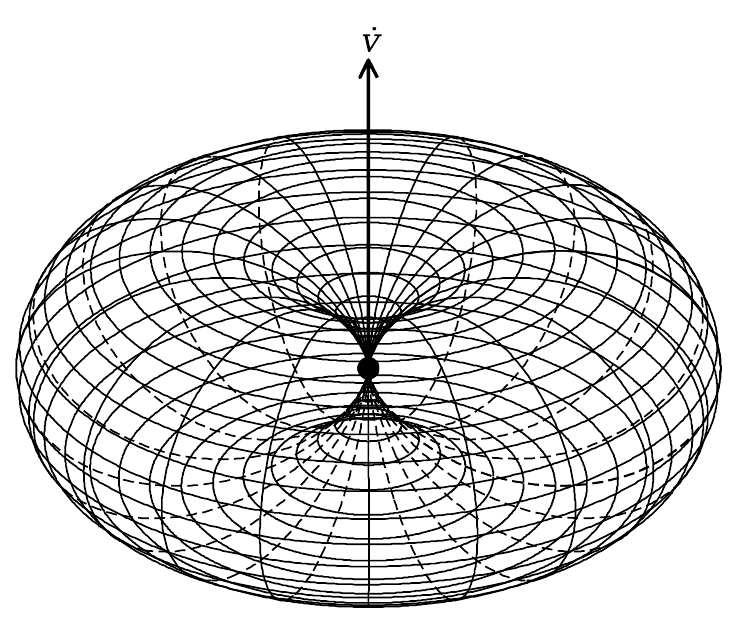}{0.95} \caption{The power pattern of Larmor radiation from a charged particle shown for an acceleration vector $\dot{\vec v}$ tilted $60^\circ$ from the line of sight. The power received in any direction is proportional to the component of $\dot{\vec v}$ perpendicular to the line of sight.}
\label{fig:Larmor2} 
\end{figure}

We may now calculate the energy flux carried by this disturbance. The energy flux transported by electromagnet fields is given by the Poynting vector 
\begin{equation}
\vec S= c \ \vec E \times \vec B \, .
\end{equation}
The magnetic field is equal and perpendicular to the electric field in electromagnetic radiation:
\begin{equation}
\vec B = \hat n \times \vec E \, .
\end{equation}
The rate loss of energy (total power radiated) through the solid angle $d\Omega$ at distance $r$ from the charge is therefore
\begin{eqnarray}
P & = & \frac{dE}{dt} d \Omega  =  |\vec S| \ r^2 d \Omega \nonumber \\
& = & \frac{q^2 \dot v^2}{4 \pi c^3} \int_{\phi = 0}^{2 \pi} \int_{\theta =0}^{\pi} \frac{\sin^2 \theta}{r^2} r^2 \sin \theta \ d \theta \ d \phi \nonumber \\
 & = & \frac{q^2 \dot v^2}{2 c^3} \int_{\theta = 0}^\pi \sin^3 \theta d \theta \, .
\label{SotR}
\end{eqnarray}
Evaluating this integral gives $\int_{\theta =0}^\pi \sin^3 \theta \, d\theta = 4/3$ so the total power emitted is
\begin{equation}
P = \frac{2}{3} \frac{q^2 \dot v^2}{c^3}  \, .
\end{equation}
This is the so-called Larmor formula. It states that any charged particle radiates when accelerated and that the total radiated power is proportional to the square of the acceleration. 
The charge radiates with a dipolar  power pattern that looks like a doughnut whose axis is parallel to $\dot{\vec v}$, see Fig.~\ref{fig:Larmor2}.

\section{Magnetic dipole moment}
\label{appD}

As you very well know, a {\it magnetic dipole} is created by a flow of electric charge around an infinitesimal loop. The {\it magnetic dipole moment} $\vec \mu$ is a measure of the strength of the magnetic dipole, or equivalently to its ability to align with a given external magnetic field $\vec B$.  For a planar loop encircling an area $A$, the magnetic moment is
\begin{equation}
\vec \mu = i A \hat n \,,
\end{equation}
where $i$ is the current and $\hat n$ is the unit vector perpendicular to the loop plane. 
The energy of the unperturbed system depends on the angle $\theta$ between the magnetic moment and the external magnetic field, 
\begin{equation}
U = - \vec \mu \cdot \vec B = - |\vec \mu| B_0 \cos \theta \,,
\end{equation}
where  $\vec B = B_0 \hat k$. The torque on the magnetic moment is given by (\ref{torque-equation}).

\begin{figure}[tbp] \postscript{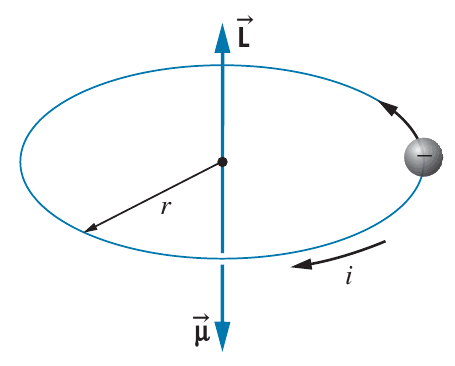}{0.95} \caption{A circulating negative charge is represented as a current loop. Because the charge is negative, $\vec L$ and $\vec \mu$ point in opposite directions.}
\label{fig:mu} 
\end{figure}

In Fig.~\ref{fig:mu} we show a particle with mass $m$ and charge $q (< 0)$ which is kept moving in a classical circular orbit by a central field $V (r)$. This constitutes a current
loop of radius $r$, velocity $v$ and revolution frequency $\nu = v/(2 \pi r)$, which leads to a current  $i = q \nu$ and a magnetic moment
\begin{equation}
|\vec \mu| = i A = q \frac{v}{2 \pi r} \pi r^2 = \frac{1}{2} q rv = \frac{1}{2} |\vec r \times \vec v| \, .
\end{equation}
Hence the magnetic moment due to the motion of the charge is proportional to the orbital angular momentum 
\begin{equation}
\vec \mu_L = \frac{q}{2m} \vec L \, .
\label{boardgear3}
\end{equation}
Note that for this kind of classical motion the angular momentum $\vec L$ and the magnetic moment $\vec \mu_L$ are not constants of motion when $\vec B$ differs from zero. Indeed, using
Newton's second law and (\ref{torque-equation}) we have that
\begin{equation}
\frac{d\vec L}{dt} = \vec \tau = \vec \mu_L \times \vec B = - \frac{q}{2m} \vec B \times \vec L \,  .
\end{equation}
Now, using (\ref{boardgear3}) we obtain
\begin{equation}
\frac{d \vec \mu_L}{dt} = \frac{q}{2m} \frac{d\vec L}{dt} = -  \left(\frac{q}{2m} \right)^2 \vec B \times \vec L = - \frac{q}{2m} \vec B \times \vec \mu_L \, .
\label{glade}
\end{equation}
Note that the magnitude of the magnetic moment does not change since
\begin{eqnarray}
\frac{d|\vec L|^2}{dt} & = & \frac{d (\vec L \cdot \vec L)}{dt} = 2 \vec L \cdot \frac{d\vec L}{dt} = 2 \vec L \cdot (\vec \mu_L \times \vec B) \nonumber \\
& = & 2 \vec L \cdot \left(-\frac{q}{2m}  \vec B \times \vec L \right)= 0 \, .
\end{eqnarray}
Similarly the angle between $\vec \mu_L$ and $\vec B$ (i.e., the projection of $\vec \mu_L$ along the $z$ axis) does not change, since
\begin{eqnarray}
\frac{d (\vec B \cdot \vec L )}{dt} & = & 0 + \vec B \cdot \frac{d\vec L}{dt} = \vec B \cdot (\vec \mu_L \times \vec B) \nonumber \\
& = & -  \frac{q}{2m}  \vec B \cdot (\vec L \times \vec B) = 0 \, .
\end{eqnarray}
The equations of motion for the magnetic moment follow from (\ref{glade})
\begin{eqnarray}
\frac{d\mu_{L,x}}{dt} & = & - \omega_0 \ \mu_{L,y} \,, \nonumber \\
\frac{d\mu_{L,y}}{dt} & = & \omega_0 \ \mu_{L,x} \, , \nonumber \\
\frac{d\mu_{L,z}}{dt} & = & 0 \, ,
\end{eqnarray} 
where  $\omega_0 = - q \ B_0 /(2m)$
is the Larmor frequency. For the initial condition $\vec \mu_L (0) = \mu_{L,x} (0) \hat \imath + \mu_{L,z} (0) \hat k$, i.e. $\theta \neq 0$, the solutions are
\begin{eqnarray}
\mu_{L,x} (t) & = & \mu_{L,x}(0) \cos (\omega_0 t) \nonumber \\
 \mu_{L,y} (t) & = & \mu_{L,x}(0) \sin (\omega_0 t) \, .
\end{eqnarray}
That is, the magnetic moment precesses around the $z$-axis at the Larmor frequency.

\section{Pauli exclusion principle}
\label{appE}

Consider the simplest case of a two-particle system. The wave function is $\psi(\vec r_1,\vec r_2)$. If we assume that there is no interaction between the two particles, we will be able to describe the states using separation of variables: 
\begin{equation}
\psi (\vec r_1, \vec r_2) = \psi_a(\vec r_ 1) \psi_b (\vec r_2)
\end{equation}
 where $a$ and $b$ label two different single-particle states. Implicit in this expression is the assumption that we can distinguish the two particles by some mean and link particle one to the position $1$ and the state $a$. However, if we consider two identical particles (such as two electrons, two photons, two neutrons) there is no physical mean to distinguish them. Even if we try to measure them in order to keep track of which one is which, we know that in the process we destroy the state (by the wavefunction collapse) so not even this is a possibility.  

In quantum mechanics identical particles are fundamentally indistinguishable. Then the expression above does not correctly describe the state anymore. In order to faithfully describe a state in which we cannot know if particle $a$ or $b$ is at $r_1$ or $r_2$, we can take a linear combination of these two possibilities: 
\begin{equation}
\psi(\vec  r_1, \vec r_2) = A_1 \ \psi_a(\vec r_1) \psi_b(\vec r_2)+A_2\psi_b(\vec r_1) \psi_a(\vec r_2).
\end{equation}  
Now, since the two possibilities have the same probability, we have $|A_1| = |A_2| = 1/sqrt{2}$. Then there are two possible combinations:
\begin{equation}
\psi(\vec  r_1, \vec r_2) = \frac{1}{\sqrt{2}} [\psi_a(\vec r_1) \psi_b(\vec r_2) \pm \psi_b(\vec r_1) \psi_a(\vec r_2)].
\end{equation}  
These two combinations describe two types of particle. The combination with the plus sign describes bosons, particles that are invariant under exchange of a particle pair. The combination with the minus sign describes fermions.  All particles with integer spin are bosons, whereas all particles with half-integer spin are fermions.

We can define an operator $\hat {\rm P}$ that interchanges the two particles: 
\begin{equation}
\hat {\rm P} [\psi(\vec r_1, \vec r_2)] = \psi (\vec r_2,\vec r_1)
\end{equation}
Since of course $\hat {\rm P} [\hat {\rm P} [\psi(\vec r_1, \vec r_2)]] = \psi (\vec r_1, \vec r_2)$, we have that $\hat {\rm P}^2= 1$. Then the eigenvalues of $\hat {\rm P}$ must be $\pm 1$. If two particles are identical, then the hamiltonian is invariant with respect to their exchange and $[H,{\rm P}] = 0$. Then we can find energy eigenfunctions that are common eigenfunctions of the exchange operator, or 
$\psi (\vec r_1, \vec r_2) = \pm \psi (\vec r_2, \vec r_1)$. Then if the system is initially in such a state, it will be always be in a state with the same exchange symmetry. For the considerations above, however, we have seen that the wavefunction is not only allowed, but it must be in a state with a definite symmetry:
\begin{equation}
\psi(\vec r_1, \vec r_2) = \left\{\begin{array}{c l}
\phantom{-} \psi (\vec r_2, \vec r_1) & ~~~{\rm bosons} \\
- \psi (\vec r_2, \vec r_1) & ~~~{\rm fermions} \end{array} \right. \, .
\end{equation}

From the form of the allowed wave function for fermions, it follows that two fermions cannot occupy the same state. Assume that $\psi_a(\vec r) = \psi_b(\vec r)$, 
then we always have that 
\begin{equation}
\psi_f(\vec r_1, \vec r_2) = \frac{1}{\sqrt{2}} [\psi_a(\vec r_1) \psi_b (\vec r_2) - \psi_b (\vec r_1 \psi_a (\vec r_2) = 0 \, .
\end{equation}
This is the well-known Pauli exclusion principle. Notice that of course it applies to any fermions. For example, it applies to electrons, and this is the reason why electrons do not pile up in the lowest energy level of the atomic structure, but form a shell model. 

\section{A little group theory}

\label{appF}

A group is a set of elements $S$ plus a compostion rule $\otimes$, such that: {\it (i)}~Combining two elements under the rule gives another element of the group, i.e.  if $E, E' \in S$ then \mbox{$E \otimes E' = E''$}, with $E'' \in S$. {\it (ii)}~The composition rule is associative $E\otimes (E'\otimes E'') = (E \otimes E') \otimes E''$. {\it (iii)}~There is an indentity element $\mathds{1}$, such that $E \otimes \mathds{1} = \mathds{1} \otimes E = E$ {\it (iv)}~Every element has a unique inverse $E^{-1}$ such that  \mbox{$E \otimes E^{-1} = E^{-1} \otimes E = \mathds{1}$}. 

If the elements of a group are differentiable with respect to their parameters, the group is a Lie group. For a Lie group, any element can be written in the form
\begin{equation}
E(\theta_1, \theta_2, \cdots , \theta_n) = {\rm exp} \left(\sum_{i=1}^n i \theta_i F_i\right) \, .
\end{equation}
The quantities $F_i$ are the generators of the group. The quatities $\theta_i$ are the parameters of the group: a set of $i$ real numbers that are needed to specify a particular element of the group. Note that the number of generators and parameters are the same. There is one generator for each parameter.

The group $U(1)$ is the set of all one dimensional, complex unitary matrices. The group has one generator $F = 1$, and one parameter, $\theta$. It simply produces a complex phase change
\begin{equation}
E(\theta) = e^{-i \theta F} = e^{-i \theta} \, .
\end{equation}
$U(1)$ is a Lie group,
\begin{equation}
\frac{dE}{d \theta} = iE \, .
\end{equation}
Since the generator $F$, commutes with itself, the
group elements also commute
\begin{equation}
E(\theta_1) \otimes E(\theta_2) = e^{-i\theta_1} e^{-i \theta_2} = e^{-i\theta_2} e^{-i \theta_1} \, .
\end{equation}
Such groups are called Abelian groups.

The group $U(2)$ is the set of all two dimensional, complex unitary matrices.  An  complex $n \times n$ matrix has $2n^2$ real parameters. The unitary condition constraint removes $n^2$ of these.  The group $U(2)$ then has four generators and four parameters
\begin{equation}
E(\theta_0,\theta_1,\theta_2,\theta_3) = e^{-i\theta_j F_j}
\end{equation}
where $j = 0,1,2,3$. The generators are: $F_i = \sigma_i/2$, where $\sigma_0$ denotes  the identity matrix. The operators represented by the elements of $U(2)$ act on two dimensional complex vectors.  The operations generated by $F_0 = \sigma_0/2$ simply change the complex phase of both components of the vector by the same amount. In general we are not so interested in these operations.

The group $SU(2)$ is the set of all two dimensional, complex unitary matrices with unit determinant. The unit determinant constraint removes one more parameter. The group $SU(2)$ then has three generators and three parameters
\begin{equation}
E(\theta_1,\theta_2,\theta_3) = e^{-i \theta_j F_j} \,,
\end{equation}
where $j = 1, 2, 3$. The generators of $SU(2)$ are a set of three linearly
independent, traceless $2\times2$ hermitian matrices: $F_j = \tau_j/2$, where $\tau_j = \sigma_j$ with $j =1,2,3$. 
Since the generators do not commute with one another,
$[\sigma_i,\sigma_j] = 2 i \epsilon_{ijk} \sigma_k$, this is a non-Abelian group. The Pauli matrices obey the following anticommutation relatios $\{\sigma_i, \sigma_j\} = 2 \delta_{ij} \mathds{1}$.

The group $SU(3)$ is the set of all three dimensional, complex unitary matrices with unit determinant.
This set has $2 (3)^2 - (3)^2 -1 = 8$ parameters and generators
\begin{equation}
E(\theta_1,\theta_2, \cdots, \theta_8) = e^{-i  \theta_j F_j} \,,
\end{equation}
where $j = 1,2, \cdots, 8$. The generators of $SU(3)$ are a set of eight linearly
independent, traceless $3\times3$ hermitian matrices.
Since there are eight generators, the  $SU(3)$ elements represent rotations of complex three component vectors in an eight dimensional space. The generators are
\begin{eqnarray}
\lambda_1 = \left(\begin{array}{ccc} 0 & 1 & 0 \\ 1 & 0 & 0 \\
0 & 0 & 0 \end{array} \right) \quad && \lambda_2 = \left(\begin{array}{ccc} 0 & -i & 0 \\ i & 0 & 0 \\
0 &0 & 0 \end{array} \right) \nonumber \\
\lambda_3 = \left(\begin{array}{ccc} 1 & 0 & 0 \\ 0& -1 & 0 \\
0 & 0 & 0 \end{array} \right) \quad && \lambda_4 = \left(\begin{array}{ccc} 0 & 0 & 1 \\ 0 & 0 & 0 \\
1 & 0 & 0 \end{array} \right) \nonumber \\
\lambda_5 = \left(\begin{array}{ccc} 0 & 0 & -i \\ 0 & 0 & 0 \\
i & 0 & 0 \end{array} \right) \quad && \lambda_6 = \left(\begin{array}{ccc} 0 & 0 & 0 \\ 0 & 0 & 1 \\
0 & 1 & 0 \end{array} \right) \nonumber \\
\lambda_7 = \left(\begin{array}{ccc} 0 & 0 & 0 \\ 0 & 0 & -i \\
0 & i & 0 \end{array} \right) \quad && \lambda_8 = \frac{1}{\sqrt{3}} \left(\begin{array}{ccc} 1 & 0 & 0 \\ 0 & 1 & 0 \\
0 & 0 & 0 \end{array} \right) \, .
\end{eqnarray}
The structure of $SU(3)$ is: $[\lambda_a,\lambda_b] = 2 i f_{abc} \lambda_c$, where $f_{123} = 1$, $f_{458} = f_{678} = \sqrt{3}/2$, $f_{147} = f_{516} = f_{246} = f_{257} = f_{345} = f_{637} = 1/2$; $f_{abc}$ is totally antisymmetric.

\section{Wu experiment}
\label{appG}

In the Wu experiment, $^{60}$Co beta decays into $^{60}$Ni$^*$ (an excited form of Ni), which de-excites via gamma radiation to the ground state of Ni~\cite{Lee:1956qn,Wu:1957my}. From the angular distribution of the gamma rays, one can obtain the polarization of the original excited state of nickel. The result, roughly speaking, is that if the original Co (spin $J=5$) is fully polarized $(J_z=5)$, then Ni$^*$ has $J=4, J_z=4$. Again roughly speaking, we can regard the transition as if a particle with $J=1, J_z=1$ were to decay into a back-to-back pair $e^-\ \bar\nu_e$, with the electron emitted at an angle $\theta$ with respect to the original spin of the decaying spin-1 ``particle.''

Consider an effective
  spin-1 particle decaying into an $e$ and a $\bar \nu_e$. We have a
  Hamiltonian that interacts with  a spin-1 particle at rest  and as a
  consequence it decays back-to-back. If the original state of the
decaying spin -1 particle is $\ket{\Lm}$, then the {\it colinear} amplitude for decay
to an angle $\theta$ for an electron with helicity $\lme,$
antineutrino helicity $\lmnu$ is 
\begin{eqnarray}
\mathscr{M}^{\Lm}_{\lme \lmnu}(\theta) \!\!& = &\!\!\!  \amp{\lme \lmnu \theta}{H}{\Lm}
= \sum_{\Lmp} \amp{\lme \lmnu 0}{H} {\Lmp}  \langle\Lmp|\Lm \rangle \nonumber \\
& = &\!\!\! \amp{\lme \lmnu 0}{H} {\lme-\lmnu} d^1_{\lme-\lmnu, \Lm}(\theta),
\end{eqnarray} where conservation of angular momentum along the $z'$ axis (the
direction of the electron) implies $\Lmp = \lme - \lmnu$. For given
$\Lm$, say $\Lm=1,$ there are four matrix elements: \bea
\mathscr{M}^{1}_{\thalf\ \thalf}(\theta) &=&           a\ d^1_{0\ 1}(\theta)\\
\mathscr{M}^{1}_{-\thalf\ -\thalf}(\theta) &=& \bar a\ d^1_{0\  1}(\theta)\\
\mathscr{M}^{1}_{\thalf\ -\thalf}(\theta) &=&          b\ d^1_{1\ 1}(\theta)\\
\mathscr{M}^{1}_{-\thalf\ \thalf}(\theta)&=& \bar b\ d^1_{-1\ 1}(\theta) ,\eea
where  for example, the collinear matrix element $b = \amp{\thalf\
  -\thalf\ 0}{H} {1}$, $\bar b = \amp{-\thalf\
  +\thalf\ 0}{H} {-1}$, etcetera. Squaring and adding helicity
amplitudes, we find the decay rate for electrons into angle $\theta$
with respect to the initial polarization direction \begin{eqnarray}
\frac{d\Gamma}{d\Omega} & \propto & \thalf\ (|a|^2 + |\bar a|^2) \sin^2{\theta} \nonumber \\
& + & \tfrac{1}{4} (|b|^2 + |\bar b|^2) (1 + \cos^2{\theta}) \nonumber \\
& + &  \thalf(|b|^2 - |\bar b |^2) \cos{\theta} \, .
\end{eqnarray} 
The decay will be up-down symmetric iff $|b| = |\bar b|.$ From the defining equations
above, we can see that $\bar b$ is the mirror image of $b$ i.e.,
it is $b$ with all the spins around the $z'$ axis reversed. Thus, if
an asymmetry is observed, parity is violated. Moreover, the asymmetry
is predicted to go as $\cos{\theta}.$ 
For weak interaction, we have maximal parity violation as
  $b=0$. If $|b| = |\bar b|$ then there is no dependence with $\cos
  \theta$ and parity is conserved.

\section{Electroweak symmetry breaking}
\label{appH}

Herein we demonstrate that the vacuum (\ref{vacuum}) breaks the gauge 
symmetry.  The vacuum state $\vev{\phi}$ is invariant under a symmetry 
operation $\exp{(i \theta_j F_j)}$ corresponding to the 
generator $F_j$ provided that $\exp{(i \theta_j F_j)}\langle {\phi} \rangle_0 = \langle \phi \rangle_0$, that is, if $F_j \langle \phi \rangle_0 = 0$ (see Appendix~\ref{appF}).  
We easily compute that 
\begin{eqnarray}
   \frac{1}{2} \tau_{1}\vev{\phi} & = & \frac{1}{2} \left( 
    \begin{array}{cc}
	0 & 1  \\
	1 & 0
    \end{array}
    \right) \left( 
    \begin{array}{c}
	0  \\
	v/\sqrt{2}
    \end{array}
    \right) \nonumber \\
&  =  & \frac{1}{2} \left( 
    \begin{array}{c}
	v/\sqrt{2}  \\
	0
    \end{array}
    \right) \neq 0 \quad\hbox{broken!}
    \nonumber  \\
   \frac{1}{2} \tau_{2}\vev{\phi} & = & \frac{1}{2} \left( 
    \begin{array}{cc}
	0 & -i  \\
	i & 0
    \end{array}
    \right) \left( 
    \begin{array}{c}
	0  \\
	v/\sqrt{2}
    \end{array}
    \right) \nonumber \\ & = & \frac{1}{2} \left( 
    \begin{array}{c}
	-iv/\sqrt{2}  \\
	0
    \end{array}
    \right) \neq 0 \quad\hbox{broken!}
    \nonumber  \\
   \frac{1}{2} \tau_{3}\vev{\phi} & = & \left( 
    \begin{array}{cc}
	1 & 0  \\
	0 & -1
    \end{array}
    \right) \left( 
    \begin{array}{c}
	0  \\
	v/\sqrt{2}
    \end{array}
    \right) \nonumber \\ & = & \frac{1}{2} \left( 
    \begin{array}{c}
	0  \\
	-v/\sqrt{2}
    \end{array}
    \right) \neq 0 \quad\hbox{broken!}
    \nonumber  \\
    Y\vev{\phi} & = & Y_{\phi}\vev{\phi} = +\frac{1}{2} \vev{\phi}  \nonumber \\ & = & \frac{1}{2} \left( 
    \begin{array}{c}
	0  \\
	v/\sqrt{2}
    \end{array}
    \right) \neq 0 \quad\hbox{broken!} 
    \label{eq:brisure}
\end{eqnarray}
However, if we examine the effect of the electric charge operator $Q$ 
on the (electrically neutral) vacuum state, we find that
\begin{eqnarray}
    Q \vev{\phi} & =&  \left(\frac{1}{2} \tau_{3} + Y \right) \vev{\phi} = 
     \left( 
    \begin{array}{cc}
	Y_{\phi}+\frac{1}{2} & 0  \\
	0 & Y_{\phi}- \frac{1}{2}
    \end{array}
    \right) \vev{\phi}
    \nonumber  \\
     & = & \left( 
     \begin{array}{cc}
	 1 & 0  \\
	 0 & 0
     \end{array}
      \right) \left(     \begin{array}{c}
	0  \\
	v/\sqrt{2}
    \end{array}
\right)  = \left( 
     \begin{array}{c}
	 0  \\
	 0
     \end{array}
     \right)~\hbox{{\rm unbroken!}} 
    \label{eq:Qok}
\end{eqnarray}
The original four generators are all broken, but the electric charge is
not.  It appears that we have accomplished our goal of breaking
$SU(2)_{L} \otimes U(1)_{Y} \to U(1)_{\mathrm{em}}$.  We expect the
photon to remain massless, and expect the gauge bosons that correspond
to the generators $\tau_{1}$, $\tau_{2}$, and $\kappa \equiv
(\frac{1}{2} \tau_{3} - Y)$ to acquire masses.


\end{document}